\documentclass[a4paper,12pt]{article}
\usepackage[dvipdfmx]{graphicx}
\usepackage{amsfonts,amsthm,amsmath,amssymb,graphicx,float,mathtools}
\numberwithin{equation}{section}
\usepackage{color}
\usepackage{bbm} 
\usepackage{tensor} 
\usepackage{verbatim} 
\usepackage[
	  pagebackref=false,
	  colorlinks=true,
      linkcolor=blue,
      urlcolor=blue,
      filecolor=black,
      citecolor=red,
      pdfstartview=FitV,
      pdftitle={},
        pdfauthor={},
        pdfsubject={},
        pdfkeywords={},
        pdfpagemode=None,
        bookmarksopen=true
      ]{hyperref}

\begin{document}

\newtheorem{definition}{Definition}[section]
\newcommand{\be}{\begin{equation}}
\newcommand{\ee}{\end{equation}}
\newcommand{\bea}{\begin{eqnarray}}
\newcommand{\eea}{\end{eqnarray}}
\newcommand{\LE}{\left[}
\newcommand{\R}{\right]}
\newcommand{\nn}{\nonumber}
\newcommand{\Tr}{\text{Tr}}
\newcommand{\N}{\mathcal{N}}
\newcommand{\G}{\Gamma}
\newcommand{\vf}{\varphi}
\newcommand{\LL}{\mathcal{L}}
\newcommand{\Op}{\mathcal{O}}
\newcommand{\HH}{\mathcal{H}}
\newcommand{\arctanh}{\text{arctanh}}
\newcommand{\up}{\uparrow}
\newcommand{\down}{\downarrow}
\newcommand{\ket}[1]{\left| #1 \right>}
\newcommand{\bra}[1]{\left< #1 \right|}
\newcommand{\ketbra}[1]{\left|#1\right>\left<#1\right|}
\newcommand{\rd}{\partial}
\newcommand{\de}{\partial}
\newcommand{\ba}{\begin{eqnarray}}
\newcommand{\ea}{\end{eqnarray}}
\newcommand{\db}{\bar{\partial}}
\newcommand{\we}{\wedge}
\newcommand{\ca}{\mathcal}
\newcommand{\lr}{\leftrightarrow}
\newcommand{\f}{\frac}
\newcommand{\s}{\sqrt}
\newcommand{\vp}{\varphi}
\newcommand{\hvp}{\hat{\varphi}}
\newcommand{\tvp}{\tilde{\varphi}}
\newcommand{\tp}{\tilde{\phi}}
\newcommand{\ti}{\tilde}
\newcommand{\ap}{\alpha}
\newcommand{\pr}{\propto}
\newcommand{\mb}{\mathbf}
\newcommand{\ddd}{\cdot\cdot\cdot}
\newcommand{\no}{\nonumber \\}
\newcommand{\la}{\langle}
\newcommand{\lb}{\rangle}
\newcommand{\ep}{\epsilon}
 \def\we{\wedge}
 \def\lr{\leftrightarrow}
 \def\f {\frac}
 \def\ti{\tilde}
 \def\ap{\alpha}
 \def\pr{\propto}
 \def\mb{\mathbf}
 \def\ddd{\cdot\cdot\cdot}
 \def\no{\nonumber \\}
 \def\la{\langle}
 \def\lb{\rangle}
 \def\ep{\epsilon}
\newcommand{\mcl}{\mathcal}
 \def\g{\gamma}
\def\Tr{\text{tr}}

\begin{titlepage}
\thispagestyle{empty}

\begin{flushright}

\end{flushright}
\bigskip

\begin{center}
  \noindent{\large \textbf{Signature of quantum chaos in operator entanglement in 2d CFTs}}\\
\vspace{2cm}

Laimei Nie $^{a}$,
Masahiro Nozaki $^{a,b}$,
Shinsei Ryu $^{a,c}$, and 
Mao Tian Tan $^{c}$

\vspace{1cm}

{\it $^{a}$Kadanoff Center for Theoretical Physics, University of Chicago,\\
Chicago, Illinois 60637, USA \\}

{\it $^{b}$iTHEMS Program, RIKEN, Wako, Saitama 351-0198, Japan}

{\it $^{c}$James Franck Institute, University of Chicago, Chicago, Illinois
  60637, USA \\}

\vskip 4em
\end{center}
\begin{abstract}
We study operator entanglement measures of 
the unitary evolution operators of (1+1)-dimensional
conformal field theories (CFTs),
aiming to uncover their scrambling and chaotic behaviors.
In particular, we compute
the bi-partite and tri-partite mutual information
for various configurations
of input and output subsystems, 
and as a function of time.
We contrast three different CFTs: 
the free fermion theory,
compactified free boson theory at various radii, 
and CFTs with holographic dual.
We find that the bi-partite mutual information
exhibits distinct behaviors for these different CFTs, reflecting the different information
scrambling capabilities of these unitary operators; 
while a quasi-particle picture can describe well
the case the free fermion and free boson
CFTs,
it completely fails for 
the case of holographic CFTs.
Similarly, the tri-partite mutual information
also distinguishes the unitary evolution operators
of different CFTs.
In particular, 
its late-time behaviors,
when the output subsystems are semi-infinite,
are quite distinct for these theories.
We speculate that 
for holographic theories
the late-time saturation value of the tri-partite mutual
information takes the largest possible negative value and saturates the lower bound
among quantum field theories.

\end{abstract}
\end{titlepage} 
\tableofcontents

\section{Introduction}

\subsection{Backgrounds and motivations}

Solving far-from-equilibrium dynamics
in quantum many-body systems is a
daunting task in general.
This can be compared, for example, with the problem of finding a ground state of
a many-body Hamiltonian, which is time-independent and local. 
While the many-body Hilbert space is exponentially large,
ground states of physically sensible Hamiltonians
are expected to live on a small ``corner'' of the Hilbert space,
which consists of relatively low-entangled states,
e.g., those which obey the area law of bi-partite entanglement entropy. 
(See, e.g., \cite{2010RvMP...82..277E}.)
On the other hand,
in non-equilibrium quantum problems such as quantum quenches,  
even when we start from a low-entangled initial state,
quantum entanglement is generated during the evolution, and we are eventually
kicked out of the corner.

Nevertheless, in systems consisting of a large number of
degrees of freedom with reasonably strong interactions,
universal behaviors can emerge and efficient descriptions may be possible.
For example, many-body quantum systems can thermalize at late times,
even when their dynamics is governed by a unitary time-evolution operator
\cite{PhysRevA.43.2046, PhysRevE.50.888, 2008Natur.452..854R,
2011RvMP...83..863P, 2016AdPhy..65..239D,2016RPPh...79e6001G,1742-5468-2016-6-064001}.
Namely,
the reduced density matrix for a given local region
is given by the one describing
thermal (or generalized Gibbs) ensembles. 
The memory of the initial state that we started with may be completely lost beyond some time scale, at least locally;
The quantum information of the initial state is scrambled and spread non-locally over large length scales. 
In order for complete thermalization to occur, systems at least has
to be non-integrable or even chaotic.  
Once the system thermalizes, its dynamics may be described by hydrodynamics. 

In addition to the motivations from the many-body quantum physics,
scrambling and quantum chaos also have
an intimate connection to the physics of black holes 
\cite{
2007JHEP...09..120H,
2008JHEP...10..065S,
2014JHEP...03..067S,
2013JHEP...04..022L,
2015JHEP...03..051R,
2017JHEP...05..118C,
2015PhRvL.115m1603R,
2014arXiv1409.1231H,
2014JHEP...12..046S}.
For example, in AdS/CFT, the geometry corresponding to
highly excited states in CFTs is expected to be a black hole in the bulk.
How can CFTs, which are unitary, 
encode the thermal behaviors of the black hole?
Scrambling in the dynamics of holographic CFTs
is expected to be a key ingredient to resolve this puzzle. 

One way to characterize chaotic behaviors is to use out-of-time order correlators (OTOC) 
\cite{1969JETP...28.1200L},
which have been revived and discussed,
e.g.,
in recent studies on the Sachdev-Ye-Kitaev and related models
\cite{1993PhRvL..70.3339S, KitaevSYK, 2016JHEP...08..106M,2018JHEP...06..122R,2018arXiv180606840S,2018PhRvB..98n4304K}.
In the semiclassical sense, one can interpret that OTOCs quantify 
the exponential separation of nearby classical trajectories in the phase space.
It is expected to show the exponential growth (decrease) in chaotic systems,
characterized
by the so-called the Lyapunov exponent.

The exponential behaviors in the OTOC happens relatively early time
(the time before or around the scrambling time).
On the other hand, for a later time (time scale which is comparable to the inverse
system size),
the chaotic behaviors may also manifest in the level statistics of energy levels.
In chaotic (non-integrable) systems, energy levels are expected to repel.
For the case of single-particle chaos, it is accepted
the level statistics of chaotic systems are described by the random matrix
theory (the Bohigas-Giannoni-Schmidt conjecture)
\cite{PhysRevLett.52.1}.

In this paper, we will discuss quantum chaos 
 in (1+1)-dimensional conformal field theories (CFTs)
by studying various quantum entanglement measures
defined for the unitary time-evolution operator $U(t) = \exp(-i t H)$ of CFTs. 
Specifically, we use the operator-state mapping
\cite{Jamiolkowski,Choi}
to map the unitary evolution operator $U(t)$, acting on the CFT Hilbert space
$\mathcal{H}$, to a corresponding state $|U(t)\rangle$ in the doubled Hilbert space
$\mathcal{H}_1\otimes \mathcal{H}_2 \equiv \mathcal{H}_{{\it tot}}$, where $\mathcal{H}_{1,2}$ are two copies of the original Hilbert space $\mathcal{H}$:
Representing the (regularized) time-evolution operator using the spectral
decomposition,
\begin{align} \label{op}
U_{\epsilon}(t)=e^{H(-\epsilon-it)} = \sum_a e^{-E_a(\epsilon+it)}\ket{a}\bra{a},
\end{align}
where $|a\rangle$ is an eigenstate of $H$ with energy $E_a$,
the mapped state $|U(t)\rangle$ in $\mathcal{H}_1 \otimes \mathcal{H}_2$
is then given by simply ``flipping'' the second bra $\bra{a}$ to the ket $\ket{a}$ 
\begin{align} \label{dsuni}
  \ket{U_{\epsilon}(t)} =\mathcal{N} \sum_a e^{-itE_a}e^{-\epsilon E_a}\ket{a}_{1}\ket{a}_{2},
\end{align}
where $\mathcal{N}$ is a normalization constant.
We often refer to $\mathcal{H}_{1/2}$ as the input/output Hilbert spaces, respectively.
(See Sec.\ \ref{Operator entanglement: general consideration}
for more details.)

In much the same way as we do for ordinary states in the original Hilbert space
$\mathcal{H}$,
we can introduce various quantum entanglement measures
for the state $|U(t)\rangle$
in the doubled Hilbert space $\mathcal{H}_1 \otimes \mathcal{H}_2$
to quantify its entanglement properties.
%
For example,
let us consider a bipartitioning of the Hilbert space
$\mathcal{H}_{{\it tot}}$ into $\mathcal{H}_A$ and $\mathcal{H}_{\bar A}$.
Then, the $n$-th R\'enyi operator entropy of subsystem $A$
is defined by
\begin{align}
S^{(n)}_A = \frac{1}{1-n} \log [\mathrm{Tr}_A (\rho^n_A)],
\end{align}
where 
$\rho_A = \mathrm{Tr}_{\bar A}\,
|U_{\epsilon}(t)\rangle\langle  U_{\epsilon}(t)|$
is the reduced density matrix for the subsystem $A$.
Taking the limit $n\to 1$,
we obtain the von-Neumann operator entanglement entropy.
Here, $\mathcal{H}_A$ can lie entirely in $\mathcal{H}_1$ or $\mathcal{H}_2$,
or it can overlap both with $\mathcal{H}_1$ and $\mathcal{H}_2$.  

Using (R\'enyi) operator entanglement entropy $S^{(n)}_A$, we further introduce:

\begin{itemize}
  \item
The $n$-th {\it bi-partite operator mutual information} (BOMI)
\begin{align}
I^{(n)}(A, B) = S^{(n)}_{A} + S^{(n)}_{B} - S^{(n)}_{A \cup B},
\end{align}
for two sub Hilbert spaces $\mathcal{H}_{A}, \mathcal{H}_{B} \subset \mathcal{H}_{{\it tot}}$,
and 

\item
The $n$-th {\it tri-partite operator mutual information} (TOMI)
\begin{align}
I^{(n)}(A, B_1, B_2) = I^{(n)}(A,B_1) + I^{(n)}(A,B_2) - I^{(n)}(A, B_1 \cup B_2),
\end{align}
for three sub Hilbert spaces $\mathcal{H}_{A}, \mathcal{H}_{B_1}, \mathcal{H}_{B_2} \subset \mathcal{H}_{{\it tot}}$.
\end{itemize}
Once again,
the sub Hilbert spaces $\mathcal{H}_{A}, \mathcal{H}_{B_1}$, etc.
can lie entirely in $\mathcal{H}_1$ or $\mathcal{H}_2$,
or they can overlap both with $\mathcal{H}_1$ and $\mathcal{H}_2$.

\begin{figure}
  \begin{center}
    \includegraphics[width=14.5cm]{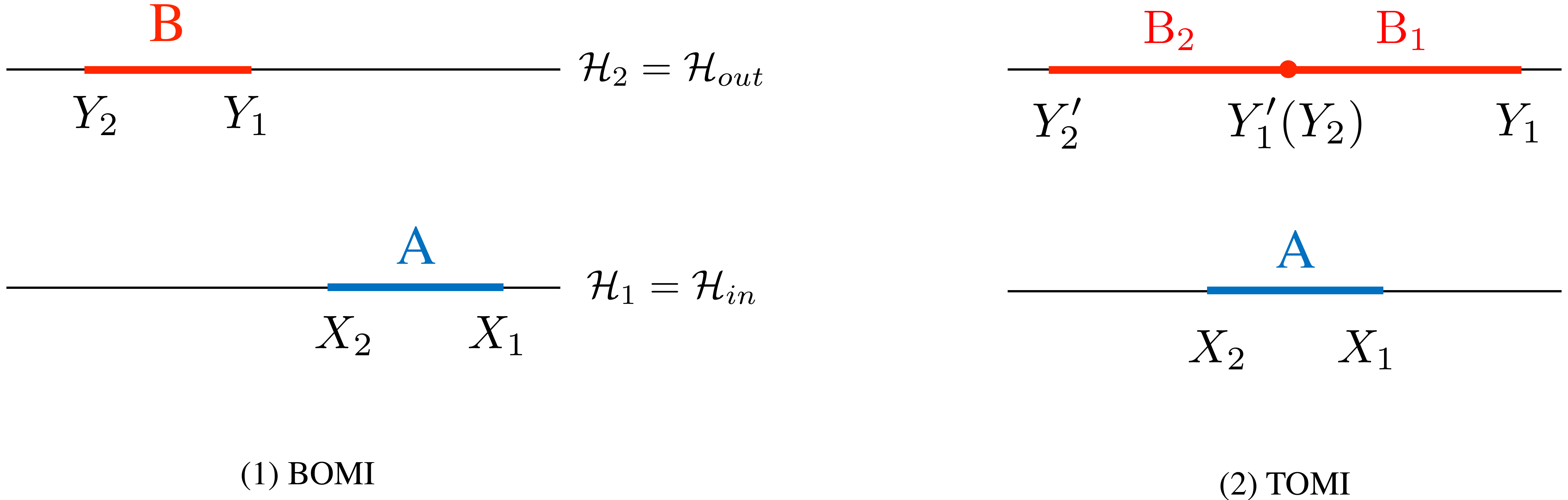} 
    \caption{General set up of operator mutual information. $\mathcal{H}_{1,2}$ are the spatially 1-dimensional doubled Hilbert spaces represented by the horizontal thin lines. $X_{1,2}, Y_{1,2}, Y'_{1,2}$ are spatial coordinates of the boundaries of the subsystems, which are marked by blue (input) and red (output) intervals.
    (1) Bi-partite case. (2) Tri-partite case, note that $B_1, B_2$ do not need to be adjacent in general. }
    \label{config}
       \end{center}
\end{figure}

These operator entanglement measures allow us to focus on the properties of the evolution operator itself and study their information scrambling and chaotic behaviors, independent of the choice of initial states of the time-evolution. 
For previous studies on the operator entanglement 
entropy of unitary evolution operators, see, for example, 
\cite{2015JHEP...08..011C, 2016JHEP...02..004H, 2017JPhA...50w4001D, 2017arXiv171206054C,2018arXiv180300089J}.
In particular, Ref.\ \cite{2016JHEP...02..004H}
proposed the negativity of TOMI of
unitary evolution operators
(or unitary channels)
as a general diagnostic of scrambling. 
When $\mathcal{H}_{A} \subset \mathcal{H}_1$
and $\mathcal{H}_{B} \subset \mathcal{H}_2$
(Fig.~\ref{config}(1)),
roughly speaking,
$I^{(n)}(A, B)$
describes the localization of information sent by the unitary channel from
$A$ to $B$:
$I^{(n)}(A, B)$ is always non-negative.
$I^{(n)}(A, B) > 0$ means that
we can mine information of $A$ from $B$ locally. 
On the other hand,
when $\mathcal{H}_A \subset \mathcal{H}_1$
and $\mathcal{H}_{B_{1,2}} \subset \mathcal{H}_2$
(Fig.~\ref{config}(2)),
$I^{(n)}(A, B_1,B_2)$
describes the delocalization of information scrambled by the unitary channel:
$I^{(n)}(A, B_1, B_2) $ can be positive, negative, or zero.
In particular, $I^{(n)}(A, B_1, B_2) < 0$
means that some information of $A$ is hidden in the entire output subsystem
$B_1\cup B_2$ and cannot be mined by simply measuring $B_1$ or $B_2$ separately.

%
%

\subsection{Summary of main results}


Chaotic behaviors of some sort are expected to occur in CFTs
with large number of degrees of freedom per local Hilbert space,
i.e., CFTs with large central charge $c \gg 1$,
and when interactions are sufficiently strong
such that these degrees freedom are ``mixed well enough''
under time evolution.
Despite the fact that a definitive characterization of chaos in generic quantum systems remains an open problem, 
we expect those CFTs which have a holographic dual description
in terms of classical gravity are strongly (maximally) chaotic.
One of our primary goals in this paper is to see if we are able to characterize
chaotic behaviors of these CFTs by using operator entanglement measures.
This is complimentary to and should be contrasted with
prior works which tried to detect
the signature of quantum chaos in CFTs by computing OTOCs
\cite{2016JHEP...08..129G,2016PTEP.2016k3B06C,2017PhRvD..96d6020C,2016JHEP...10..069P},
the relative entropy
\cite{2018JHEP...07..002N},
and others
\cite{2017JHEP...08..075D,2017arXiv171206054C}.
(We compare these different indicators of
scrambling and chaos in Sec.\ \ref{Conclusion}.)


In this paper, we compute the operator bi-partite and tri-partite mutual
information for various setups,
for the free fermion CFT $(c=1)$,
the compactified free boson CFT ($c=2$, complex boson)
at various radii,
and holographic CFTs $(c\gg 1)$.
The evolution operators in these CFTs
are expected to show the different degrees of
information scrambling capabilities,
ranging from
the complete absence of scrambling in the free fermion CFT
to 
the maximal scrambling in holographic CFTs.
The compactified free boson CFT, depending on its compactification radius, 
is expected to lie somewhere in between these two extreme cases.
(See Fig~\ref{fig:ScramblingAbility}.) 
Our findings in this work,
a quick summary and sampling of which
are presented below,
support these expectations.

\begin{figure}[t]
  \begin{center}
    \includegraphics[width=9.5cm]{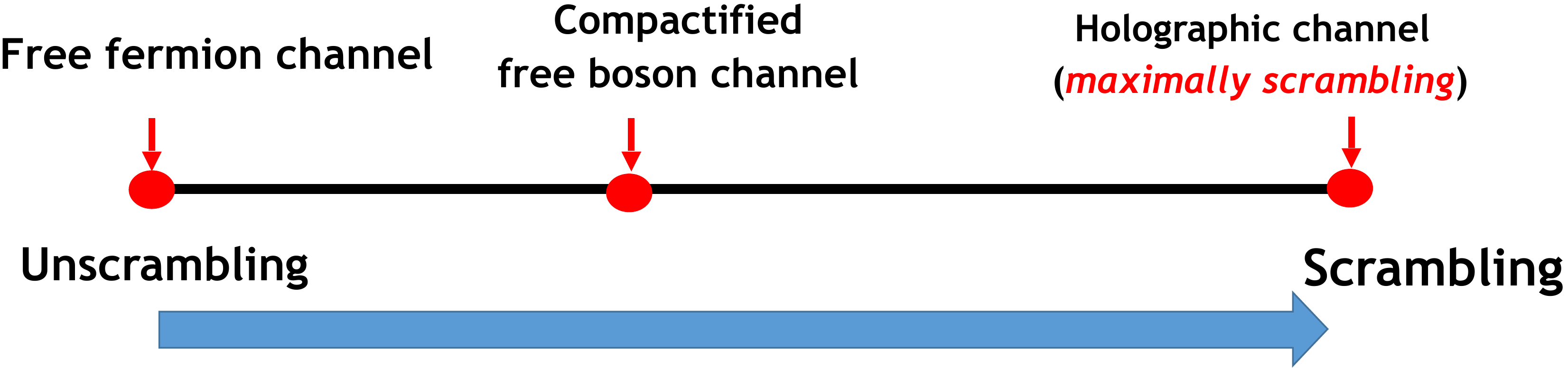} 
     \caption{Information scrambling abilities of three
       types of CFTs.}
     \label{fig:ScramblingAbility}
    \end{center}
\end{figure}


\begin{figure}[h]
  \begin{center}
    \includegraphics[width=18cm]{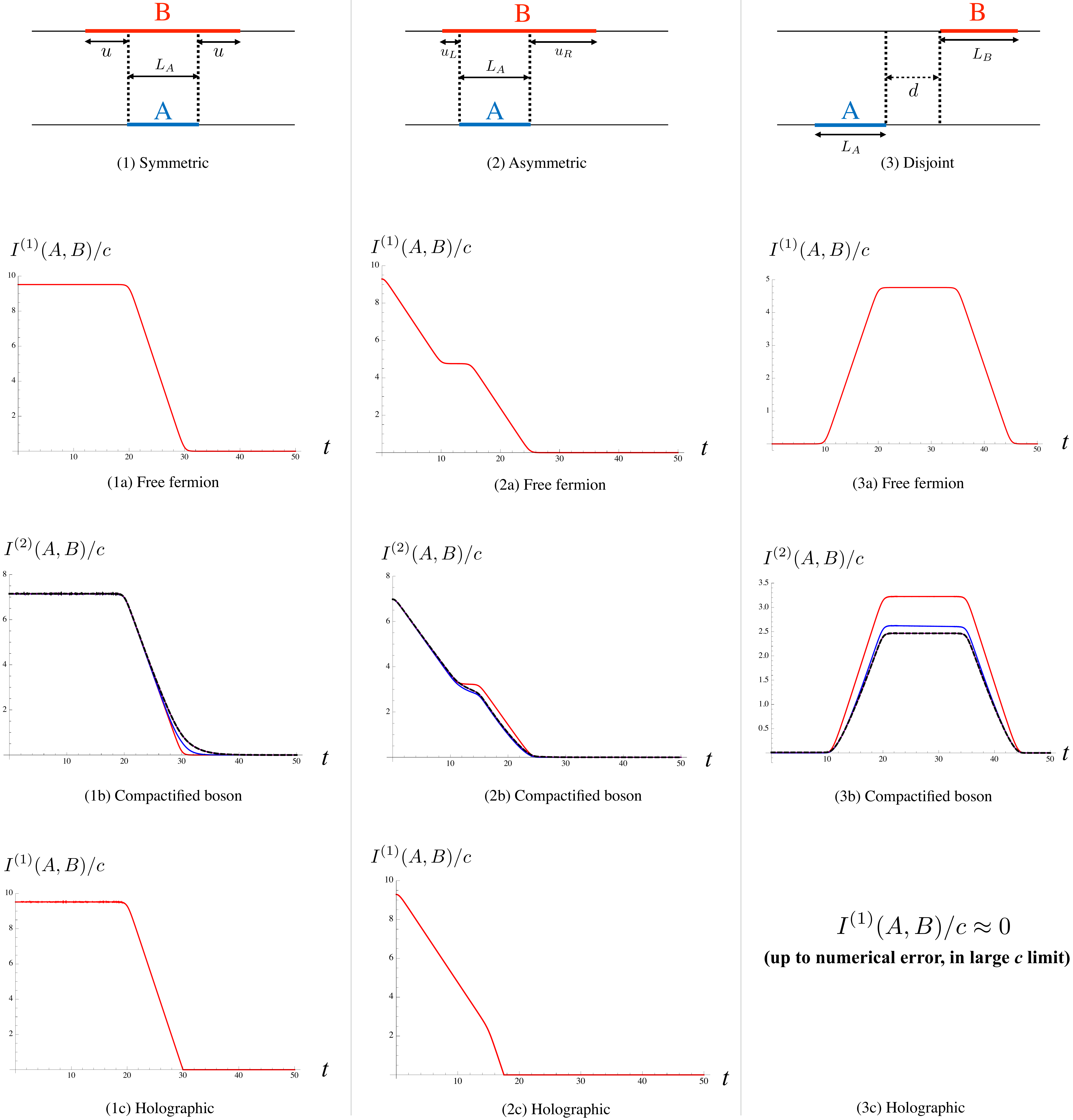} 
    \caption{\label{fig:bOpMIchart}
      BOMI per central charge for free fermion, compactified boson, and holographic CFTs, under different setups (1)(2)(3). The regulator $\epsilon = 1.1$ for all the plots. \textbf{(1a), (1b), (1c):} $(L_A , u)= (10, 20)$. \textbf{(2a), (2b), (2c):} $(L_A, u_L, u_R) = (10, 0,15).$  \textbf{(3a), (3b), (3c):} $(L_A, L_B, d) = (10,25,10).$
      \textbf{(1b), (2b), (3b):} {\color{red} red: $R^2 = \eta = 1$} ($R$ is the compactification radius; see Sec. \ref{sec: CB}), {\color{blue} blue:  $\eta = \pi$}, {\color{magenta} purple: $\eta=6$}, black dash: $\eta=1/6$, which is dual to $\eta=6$, so purple curve and black dash are not visually distinguishable. Note that due to technical difficulty of analytic continuation to $n=1$ here we are only showing $I^{(2)}$ for the compactified boson case.}
  \end{center}
\end{figure}

\clearpage 
\subsubsection{Bi-partite operator mutual information (BOMI)}

For BOMI,
we mainly consider three setups here, ``symmetric", ``asymmetric"
and ``disjoint", as shown in Fig.~\ref{fig:bOpMIchart}.
As seen in Fig. \ref{fig:bOpMIchart}, while  
the early and late-time evolution of BOMI is
rather insensitive to types of CFTs,
in intermediate time regimes,
it shows distinct behaviors. 
For example, for the case of asymmetric
subsystems $A$ and $B$, 
BOMI for the free fermion and compactified free boson CFTs
develops a plateau,
while BOMI for holographic channels
does not show such a behavior.
Also, for the disjoint case,
BOMI for the free fermion and compactified boson CFTs
shows a peak or bump,
while BOMI for holographic CFTs identically vanishes,
$I^{(1)}(A, B) = 0$,
indicating that one cannot mine information of input $A$ from $B$
due to the fact that $B$ is only a local portion of the output and has no overlap with $A$.
See Sec.\ \ref{Comparison of BOMI for various CFTs}
for more detailed descriptions and discussions.

As we will demonstrate in Sec.\ \ref{Comparison of BOMI for various CFTs},
the free fermion case can be explained using quasi-particle picture
\cite{Calabrese2006PRL,2005JSMTE..04..010C}.
This is also the case for the compactified boson theory,
except for some curves (peaks, plateaus) are rounded
(especially for non-self-dual cases in Fig.~\ref{fig:bOpMIchart}(1b)(2b)(3b)),
which presumably mean
that the quasi-particles are less sharply defined compared to free fermion case.
The holographic case has a prominent deviation from the free fermion case,
and hence from the quasi-particle picture.

\subsubsection{Tri-partite operator mutual information (TOMI)}

\begin{figure}[h]
  \begin{center}
    \includegraphics[width=5.5cm]{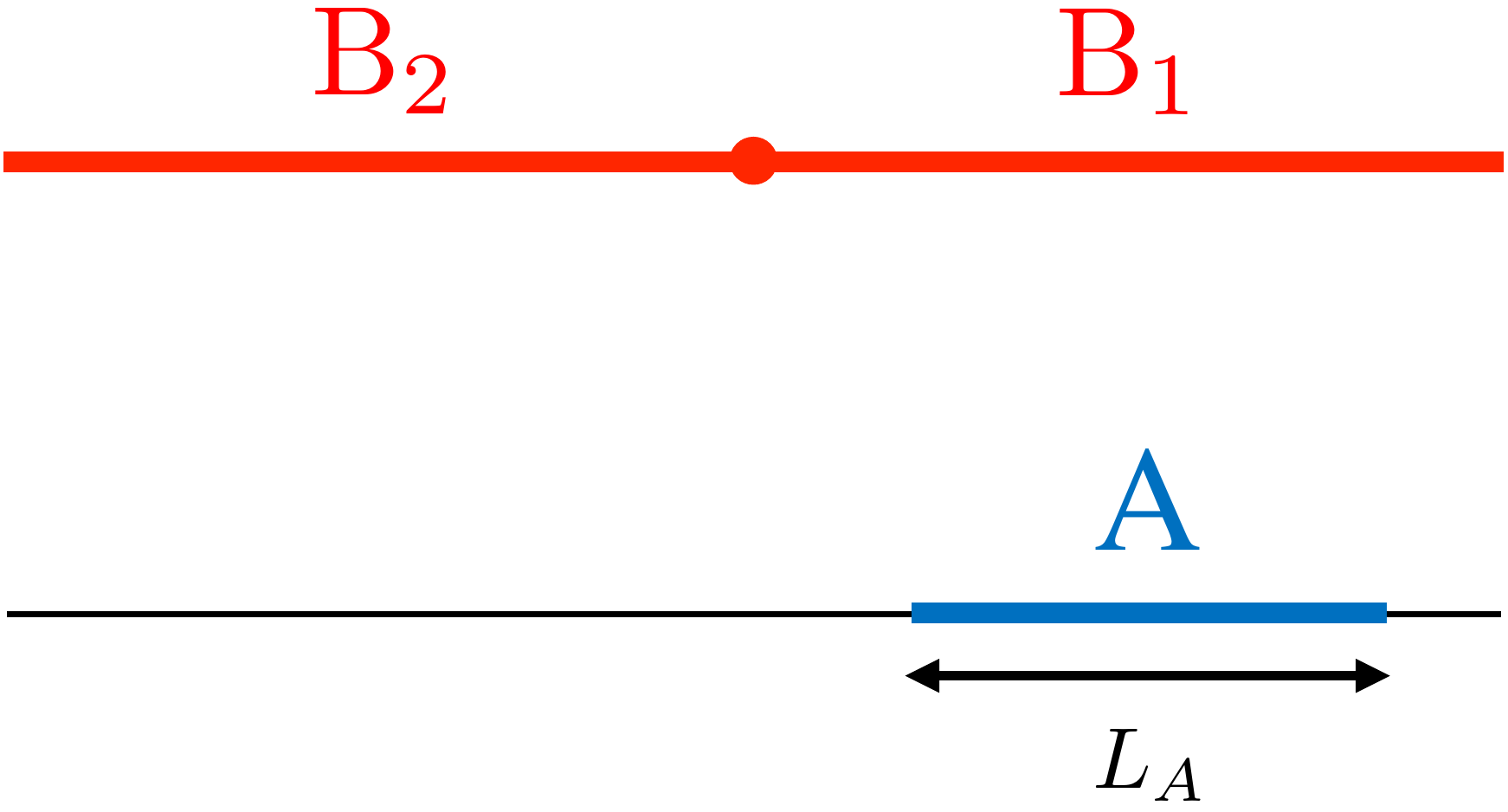} 
     \caption{Setup for TOMI with infinite output subsystem with $B_1, B_2$ being semi-infinite.} 
    \label{fig:tOpMIconfi}
       \end{center}
\end{figure}

\begin{figure}[h]
  \begin{center}
    \includegraphics[width=16.5cm]{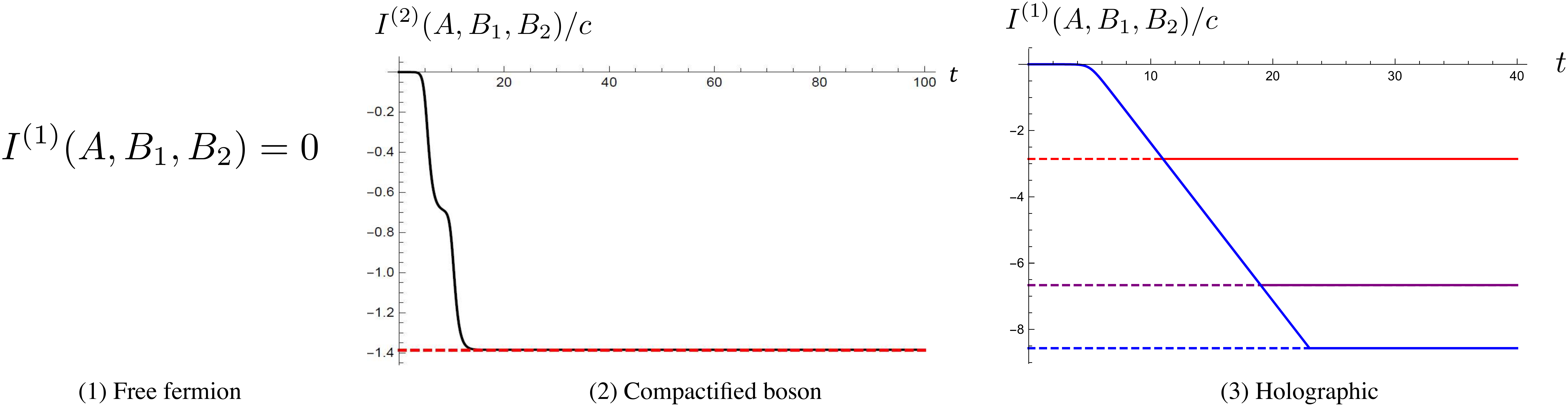} 
    \caption{TOMI per central charge
      for the configuration in Fig.\ \ref{fig:tOpMIconfi}
      for three different CFTs. $\epsilon = 1.1$ for all the curves.
      \textbf{(1)}
      Free fermion.
      \textbf{(2)}
      Compactified boson at self-dual radius $R^2 = \eta = 1$, with $L_A = 5$ (see Sec. \ref{sec: CB} for definitions of $R$ and $\eta$). The red dashed line indicates the late-time saturation value which is equal to $\log (1/4)$.
      \textbf{(3)}
      Holographic CFTs, {\color{red} red: $L_A = 3$},   {\color{magenta} purple: $ L_A = 7$},  {\color{blue} blue: $L_A = 9$}. The three curves overlap with each other for small $t$. The dashed lines are guided to the eyes for the late-time saturation values $-2S^R_A$, where $S_A^R \equiv \pi c L_A /( 6\epsilon)$ (in the limit $L_A \gg \epsilon$) is a regularized entanglement entropy of region $A$. Details of the analysis can be found in Sec. \ref{sec:TOMIinfinite}.
     }
    \label{fig:tOpMIchart}
     \end{center}
\end{figure}

For TOMI,
we study both the case of finite output subsystem (Fig.~\ref{config}(2)),
and of infinite output subsystems (Fig.~\ref{fig:tOpMIconfi})
where $Y_1, Y_2'$ in Fig.~\ref{config}(2)
are sent to $+\infty, -\infty$ respectively.
As an illustration, we present
in Fig.~\ref{fig:tOpMIchart}
our results for the case of infinite output subsystem.

For the free fermion CFT
(Fig.~\ref{fig:tOpMIchart}(1)),
TOMI vanishes identically: $I^{(1)} (A, B_1,B_2) = 0$;
Due to the lack of information scrambling, the information that can be mined
locally from $B_1$ and $B_2$ sums up to equal to the amount that can be mined
globally from $B_1 \cup B_2$.
On the other hand, 
for the compactified boson (self-dual or not)
(Fig.~\ref{fig:tOpMIchart}(2))
TOMI is non-zero, and saturates to a negative value at late times.
This is in contrast with BOMI,
which shows similar behaviors
both for the free fermion and compactified boson CFTs,
Figs.~\ref{fig:bOpMIchart} (2a) and (2b).
The late-time saturation value of TOMI
depends on the radius of the compactified boson,
and also on the size of the input subsystem ($=L_A=X_1-X_2$).
Generically, for irrational radii,
the late-time saturation value of TOMI diverges
logarithmically 
to $-\infty$
as we send $L_A\to \infty$. On the other hand, for
rational radii, 
the late-time saturation value of TOMI is finite
and given by the quantum dimension of the twist operator.
We will give a detailed analysis of the late-time behavior of TOMI 
in Sec.\ \ref{The compactified boson theory TOMI}.
Finally, for holographic CFTs (Fig.~\ref{fig:tOpMIchart}(3)),
TOMI saturates to a negative value at late times.
The saturation value grows linearly with $L_A$,
and given by the properly regularized entanglement
entropy for the region $A$; See~\eqref{late time const} and~\eqref{reg SA}.
The late-time saturation value of
the negativity of TOMI diverges mostly strongly 
for holographic CFTs (linearly divergent in $L_A$),
and we suspect that holographic CFTs, among sensible
local quantum field theories, provide
the lower bound of the late-time value of TOMI.

\subsection{Organization of the paper}

The rest of the paper is organized as follows:
In Sec.\ \ref{Operator entanglement: general consideration},
we collect necessary background materials, in particular,
the definitions of 
various operator entanglement measures,
and the corresponding twist operator formalism
in CFTs.
In Sec.\ \ref{Bi-partite operator mutual information}, 
we present the details of our analysis for BOMI
for the free fermion, compactified free boson,
and holographic CFTs.
When possible, we also establish the quasi-particle picture
explaining the behaviors of BOMI. 
In Sec.\ \ref{tri-partite operator mutual information},
our results for TOMI are presented for various configurations
of input/output subsystems.
In particular, when
the output subsystem is infinite,
we discuss in detail the late-time behaviors of TOMI,
which are quite distinct for
the free fermion (zero),
compactified free boson (negative but magnitude is not large),
and 
holographic (large negative and saturates the lower bound)
CFTs.
Finally in Sec.\ \ref{Conclusion},
we summarize our results and give further discussions,
including a possible ``hydrodynamic''
information flow picture that can explain the
time-dependent behaviors of the operator entanglement measure
in holographic CFTs.

\section{Operator entanglement: general consideration}
\label{Operator entanglement: general consideration}

\subsection{Operator-state mapping and operator entanglement}

Here we present the general setup of evaluating operator entanglement and mutual information in the context of CFT. 
Consider a Hilbert space $\mathcal{H}$,
and Hamiltonian $H$.
The time-evolution operator regularized by energy cutoff $\epsilon^{-1}$ and its spectral decomposition are given by
\be \label{op}
U_{\epsilon}(t) = e^{H(-\epsilon-it)} = \sum_a e^{-E_a(\epsilon+it)}\ket{a}\bra{a},
\ee
where $|a\rangle$ is an eigenstate of $H$
with energy $E_a$.
As mentioned in the previous section, in order to study the operator entanglement of
the unitary evolution operator $U_{\epsilon}(t)$ 
we first use the operator-state mapping
\cite{Jamiolkowski,Choi}
to introduce a corresponding quantum state
$|U_{\epsilon}(t)\rangle$,
which lives in the doubled Hilbert space
$\mathcal{H}_{{\it tot}}=\mathcal{H}_1\otimes\mathcal{H}_2$,
where $\mathcal{H}_{1}=\mathcal{H}_2=\mathcal{H}$ are two identical copies
of the original Hilbert space $\mathcal{H}$. The total Hamiltonian acting on $\mathcal{H}_{tot}$ is denoted by 
\be
H_{{\it tot}} \equiv H_1 \otimes \mathbbm{1} +   \mathbbm{1} \otimes  H_2, 
\ee
where $H_1 = H$ and $H_2 = H$ act on $\mathcal{H}_{1}$ and $\mathcal{H}_2$ respectively, and $ \mathbbm{1}$ is the identity operator.
The mapping from an operator to a state is then given by
$
\bra{a} \rightarrow \ket{a}.
$
Under this map, the operator in (\ref{op}) is mapped to a
thermofield double-like state:
\begin{align} \label{dsuni}
  \ket{U_{\epsilon}(t)}&=
                         \sum_a e^{-itE_a}e^{-\epsilon E_a}\ket{a}_{1}\ket{a}_{2}
                         \nonumber
  \\
&=e^{\f{-it}{2}(H_{1}+H_{2})}
       e^{-\f{\epsilon}{2} (H_{1}+H_{2})}\sum_a\ket{a}_{1}\ket{a}_{2}
  \nonumber \\
& \equiv e^{\f{-it}{2}(H_{1}+H_{2})}\ket{{\it TFD}(t=0)}_{\epsilon},
\end{align}
where
$
\ket{{\it TFD}(t=0)}_{\epsilon}
=
e^{ - \frac{\epsilon}{2}(H_1+H_2)}
\sum_a |a\rangle_1 |a\rangle_2
$
is the (unnormalized) thermofield double state
at inverse temperature $\epsilon$ (for notational convenience we have written $H_{tot} = H_1 + H_2$).
If we consider the operator in (\ref{op}) in holographic CFT, its gravity dual
is an eternal black hole \cite{HM} where time directions in both sides is the same
as in Fig. \ref{f1} in Appendix \ref{Holographic computation}. 
Thus the eternal black hole is dual to the {\it operator} in (\ref{op}). 

Coming back to the operator-state mapping, we proceed to the evaluation of (R\'enyi) entanglement entropy by considering the normalized state
\begin{align}
  \label{opeerator state U}
  \ket{U_{\epsilon}(t)}=
  \mathcal{N}\, e^{-\left(\f{it +\epsilon}{2}\right)H_{{\it tot}}}
\sum_a\ket{a}_1\ket{a}_2,
\end{align}
where the normalization factor $\mathcal{N}$ is given by 
\begin{align}
\mathcal{N}^{-2} = {\Tr_{\mathcal{H}_1}e^{-2\epsilon H_1}}.
\label{NE}
\end{align}
The corresponding (pure state) density matrix $\rho$ then reads
\begin{align}
  \rho&=\ket{U_{\epsilon}(t)}\bra{U_{\epsilon}(t)}
  \nonumber \\
&=\mathcal{N}^2\, e^{-\left(\f{\epsilon+i t}{2}\right)H_{{\it tot}}}
\sum_{a,b}\ket{a}_1  \tensor[_1]{\bra{b}}{}   \otimes \ket{a}_2  \tensor[_2]{\bra{b}}{}
e^{-\left(\f{\epsilon-i t}{2}\right)H_{{\it tot}}}.
\end{align}
It will be convenient to work with the corresponding Euclidean reduced density $\rho^E$ 
\begin{align}
\rho^E=\mathcal{N}
^2\, e^{\f{\tau_1}{2} H_{{\it tot}}}
\sum_{a,b}\ket{a}_1  \tensor[_1]{\bra{b}}{}   \otimes \ket{a}_2  \tensor[_2]{\bra{b}}{}  e^{-\f{\tau_2}{2}H_{{\it tot}}},
\end{align}
and later analytically continue $\tau_1$ and $\tau_2$:
\be
\tau_1\to -\epsilon-i t,
\quad
\tau_2 \to \epsilon -i t.
\ee
A generic matrix element of $\rho^E$ is given by 
\be
\begin{split}
&\tensor[_1]{\bra{\psi_1}}{} \tensor[_2]{\bra{\psi_2}}{} \rho^E \ket{\psi'_2}_2 \ket{\psi'_1}_1 \\
&=\mathcal{N}^2\sum_{a,b} \tensor[_1]{\bra{\psi_1}}{} e^{\f{\tau_1}{2}H_1}\ket{a}_1   \tensor[_1]{\bra{b}}{}   e^{-\f{\tau_2}{2}
H_1}\ket{\psi'_1}_1  \tensor[_2]{\bra{\psi_2}}{} e^{\f{\tau_1}{2}H_2}\ket{a}_2    \tensor[_2]{\bra{b}}{}  e^{-\f{\tau_2}{2}H_2}\ket{\psi'_2}_2 
\end{split}
\ee
where $\big\{ |\psi_i \rangle_i, |\psi'_i \rangle_i \big\}$ are states living
in the Hilbert space $\mathcal{H}_i \ \ (i = 1,2)$. 

We introduce new kets $| \psi_2^* \rangle_2$ which are
the complex conjugate of $|\psi_2 \rangle_2$:
\be
\tensor[_2]{ \bra{\psi_2}}{} e^{\frac{\tau_1 H_2}{2}} \ket{a}_2 = \tensor[_2]{\bra{\psi_2}}{} a \rangle_2 e^{\frac{\tau_1 E_a}{2}} = \Big[  \tensor[_2]{\bra{a}}{} \psi_2 \rangle_2 \Big]^* e^{\frac{\tau_1 E_a}{2}}  = \tensor[_2]{\bra{a}}{} \psi_2^* \rangle_2 e^{\frac{\tau_1 E_a}{2}}  = \tensor[_2]{\bra{a}}{} e^{\frac{\tau_1 H_2}{2}}   \ket{\psi^*_2}_2,
 \label{cc}
\ee
and similarly,
\be
\tensor[_2]{\bra{b}}{}e^{\frac{-\tau_2 H_2}{2}} \ket{\psi'_2}_2 =  \tensor[_2]{\bra{\psi'^*_2}}{} e^{\frac{-\tau_2 H_2}{2}}   \ket{b}_2.
\ee

\noindent Using these dual kets, the matrix elements can be rewritten as
\begin{align}
  \tensor[_1]{\bra{\psi_1}}{} \tensor[_2]{\bra{\psi_2}}{} \rho^E \ket{\psi'_2}_2 \ket{\psi'_1}_1
&=\mathcal{N}^2\sum_{a,b}
\langle {\psi}'^*_2|e^{-\f{\tau_2}{2}H}\ket{b}   \bra{b}e^{-\f{\tau_2}{2}H}\ket{\psi'_1}
                                                                                                                                       \bra{\psi_1} e^{\frac{\tau_1}{2} H} \ket{a} \bra{a} e^{\frac{\tau_1}{2} H} \ket{\psi_2^*}
  \nonumber \\
&=\mathcal{N}^2
\langle {\psi}'^*_2 |e^{-\tau_2 H}\ket{\psi'_1}\bra{\psi_1}
e^{\tau_1H}
|{\psi}_2^* \rangle, 
\label{rhoEelement}
\end{align}
where without introducing ambiguity we have dropped the Hilbert space indices of
the states.

Next, by bipartitioning the Hilbert space
$\mathcal{H}_{{\it tot}}$
into $\mathcal{H}_C \equiv \mathcal{H}_{A \cup B}$ and $\mathcal{H}_{\bar C}$
and taking the partial trace over $\mathcal{H}_{\bar C}$,
we will be able to obtain the reduced density matrix $\rho_C$. Similar to the
usual case of evaluating R\'enyi entropy of a state (which corresponds to a
single Hilbert space),
the $n$-th order R\'enyi entropy $S^{(n)}_C$ of the time-evolution operator
can be obtained via path integral on an $n$-sheeted Riemann surface,
albeit in the doubled Hilbert spaces. In particular, in (1+1)d CFT it can be formulated in terms of the correlation functions of twist and anti-twist operators. We will elaborate on the procedure in the next section.

\subsection{Twist operator formalism}

\begin{figure}[t]
  \begin{center}
   \includegraphics[width=100mm]{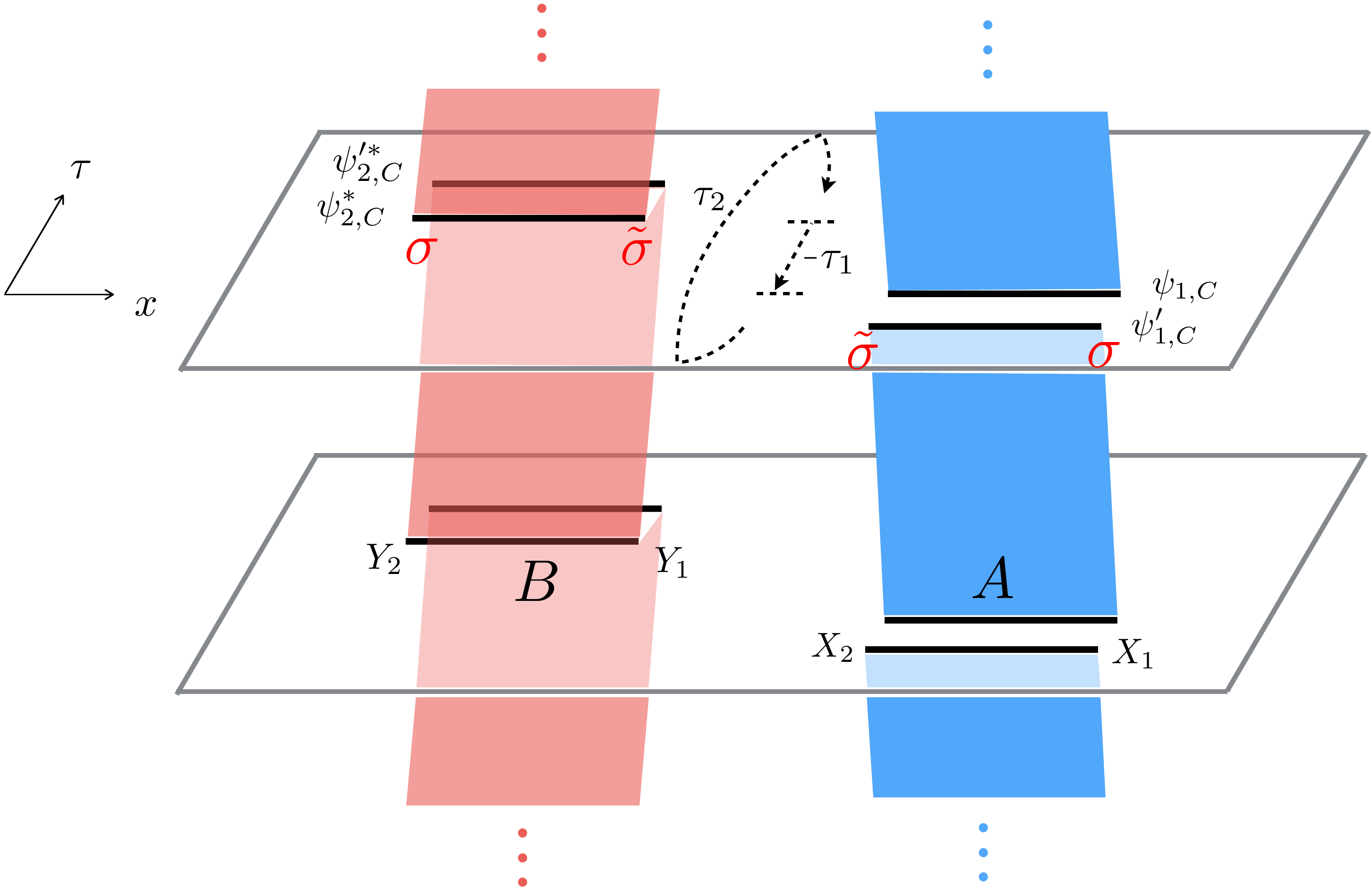}
  \end{center}
  \caption{
  \label{fig:twistoperators}
    Path integral representation of $\Tr_C (\rho_C)^n$, where $C \equiv A \cup B$. 
  Each plane is periodic along $\tau$ direction and represents one $\rho_C$ (here only showing two as an illustration), and black lines mark the subsystems $A,B$ with boundary conditions $\psi_{1,C}, \psi'_{1,C}, \psi^*_{2,C}, \psi'^*_{2,C}$ 
  that are to be identified with corresponding lines on adjacent planes. The identification procedure is indicated by red and blue shadows.
  Notice that the positions of the twist and anti-twist operators ($\sigma$ and
  $\tilde \sigma$) are switched for $B$, due to the direction through which
  adjacent sheets are connected.
}
\end{figure}

We now describe how we can compute 
operator entanglement entropy 
in quantum field theories. 
The reduced density matrix $\rho_C = \rho_{A\cup B}$ is
obtained by tracing out regions outside of $C$ (which we call $\bar C$): $\rho_C = \Tr_{\bar C} \rho^E$.
In terms of matrix elements,
\begin{align}
&\bra{\psi_{1,C}} \bra{\psi_{2,C}}  \rho_C \ket{\psi'_{2,C}} \ket{\psi'_{1,C}} \nonumber\\
&= \mathcal{N}^2  \int \mathcal{D} \psi_{1,\bar C} \mathcal{D} \psi^*_{2,\bar C} \langle \psi'^*_{2,C} , \psi^*_{2,\bar C} | e^{-\tau_2 H} | \psi'_{1,C}, \psi_{1,\bar C} \rangle \langle \psi_{1,C},\psi_{1,\bar C} | e^{\tau_1 H} | \psi^*_{2,C}, \psi^*_{2,\bar C} \rangle
\end{align}
where we have rewritten $\ket{\psi_1}$ in Eq.~\eqref{rhoEelement} as $\ket{\psi_{1,C},\psi_{1,\bar C}}$, $\ket{\psi_2}$ as $\ket{\psi_{2,C},\psi_{2,\bar C}}$ etc. 
The $n$-th moment of
the reduced density matrix $\Tr_C\left(\rho_C\right)^n$
is then given by the thermal expectation value of twist and anti-twist operators
($\sigma_n$ and $\tilde \sigma_n$) appropriately inserted at the boundaries of
$A$ and $B$, as shown in Fig.~\ref{fig:twistoperators}
\cite{2004JSMTE..06..002C, Morrison:2012iz}:
\be
\begin{split}
  \Tr_C\left(\rho_C\right)^n&= C_0 \cdot \left(\mathcal{N}^2\right)^n
  \Tr_{\mathcal{H}}
  \left[e^{-\tau_2 {H}^{(n)}}\sigma_n(X_1)\tilde{\sigma}_n(X_2)
    e^{\tau_1{H}^{(n)}}\tilde{\sigma}_n(Y_1)\sigma_n(Y_2)\right]
  \\
  &= C_0 \cdot\left(\mathcal{N}^2\right)^n
  \Tr_{\mathcal{H}}
  \left[e^{-{H}^{(n)}(\tau_2-\tau_1)}\sigma_n(X_1)\tilde{\sigma}_n(X_2)  \tilde{\sigma}_n(Y_1, \tau_1) \sigma_n(Y_2,
   \tau_1) \right]
  \\
  &= C_0 \cdot \left \langle \sigma_n(X_1)\tilde{\sigma}_n(X_2)   \sigma_n(Y_2, \tau_1) \tilde{\sigma}_n(Y_1, \tau_1)  \right \rangle_{2\epsilon}
\end{split}
\label{TrrhoAn}
\ee
where $H^{(n)} \equiv \sum\limits_{a=1}^n H[\psi^{(a)}]$ is the Hamiltonian of the $n$-copied system ($a$ is the replica index), and we have used $\tau_2 - \tau_1 = 2\epsilon$ together with Eq.~\eqref{NE} which gives $(\mathcal{N}^2)^n =\big(1/ \Tr_{\mathcal{H}_1}e^{-2\epsilon H_1} \big)^n = 1 / \Tr_{\mathcal{H}} \big[e^{-2\epsilon H^{(n)}} \big] $.
$\left \langle \cdots \right \rangle_{2\epsilon}$ is the thermal expectation value,  defined as
\be
\left \langle \cdots \right \rangle_{2\epsilon} \equiv \frac{1}{\mathcal{Z}_{2\epsilon}} \Tr_{\mathcal{H}} \Big[ e^{-2\epsilon H^{(n)}}  \cdots \Big], \ \ \mathcal{Z}_{2\epsilon} \equiv \Tr_{\mathcal{H}} \Big[  e^{-2\epsilon H^{(n)}} \Big].
\ee
$C_0$ is a proportionality constant which together with $C_0'$ introduced
in~\eqref{Sa12} below, will be determined by the requirement that the bi-partite
operator mutual information when $A,B$ do not overlap (disjoint case) vanishes at $t=0$ as $\epsilon \to
0$~\cite{Morrison:2012iz}. Notice that the positions of $\sigma_n$ and
$\tilde \sigma_n$ are interchanged for $B$, a reminiscent of the setup of
entanglement negativity calculation (in which case $A$ and $B$ are on the
same imaginary time)
\cite{Morrison:2012iz, CalabreseNegativity}.

Thus, the $n$-th R\'enyi entanglement entropy
and the von Neumann entanglement entropy
for $C = A \cup B$
are given by 
\begin{align}\label{sna}
 & S^{(n)}_{A \cup B}
=\f{1}{1-n}\log{\Big[ 
C_0 \cdot \left \langle \sigma_n(X_1)\tilde{\sigma}_n(X_2)  \sigma_n(Y_2, \tau_1) \tilde{\sigma}_n(Y_1, \tau_1) \right \rangle_{2\epsilon}\Big]},
                          \nonumber \\
& S_{A \cup B} =\lim_{n\rightarrow 1} S^{(n)}_{A \cup B}.
\end{align}
%

\noindent  Meanwhile, the derivation of $S^{(n)}_{A,B}$ is less involved
than $S^{(n)}_{A\cup B}$
as it essentially returns to the single Hilbert space case \cite{2004JSMTE..06..002C}:
\be
\label{Sa12}
\begin{split}
  &S^{(n)}_{A}=\f{1}{1-n}\log{\Big[ 
   C_0' \cdot \left\langle\sigma_n(X_1)\tilde{\sigma}_n(X_2)\right\rangle_{2\epsilon}\Big]},
  \\
  &S^{(n)}_{B}=\f{1}{1-n}\log{\Big[
     C_0' \cdot  \left\langle \tilde \sigma_n(Y_1){\sigma}_n(Y_2)\right\rangle_{2\epsilon}\Big]}, \\
\end{split}
\ee
where $C'_0$ is a constant independent of the coordinates.
\subsection{$S^{(n)}_{A}$ and $S^{(n)}_{B}$ in 2d CFT}

We now turn to the
calculation of the operator mutual information $I^{(n)}(A,B)$
in two-dimensional CFT.
Let us first consider $S^{(n)}_{A}$ and $S^{(n)}_{B}$
given by \eqref{Sa12}.
Since these quantities
are expressed in terms of two-point correlation functions of $\sigma_n$ and $\tilde{\sigma}_n$,
they are rather insensitive
to details of theories and can depend only on the central charge. 

Generally speaking, the two-point function 
$\left\langle\sigma_n(w_1, \bar{w}_1)\tilde{\sigma}_n(w_2,
  \bar{w}_2)\right\rangle_{2\epsilon}$
on a cylinder
can be expressed in terms of
the corresponding correlation function
on an infinite plane
via a conformal transformation:
\be
z=e^{\f{\pi}{\epsilon}w}
\label{conformalmap}
\ee
where $\epsilon$ is half of the inverse temperature. The two-point function then becomes
\be
\begin{split}
\left\langle\sigma_n(w_1, \bar{w}_1)\tilde{\sigma}_n(w_2, \bar{w}_2)\right\rangle_{2\epsilon} &=\prod_{i=1}^2\left(\f{dz}{dw}\right)^{h_n}_{z_i}\left(\f{d\bar{z}}{d\bar{w}}\right)^{\bar{h}_n}_{\bar{z}_i}\left\langle\sigma_n(z_1, \bar{z}_1)\tilde{\sigma}_n(z_2, \bar{z}_2)\right\rangle_{\text{plane}} \\
&=\prod_{i=1}^2\left(\f{dz}{dw}\right)^{h_n}_{z_i}\left(\f{d\bar{z}}{d\bar{w}}\right)^{h_n}_{\bar{z}_i}\f{1}{\left|z_{12}\right|^{4h_n}},\\
\end{split}
\ee
where $h_n = \bar{h}_n = \frac{c}{24}(n - \frac{1}{n})$ is the conformal dimension of $\sigma_n$ and $\tilde \sigma_n$, and we have absorbed a possible proportionality constant in the two-point function into $C'_0$ introduced in~\eqref{Sa12}.
In our specific setup, 
\be
\omega_1 = X_1, \ \ \omega_2 = X_2  \ \   \mbox{ for } S^{(n)}_{A},
\ee
\be
\omega_1 = Y_1, \ \ \omega_2 = Y_2  \ \   \mbox{ for } S^{(n)}_{B},
\ee
hence the relevant two point functions are given by
\be
\begin{split}
  \left\langle\sigma_n(X_1)\tilde{\sigma}_n(X_2)\right\rangle_{2\epsilon}
  &=\left(\f{\pi}{2\epsilon}\right)^{4h_n}\f{\delta^{4h_n}  }{\left| \sinh{\f{\pi (X_1-X_2)}{2 \epsilon}  }\right|^{4h_n}}, \\
  \left\langle \tilde{\sigma}_n (Y_1 )\sigma_n(Y_2 )\right\rangle_{2\epsilon}
  &=\left(\f{\pi}{2\epsilon}\right)^{4h_n}\f{\delta^{4h_n} }{\left|\sinh{\f{\pi (Y_1-Y_2)}{2 \epsilon}}\right|^{4h_n}},\\
\end{split}
\label{eq:TwoPointTwist}
\ee
where we have put in a lattice cutoff $\delta$ to ensure that the two-point
functions at the same point, e.g.
$\left\langle\sigma_n(X_1)\tilde{\sigma}_n(X_1)\right\rangle_{2\epsilon}$, do
not diverge~\cite{Morrison:2012iz}.
From now on, the distances $X_1 - X_2$ etc. and $\epsilon$ will be all measured in unit of $\delta$.  
$S^{(n)}_{A}+S^{(n)}_{B}$ then follows as
\begin{align}
  \label{SA1SA2}
  S^{(n)}_{A}+S^{(n)}_{B}
  &= \f{2}{1-n}
    \log\big[C_0' \big] 
    \nonumber \\
  &\quad +
    \frac{1}{1-n}
    \log
    \left[ 
    \left(\f{\pi}{2\epsilon}\right)^{8h_n}
     \f{ 1 }{\left|\sinh{\f{\pi (X_1-X_2)}{2 \epsilon}}
  \sinh{\f{\pi (Y_1-Y_2)}{2 \epsilon}}\right|^{4h_n}}\right].
\end{align}

\subsection{$S^{(n)}_{A \cup B}$ in 2d CFT}

Let us move on to $S^{(n)}_{A\cup B}$ given by (\ref{sna}).
Expressed in terms of the four-point function of twist and anti-twist operators,
$S^{(n)}_{A\cup B}$ depends not only on the central charge, but also the operator content (i.e. conformal blocks) of the theory.
Using the same conformal map \eqref{conformalmap}, 
the four-point function $\left \langle
  \sigma_n(X_1)\tilde{\sigma}_n(X_2) \tilde{\sigma}_n(Y_1, \tau_1)\sigma_n(Y_2,
  \tau_1)\right \rangle_{2\epsilon}$
can be expressed in terms of the
four point function on the plane as 
\begin{align}
  \label{master formula}
  &
\left \langle\sigma_n(X_1)\tilde{\sigma}_n(X_2) \tilde{\sigma}_n(Y_1,
  \tau_1)\sigma_n(Y_2, \tau_1)\right \rangle_{2\epsilon}
    \nonumber \\
  &=\prod^4_{i=1}\left(\f{dz}{dw}\right)^{h_n}_{z_i}\left(\f{d\bar{z}}{d\bar{w}}\right)^{h_n}_{\bar{z}_i}\left \langle\sigma_n(z_1,\bar{z}_1)\tilde{\sigma}_n(z_2,\bar{z}_2)\sigma_n(z_3,\bar{z}_3)\tilde{\sigma}_n(z_4,\bar{z}_4)\right \rangle_{\text{plane}}
    \nonumber \\
 &= \left(\f{\pi}{\epsilon}\right)^{8h_n}e^{\f{2h_n\pi}{\epsilon}\left(X_1+X_2+Y_1+Y_2\right)} \cdot \left \langle\sigma_n(z_1,\bar{z}_1)\tilde{\sigma}_n(z_2,\bar{z}_2)\sigma_n(z_3,\bar{z}_3)\tilde{\sigma}_n(z_4,\bar{z}_4)\right \rangle_{\text{plane}},
\end{align}
where
$w_1 = X_1$, $w_2 = X_2$, $w_3 = Y_2+i\tau_1$, $w_4 = Y_1+i\tau_1$
and $z_i=e^{\frac{\pi}{\epsilon}w_i}$.
By yet another conformal map,
$\eta(z) = \frac{z_{34}}{z_{31}}\frac{z-z_1}{z-z_4}$,
which sends $z_1 \to 0$,
$z_2 \to x$,
$z_3 \to 1$,
and 
$z_4 \to \infty$, 
and introducing 
\begin{align}
  G_n(x,\bar{x})=\lim\limits_{\eta_4\rightarrow \infty} |\eta_4|^{4h_n}\langle \sigma_n(0) \tilde \sigma_n(x,\bar{x})\sigma_n(1)\tilde \sigma( \eta_4 )
  \rangle,
\end{align}
then our four point function becomes
\begin{equation}
\langle \sigma_n(X_1) \tilde \sigma_n(X_2) \sigma_n(Y_2,\tau_1) \tilde \sigma_n(Y_1,\tau_1) \rangle
=\left(\frac{\pi}{2\epsilon} \right)^{8h_n}
\frac{x^{2h_n}\bar{x}^{2h_n} G_n(x,\bar{x})}
{\left|\sinh\frac{\pi(X_1-X_2)}{2\epsilon}\sinh\frac{\pi(Y_2-Y_1)}{2\epsilon} \right|^{4h_n}},
\end{equation}
where we have absorbed a possible proportionality constant into $C_0$ introduced in~\eqref{TrrhoAn}.
Here, the cross ratios are
\be \label{crossratio}
\begin{split}
&x=\f{\sinh{\left[\f{\pi}{2\epsilon}\left(X_1-X_2\right)\right]}\sinh{\left[\f{\pi}{2\epsilon}\left( Y_2-Y_1 \right)\right]} }{\sinh{\left[\f{\pi}{2\epsilon}\left(X_1-Y_2-i \tau_1\right)\right]}\sinh{\left[\f{\pi}{2\epsilon}\left(X_2-Y_1-i \tau_1\right)\right]} }, \\
&\bar{x}=\f{\sinh{\left[\f{\pi}{2\epsilon}\left(X_1-X_2\right)\right]}\sinh{\left[\f{\pi}{2\epsilon}\left(Y_2-Y_1\right)\right]} }{\sinh{\left[\f{\pi}{2\epsilon}\left(X_1-Y_2+i \tau_1\right)\right]}\sinh{\left[\f{\pi}{2\epsilon}\left(X_2-Y_1+i \tau_1\right)\right]} }. \\
\end{split}
\ee
Then, $S^{(n)}_{A \cup B}$ is given by
\begin{equation}
  S_{A\cup B}^{(n)}=\frac{1}{1-n} \log\left[C_0 \cdot \left(\frac{\pi}{2\epsilon} \right)^{8h_n}
    \frac{x^{2h_n}\bar{x}^{2h_n}G_n(x,\bar{x})}{\left|\sinh\frac{\pi(X_1-X_2)}{2\epsilon}\sinh\frac{\pi(Y_2-Y_1)}{2\epsilon} \right|^{4h_n}}\right].
\end{equation}

Combine this with the previous expression for the operator (R\'enyi) entanglement entropy of $A$ and $B$ separately to get the mutual information
\begin{equation}
I^{(n)}(A,B)=\frac{1}{1-n}\log \left[\frac{ (C'_0)^2/C_0}{x^{2h_n}\bar{x}^{2h_n}G_n(x,\bar{x})} \right].
\end{equation}
In particular, 
the time dependent piece is
\begin{equation}
\tilde I^{(n)}(A,B)(t) = \frac{1}{n-1}\log\left[x^{2h_n}\bar{x}^{2h_n}G_n(x,\bar{x})\right].
\end{equation}

To discuss the evolution as a function of real time $t$,
we need to perform the analytic continuation $\tau_1=-\epsilon-i t$
in the above expressions.
Under the analytic continuation,
the cross ratios become
\begin{align}
\label{cora}
  &x=\f{\sinh{\left[\f{\pi}{2\epsilon} \left(X_1-X_2\right)\right]}\sinh{\left[\f{\pi}{2\epsilon}\left(Y_1-Y_2\right)\right]} }{\cosh{\left[\f{\pi}{2\epsilon}\left(X_1-Y_2-t\right)\right]}\cosh{\left[\f{\pi}{2\epsilon}\left(X_2-Y_1-t\right)\right]}},
  \nonumber \\
&\bar{x}=\f{\sinh{\left[\f{\pi}{2\epsilon} \left(X_1-X_2\right)\right]}\sinh{\left[\f{\pi}{2\epsilon}\left(Y_1-Y_2\right)\right]} }{\cosh{\left[\f{\pi}{2\epsilon}\left(X_1-Y_2+t\right)\right]}\cosh{\left[\f{\pi}{2\epsilon}\left(X_2-Y_1+t\right)\right]}}.
\end{align}
%
%

\section{Bi-partite operator mutual information}
\label{Bi-partite operator mutual information}

Using the twist operator formalism introduced in
the previous section, here we compute BOMI $I^{(n)}(A, B)$
for three different types of CFTs;
the free fermion CFT ($c=1$),
the compactified free boson ($c=2$),
and holographic CFTs ($c \gg 1$).

\subsection{The free fermion CFT}
Let us start with the free fermion CFT. 
To evaluate BOMI
$
I^{(n)} (A,B)
$
for the free fermion theory,
we recall ~\eqref{sna}.
This four-point function of the twist operators,
can be found, e.g., in Eq.\ (2.57) in \cite{Morrison:2012iz},
which, 
with the choice
$u_1 = X_1$, $v_1 = X_2$, $u_2 = Y_1 + i\tau_1$, $v_2 = Y_2 + i\tau_1$,
reads
\begin{align}
&\langle \sigma_n(X_1) \tilde{\sigma}_n(X_2)  \sigma_n(Y_2,\tau_1) \tilde{\sigma}_n(Y_1,\tau_1)\rangle_{2\epsilon} \nonumber\\
&= \left(\frac{\pi^2 
\sinh\frac{\pi(Y_2+i\tau_1-X_1)}{2\epsilon} \sinh\frac{\pi(Y_1+i\tau_1-X_2)}{2\epsilon}}{4\epsilon^2   \sinh\frac{\pi(Y_2-Y_1)}{2\epsilon}\sinh\frac{\pi(Y_2+i\tau_1-X_2)}{2\epsilon} \sinh\frac{\pi(Y_1+i\tau_1-X_1)}{2\epsilon}\sinh\frac{\pi(X_2-X_1)}{2\epsilon} }\right)^{2h_n}  \nonumber\\
& \qquad \times  \left(\frac{\pi^2 
\sinh\frac{\pi(Y_2-i\tau_1-X_1)}{2\epsilon} \sinh\frac{\pi(Y_1-i\tau_1-X_2)}{2\epsilon}}{4\epsilon^2   \sinh\frac{\pi(Y_2-Y_1)}{2\epsilon}\sinh\frac{\pi(Y_2-i\tau_1-X_2)}{2\epsilon} \sinh\frac{\pi(Y_1-i\tau_1-X_1)}{2\epsilon}\sinh\frac{\pi(X_2-X_1)}{2\epsilon} }\right)^{2h_n}, 
\end{align}
After analytic continuation $\tau_1 = -\epsilon-i t$, we arrive at
\begin{align}
&\langle \sigma_n(X_1) \tilde{\sigma}_n(X_2)  \sigma_n(Y_2,  t) \tilde{\sigma}_n(Y_1,  t) \rangle_{2\epsilon} \nonumber\\
&=  \left(\frac{\pi^2 \cosh\frac{\pi(Y_2-X_1+t)}{2\epsilon} \cosh\frac{\pi(Y_1-X_2+t)}{2\epsilon}}{4\epsilon^2   \sinh\frac{\pi(Y_2-Y_1)}{2\epsilon}\cosh\frac{\pi(Y_2-X_2+t)}{2\epsilon} \cosh\frac{\pi(Y_1-X_1+t)}{2\epsilon}\sinh\frac{\pi(X_2-X_1)}{2\epsilon} }\right)^{2h_n} \\ \nonumber
&\qquad \times  \left(\frac{\pi^2 \cosh\frac{\pi(Y_2-X_1-t)}{2\epsilon} \cosh\frac{\pi(Y_1-X_2-t)}{2\epsilon}}{4\epsilon^2   \sinh\frac{\pi(Y_2-Y_1)}{2\epsilon}\cosh\frac{\pi(Y_2-X_2-t)}{2\epsilon} \cosh\frac{\pi(Y_1-X_1-t)}{2\epsilon}\sinh\frac{\pi(X_2-X_1)}{2\epsilon} }\right)^{2h_n}.
\end{align}
The $n$-th R\'enyi entropy for the subregion $A\cup B$ is then given by
\begin{align}
S_{A\cup B}^{(n)} =&\frac{1}{1-n}\log \Bigg[ C_0 \cdot 
\left(\frac{\pi}{2\epsilon} \right)^{8h_n}\frac{1    }{\left[\sinh\frac{\pi(Y_2-Y_1)}{2\epsilon}\sinh\frac{\pi(X_2-X_1)}{2\epsilon} \right]^{4h_n}} \\ \nonumber
&\times \left(\frac{ \cosh\frac{\pi(Y_2-X_1+t)}{2\epsilon} \cosh\frac{\pi(Y_1-X_2+t)}{2\epsilon} \cosh\frac{\pi(Y_2-X_1-t)}{2\epsilon} \cosh\frac{\pi(Y_1-X_2-t)}{2\epsilon} }{ \cosh\frac{\pi(Y_2-X_2+t)}{2\epsilon} \cosh\frac{\pi(Y_1-X_1+t)}{2\epsilon} \cosh\frac{\pi(Y_2-X_2-t)}{2\epsilon} \cosh\frac{\pi(Y_1-X_1-t)}{2\epsilon}} \right)^{2h_n}\Bigg].
\end{align}
Combining with Eq.~\eqref{SA1SA2},
$I^{(n)}(A,B)(t) = S_{A}^{(n)}+S_{B}^{(n)}-S_{A\cup B}^{(n)}$
is given by
\begin{align}
  I^{(n)}(A,B)(t)
   &=
\tilde I^{(n)}(A,B)(t)
     +
     \frac{1}{1-n}\log \Big[ \frac{(C'_0)^2 }{C_0   }   \Big],
\end{align}
where
$\tilde I^{(n)}(A,B)(t)$ is the time dependent part
(see below for its expression). 
The constant term can be fixed as follows. When $A$ and $B$ do not have any
overlap, at $t = 0$, as the inverse temperature $2\epsilon \to 0$
the operator mutual information $I^{(n)}(A,B)(0)$ should approach zero,
since the
correlations between the two Hilbert spaces become extremely short ranged in the
infinite temperature limit \cite{Morrison:2012iz}.
For non-overlapping $A$ and $B$ we have
$\tilde I^{(n)}(A, B)(0) = 0$,
thus $\frac{1}{1-n} \log \Big[  \frac{(C_0')^2 }{C_0  } \Big] = 0$, and we eventually have
\begin{align}
\label{fOMI}
  &I^{(n)}(A,B)(t) = \tilde I^{(n)}(A,B)(t)
    \nonumber\\
  \quad 
&= \frac{c}{12}\frac{n+1}{n}\log \left(\frac{ \cosh\frac{\pi(Y_2-X_1+t)}{2\epsilon} \cosh\frac{\pi(Y_1-X_2+t)}{2\epsilon} \cosh\frac{\pi(Y_2-X_1-t)}{2\epsilon} \cosh\frac{\pi(Y_1-X_2-t)}{2\epsilon} }{ \cosh\frac{\pi(Y_2-X_2+t)}{2\epsilon} \cosh\frac{\pi(Y_1-X_1+t)}{2\epsilon} \cosh\frac{\pi(Y_2-X_2-t)}{2\epsilon} \cosh\frac{\pi(Y_1-X_1-t)}{2\epsilon}} \right).
\end{align}

\subsection{The compactified boson CFTs}
\label{sec: CB}
Here, we compute the operator mutual information
$I^{(2)}(A, B)$
in the $c=2$ free boson theory compactified on $S^1\times S^1$
with radius $R$.
We denote 
\begin{align}
\label{eq:eta}
  \eta = R^2.
\end{align}
When rational, $\eta$ can be written as
\begin{align}
\label{eq:ppprime}
  \eta = \frac{p}{p'},
  \quad
  p, p': \mbox{relatively coprime integers}
\end{align}
whereas when irrational, $\eta$ cannot be written as $p/p'$.
In this notation, the self-dual point corresponds to
$\eta=1$.

To compute $S^{(n)}_{A\cup B}$, we need the four point function for the twist operator in the compactified boson theory. From \cite{2017PhRvD..96d6020C} equation (2.3), we have the following expression for the four-point function
\begin{equation}
  \langle\sigma_n(z_1,\bar{z}_1)\tilde \sigma_n(z_2,\bar{z}_2)\sigma_n(z_3,\bar{z}_3)\tilde \sigma_n(z_4,\bar{z}_4) \rangle = \left|z_{12}z_{34}\right|^{-4h_n}
  \left|1-x \right|^{-4h_n}F_n(x,\bar{x}).
\end{equation}
For $n=2$, $F_2$ is given by
\begin{equation}\label{F2}
F_2(x,\bar{x})=\frac{\Theta(0|T)^2}{f_{1/2}(x)f_{1/2}(\bar{x})}
\end{equation}
where $\Theta(0|T)$ is the Siegel Theta function with modular matrix $T$ which is a function of the modular parameters $\tau$ and $\bar{\tau}$ and 
\begin{align}
  f_{1/2}(x) &= _2F_1\left(\frac{1}{2},\frac{1}{2},1,x\right),
               \quad
f_{1/2}(\bar{x}) = _2F_1\left(\frac{1}{2},\frac{1}{2},1,\bar{x}\right).
\end{align}
The Siegel theta function $\Theta(0|T)$ is given explicitly as (see (3.1) in \cite{2017PhRvD..96d6020C}),
\begin{align}
  \Theta(0|T)
  &= \sum_{\mu, \nu \in \mathbb{Z}}
    \exp
    \left[\frac{\pi i \tau}{2}\left(\nu \sqrt{\eta}+\frac{\mu}{\sqrt{\eta}}
    \right)^2 \right]
    \exp\left
    [-\frac{\pi i \bar{\tau}}{2}\left(\nu \sqrt{\eta}-\frac{\mu}{\sqrt{\eta}}
    \right)^2 \right]
  \nonumber \\
&=  \sum_{\mu, \nu \in \mathbb{Z}} \exp\left[i \pi 
\begin{pmatrix}
\nu &  \mu
\end{pmatrix}
      T
\begin{pmatrix}
\nu \\
\mu
\end{pmatrix}
\right]
\end{align}
where the modular matrix $T$ is
\begin{align}
T&=
\begin{pmatrix}
\eta \frac{\tau-\bar{\tau}}{2} &  \frac{\tau+\bar{\tau}}{2} \\ \nonumber
\frac{\tau+\bar{\tau}}{2} & \frac{1}{\eta} \frac{\tau-\bar{\tau}}{2}
\end{pmatrix},
   \quad
  \tau = i\frac{K(1-x)}{K(x)},
         \quad
         \bar{\tau} = -i\frac{K(1-\bar{x})}{K(\bar{x})}.
\end{align}
Note that after analytic continuation, $x \neq \bar{x}$,
so we have the two modular parameters given by (3.4) of \cite{2017PhRvD..96d6020C}.
This modular matrix has the same form as (147) in \cite{CalabreseNegativity}.
\footnote{
  This expression corresponds to \texttt{SiegelTheta[T,0]}
  in {\it Mathematica}.}

Now we have all the ingredients for $F_2(x,\bar{x})$.
From \eqref{master formula},
\begin{align}
&\langle\sigma_n(X_1)\tilde \sigma_n(X_2)\sigma_n(Y_2,\tau_1)\tilde \sigma_n(Y_1,\tau_1) \rangle  \\ \nonumber
  &= \left(\frac{\pi}{2\epsilon} \right)^{8 h_n}
    \frac{F_n(x,\bar{x})}{\left[\sinh \frac{\pi (X_1-X_2)}{2\epsilon} \sinh \frac{\pi (Y_2-Y_1)}{2\epsilon} \right]^{4h_n}(1-x)^{2h_n}(1-\bar{x})^{2h_n}}.
\end{align}
Combined with the sum of the entropies of $S_{A}^{(n)}+S_{B}^{(n)}$
in \eqref{SA1SA2},
the bi-partite mutual information
$I^{(n)}(A,B)=S_{A}^{(n)}+S_{B}^{(n)}-S_{A \cup B}^{(n)}$
is given by
\begin{align}
  I^{(n)}(A,B)
  &= \frac{1}{1-n}\log \left[ \frac{(C'_0)^2}{C_0}   \right]
    +
    \tilde{I}^{(n)}(A,B)(t) 
\end{align}
where the time dependent piece is 
\begin{equation}
  \tilde{I}^{(n)}(A,B)(t) =\frac{1}{1-n}
  \log\frac{(1-x)^{2h_n}(1-\bar{x})^{2h_n}}{F_n(x,\bar{x})}
\end{equation}
with the analytically continued cross ratios given in \eqref{cora}.
For $n=c=2$,
the conformal dimension of the twist operator
is $h_2
={1}/{8}$, and the 2nd R\'enyi  mutual information is
\begin{equation}\label{CBI2}
  \tilde{I}^{(2)}(A,B)(t) =
  \log \frac{F_2(x,\bar{x})}{(1-x)^{1/4}(1-\bar{x})^{1/4}}.
\end{equation}
To fix the time-independent piece of BOMI, look at the rightmost panel of Fig.\ \ref{BOMI fb}
which corresponds to the case of non-overlapping subregions.
For set-ups with non-overlapping $A$ and $B$, 
we see that $\tilde{I}^{(2)}(A,B)(0)= 0$.
As we noted in the previous subsection, when $\epsilon \rightarrow 0$, $I^{(2)}(A,B)(0)$ must vanish, so we again conclude that the time-independent term $\frac{1}{1-n}\log \left[ \frac{(C'_0)^2}{C_0}   \right]=0$.

\subsection{Holographic CFTs}

Using the twist operator formalism,
BOMI for the evolution operator of holographic CFTs
can be computed by using the semiclassical conformal blocks
of the twist operators,
obtained by the monodromy method\textcolor{blue}{,}
following Refs.\ \cite{2013arXiv1303.6955H,2010PhRvD..82l6010H,2013arXiv1303.7221F}.
Alternatively, we can also use the holographic entanglement entropy formula
\cite{2006PhRvL..96r1602R,2006JHEP...08..045R}
following \cite{HM}.
(See Appendix \ref{Holographic computation}.)
Both computations agree, and give the four point function  
\begin{align}
  G_n(x,\bar{x})
  \approx \min \left\{x\bar{x},(1-x)(1-\bar{x}) \right\}^{-\frac{c}{6}(n-1)},
\end{align}
for $n\rightarrow 1$.
This gives us, for the time-dependent part of BOMI, 
\begin{equation}
  \label{Holographic BOMI}
\tilde{I}^{(1)}(A,B)(t) = \frac{c}{6} \log \left[\frac{x \bar{x}}{\min \left\{x\bar{x},(1-x)(1-\bar{x}) \right\}} \right].
\end{equation}
It turns out that BOMI between two disjoint subregions for a holographic CFT vanishes, so we conclude that the time-independent part of BOMI for a holographic CFT vanishes, just like in the previous two examples.
\subsection{Comparison of BOMI for various CFTs}
\label{Comparison of BOMI for various CFTs}

Having computed BOMI for the free fermion theory,
the compactified free boson theory, and holographic CFTs,
we now discuss in more details the behaviors of BOMI in these theories. 

BOMI $I^{(n)}(A,B)$ can be studied for various configurations of output and
input subsystems $A$ and $B$.
As in Fig. \ref{fig:bOpMIchart} (1) - (3),
of our primary interest 
are the following three configurations:
\begin{itemize}
  \item
{\bf Symmetric,} 
  where the middle points of the input and output subsystems spatially coincide with each other 
  while locating on the different time slices.
  \item
{\bf Asymmetric,} 
where the middle points of the input and output subsystems are at different spatial locations; 
  \item
{\bf Disjoint,} where the input subsystem does not have any overlap with the output subsystem.
\end{itemize}

We will contrast the differences between
different CFTs with different levels of information scrambling capabilities.
As we will see, BOMI in 
the free fermion and compactified free boson theories,
albeit some minor differences, shows rather similar behaviors.
We can loosely call the time-evolution operators in
these systems ``integrable channels'',
as their BOMI can be interpreted in terms of the quasi-particle picture.  
On the other hand, the dynamics of
BOMI in holographic CFTs behaves quite differently from integrable channels.
We will call the time-evolution operators in holographic theories 
``holographic channels'' or ``chaotic channels''.

\begin{figure}[t]
  \begin{center}
   \flushleft{Symmetric:}\\
    \begin{tabular}{lc}
   \includegraphics[width=70mm]{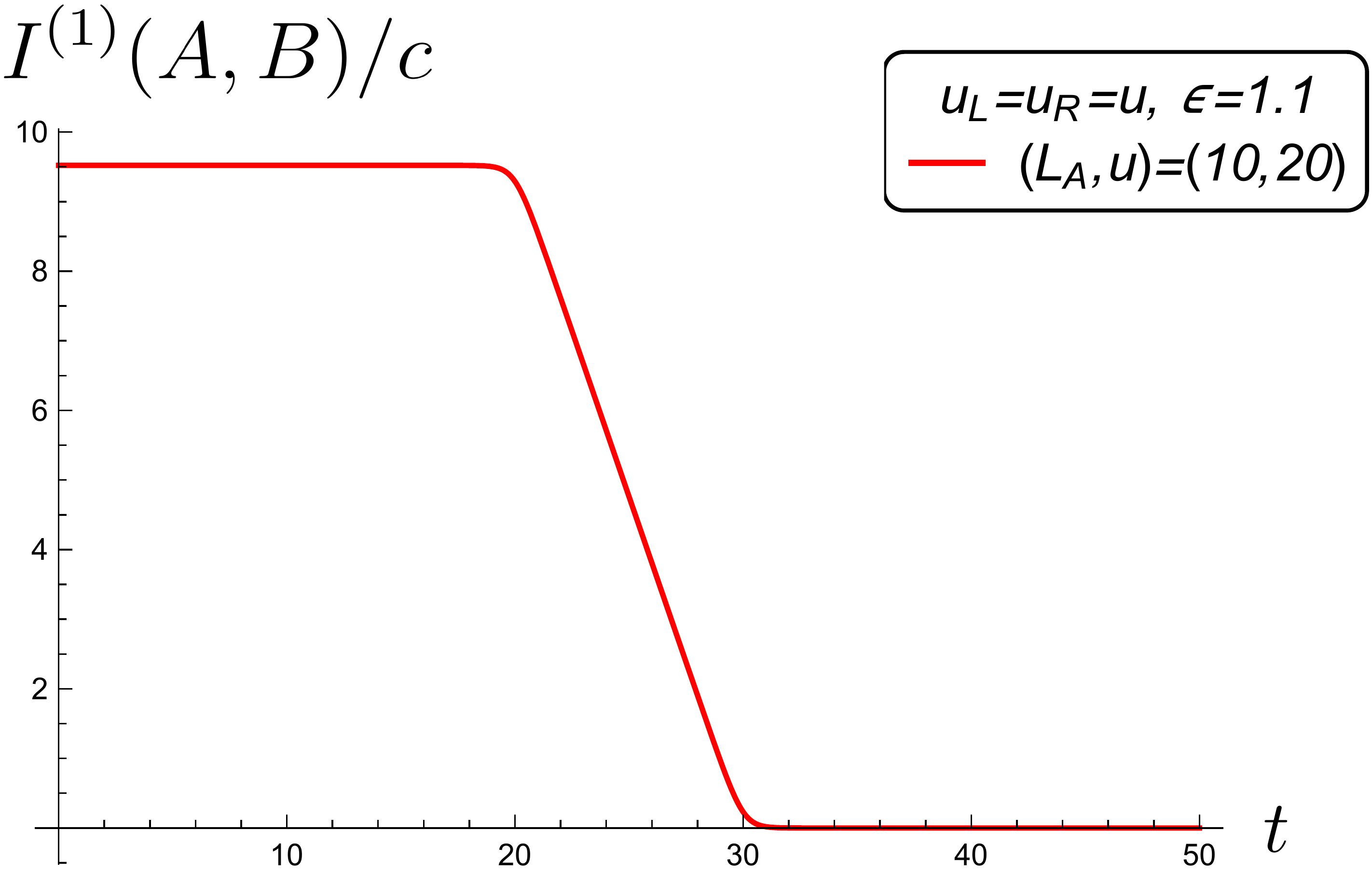}
     &
   \includegraphics[scale=0.3]{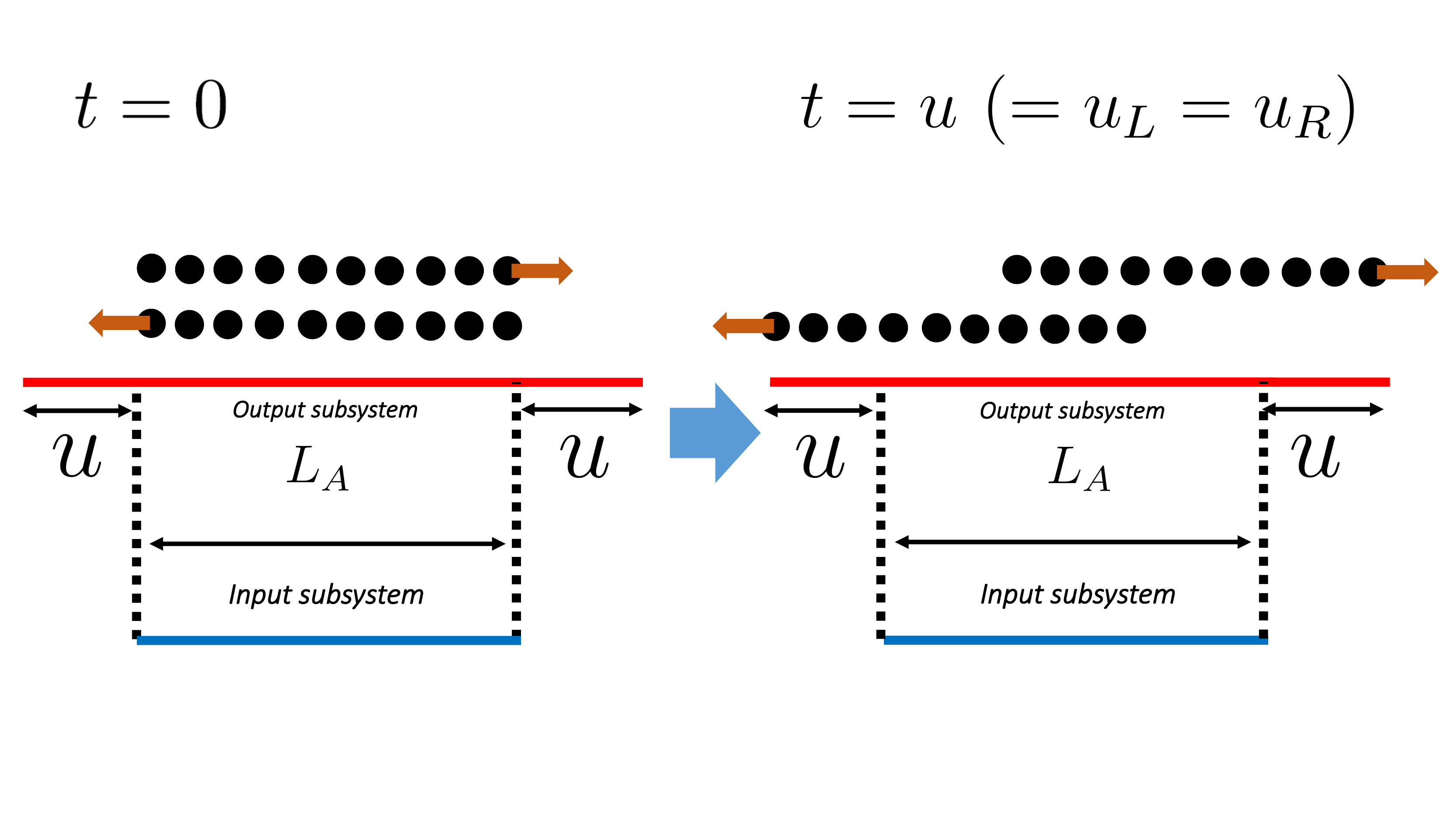}
      \end{tabular}
   \\
   \flushleft{Asymmetric:}\\
   \begin{tabular}{lc}
     \includegraphics[width=70mm]{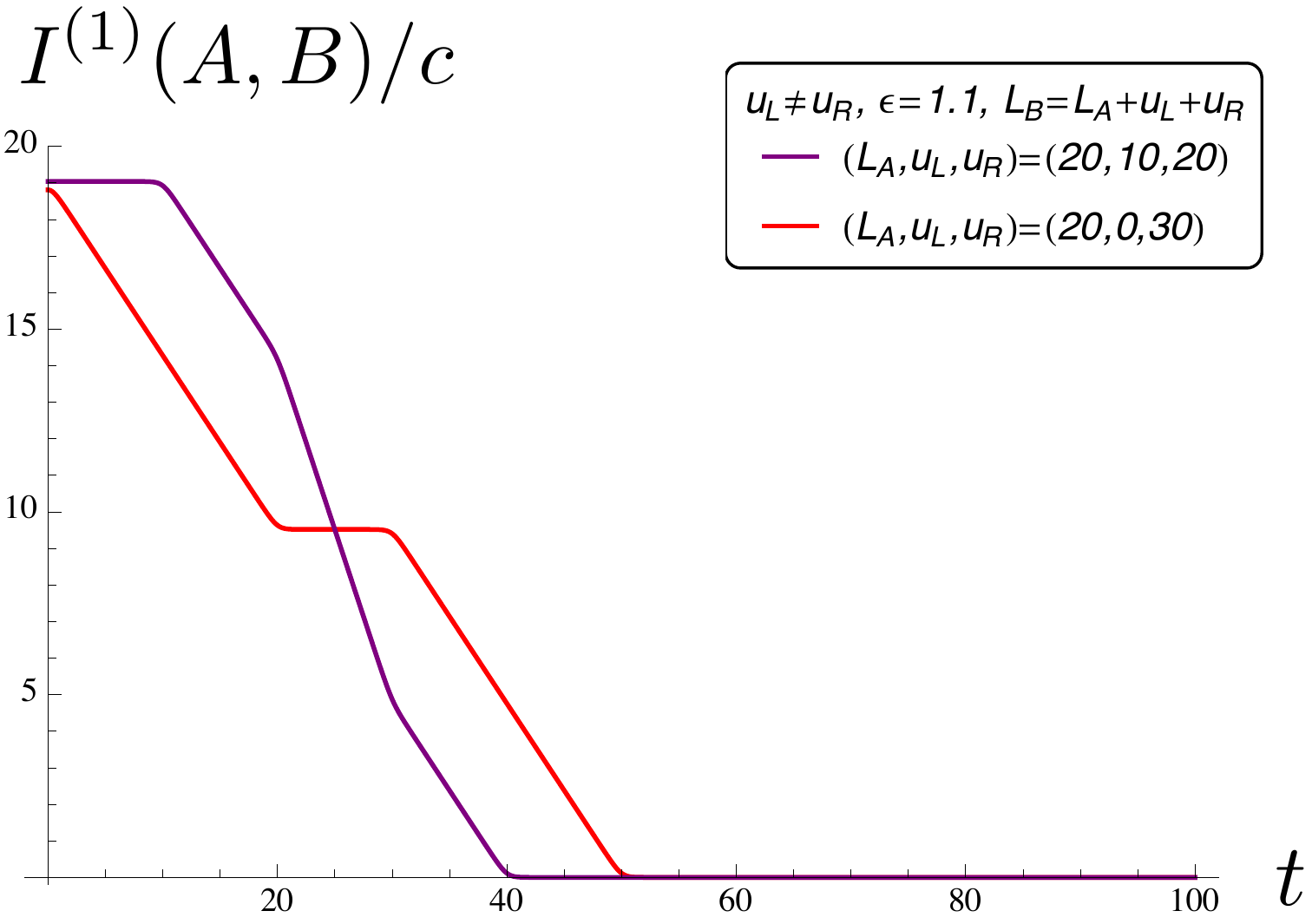}
     &
   \includegraphics[scale=0.3]{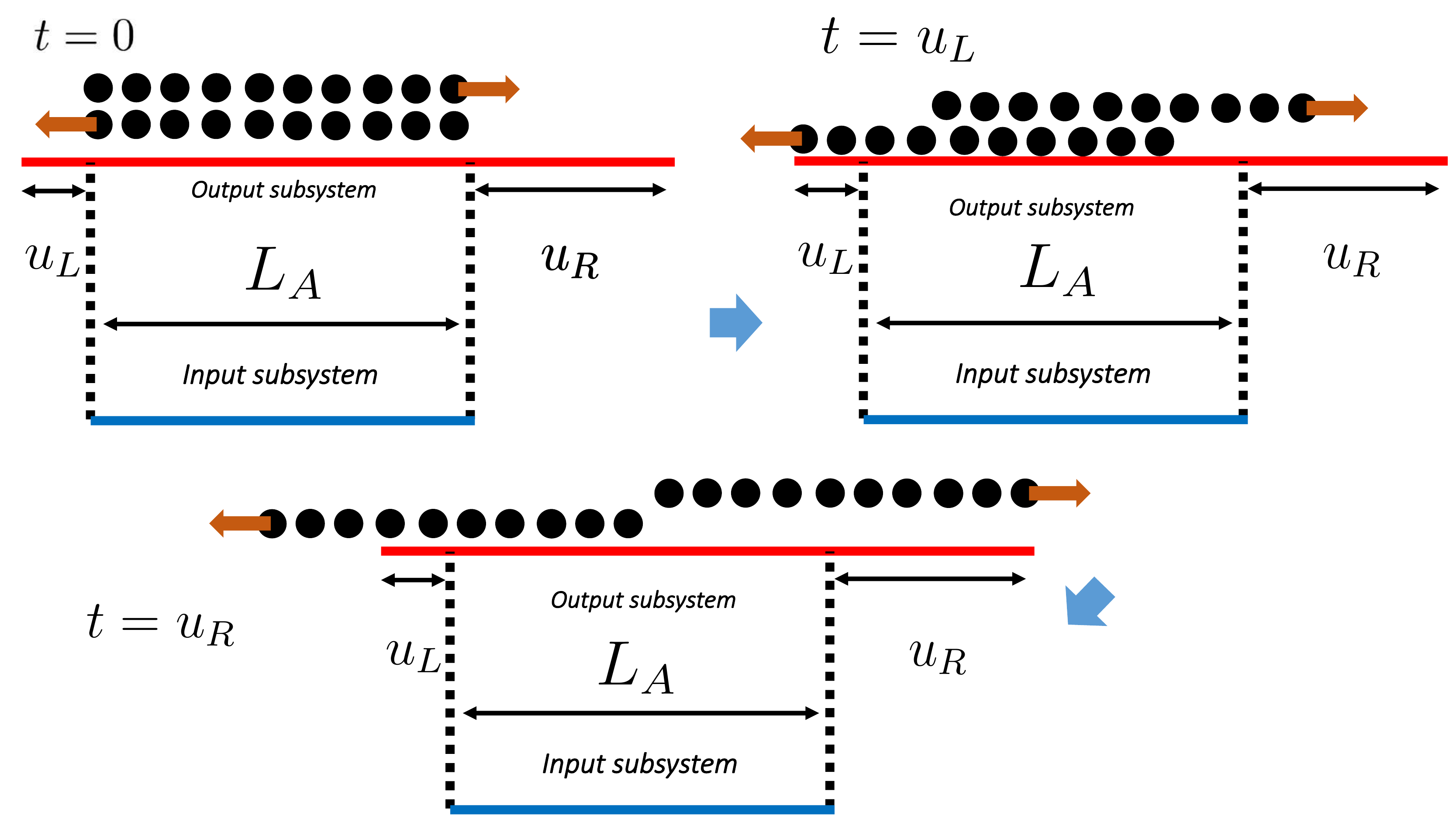}
     \end{tabular}
   \\
   \flushleft{Disjoint:}\\
   \begin{tabular}{lc}
     \includegraphics[width=70mm]{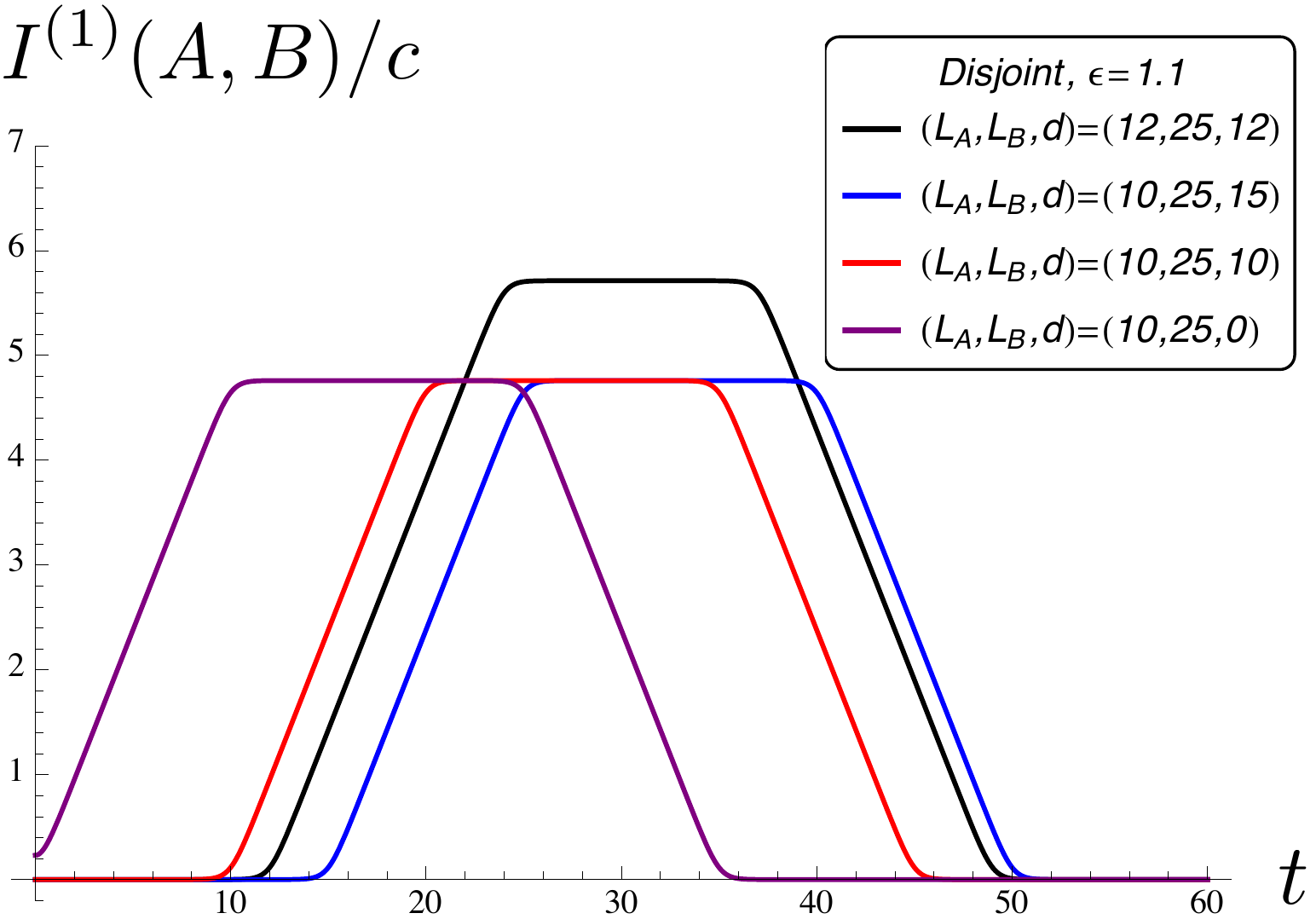}
     &
   \includegraphics[scale=0.3]{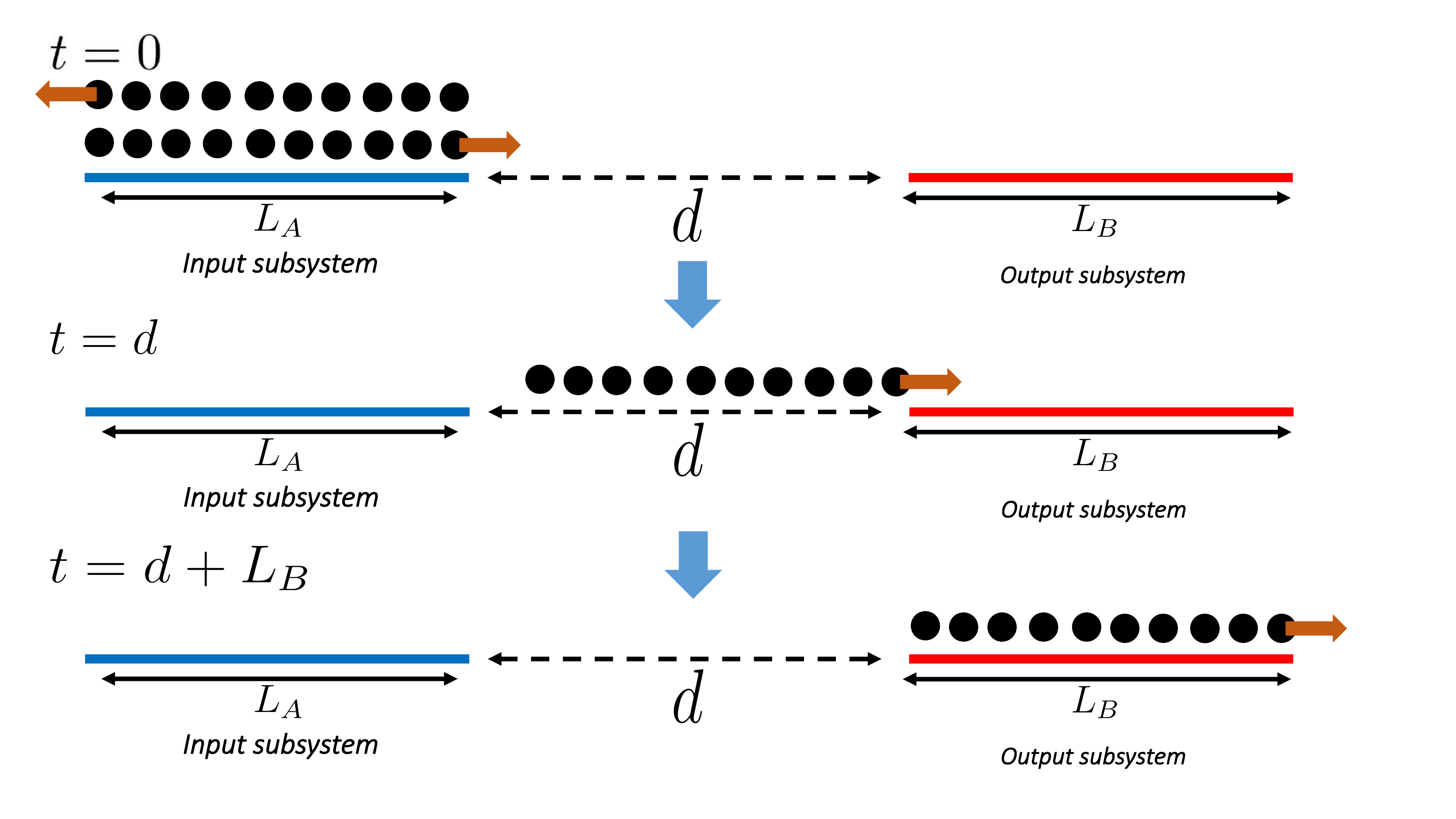}
       \end{tabular}
  \end{center}
  \caption{
    Time evolution of BOMI in the free fermion theory.
    Left column: BOMI for three different configurations:
    symmetric, asymmetric, and disjoint. We choose $\epsilon=1.1$.   The input and output subsystem sizes are $L_A$ and $L_B$, respectively. Parameters of the configurations are listed in the insets.
    Right column: corresponding quasi-particle pictures. Black dots represent the left- and right-moving quasi-particles whose moving directions are indicated by the brown arrows.
    {\bf Symmetric}:
    In this configuration, $L_B=L_A+2u$, where $u_L=u_R=u$.
    {\bf Asymmetric }:
    $L_B=L_A+u_L+u_R$ and $u_L \ne u_R$.
    {\bf Disjoint}:
$A$ is spatially far
     from $B$ and the distance between them is $d$. Note that the red curves in Fig.\ \ref{fig:bOpMIchart} (1a), (2a), (3a) are correspondingly the same as the red curves in the left column above (except that the middle panel uses twice the lengths of the setup of Fig.\ \ref{fig:bOpMIchart} (2a) for a more precise demonstration of the quasi-particle picture). 
  }
  \label{schem1}
\end{figure}

\clearpage

\subsubsection{Integrable channels}

\begin{figure}[t]
 \begin{center}
\begin{tabular}{c}
 \begin{minipage}{0.33\hsize}
   \includegraphics[width=55mm]{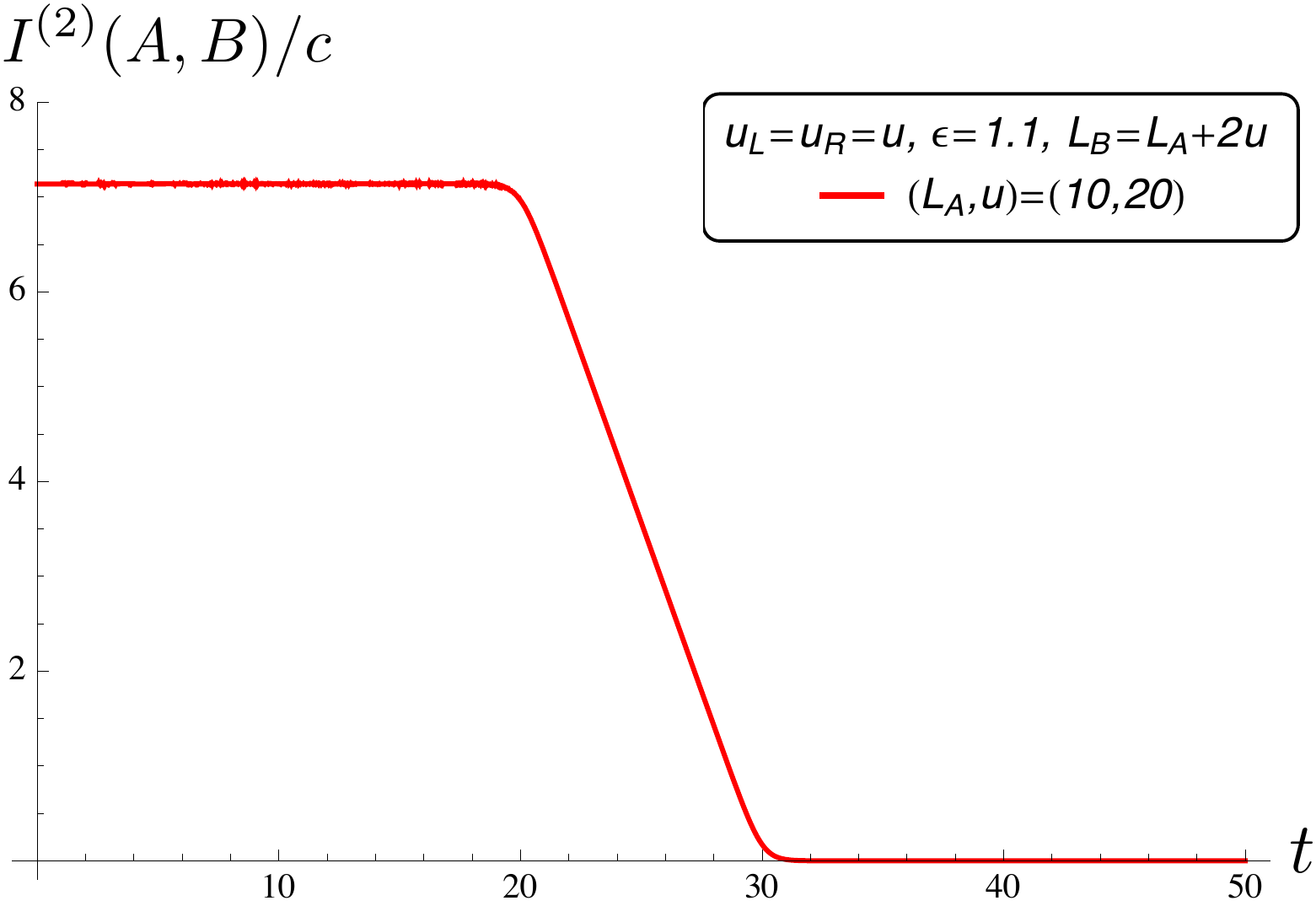}
 \end{minipage} 
 \begin{minipage}{0.33\hsize}
   \includegraphics[width=55mm]{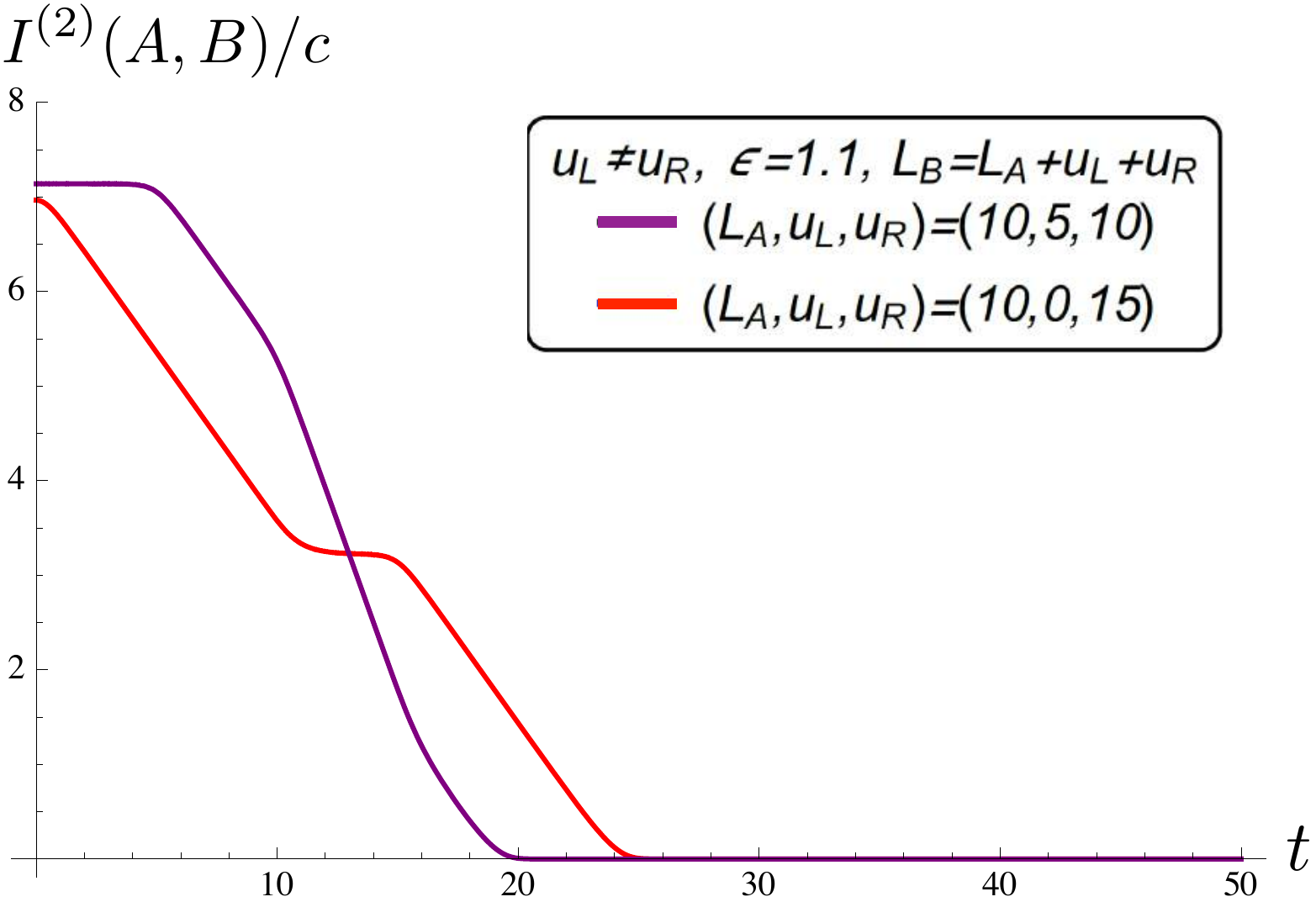}
 \end{minipage} 
 \begin{minipage}{0.33\hsize}
   \includegraphics[width=55mm]{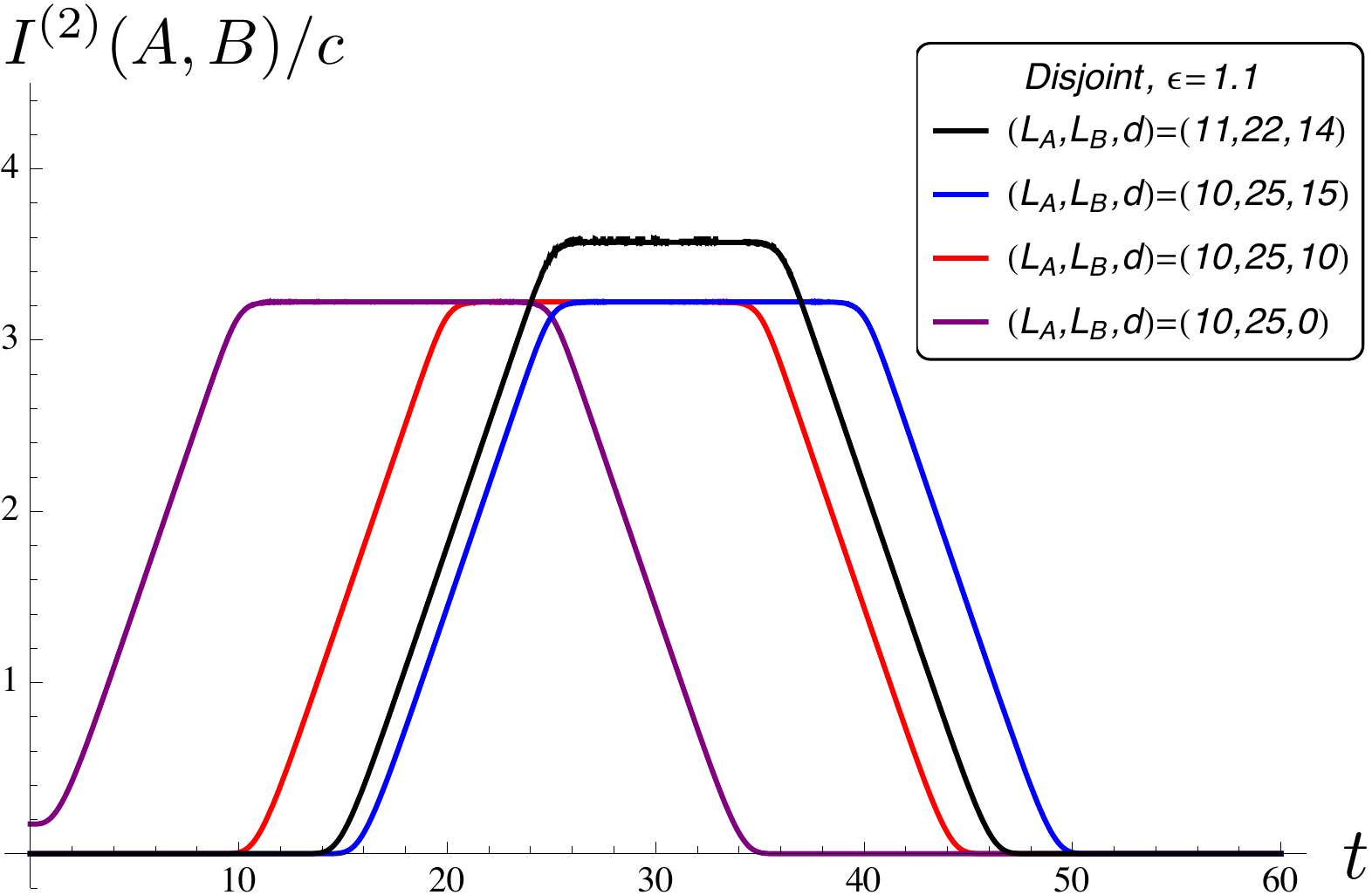}
 \end{minipage} 
  \end{tabular}
  \end{center}
  \caption{
    Time evolution of BOMI in the compactified free boson theory at self-dual radius ($\eta = R^2 = 1$) for
    the symmetric (left), asymmetric (middle), and disjoint (right) configurations.
    We choose $\epsilon = 1.1$. The parameters (shown in insets) of all three configurations are correspondingly the same as the free fermion configurations in Fig.\ \ref{schem1} (except that the middle panel of Fig.\ \ref{schem1} has twice the lengths of the setup in the middle panel here). The input and output subsystem sizes are $L_A$ and $L_B$, respectively.
    {\bf Symmetric}: $L_B = L_A + u_L + u_R = L_A + 2u$.
    {\bf Asymmetric }: $L_B = L_A + u_L + u_R$ where $u_L \ne u_R$.
    {\bf Disjoint}: $A$ does not have any spatial overlap with $B$ and $d$ is the separation between them. Note that the red curves in Fig.\ \ref{fig:bOpMIchart} (1b), (2b), (3b) are correspondingly the same as the red curves above.}
 \label{BOMI fb}
\end{figure}

As seen in Fig.\ \ref{fig:bOpMIchart},
the early- and late-time evolution of BOMI is insensitive to the types of channels.
On the other hand, 
in intermediate time regimes,
BOMI in both the asymmetric and disjoint cases 
depends on the channels.
In the asymmetric setup,
BOMI for the integrable channels 
develops a plateau with a value that depends on the compactification radius in the compactified boson case.
In contrast,  
BOMI for holographic channels  in the asymmetric configuration
does not develops a plateau, but rather drops rapidly.
For the disjoint case,
BOMI for the integrable channels shows bumps,
which is not the case
for holographic channels where correlation between $A$ and $B$ almost vanishes.
See also Figs.\ \ref{schem1} (free fermion)  and \ref{BOMI fb} (compactified boson at self-dual radius)
where we show a more comprehensive collection of data on the time dependence of BOMI
under the three configurations. 

The free fermion and compactified free boson theories
are integrable in the sense that
there are enough number of integrals of motion (conservation laws),
implying that there are states (degrees of freedom)
which retain their identities even during 
an infinitely long time evolution
(i.e., they do not decay during time evolution).
We therefore expect that the behavior of BOMI
can be interpreted in terms of the so-called
quasi-particle picture, 
following Refs.\ \cite{Calabrese2006PRL,2005JSMTE..04..010C}.
In the quasi-particle picture,
pairs of quasi-particles are produced and
propagate ballistically during the time-evolution.
These pairs are entangled and carry quantum information.
This model successfully describes the time evolution of mutual information
\cite{2017PNAS..114.7947A,2018ScPP....4...17A}
and entanglement negativity
\cite{2018arXiv180909119A}
in $2$d CFTs defined when the total Hilbert space is divided geometrically.
The quasi-particle picture can also be extended to
interacting but integrable systems
by making use of integrability -- see, for example,
\cite{2017PNAS..114.7947A,
2018ScPP....4...17A,
2013PhRvL.110y7203C,
2017JSMTE..11.3105A,
2017PhRvB..96k5421A,
2018PhRvB..97x5135A,
2018JSMTE..08.3104M,
2018arXiv180701800A,
2018JPhA...51MLT01B}.

\begin{figure}[htbp]
  \begin{center}
    \includegraphics[clip,width=9.0cm]{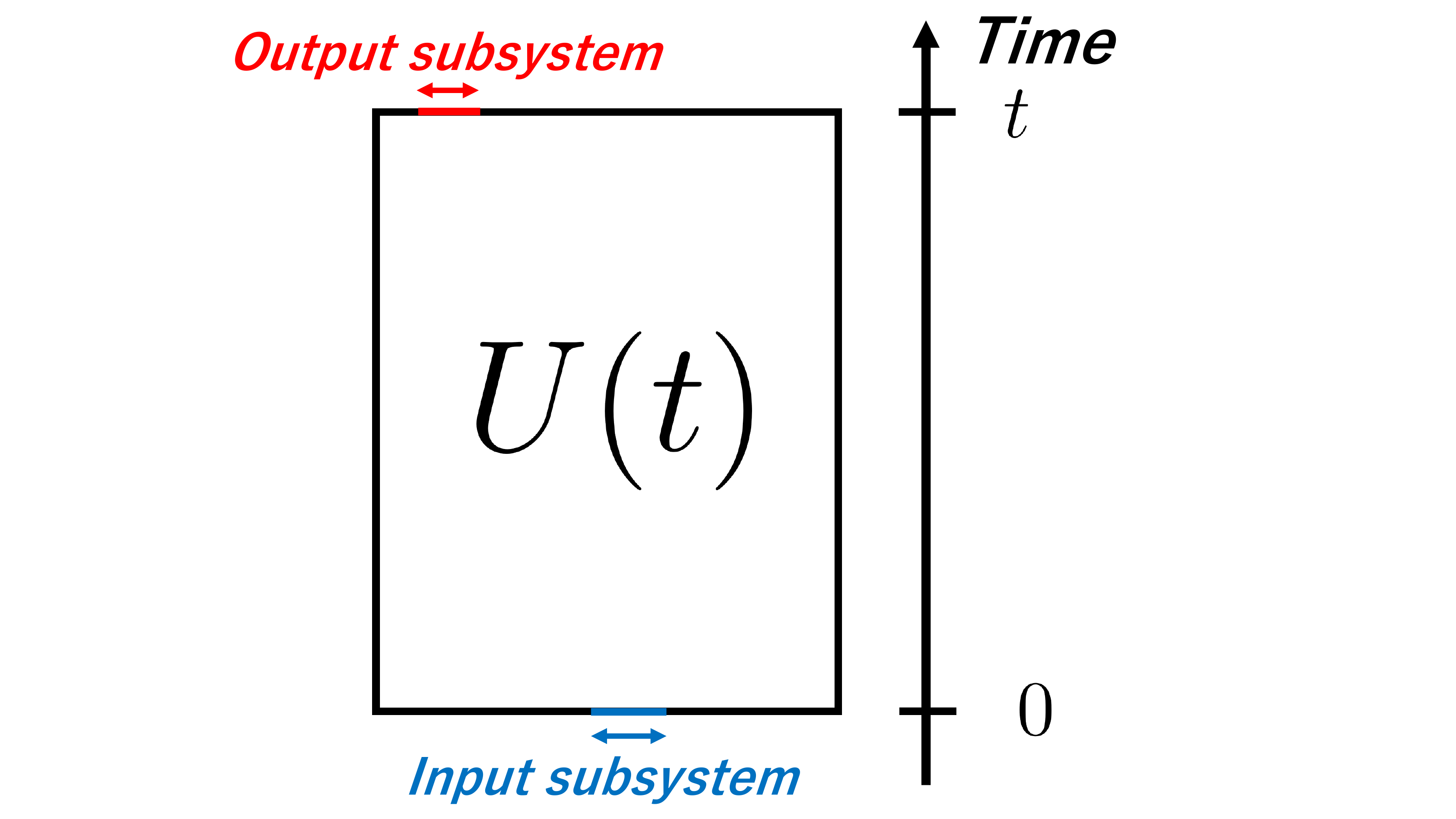}
    \caption{A sketch of unitary time evolution operator, with input and output subsystems located in the two Hilbert spaces respectively. We expect BOMI to
      measure how much information $U(t)$
      sends from the input to output subsystems.  \label{Fig:unitary}}
  \end{center}
\end{figure}

Here, we propose a quasi-particle picture which describes
the time evolution of BOMI for the integrable channels,
which is expected to characterize how much information a unitary time evolution operator $U(t)$ sends from the input subsystem to the target output subsystem as illustrated in Fig.\ \ref{Fig:unitary}.
The key ingredients of our proposed model
can be summarized as follows:
\begin{itemize}
\item The input subsystem is at $t=0$, and the output subsystem is at $t$. Pairs of particles are created everywhere in input subsystem, and move in the left and right direction with the speed of light. 
\item BOMI between input and output subsystems depends on the number of particles included in the output subsystem.
\end{itemize}

This toy model is able to describe the time evolution of BOMI 
for the free fermion channel.
Let us have a closer look at how this picture 
works for free fermion under various configurations.
\begin{itemize}
\item
  Fig. \ref{schem1} Symmetric case: the input subsystem is at the center in the output
  system.
  Here, we assume $L_A<L_B=L_A+2u$, where $L_A,L_B$ are the sizes of the input and output subsystems, respectively (see the quasi-particle picture on the right for definition of $u$).
  Since all particles sent from the input subsystem are included in the output
  system before $t=u$, BOMI maintains a plateau.
  BOMI decreases around $t=u$ because some of the quasi-particles start
  to leave the output subsystem.
  After $t=u+L_A$, all particles are outside
  of the output subsystem, so that BOMI vanishes.
  The plateau and slope depend on the regulator $\epsilon$. 

\item
 Fig. \ref{schem1} Asymmetric case:  the input subsystem is in the output subsystem,
  but is not located at the center.
  Again we assume $L_A<L_B=L_A+u_R+u_L$.
  We also assume $u_R > u_L$ for simplicity.
  BOMI does not change before $t=u_L$, and starts to decrease after $t=u_L$.
  If $u_R>u_L+L_A$ (red curve), BOMI in $u_L+L_A<t<u_R$ shows a plateau. 
  All the right-moving particles completely remain in the output subsystem until $t = u_R$ 
  after which they start to leave the right boundary.
  Since all right-moving particles are outside of the output subsystem at $t=L_A + u_R$, BOMI vanishes.
  If $u_R<u_L+L_A$ (purple curve), particles leave the output subsystem from both boundaries simultaneously during $u_R < t < u_L + L_A$, and BOMI decreases
  twice faster than $u_R - u_L < t<u_R$. 
  For $u_L + L_A< t < u_R + L_A$ all the left-moving quasi-particles have left the left boundary but there are still some of the right-moving ones in the output subsystem, and the slope of BOMI as a function of time $t$ reduces to the same as the one in the period of $u_R - u_L < t < u_R$.

\item
  Fig. \ref{schem1} Disjoint case: the input subsystem is completely outside of the output subsystem, with a distance $d$ between them.
  Since particles sent from the input subsystem are outside of the output subsystem
  before $t=d$, BOMI remains zero (except for the purple curve with $d=0$, where the nonzero value at $t=0$ is due to finite $\epsilon$).
  BOMI increases between $t=d$ and $t =d+L_A $ because the particles start to enter the
  output region.
  Similarly BOMI decreases between $t=d+L_B$ and $d+L_B+L_A$ due to particles leaving the output region. BOMI after $t=d+L_B+L_A$ simply vanishes. 
 \end{itemize}


Plotted in Fig.\ \ref{BOMI fb}
is BOMI for the compactified free boson at self-dual radius
using the same configurations with the free fermion plots.
The case of the self-dual radius can be explained rather nicely by the
quasi-particle picture.
The key difference between free fermions and compactified bosons at self-dual radius is that the graphs of the latter are slightly rounded.
The quasi-particle picture also seem to work well even away from the the
self-dual radius, although the deviation from the ideal quasi-particle picture is more pronounced, as previously shown in Fig.\ \ref{fig:bOpMIchart}.
However, the quasi-particle picture is not capable to explain the radius-dependence of BOMI in the compactified boson channel. 

\begin{figure}[h]
 \begin{center}
\begin{tabular}{c}
 \begin{minipage}{0.33\hsize}
   \includegraphics[width=55mm]{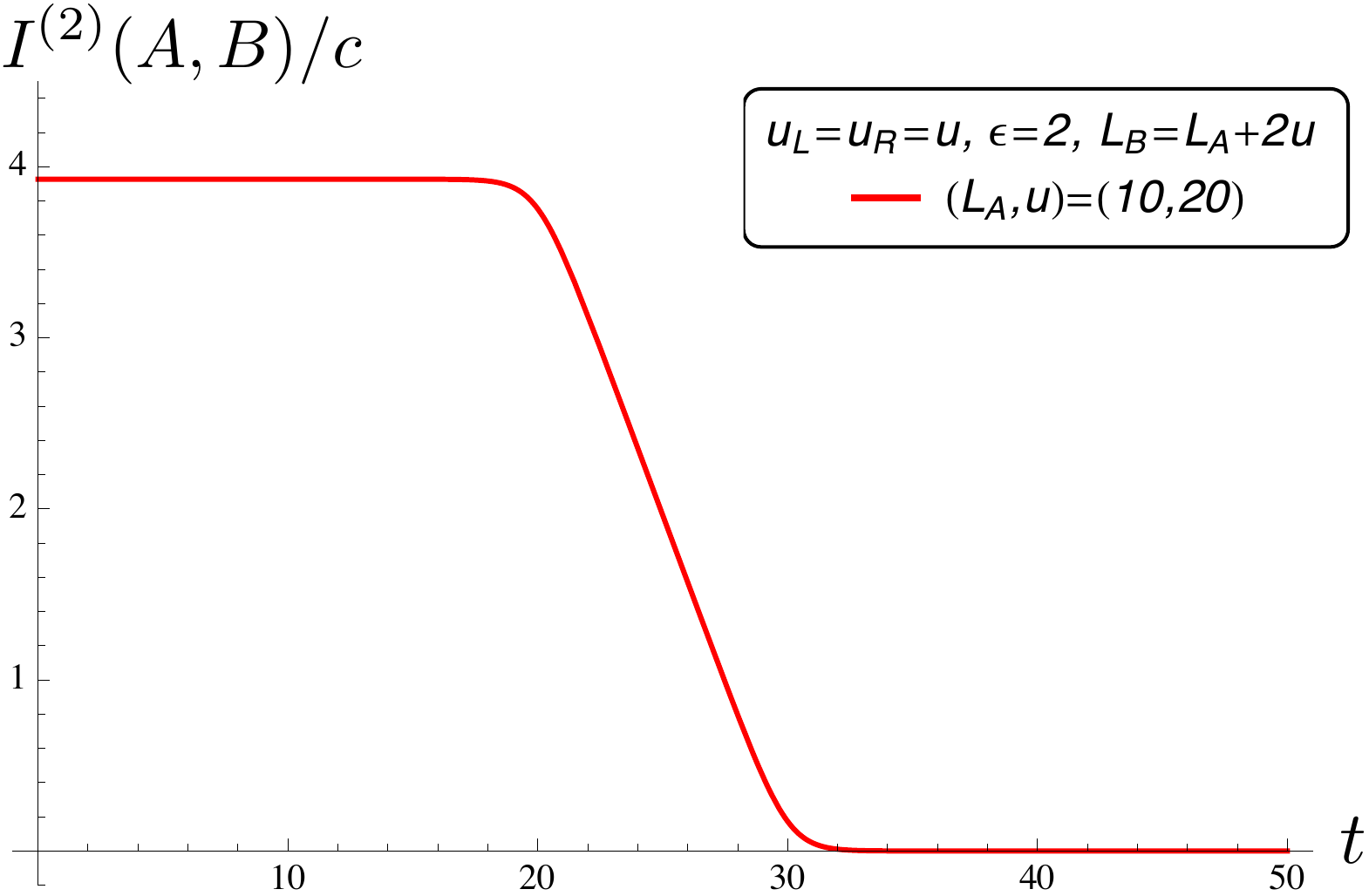}
 \end{minipage} 
 \begin{minipage}{0.33\hsize}
   \includegraphics[width=55mm]{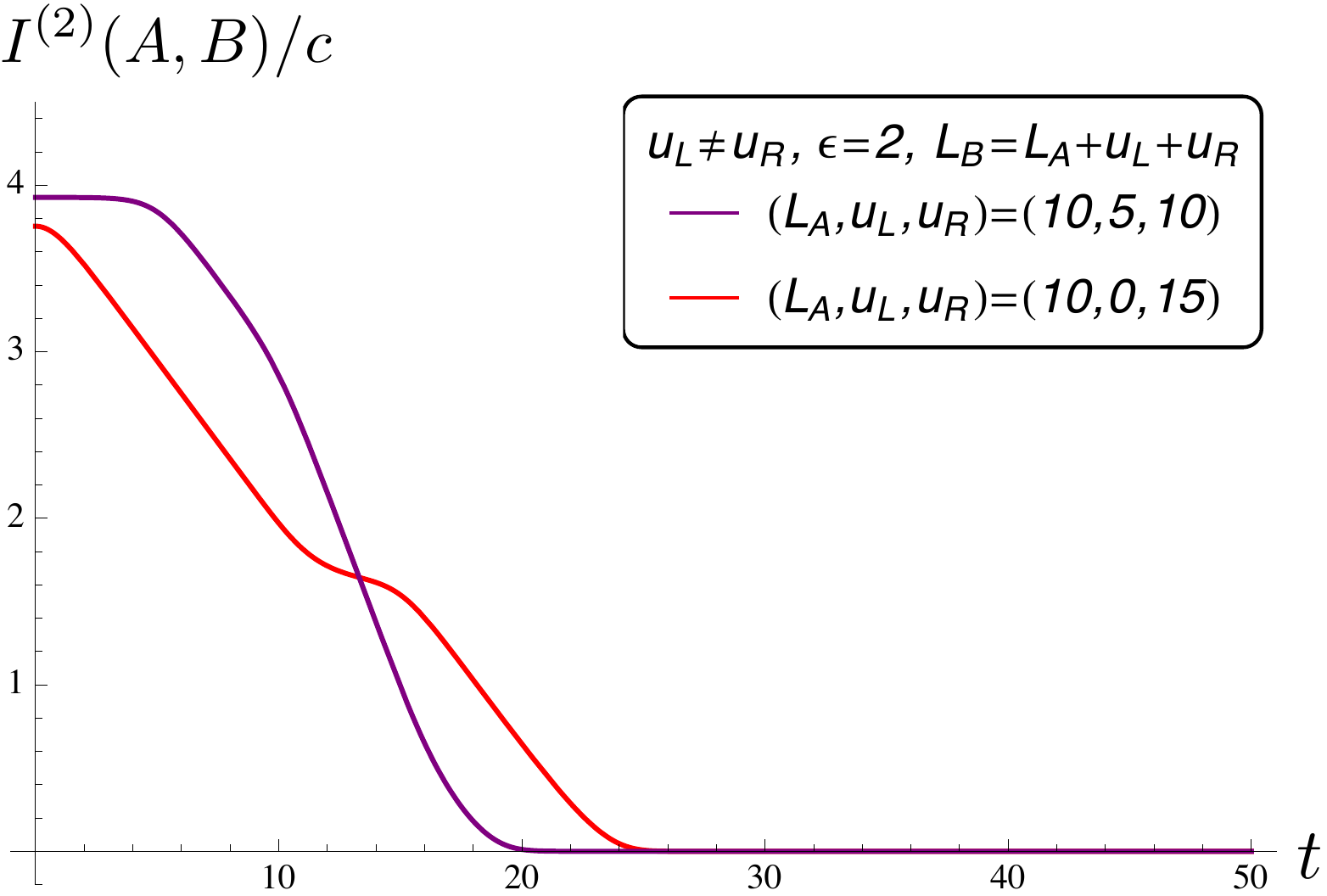}
 \end{minipage} 
 \begin{minipage}{0.33\hsize}
   \includegraphics[width=55mm]{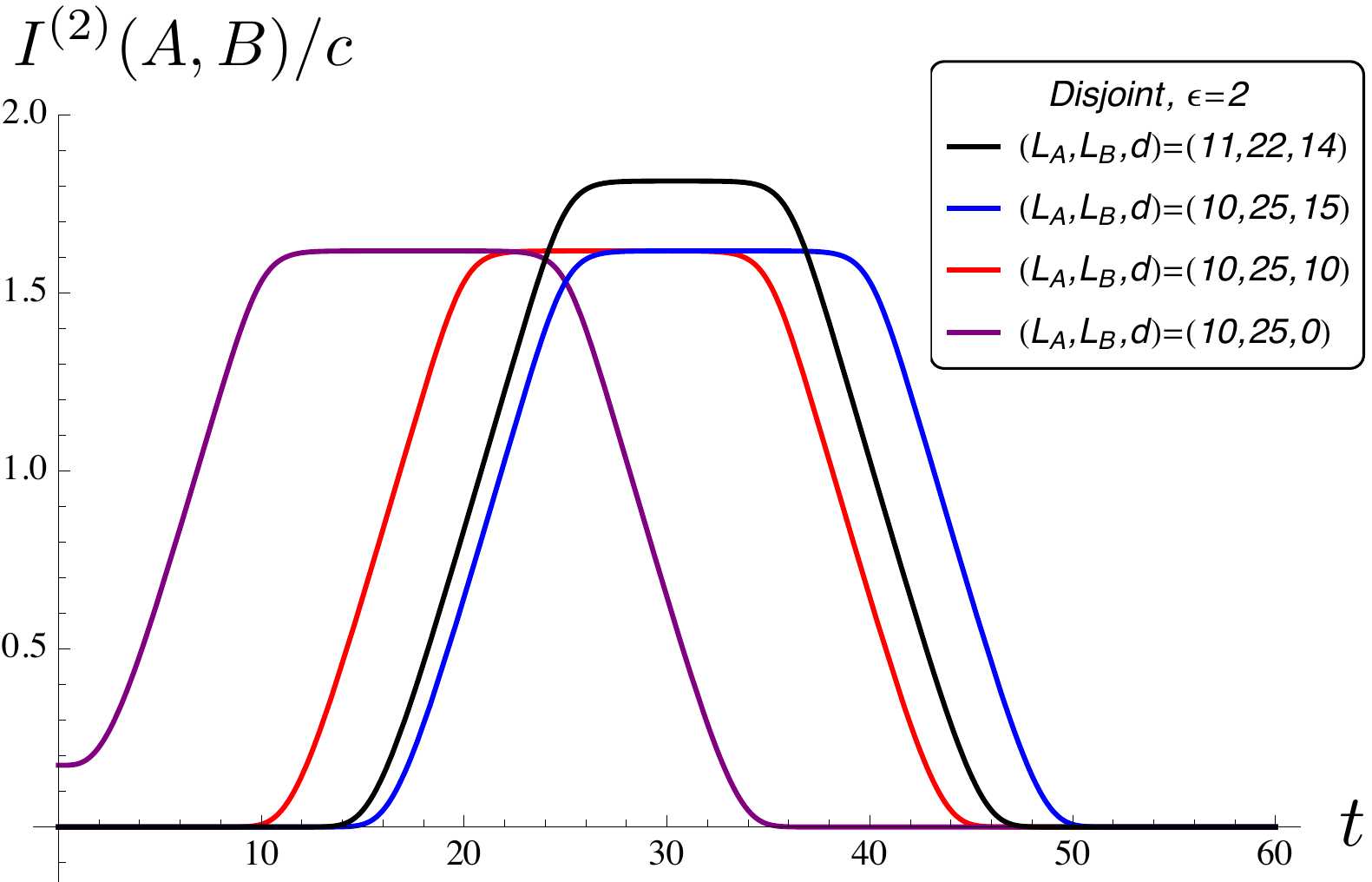}
 \end{minipage} 
  \end{tabular}
  \end{center}
  \caption{
    Time evolution of BOMI in the compactified free boson theory at self-dual radius ($\eta = R^2 = 1$) under the same configurations as Fig. \ref{BOMI fb} above, but with  $\epsilon = 2$. In the disjoint setup (right panel), the non-vanishing BOMI at $t=0$ (purple curve) is due to the relatively large value of $\epsilon$.}
     \label{BOMI fb LargerEpsilon}
\end{figure}

Figure \ref{BOMI fb LargerEpsilon} shows the time evolution of BOMI for compactified boson at self-dual radius under the same configurations as Fig.\ \ref{BOMI fb}, but with larger $\epsilon$. One can think of $\epsilon$ as being the scale related to the size of the quasi-particles,
and when $\epsilon$ is too large compared with length scales of the subsystems,
the time of entry or exit of a quasi-particle into or out of a region is uncertain, hence the more rounded curves in Fig.\ \ref{BOMI fb LargerEpsilon} in comparison with Fig.\ \ref{BOMI fb}.

Finally, it is worth pointing out the invariance of BOMI for the integrable
channels under the transformations,
$L_A  \leftrightarrow  L_B$, or $t \rightarrow -t$, or $X_1\leftrightarrow Y_1$ combined with $X_2 \leftrightarrow Y_2$, which can be readily seen from~\eqref{fOMI} and~\eqref{CBI2}.


\subsubsection{Holographic channels}

\begin{figure}[ht]
  \begin{center}
    \begin{tabular}{cc}
      Holographic: & Integrable: \\
      \includegraphics[scale=0.25]{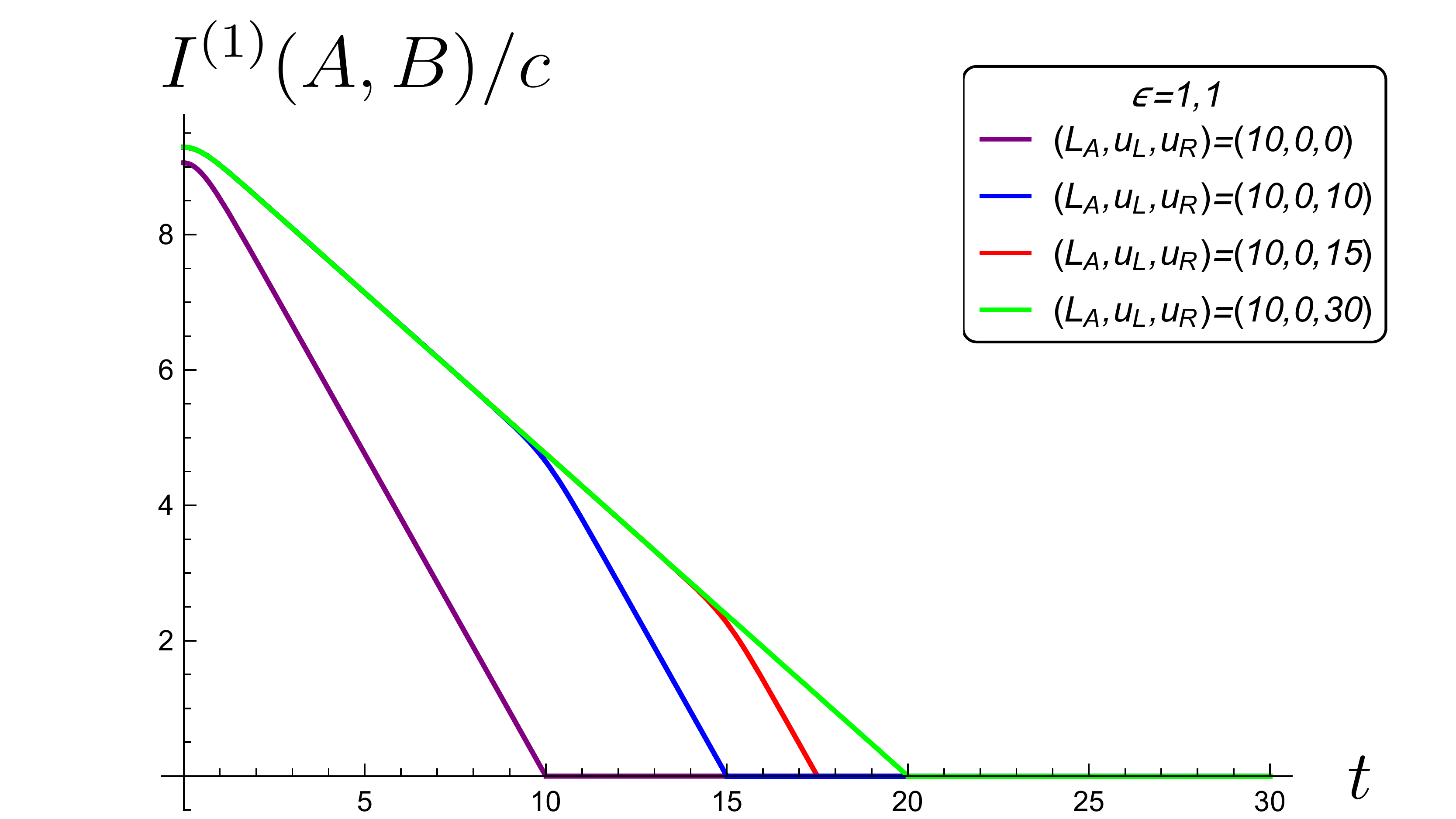}
             &
    \includegraphics[scale=0.25]{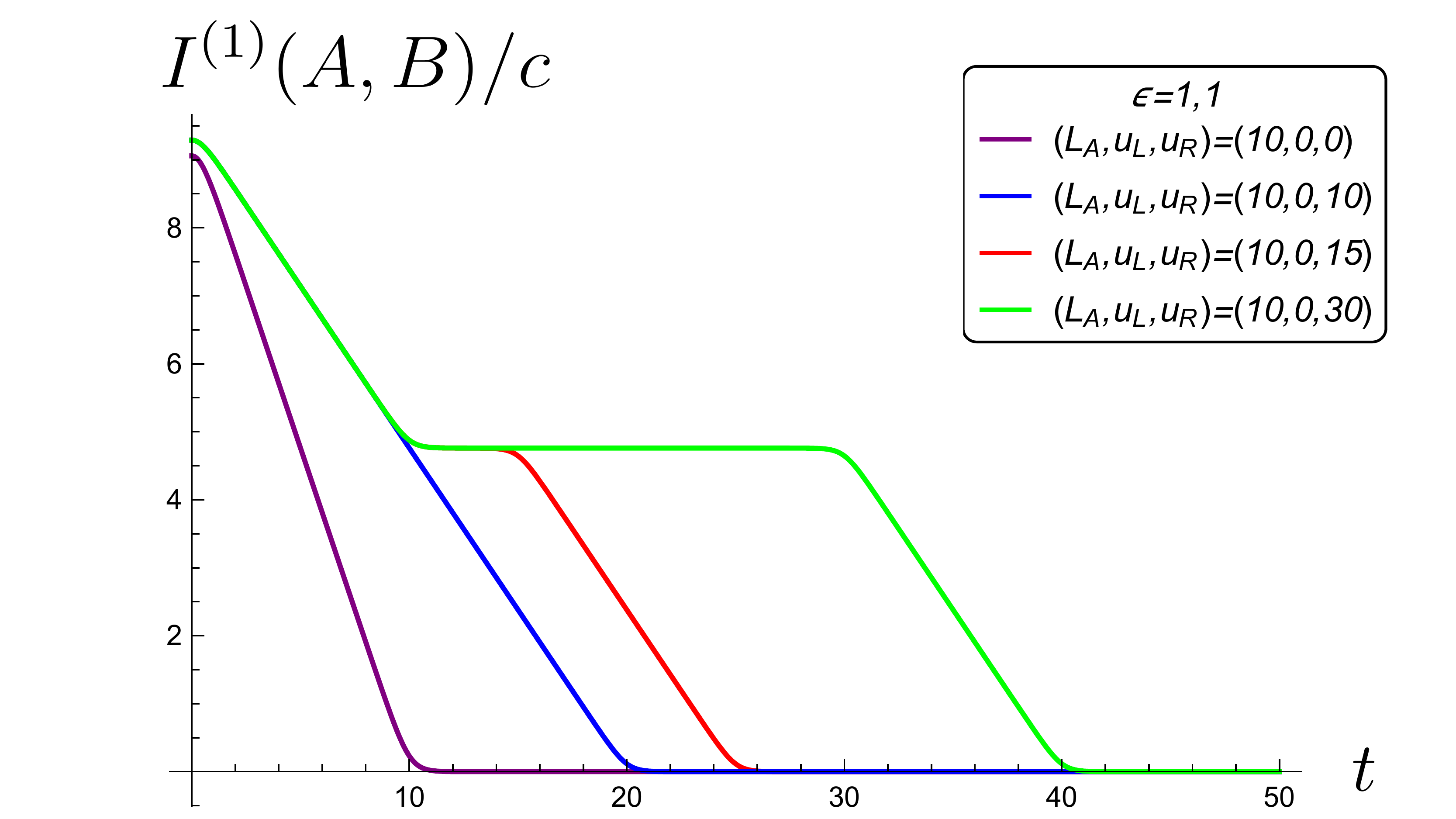}
    \end{tabular}
  \end{center}
  \caption{
    Time evolution of BOMI 
    in 
    holographic (left) and integrable (right, here showing free fermion results as an example) channels
    for asymmetric configurations with $u_L = 0$. 
    Insets list the parameters being used.
	}
	 \label{BOMI chaotic vs integrable perfectly overlapping}
\end{figure}

\begin{figure}[t]
  \begin{center}
         \begin{tabular}{cc}
         Holographic: & Integrable: \\
   \includegraphics[width=78mm]{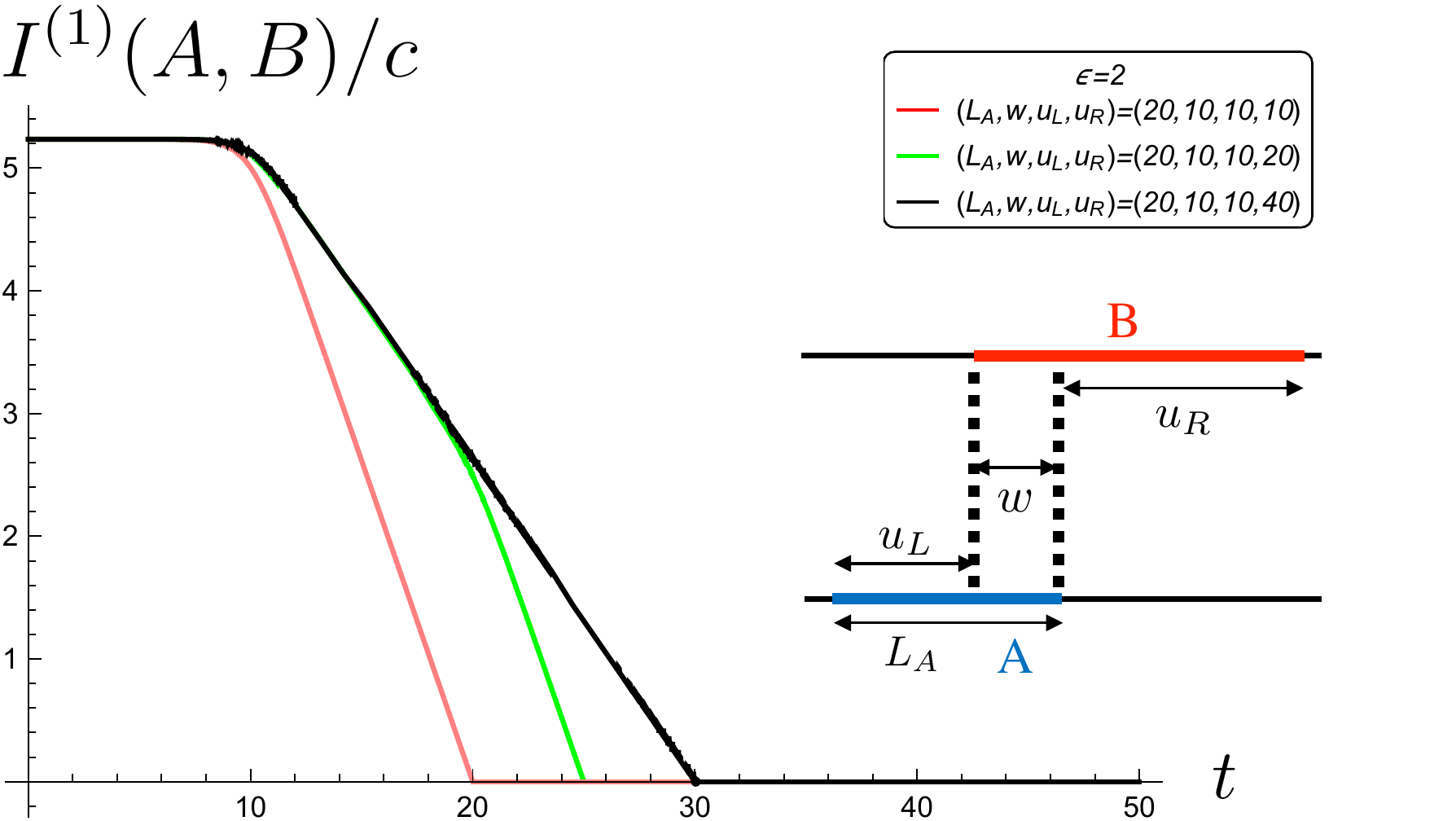}
           &
 \includegraphics[width=69mm]{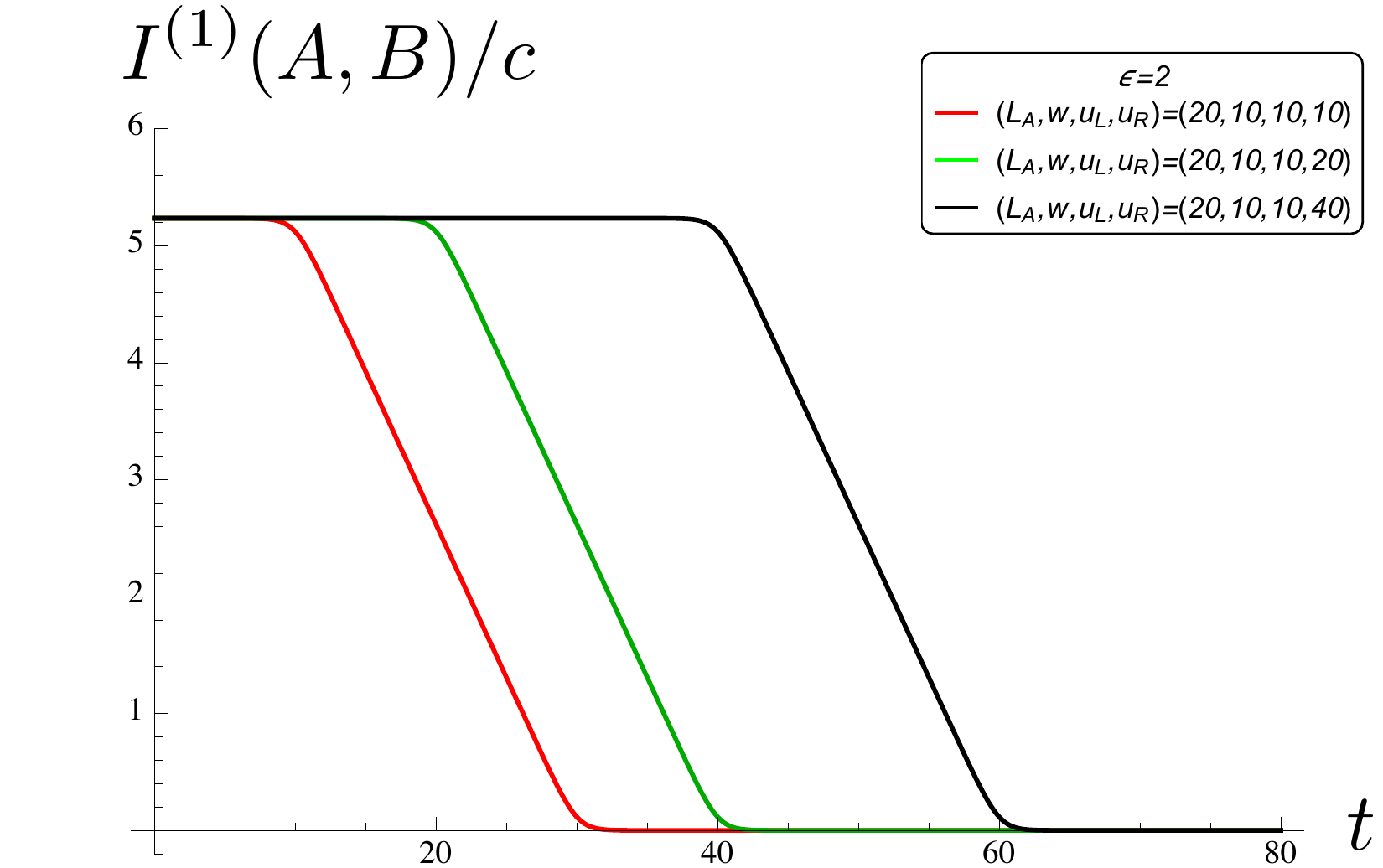}
    \end{tabular}
    \caption{
    \label{BOMI hol overlapping} 
      Time evolution of BOMI in holographic and integrable (free fermion) CFTs when input and output subsystems are generically asymmetric and partially overlapping with $L_A \le L_B =  w + u_R$ (see bottom inset of the left panel for detailed configuration and definitions of the parameters) . 
     {\bf Left}: holographic channel. We set $\epsilon = 2$ for all the curves due to numerical instability induced by $\epsilon = 1.1$. 
      {\bf Right}: free fermion channel, with parameters same as the holographic case as shown in the inset.
}
  \end{center}
\end{figure}

Let us now move on to holographic channels.
As a simple start we first discuss the case of asymmetric input and output subsystems with $u_L = 0$
(Fig.\ \ref{BOMI chaotic vs integrable perfectly overlapping}),
then provide further data on more complicated asymmetric setups
(Fig.\ \ref{BOMI hol overlapping}),
and finally touch upon the case of disjoint input and output
subsystems.

\paragraph{Asymmetric, $u_L = 0$:}

In Fig.\ \ref{BOMI chaotic vs integrable perfectly overlapping},
the time evolution of BOMI for holographic channels is
plotted for four different configurations (Left panel).
For comparison,
BOMI for integral channels (the free fermion theory)
is also plotted for the same configurations (Right panel).
Below we point out their key features and contrasts.

\begin{itemize}
\item (Purple curves)
As a special case and the simplest configuration of the (a)symmetric setup, 
  the configuration of perfectly overlapping between input and output subsystems of equal length
  ($u_R=u_L=0$) does not seem to distinguish the holographic and integrable channels, a reminiscent of the left column of Fig.\ \ref{fig:bOpMIchart} (where $u_R=u_L \ne 0$ was used).
  BOMI for both channels decreases almost linearly with time $t$ and vanishes after $t\approx L_A$. 

\item	(Red and blue curves, $2L_A > u_R$)
  The distinctions between
  holographic and integrable channels are pronounced for asymmetric configurations
  with $u_L = 0$ and $u_R \neq 0$.
  For example, 
  while for integrable channels BOMI decreases with (almost) constant slope, 
  for holographic channels
  BOMI at $t < u_R$ decreases twice as fast as for $t > u_R$. Furthermore, the red curve in the holographic channel (same as the red curve in Fig.\ \ref{fig:bOpMIchart} (2c)) lacks the plateau feature compared with the one in integrable channel (same as the red curve in Fig.\ \ref{fig:bOpMIchart} (2a)).
  
\item (Green curves, $2L_A  \le u_R$)
  For yet another case of asymmetric configuration represented by
  green curves, 
  BOMI for holographic channels decreases
  with constant slope, and vanishes at $t\simeq 2L_A$. 
  On the other hand, for integrable case,
  BOMI develops a plateau for $L_A<t<u_R$,
  monotonically decreases after $t=u_R$,
  and finally vanishes around $t = L_A + u_R$. 
\end{itemize}

\paragraph{Asymmetric, generic case:}

Figure \ref{BOMI hol overlapping} 
shows the time evolution of BOMI in holographic and free fermion channels for more general setups of the input and output subsystems. 
As illustrated in the bottom inset of the left panel of Fig.\ \ref{BOMI hol overlapping}, 
here the size of the overlapping region is $w$, the sizes of input and output subsystem are $L_A=w+u_L$ and $L_B=w+u_R$. We also assume $u_L \le u_R$.

\begin{itemize}
\item
  Left panel of  Fig.\ \ref{BOMI hol overlapping} plots the time evolution of BOMI in holographic channels for $L_A \le L_B$.
  All curves exhibit a plateau before $t=u_L$.
  After $t=u_L$, BOMI starts decrease.
  If $u_R\ge 2w+u_L$ (black curve), BOMI will monotonically decrease after the plateau ends, vanishing around $t=u_L+2w$. 
 If  $u_R\le 2w+u_L$ (green curve), the slope for $u_R< 2w+u_L$ changes around $t=u_R$, and BOMI
  starts to decrease twice as fast as $t < u_R$. Furthermore, if $u_L = u_R$, the region with twice the slope, which starts at $t=u_R$, expands to where the plateau ends, which is at $t=u_L$, leading to another monotonically decreasing BOMI (red curve).
\item
Right panel: time evolution of BOMI in free fermion channel for $L_A \le L_B$. Unlike the holographic case above, here all the important features in the plot agree well with quasi-particle picture, including the duration of the plateau, the slope during intermediate time regime, and the places where BOMI vanishes, etc..

\end{itemize}

\paragraph{Disjoint:}
Finally, when the input subsystem does not overlap the output subsystem,
BOMI of holographic channels between the subsystems simply vanishes. It should be mentioned that in~\cite{2016JHEP...02..004H} the authors studied BOMI finite quantum spin chains, and showed that the dynamics of BOMI of the time evolution operator depends on the type of channels, i.e. integrable or chaotic. \footnote{
  In \cite{2016JHEP...02..004H},
  integrable models are defined 
  by their quantum recurrence time
  which is given by a polynomial in the size of total Hilbert space, $n$.
  On the other hand, chaotic models have the recurrence time that is
$\mathcal{O}(e^{e^{n}})$. 
It is unclear how such distinction between
integrable and chaotic behaviors can be extended to 
quantum chaos in CFTs in our setup.
  The quantum chaos in quantum field theories may be defined by
  the exponential decay of OTOCs, which for $2$ dimensional holographic CFTs
  decay exponentially with a Lyapunov exponent $\lambda_L = {2\pi}/{\beta}$
  between $t= \beta$ and $t=\beta \log{c}$, where $\beta$ is an inverse
  temperature. }
In particular, it was found that if the input subsystem is spatially far from the output one, BOMI for the integrable channel shows a finite bump, while BOMI for the chaotic channel does not, consistent with what we have found in the CFT context under the ``Disjoint" configuration.

\vphantom{a}

For all the configurations listed above,
it is clear that the time dependence of BOMI
in holographic channels is significantly different from
integrable channels, and defies interpretations in terms of the quasi-particle picture.
In Sec.\ \ref{Conclusion},
we will speculate on a possible heuristic model
that can capture the dynamics of BOMI in holographic channels. 
It is also interesting to note that the distinction between
the free fermion CFT and compactified free boson theory 
is rather minor in terms of the dynamics of BOMI.
(In both cases, the quasi-particle picture applies to some extent,
although it works best for the free fermion CFT.)
In the next section, we will turn to tri-partite operator mutual information (TOMI).
As we will see, the dynamics of TOMI -- in particular,
its late-time behavior -- seems to serve as
a more quantitative measure for the information scrambling
capabilities of these channels. 

\section{Tri-partite operator  mutual information}
\label{tri-partite operator  mutual information}

In this section,
we will compute and discuss
tri-partite operator mutual information (TOMI)
\begin{align}  \label{tomi}
I^{(n)}(A,B_1,B_2)=I^{(n)}(A,B_1)+I^{(n)}(A,B_2)-I^{(n)}(A, B),
\end{align}
where $B$ is $B_1 \cup B_2$.
Intuitively, the first and second terms measure how much
information an observer in $B_1$ or $B_2$ is able to obtain,
whereas
the last term measures how much information
an observer in $B_1 \cup B_2$ is able to obtain.
If $I(A,B_1,B_2)$ is negative, the information one can get in $B_1\cup B_2$ is more
than the sum of local information one can obtain from $B_1$ and $B_2$.
Thus, the information is partially delocalized. 


As mentioned in the previous section, authors in \cite{2016JHEP...02..004H} have studied the operator mutual information in the integrable and chaotic finite spin chains.
~Besides studying BOMI they also showed that the dynamics of TOMI of the time evolution operator is capable to distinguish integrable and chaotic channels. 
Specifically, TOMI for the chaotic channel is negative, and its late-time saturation value has a larger magnitude than that of the integrable channel, which is similar to what we have found (to be introduced below) in the context of CFTs.

In this section, we will focus on how different $(1+1)d$ CFTs yield different dynamics of TOMI. 
Of particular interest to us we aim to find characteristic properties of holographic CFTs using TOMI as a criterion. 
We will study the time evolution of TOMI in
the following two setups:
\begin{itemize}
\item {\bf $B=$
    Finite output subsystem}: In this setup, $B$ is defined by a subsystem in the
  output system (see Fig.\ \ref{fig:TOMIdetailedsetup} for illustration). 
  In the long time limit the information from $A$ eventually leaves $B$. 

\item {\bf $B=$ Infinite output subsystem}: In this setup, $B$ is defined by the entire
  output system (namely, sending $Y_1 \to \infty$ and $Y_2' \to -\infty$ in Fig.\ \ref{fig:TOMIdetailedsetup}), leading to the consequence that information from $A$ encoded in $B$ is unable to escape $B$ even in the long time limit.

\begin{figure}[h]
  \begin{center}
   \includegraphics[width=60mm]{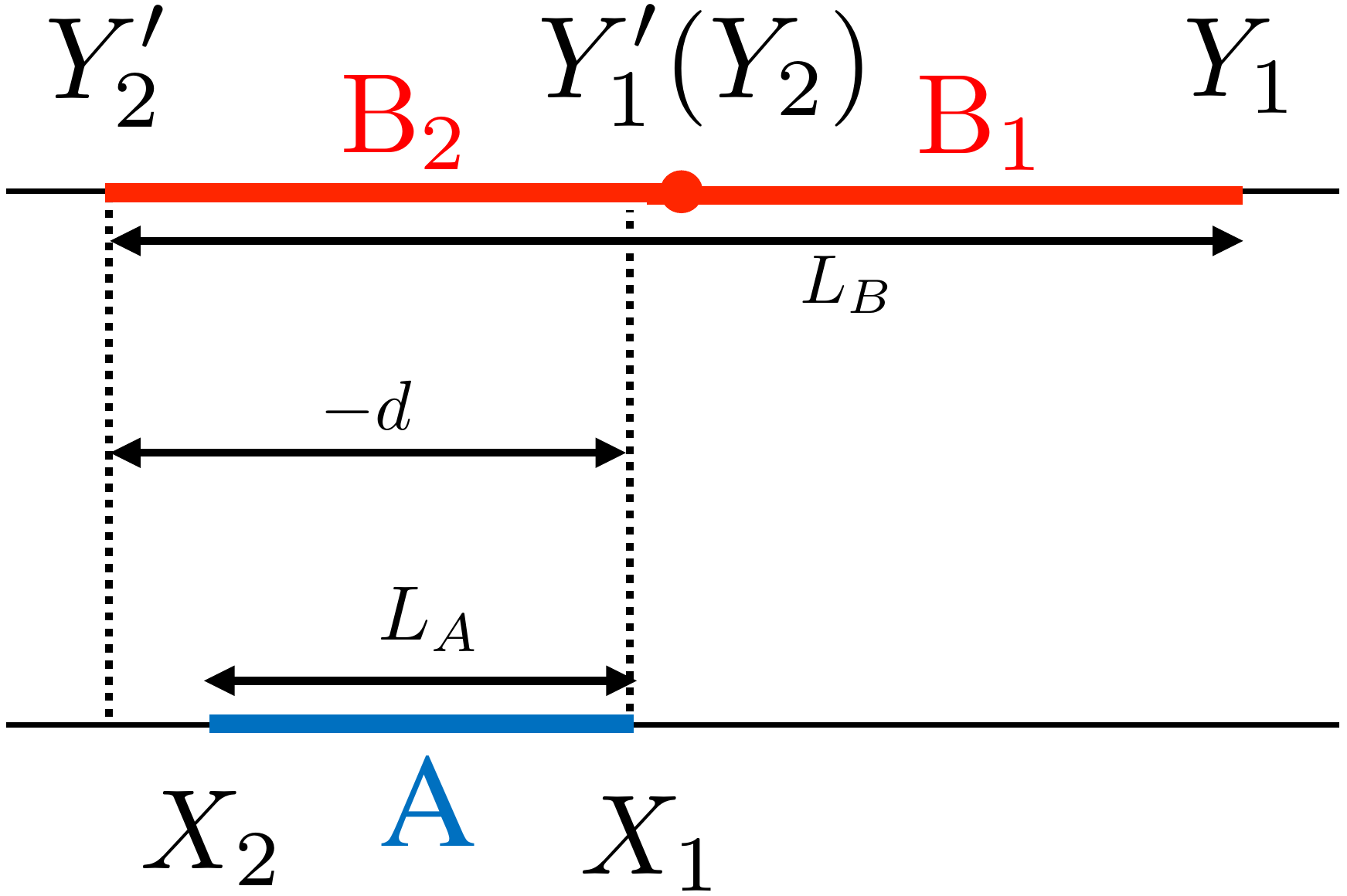}
        \caption{Setup for TOMI, where $A$ is the input subsystem, $B = B_1 \cup B_2$ is the output (sub)system. $X_{1,2}, Y_{1,2}, Y'_{1,2}$ are spatial coordinates of the boundaries of $A,B_1,B_2$. $L_A, L_B$ are the lengths for $A, B$, respectively, and $d$ equals to the position of the leftmost edge of $B_2$ minus the position of the rightmost edge of $A$.
        }
\label{fig:TOMIdetailedsetup}
  \end{center}
\end{figure}

\end{itemize}
\subsection{TOMI for finite output subsystem}

Let us start with TOMI when the output subsystem $B$ is finite.

\subsubsection*{Free fermion channel} 
Using the previously obtained expressions
(\ref{fOMI})
for BOMI, one can show that the free fermion TOMI $I^{(n)}(A,B_1, B_2) = 0$
 for all $t$. 
Once again,
the time evolution of TOMI for the finite output subsystem is interpreted in
terms of quasi-particle model.
In the model, the information sent from $A$ is described in terms of the
relativistic propagation of quasi-particles.
Then, since the number of particles in $B_1$ plus the number in $B_2$ equals to
the number in $B$,
$I^{(n)}(A,B_1,B_2)$ for the free fermion channel vanishes identically for all $t$.

\subsubsection*{Compactified boson channel}

\begin{figure}[t]
  \begin{center}
   \includegraphics[width=50mm]{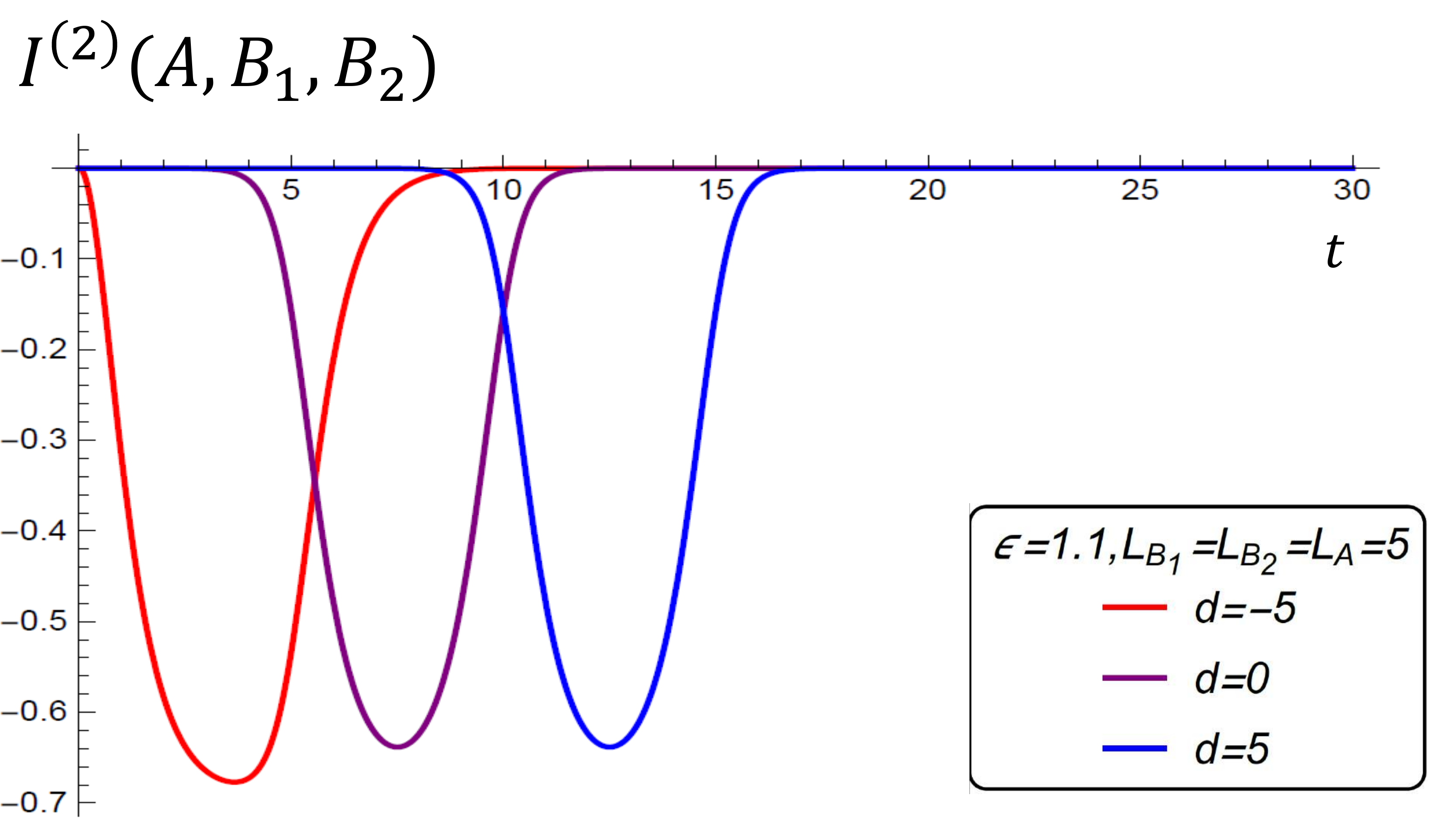}
   \includegraphics[width=50mm]{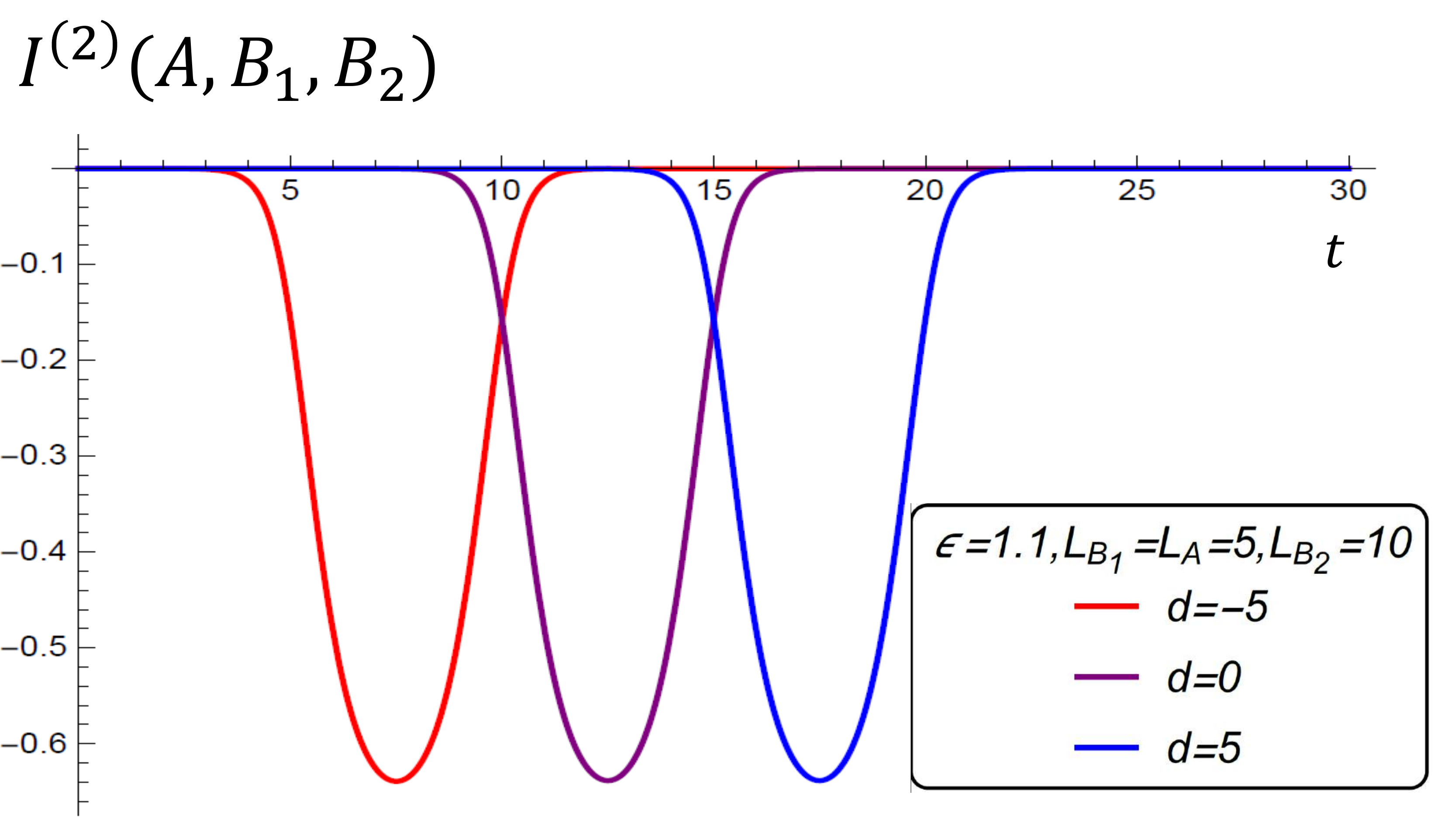}
   \includegraphics[width=50mm]{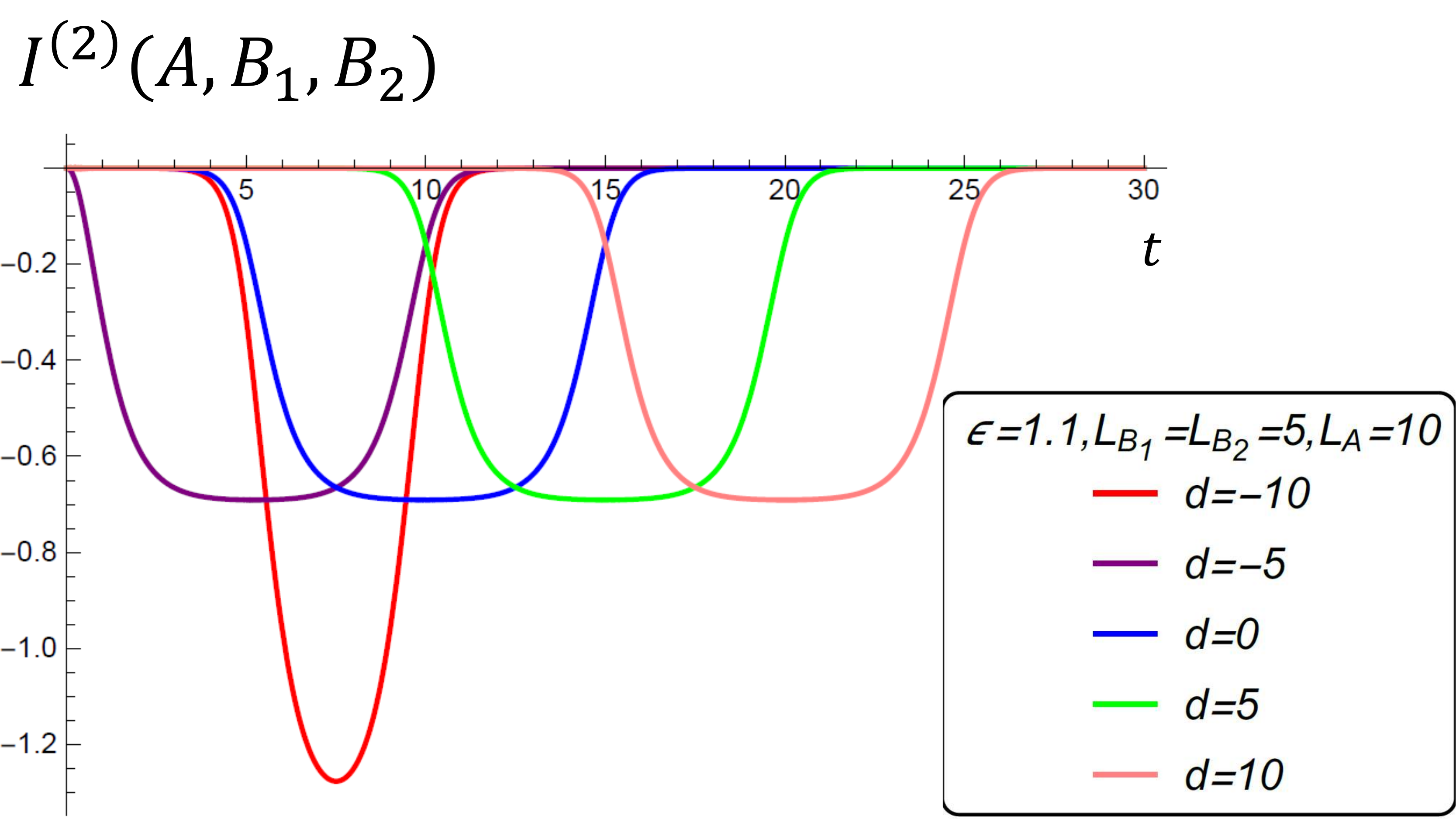}
  \end{center}
  \begin{center}
   \includegraphics[width=50mm]{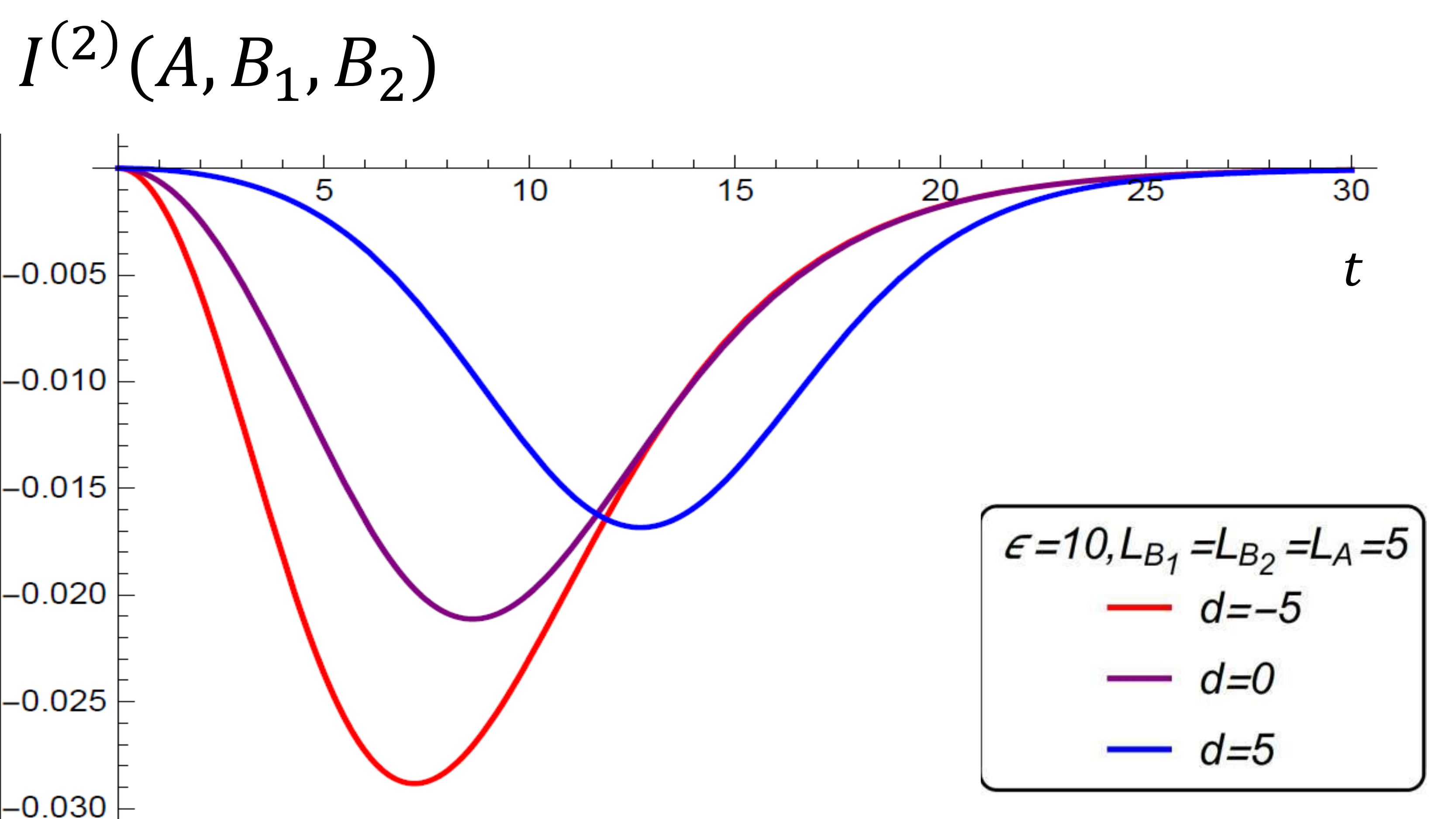}
   \includegraphics[width=50mm]{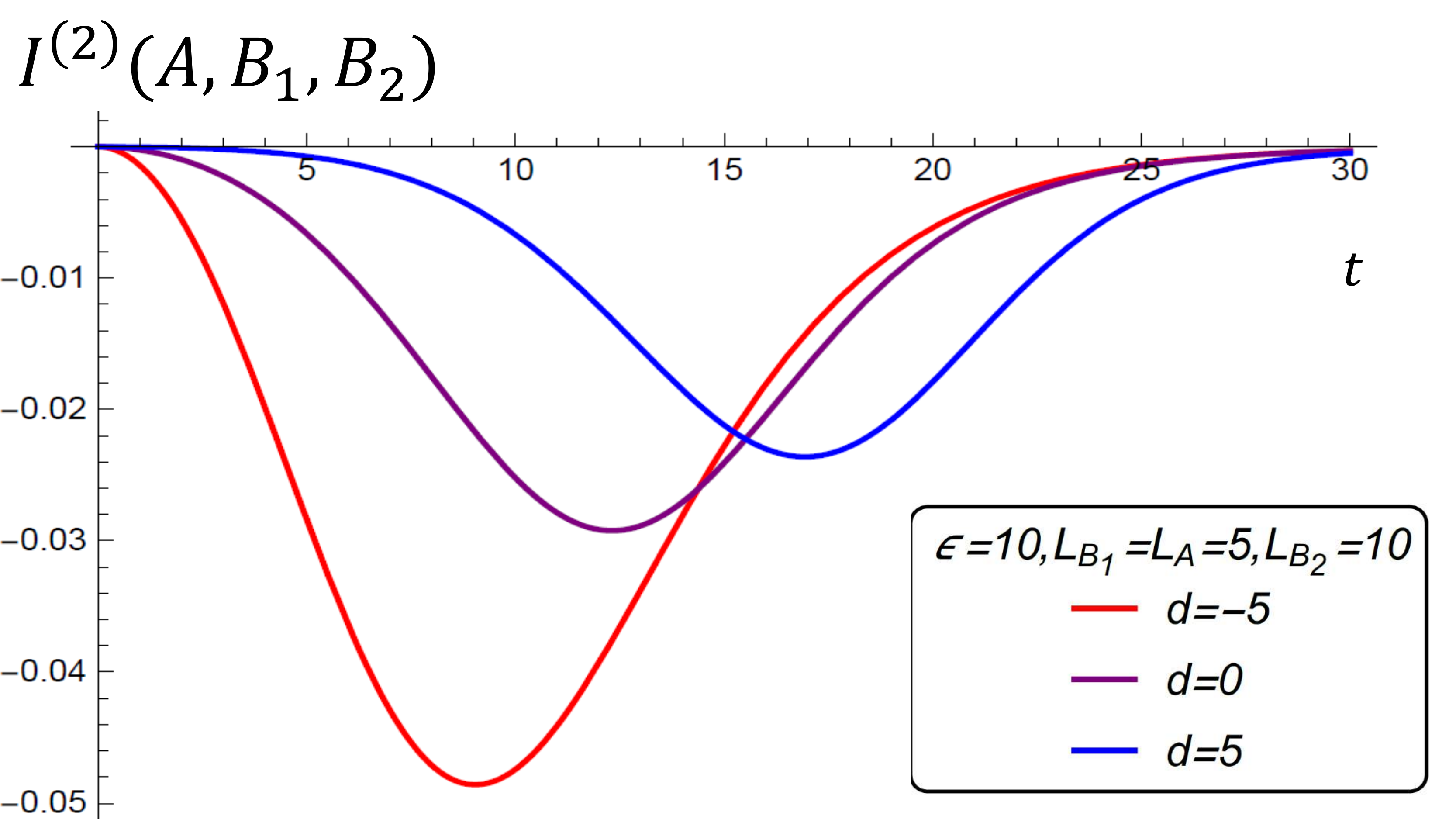}
   \includegraphics[width=50mm]{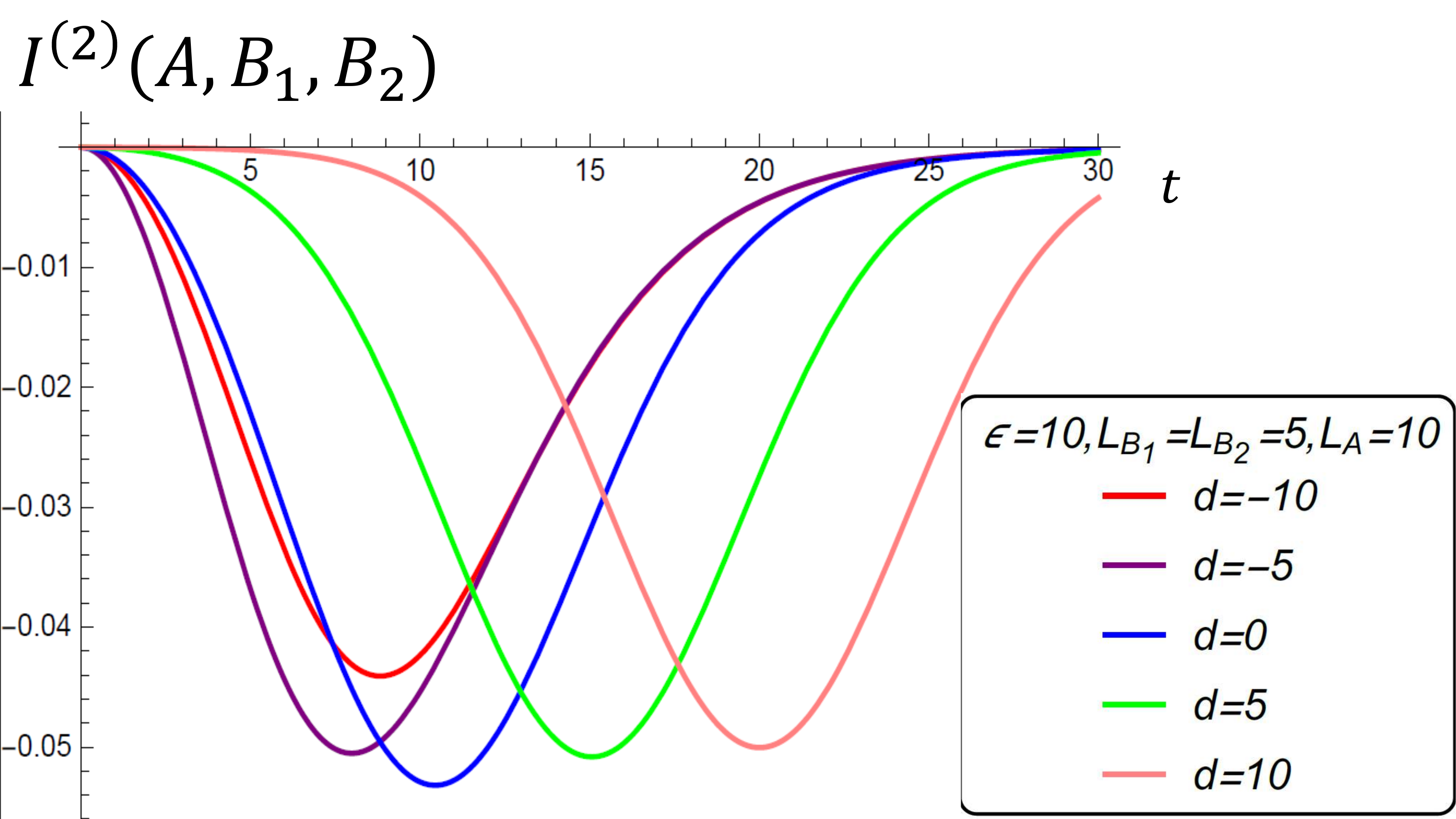}
  \end{center}
 \caption{
   Time evolution of TOMI in the compactified boson theory with finite output subsystem. 
   \textbf{Top:}
   In the top three panels, we take $\epsilon=1.1$. See insets for the parameters being used and Fig.\ \ref{fig:TOMIdetailedsetup} for their definitions.
   \textbf{Bottom:}
   The same configurations as top panels, but with $\epsilon=10$, which is comparable with the length scales $L_A$ and $d$.
 }
 \label{TOMI fb finite output}
\end{figure}

TOMI for the compactified boson theory at self-dual radius with finite output subsystem is shown in Fig.\ \ref{TOMI fb finite output}.  
(See Fig.\ \ref{fig:TOMIdetailedsetup} for detailed definitions of $L_A, L_B, d$).
As mentioned before, the key difference between BOMI for the free fermion and compactified boson theories
is that the curves of the latter are slightly rounded. 
As a result, when we compute the sums and differences in TOMI,
we do not have exact cancellations among BOMIs and instead see small bumps/peaks as shown in Fig.\ \ref{TOMI fb finite output}, despite that the individual BOMI terms behave qualitatively consistently with the quasi-particle picture. In the bottom panel of Fig.\ \ref{TOMI fb finite output}, the broadening of TOMI comes from the broadening of BOMI in Fig.\ \ref{BOMI fb LargerEpsilon}  due to larger $\epsilon$.

\subsubsection*{Holographic CFTs}
\begin{figure}[htbp]
  \begin{center}
    \includegraphics[width=50mm]{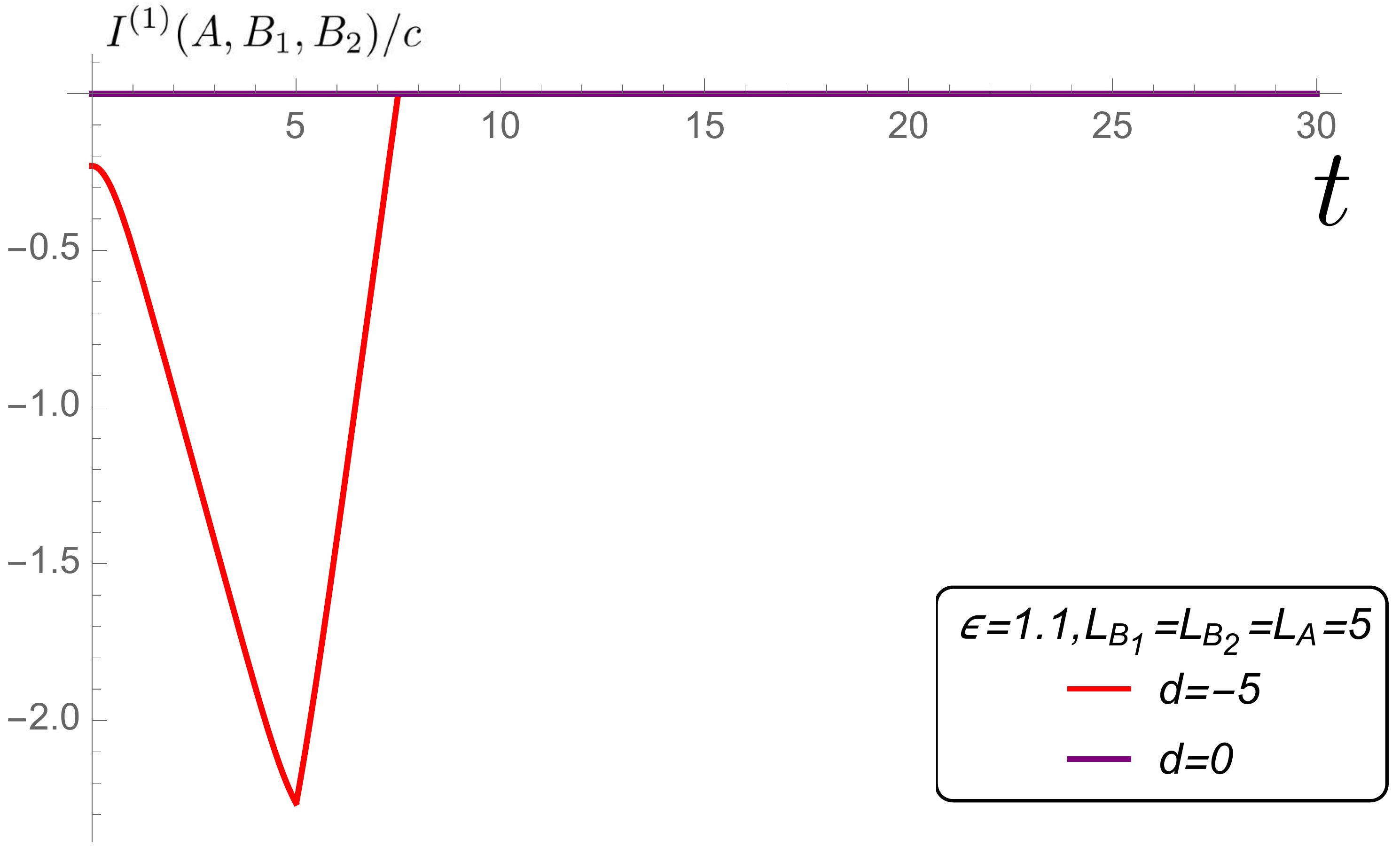}
    \includegraphics[width=50mm]{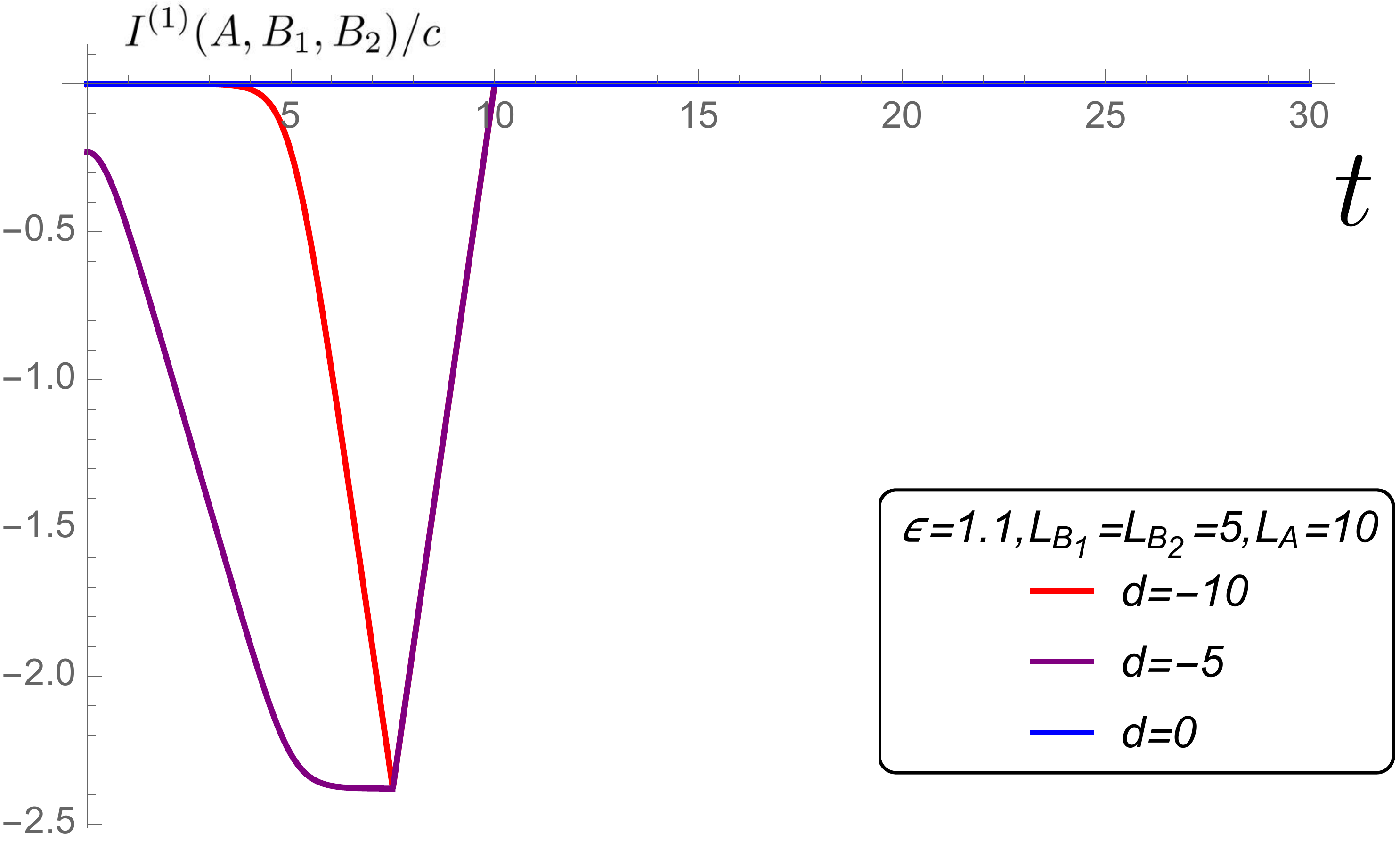}
     \includegraphics[width=50mm]{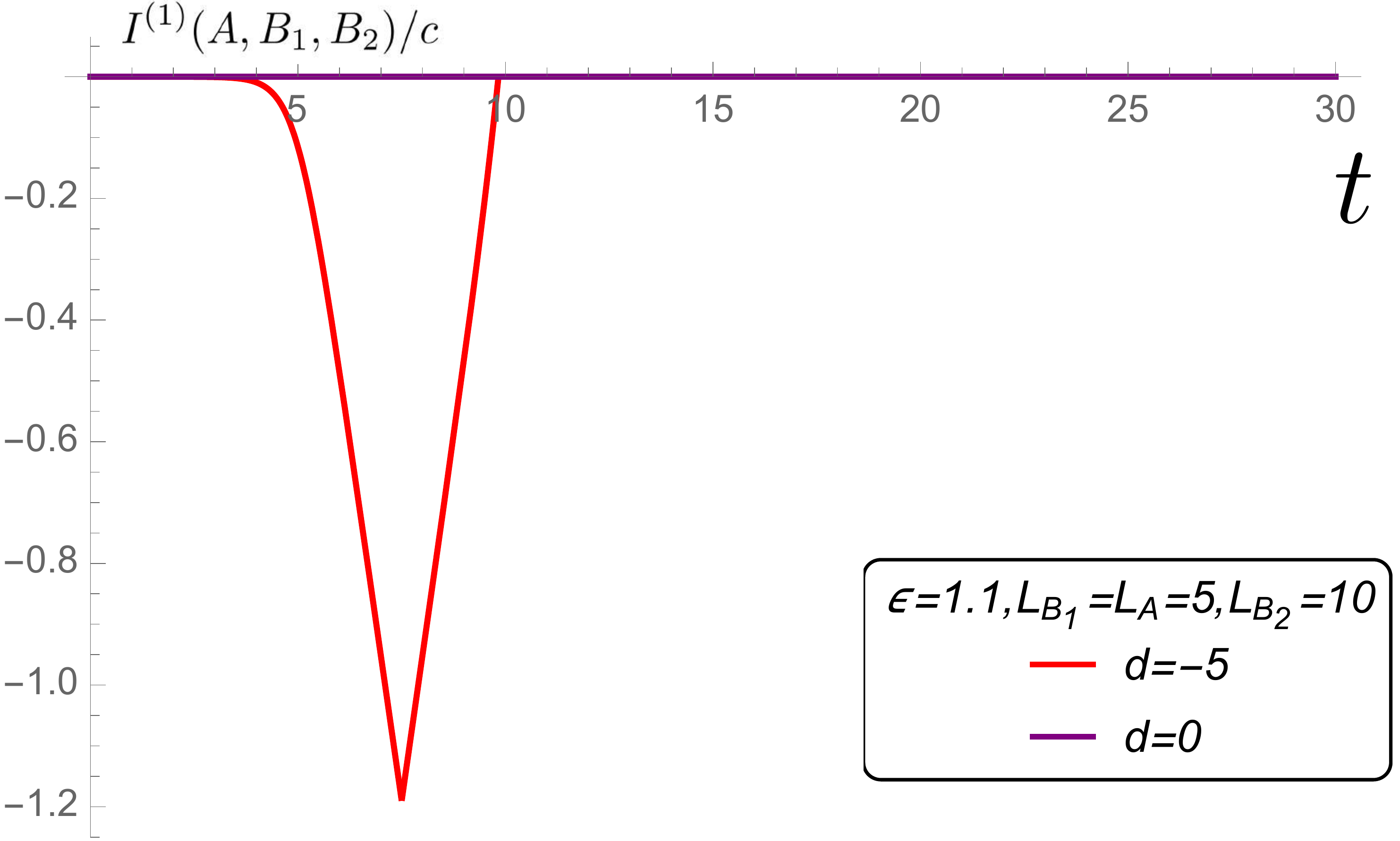}
  \end{center}
  \begin{center}
    \includegraphics[width=50mm]{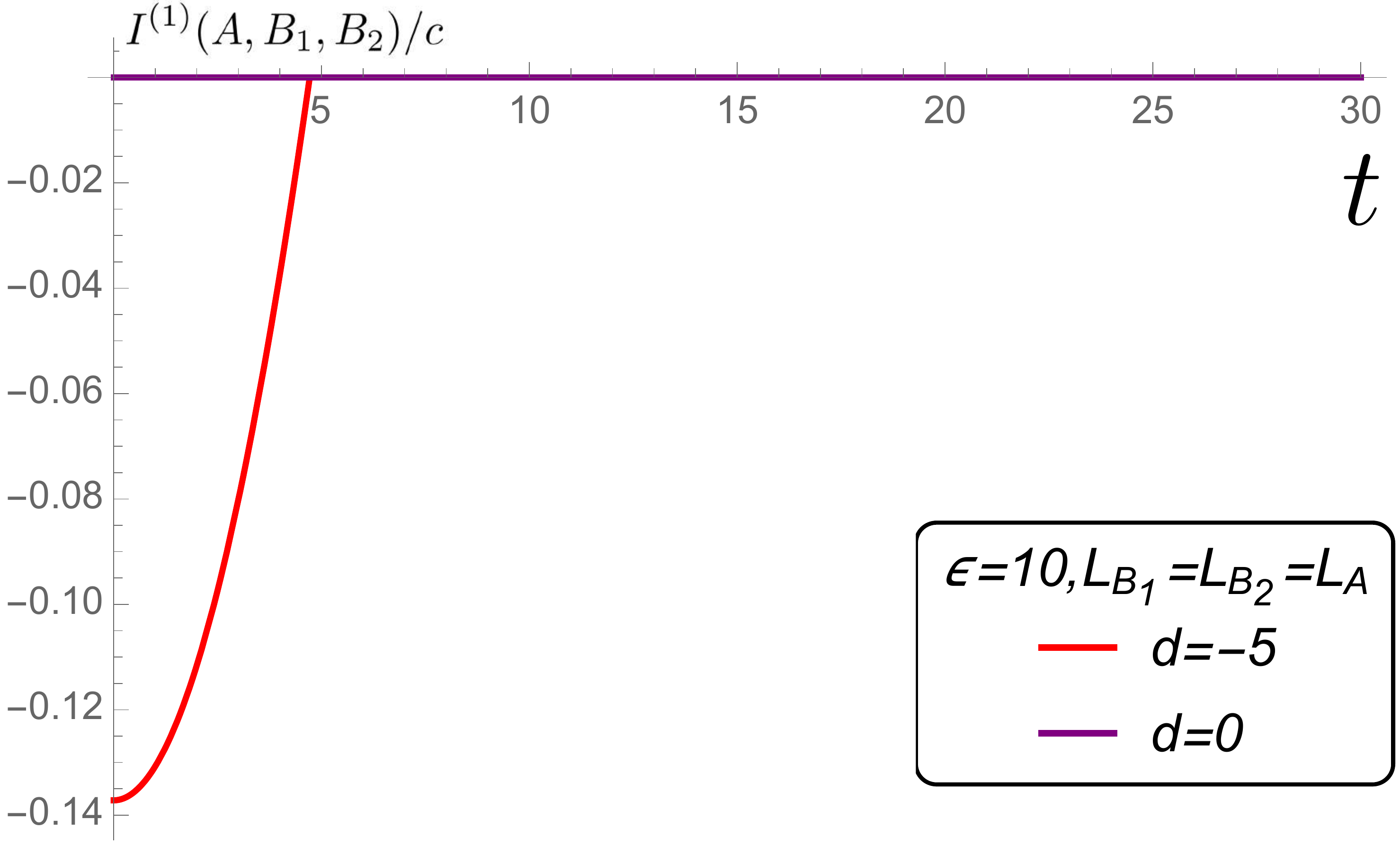}
    \includegraphics[width=50mm]{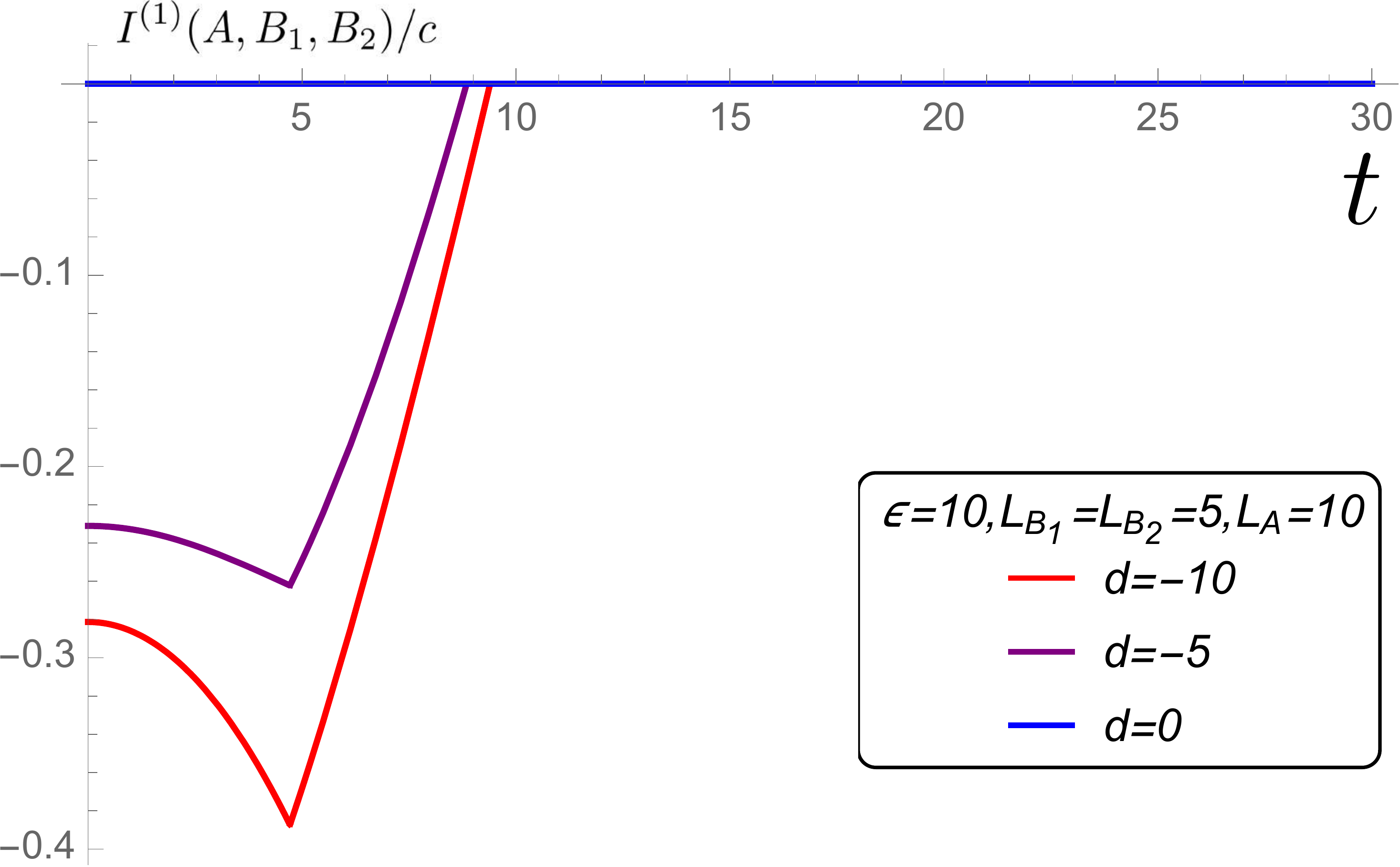}
     \includegraphics[width=50mm]{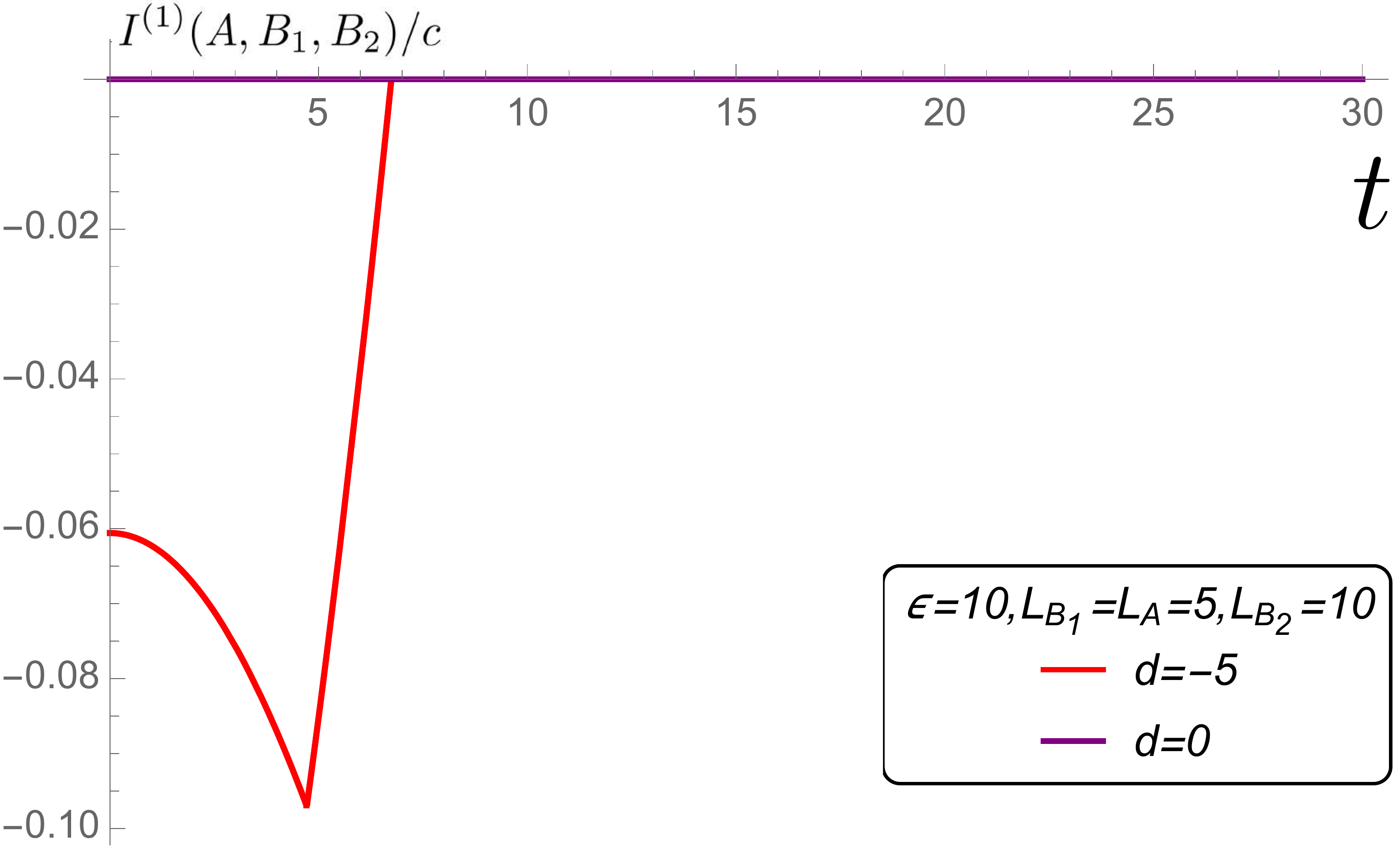}
  \end{center}
 \caption{
   Time evolution of TOMI in holographic CFTs with finite output subsystem. 
  \textbf{Top:} $\epsilon=1.1$. See insets for parameters being used and Fig. \ref{fig:TOMIdetailedsetup} for their definitions.
   \textbf{Bottom:}
  The same configurations as top panels, but with $\epsilon=10$. 
 \label{trpfcc}}
\end{figure}

Figure \ref{trpfcc} shows the time evolution of $I^{(1)}(A,B_1,B_2)$
for holographic CFTs.
The top panels 
are for $L_{B_1}, L_{B_2}, L_A \gg \epsilon$,
and the bottom panels are for $L_{B_1}, L_{B_2}, L_A \ll \epsilon$.
TOMI for holographic channels can be negative,
and the time evolution of TOMI cannot be interpreted by relativistic propagation of local objects in quasi-particle pitcure. 
Furthermore, the time evolution of TOMI for the holographic channels
is different from TOMI for compactified boson channels.
The time evolution of TOMI for $L_{B_1}, L_{B_2}, L_A \gg \epsilon$
is interpreted in terms of the properties of BOMI which we find in
Section \ref{Bi-partite operator mutual information}:
as we have seen in Figs. \ref{BOMI chaotic vs integrable perfectly overlapping} and \ref{BOMI hol overlapping}, BOMI for holographic channels shows sudden change of slopes,
as opposed to BOMI for the compactified boson theory, which leads to the sharp features in TOMI.

\subsection{TOMI for infinite output subsystem}
\label{sec:TOMIinfinite}

In this section, we take the output system $B$
to be the entire system with infinite volume.
$B_1$ and $B_2$ are semi-infinite, given by
\begin{align}
 B_1:=\left\{X| X \ge 0\right\},
  \quad
  B_2:=\left\{X| X < 0\right\}. 
\end{align}
This configuration translates
into the cross ratios, $x_{AB}$, $x_{A B_1}$, and $x_{A B_2}$,
which are needed to compute $I^{(n)}(A, B)$, $I^{(n)}(A,B_1)$ and $I^{(n)}(A,B_2)$,
respectively, as the following:
\begin{align} \label{cri}
  &x_{AB}=\bar{x}_{AB}\rightarrow 1-e^{-\f{\pi}{\epsilon}(X_1-X_2)},
  \nonumber \\
  &x_{AB_1}\rightarrow \f{\sinh{\left[\f{\pi}{2\epsilon}(X_1-X_2)\right]}}{e^{\f{\pi}{2\epsilon}(t-X_2)}\cosh{\left[\f{\pi}{2\epsilon}(X_1-t)\right]}},
    \quad
   \bar{x}_{AB_1}\rightarrow \f{\sinh{\left[\f{\pi}{2\epsilon}(X_1-X_2)\right]}}{e^{-\f{\pi}{2\epsilon}(t+X_2)}\cosh{\left[\f{\pi}{2\epsilon}(X_1+t)\right]}},
  \nonumber \\
  &x_{AB_2}\rightarrow \f{\sinh{\left[\f{\pi}{2\epsilon}(X_1-X_2)\right]}}{e^{\f{\pi}{2\epsilon}(X_1-t)}\cosh{\left[\f{\pi}{2\epsilon}(X_2-t)\right]}},
    \quad
    \bar{x}_{AB_2}
    \rightarrow \f{\sinh{\left[\f{\pi}{2\epsilon}(X_1-X_2)\right]}}{e^{\f{\pi}{2\epsilon}(t+X_1)}\cosh{\left[\f{\pi}{2\epsilon}(X_2+t)\right]}},
\end{align}
as we send the size of the subsystem infinite
($Y_1\to \infty$ and $Y_2\to -\infty$) in Fig. \ref{fig:TOMIdetailedsetup}.  

Of particular interest for us is the late-time behavior
of TOMI, i.e., when $t\pm X_1 \gg \epsilon$ and $t\pm X_2 \gg \epsilon$.
In this case, the cross ratios are given by
\be
\label{cross ratio, late time}
\begin{split}
&x_{AB}=\bar{x}_{AB}\rightarrow 1-e^{-\f{\pi}{\epsilon}(X_1-X_2)}, \\
&x_{AB_1}\rightarrow \f{2\sinh{\left[\f{\pi}{2\epsilon}(X_1-X_2)\right]}}{e^{\f{\pi}{2\epsilon}(2t-X_1-X_2)}}, ~~\bar{x}_{AB_1}\rightarrow 1-e^{-\f{\pi}{\epsilon}(X_1-X_2)}, \\
&x_{AB_2}\rightarrow 1-e^{-\f{\pi}{\epsilon}(X_1-X_2)}, ~~\bar{x}_{AB_2}\rightarrow \f{2\sinh{\left[\f{\pi}{2\epsilon}(X_1-X_2)\right]}}{e^{\f{\pi}{2\epsilon}(2t+X_1+X_2)}}. \\
\end{split}
\ee

\begin{figure}[t]
  \begin{center}
         \begin{tabular}{c}
 \begin{minipage}{0.50\hsize}
  \begin{center}
   \includegraphics[width=80mm]{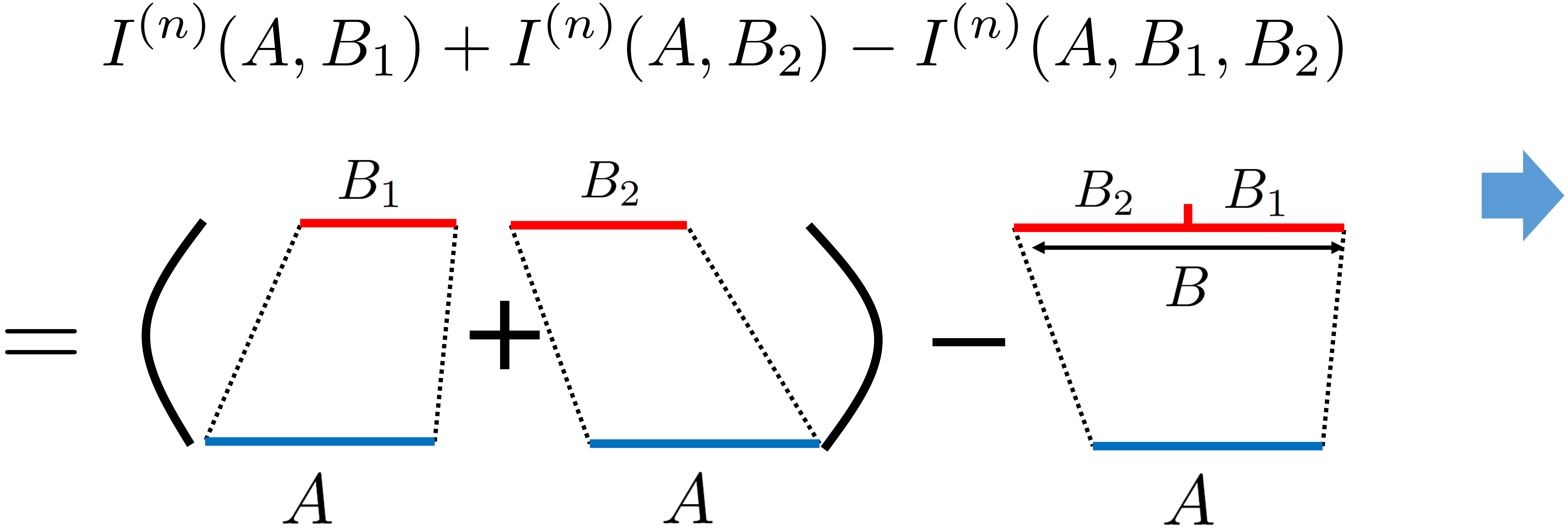}
  \end{center}
 \end{minipage}
 \begin{minipage}{0.50\hsize}
  \begin{center}
   \includegraphics[width=80mm]{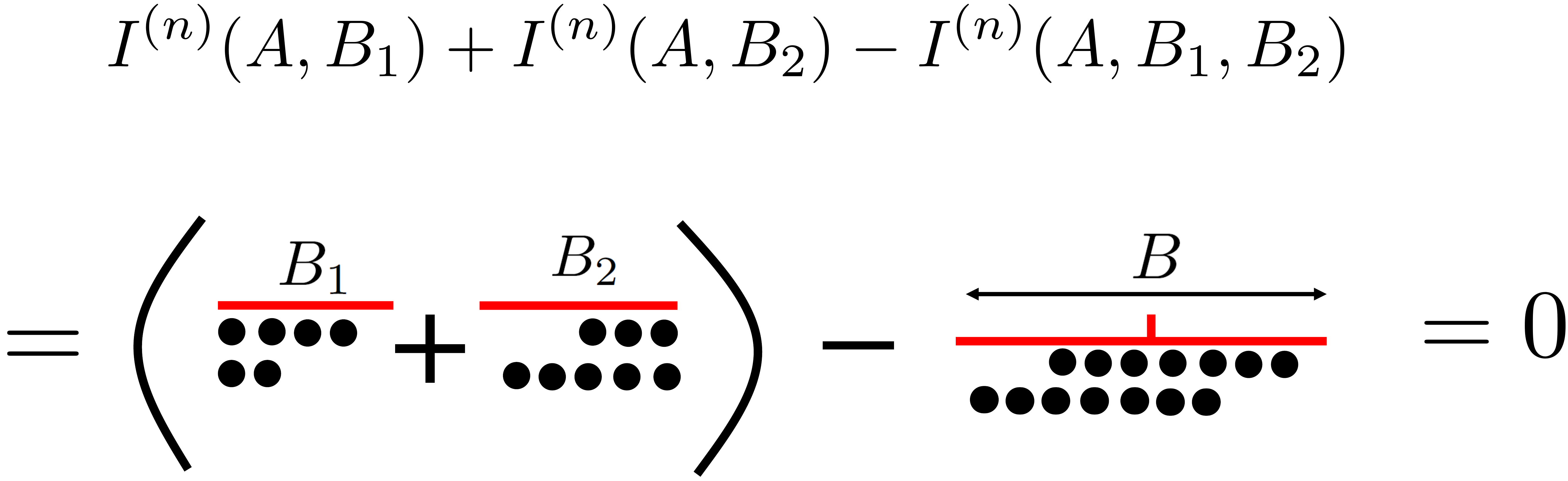}
  \end{center}
 \end{minipage}
    \end{tabular}
    \caption{Schematic explanation of
      the vanishing of TOMI for free fermion channel 
      by the quasi-particle picture. \textbf{Left}: Sketch of the setup.  \textbf{Right:} Corresponding quasi-particle picture, where the two rows of black dots represent the quasi-particles that propagate to the right (upper row) and left (bottom row) ballistically.  }
\label{TOMI_Ftoy}
  \end{center}
\end{figure}

\subsubsection{The free fermion CFT}
We recall that BOMI for the free fermion theory is given by \eqref{fOMI}.
In the limit  $Y_1 \rightarrow \infty$ and $Y'_2 \rightarrow -\infty$,
\be
 I^{(1)}(A, B) /c        \rightarrow \f{\pi}{3\epsilon}(X_1-X_2).
\ee
On the other hands, 
$I^{(1)}(A, B_1)$
and
$I^{(1)}(A, B_2)$
are given by (\ref{fOMI})
by taking $Y_1 \rightarrow \infty, Y_2 =0$,
and $Y'_2 \rightarrow -\infty,Y'_1=0$, respectively:
\begin{align}
I^{(1)}(A, B_{1,2}) /c \rightarrow
\f{\pi}{6\epsilon}(X_1-X_2)
  \pm
\f{1}{6}\log{\left[\f{\cosh{\left[\f{\pi}{2\epsilon}(X_1-t)\right]}\cosh{\left[\f{\pi}{2\epsilon}(X_1+t)\right]}}{\cosh{\left[\f{\pi}{2\epsilon}(X_2-t)\right]}\cosh{\left[\f{\pi}{2\epsilon}(X_2+t)\right]}}\right]}.
\end{align}
Thus, just like the finite output case, for infinite output subsystem $I^{(1)}(A,B_1,B_2)$ in the free fermion CFT vanishes, meaning that 
the amount of information we are able to obtain additively from $B_1$ and $B_2$
is equal to the total  amount of information sent from $A$. 

Fig.\ \ref{TOMI_Ftoy} is a more pictorial interpretation of vanishing TOMI in the free fermion CFT using the quasi-particle picture.
In the left panel, $I^{(n)}(A,B)$ measures the correlation between the input subsystem $A$, and the total output
system $B$, and 
$I^{(n)}(A,B_1)$ and $I^{(n)}(A,B_2)$ measure the correlation between $A$ and
the halves of total system, $B_1$ and $B_2$, respectively;
TOMI measures the difference
between the correlation between $A$ and $B$, and the sum of correlations between
$A$ and the halves. This amounts to counting quasi-particle numbers in the right panel of Fig.\ \ref{TOMI_Ftoy}: 
since the total number of particles is conserved in our model,
the number of particles in $B$ is equal to 
the sum of number of particles in $B_1$ and $B_2$, from which we conclude TOMI is identically zero.

\subsubsection{The compactified boson CFTs}
\label{The compactified boson theory TOMI}

Let us now discuss the case of the compactified free boson theory
at different radii. We will first present the results for self-dual radius, then move on to the more general cases where we investigate how the {\it late-time saturation value} of TOMI depends
on the radius of the compactification. We will limit our attention to the case $L_A \gg \epsilon$, i.e. the length scale corresponding to the energy cutoff is insignificant compared to the subsystem size. 

\begin{figure}[t]
  \begin{center}
   \includegraphics[width=65mm]{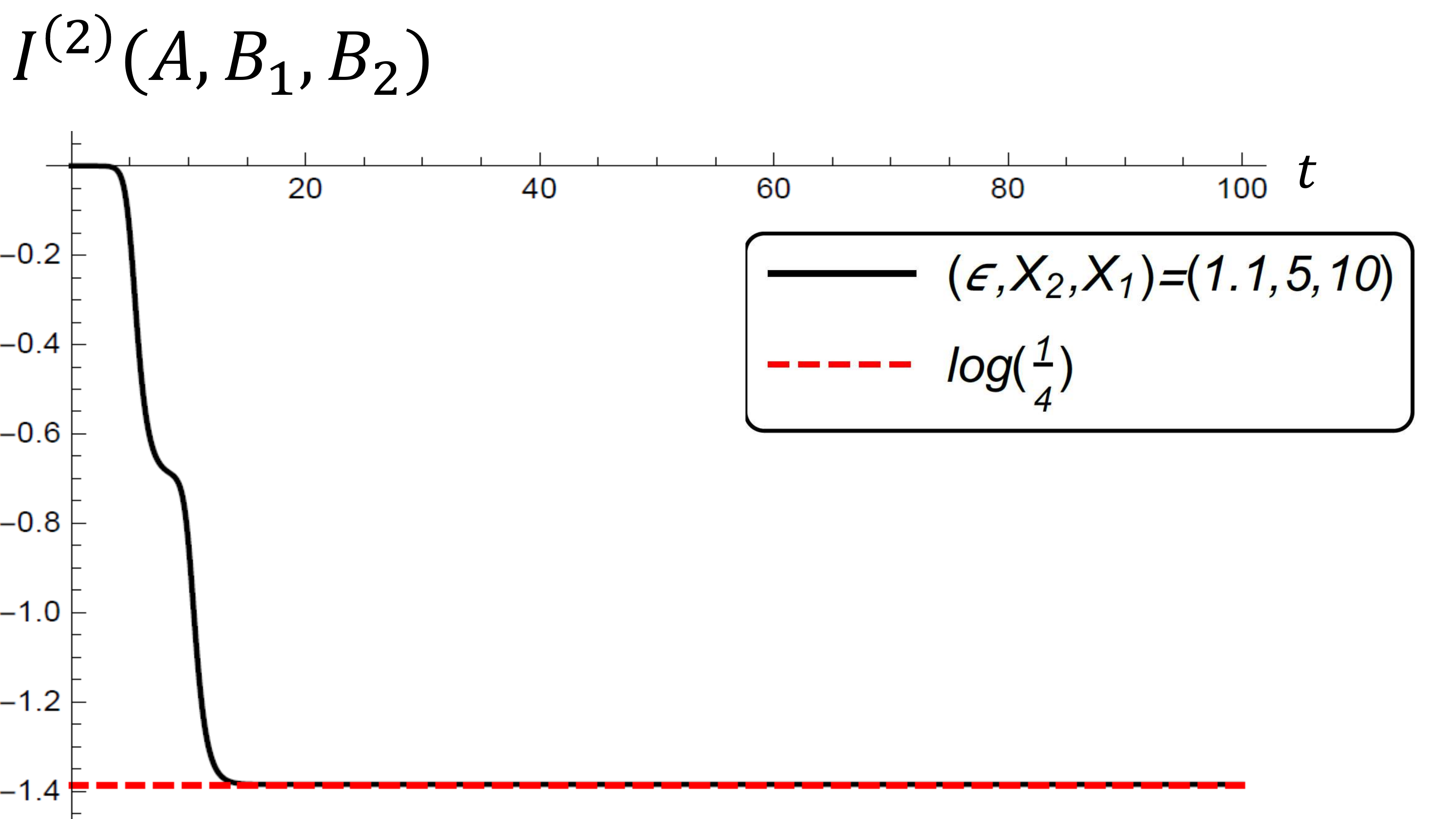}
   \includegraphics[width=65mm]{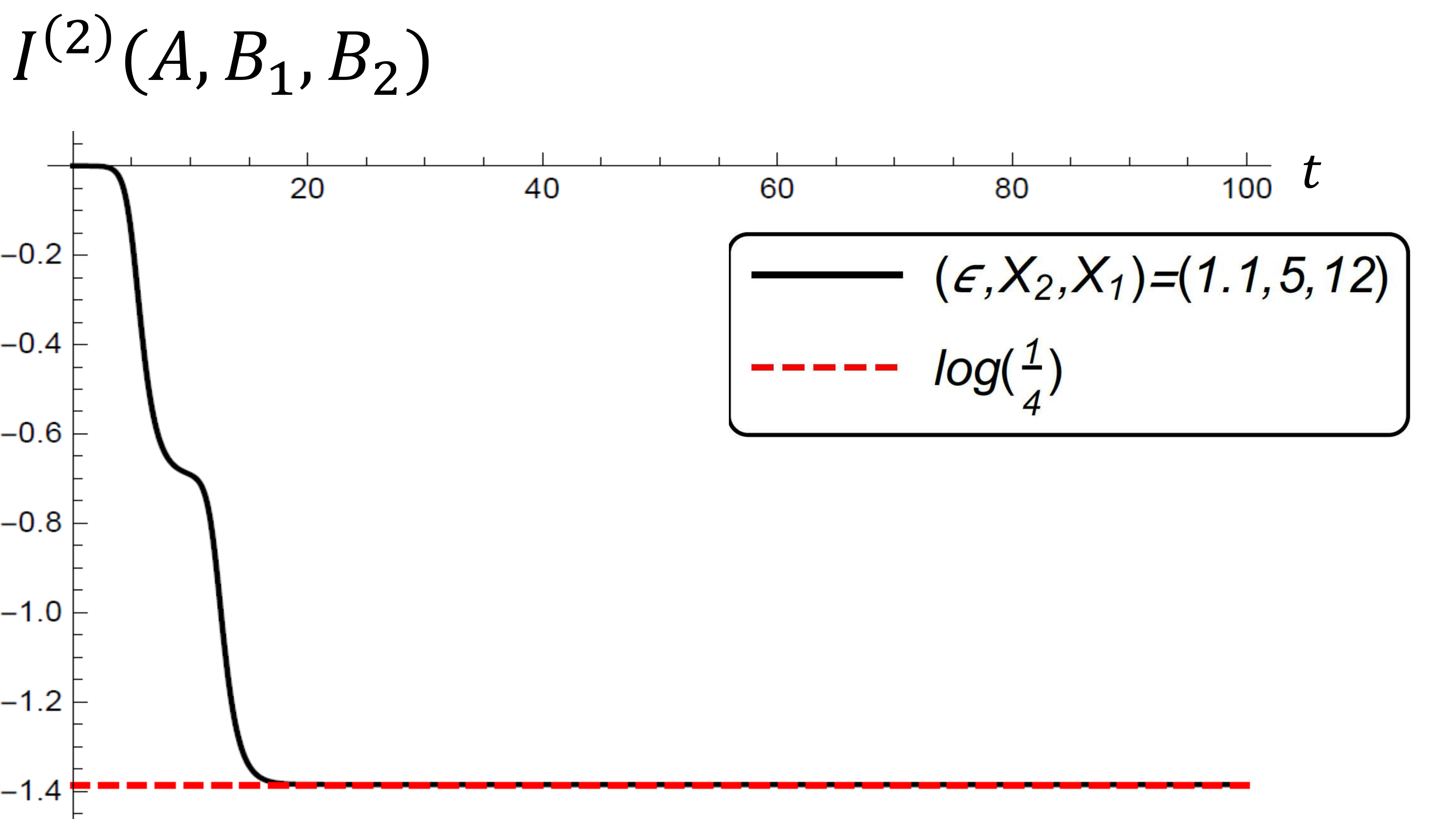}
   \\
   \includegraphics[width=65mm]{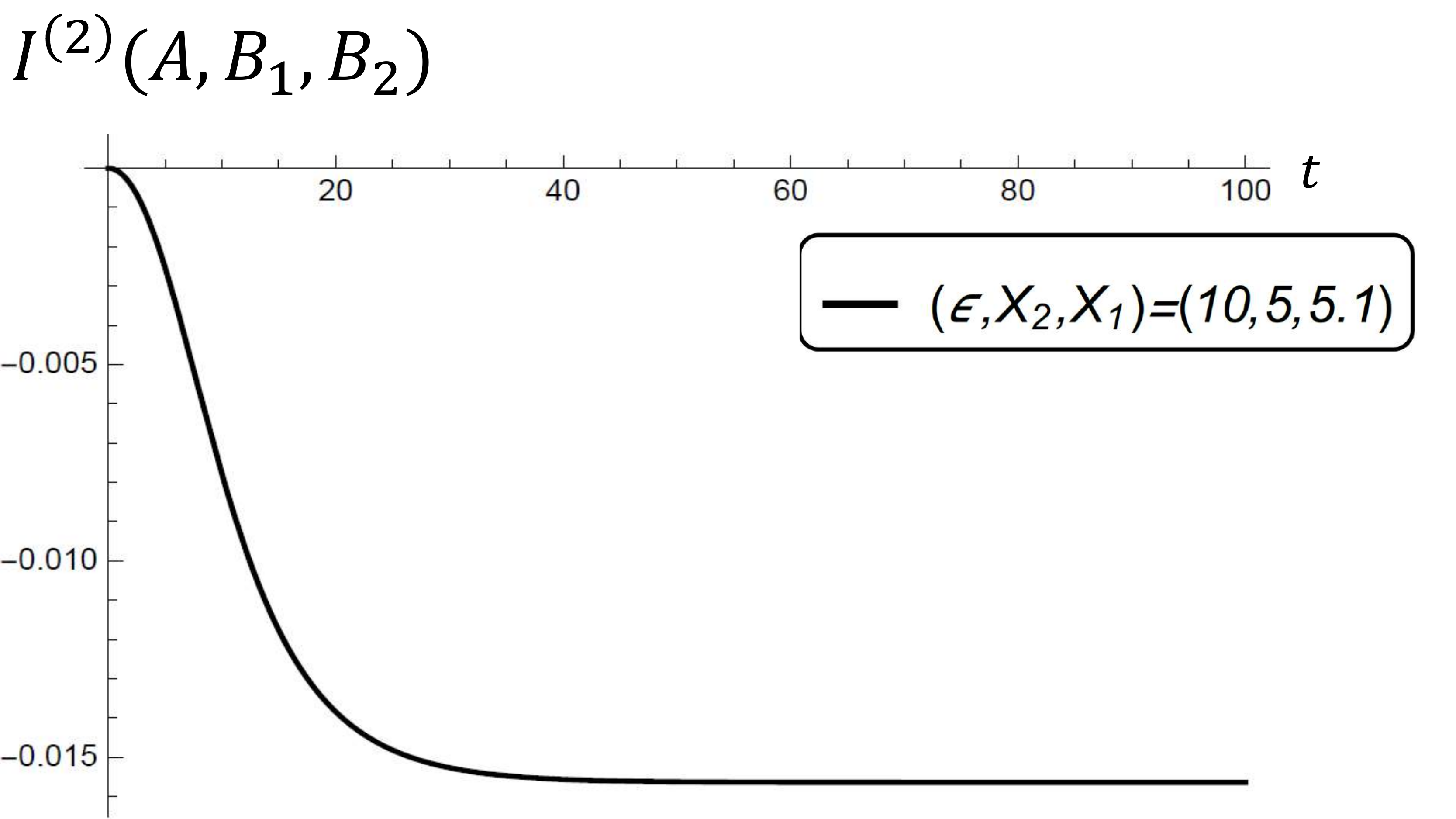}
   \includegraphics[width=65mm]{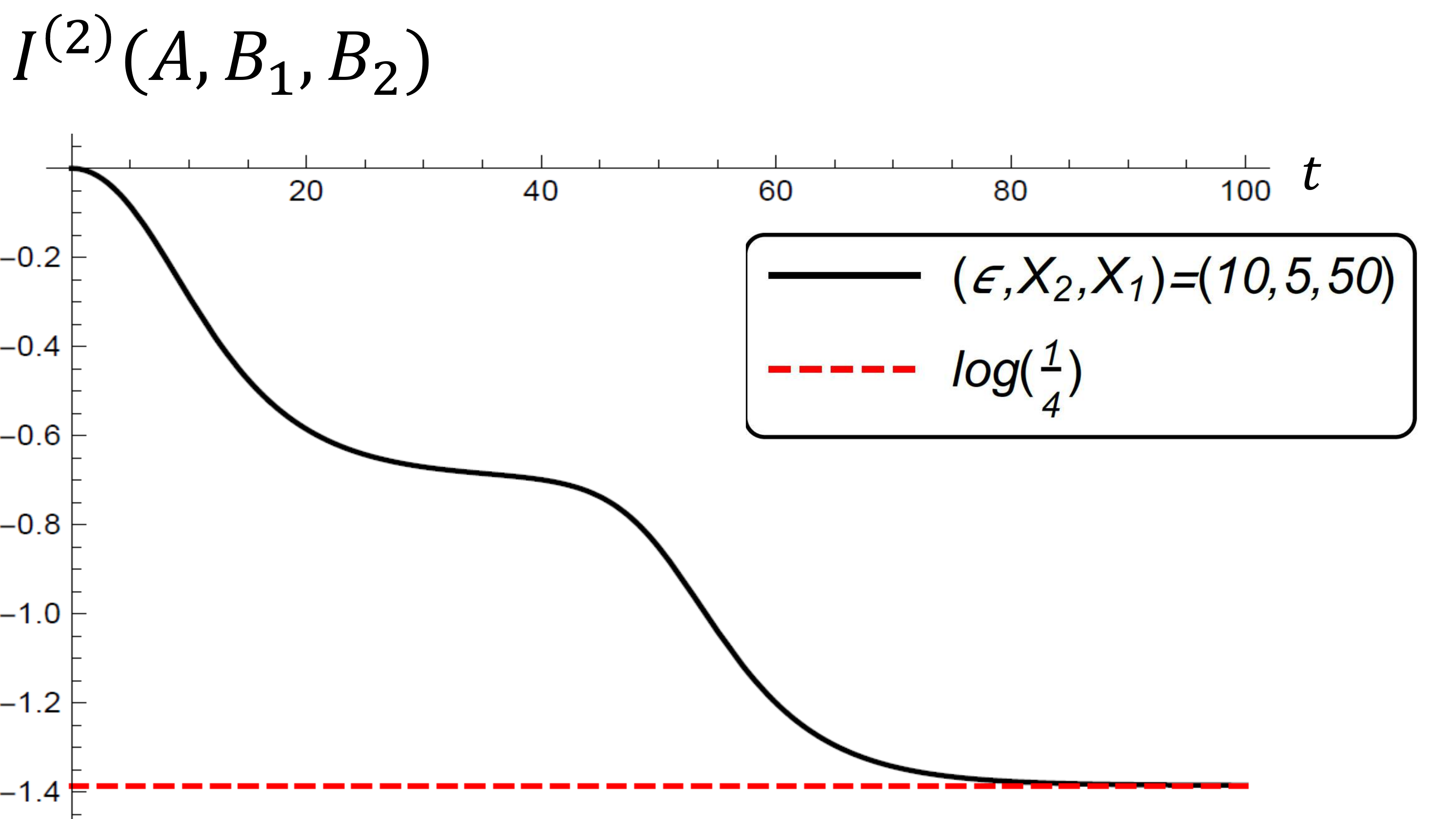}
   \end{center}
   \caption{
      Time evolution of TOMI
     $I^{(2)}(A,B_1,B_2)$ for the compactified boson theory
     at self-dual radius ($\eta  = R^2 = 1$) for different lengths of input subsystem $A$ and values of $\epsilon$. The insets list the parameters being used, with $X_1, X_2$ being the coordinates of the boundary of $A$ (see Fig. \ref{fig:TOMIdetailedsetup} for details of the setup). \textbf{Top:} $\epsilon = 1.1$. Both top left and top right panels satisfy $\epsilon  \ll L_A = X_1 - X_2$.  The red dashed line is the analytic approximation for the late time in the limit $L_A \gg \epsilon$. The top left panel is the same as Fig. \ref{fig:tOpMIchart} (2).
     \textbf{Bottom:} $\epsilon = 10$. Bottom right panel again satisfies $\epsilon \ll L_A$ and its late-time value is shown by the red dashed line. 
     As a comparison, bottom left panel is plotted with $\epsilon \gg L_A$.
       }
       \label{fomi1}
 \end{figure}

Fig. \ref{fomi1} shows the time evolution of TOMI at self-dual radius. For $L_A \gg \epsilon$ (top panels and bottom right panel), at $t>0$ TOMI becomes negative and saturates to $\log (1/4)$ denoted by red dashed lines, which as we will see is a special case of $-2\log d_{\sigma_2}$ in ~\eqref{eq:rationalEta} evaluated analytically under certain approximations, where $d_{\sigma_2}$ is the quantum dimension of twist operator $\sigma_2$ of the two-sheeted Riemann surface. For comparison we also plot TOMI in the opposite limit $L_A \ll \epsilon$ in the bottom left panel. It also saturates to a negative value, but with much smaller magnitude and a much smoother curve compared to the case $L_A \gg \epsilon$, similar to the situation in Fig. \ref{TOMI fb finite output}.
As stated before we are not interested in this limit therefore didn't evaluate the analytical expression of the late-time value.

Next, we turn our focus to the radius dependence of the dynamics of TOMI, in particular its late-time value. 
Below we summarize the analytic
approximations for the late-time value of the TOMI of the compactified boson when $L_A =X_1-X_2\gg \epsilon$. More
accurate expressions with subleading terms can be found
in Appendix \ref{Late time behavior of TOMI for the compactified boson theory},
along with the derivation of these expressions. 
The main takeaway is that the
saturation value of TOMI for the compactified boson at
rational values of radius squared is given by the quantum dimension of the twist
operators $\sigma_n$ (where $n$ is the number of replicas), while the saturation value at irrational radius
diverges as $-\log{\left(L_A/\epsilon\right)}$ as $L_A/\epsilon\to \infty$.
In the following we denote $\tilde{\eta} \coloneqq \text{Max}\left\{\eta, {1}/{\eta}\right\}$, where $\eta$ is the radius squared defined in~\eqref{eq:eta}.
(Our analysis in this section is similar in spirit to, 
and largely inspired by Refs.\ \cite{2017PhRvD..96d6020C},
where the OTOC in the orbifolded compactified free boson
theory was studied; One should keep in mind, however, that here our compactified boson CFT is not ``intrinsically" orbifolded, i.e. it is orbifolded only due to the use of replica trick.)

\begin{figure}[t]\label{eta3tri-partiteMutualInfoPlot}

 \begin{minipage}{0.32\hsize}
  \begin{center}
   \includegraphics[width=53mm]{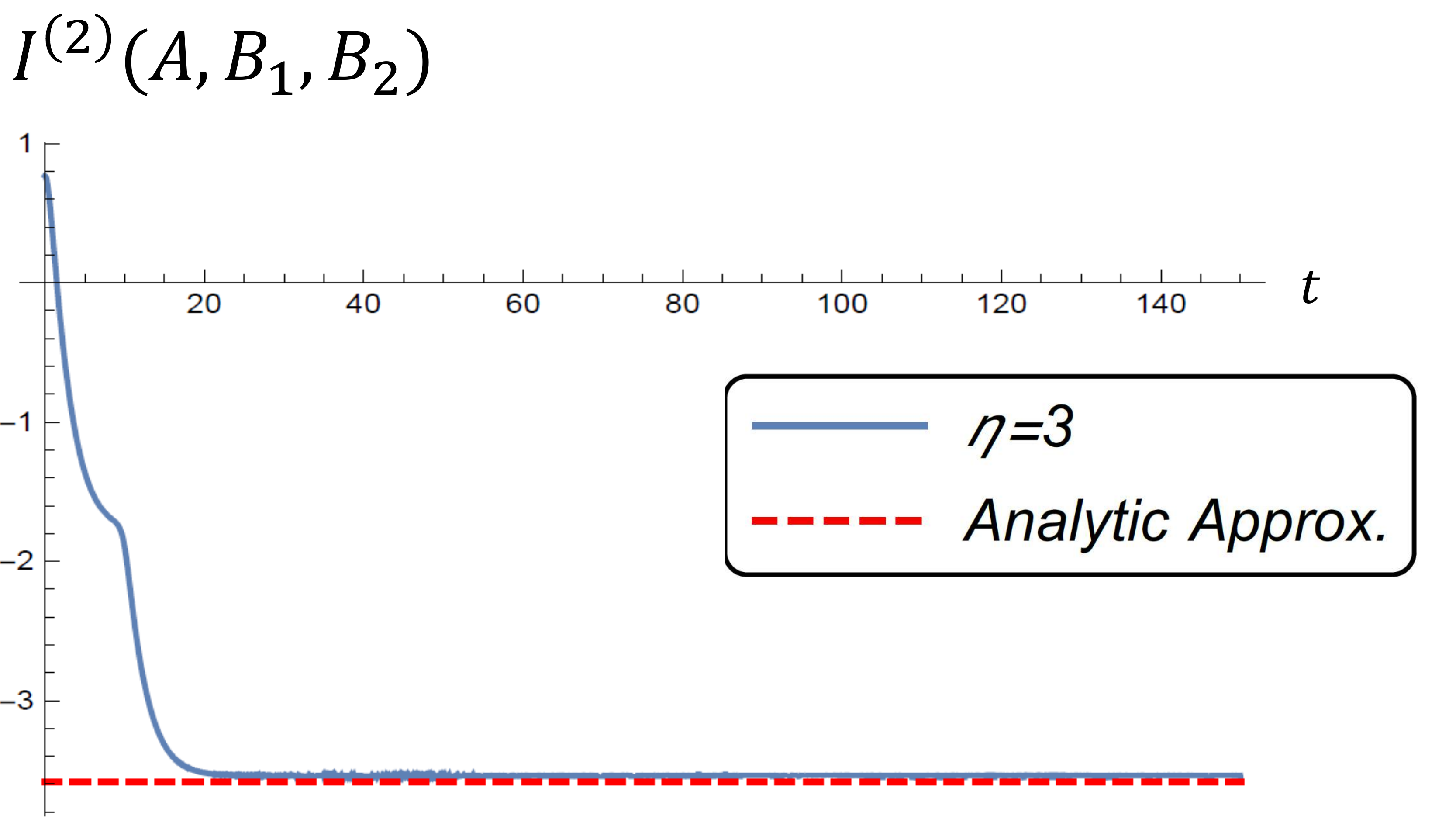}
  \end{center}
 \end{minipage}
 \begin{minipage}{0.32\hsize}
  \begin{center}
   \includegraphics[width=53mm]{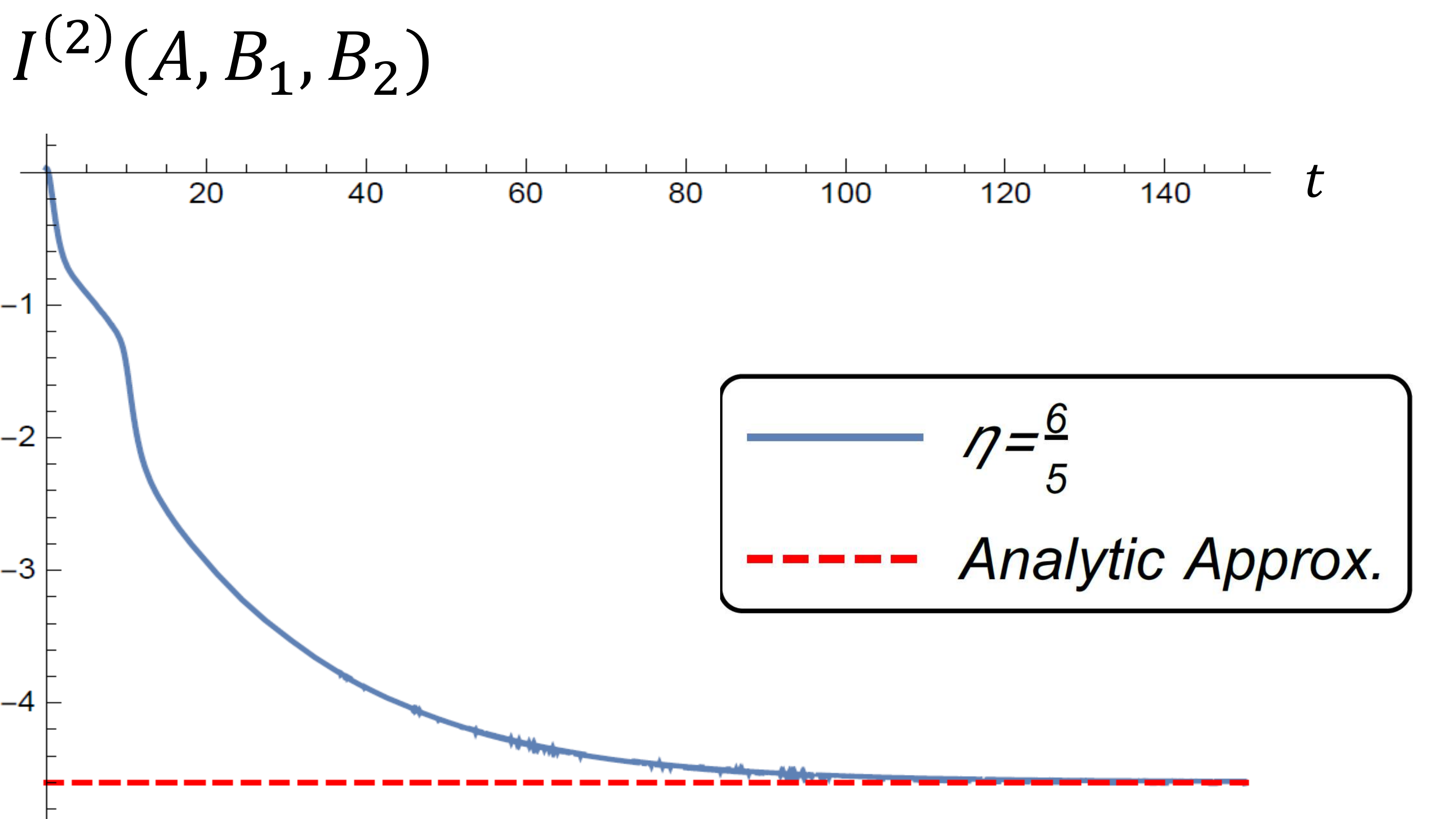}
  \end{center}
 \end{minipage}
 \begin{minipage}{0.3\hsize}
  \begin{center}
   \includegraphics[width=53mm]{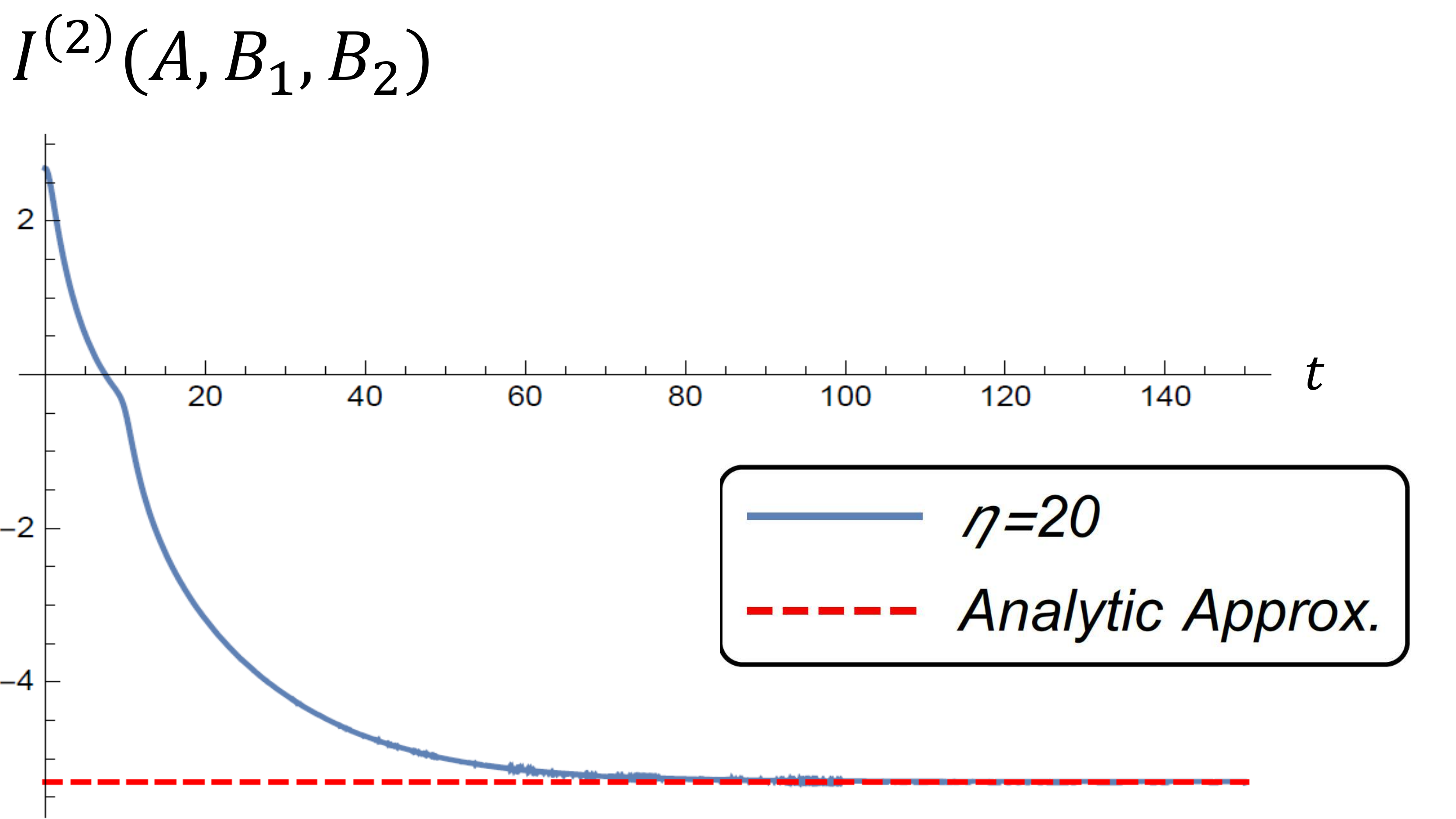}
  \end{center}
 \end{minipage}
  \vspace{0.5cm}\\
 \begin{minipage}{0.5\hsize}
  \begin{center}
   \includegraphics[width=53mm]{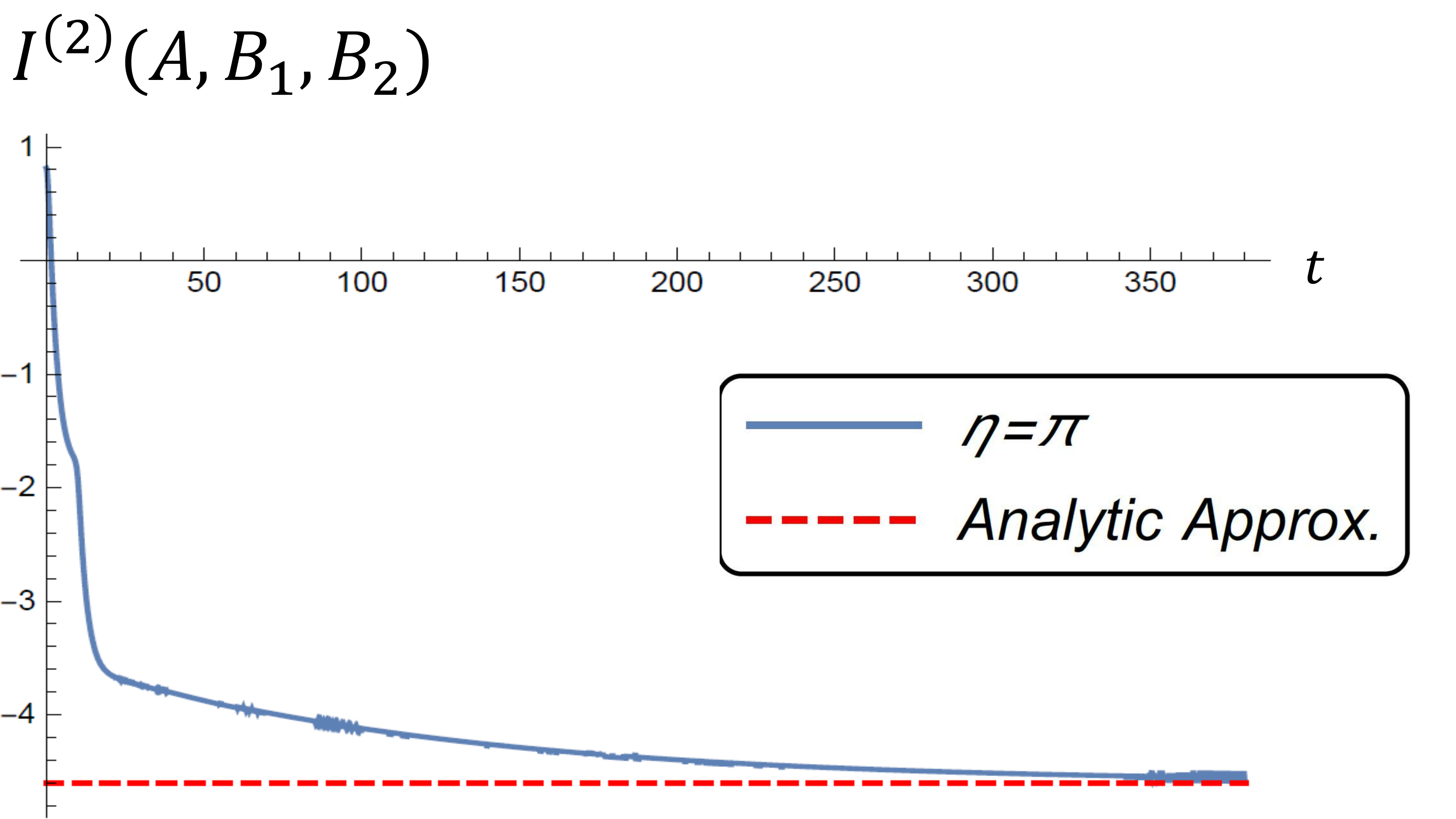}
  \end{center}
 \end{minipage}
 \begin{minipage}{0.5\hsize}
  \begin{center}
   \includegraphics[width=53mm]{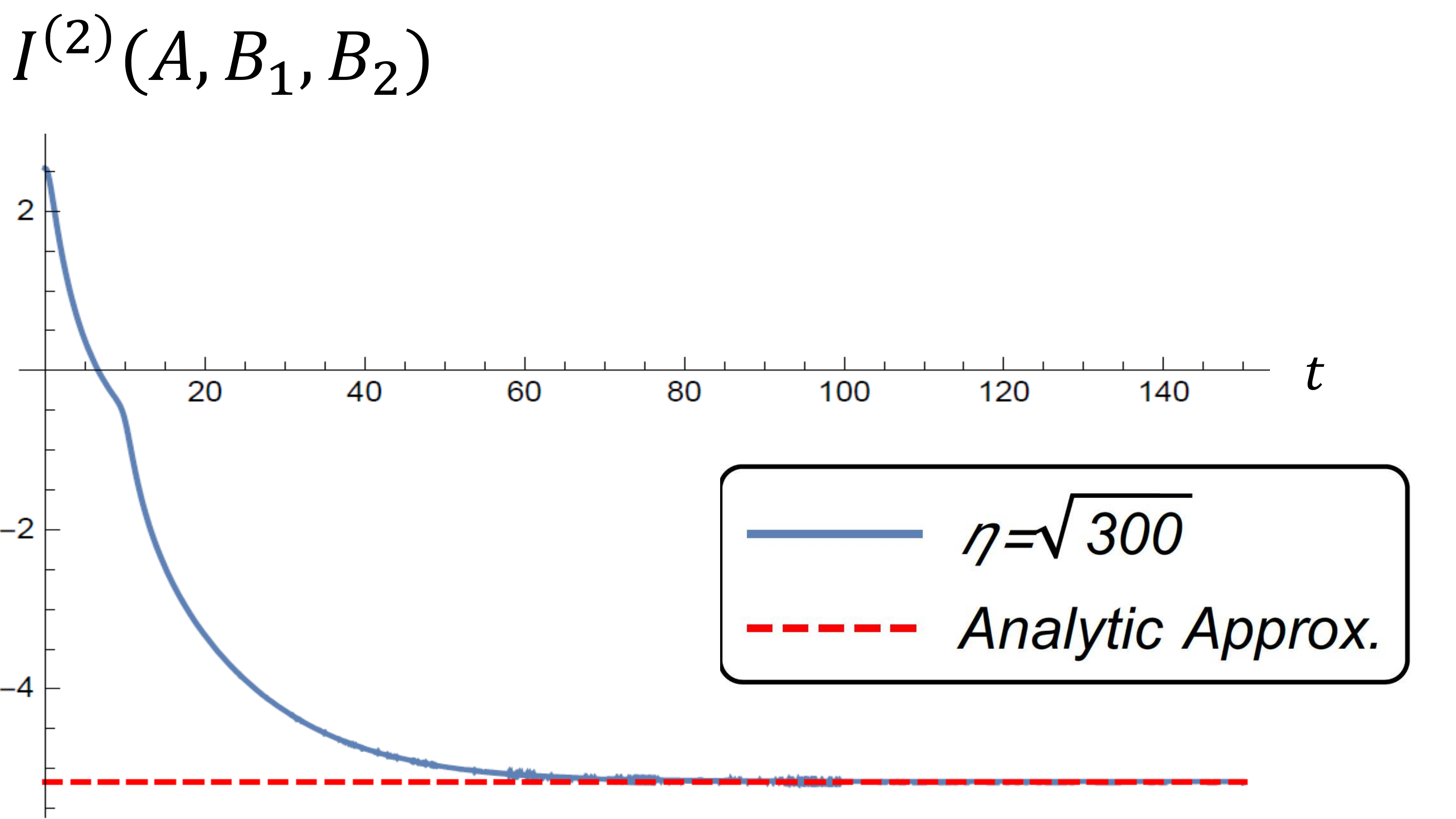}
  \end{center}
 \end{minipage}
 \caption{
   The time-dependence of 
   the tri-partite mutual information
   $I^{(2)}(A,B_1,B_2)(t)$ for
   the compacified free boson theory with
   $\eta = 3, {6}/{5}, 20, \pi,
    \sqrt{300}$, which correspond to the five cases: $\eta$ small and $2pp'$
    small, $\eta$ small and $2pp'$ large, $\eta$ large and $2pp'$ large, $\eta$
    irrational and small and lastly, $\eta$ irrational and large. The red dashed
    lines correspond to the analytically obtained values of the late-time
    behavior, which are given by 
    \eqref{2pp'etasmall},
    \eqref{2pp'largeEtaSmall},
    \eqref{2pp'largeEtaLarge},
    \eqref{EtaIrrationalSmall} and
    \eqref{EtaIrrationalLarge} respectively (note that the parameters $(L_A, \epsilon)$ used to generate the above plots do not strictly follow $L_A \gg \epsilon$, therefore we choose to plot~\eqref{2pp'etasmall} $\sim $~\eqref{EtaIrrationalLarge} instead of~\eqref{eq:rationalEta} and~\eqref{eq:irrationalEta}).
    }
  \label{TOMI CB different radii}
\end{figure}

First, when $\eta$ is rational, we find the following analytic approximations in the late-time behavior of TOMI:
\begin{equation}
I^{(2)}(A,B_1,B_2) \xrightarrow[L_A \gg\epsilon]{t \to \infty}
\begin{cases} 
      -2\log d_{\sigma_2} & 2pp' , \tilde{\eta}\ll \frac{L_A}{\epsilon} \\
      -2\log \frac{L_A}{\epsilon} & \tilde{\eta}\ll \frac{L_A}{\epsilon} \ll 2pp'\\
      -\log\left(\tilde{\eta}\frac{L_A}{\epsilon}\right) & \frac{L_A}{\epsilon} \ll 2pp', \tilde{\eta}
   \end{cases}
   \label{eq:rationalEta}
\end{equation}
where $p,p'$ are defined in~\eqref{eq:ppprime}, and 
$2p p' \equiv d_{\sigma_2}$ is the quantum dimension of the twist operator for the two-sheeted Riemann surface. See Fig.\ \ref{TOMI CB different radii} for plots of $I^{(2)}(A,B_1,B_2)$ with $\eta = 3, 6/5, 20$, which are examples of the three cases listed above.

As for irrational values of $\eta$, we obtain
\begin{equation}
I^{(2)}(A,B_1,B_2)  \xrightarrow[L_A \gg\epsilon]{t \to \infty}
\begin{cases} 
      -2\log \frac{L_A}{\epsilon} & \tilde{\eta}\ll \frac{L_A}{\epsilon} \\
      -\log\left(\tilde{\eta} \frac{L_A}{\epsilon} \right)& \tilde{\eta}\gg \frac{L_A}{\epsilon} 
   \end{cases}
   \label{eq:irrationalEta}
\end{equation}
See Fig.\ \ref{TOMI CB different radii} for examples with $\eta = \pi,\sqrt{300}$, which correspond to the two cases listed.

Observe that if we fix the radius to be finite
and take $L_A/\epsilon \gg \tilde{\eta}$, then the saturation value of $I^{(2)}(A,B_1,B_2)$ for rational $\eta$ remains at a constant determined by the quantum dimension, while the saturation values for irrational $\eta$ have the following asymptotic form
\begin{equation}
I^{(2)}(A,B_1,B_2)  \sim -\log \frac{L_A}{\epsilon}.
\end{equation}
In contrast, in the holographic case as we will see below, 
 $I^{(1)} (A, B_1, B_2)$ is negative and grows linearly in ${L_A}/{\epsilon}$ at late times:
\be
I^{(1)}(A, B_1,B_2) \sim - \frac{\pi c L_A}{6\epsilon}
\ee
In this sense, the compactified boson theory
experiences less information scrambling than the holographic case.

\begin{figure}[H]
\center
   \includegraphics[width=100mm]{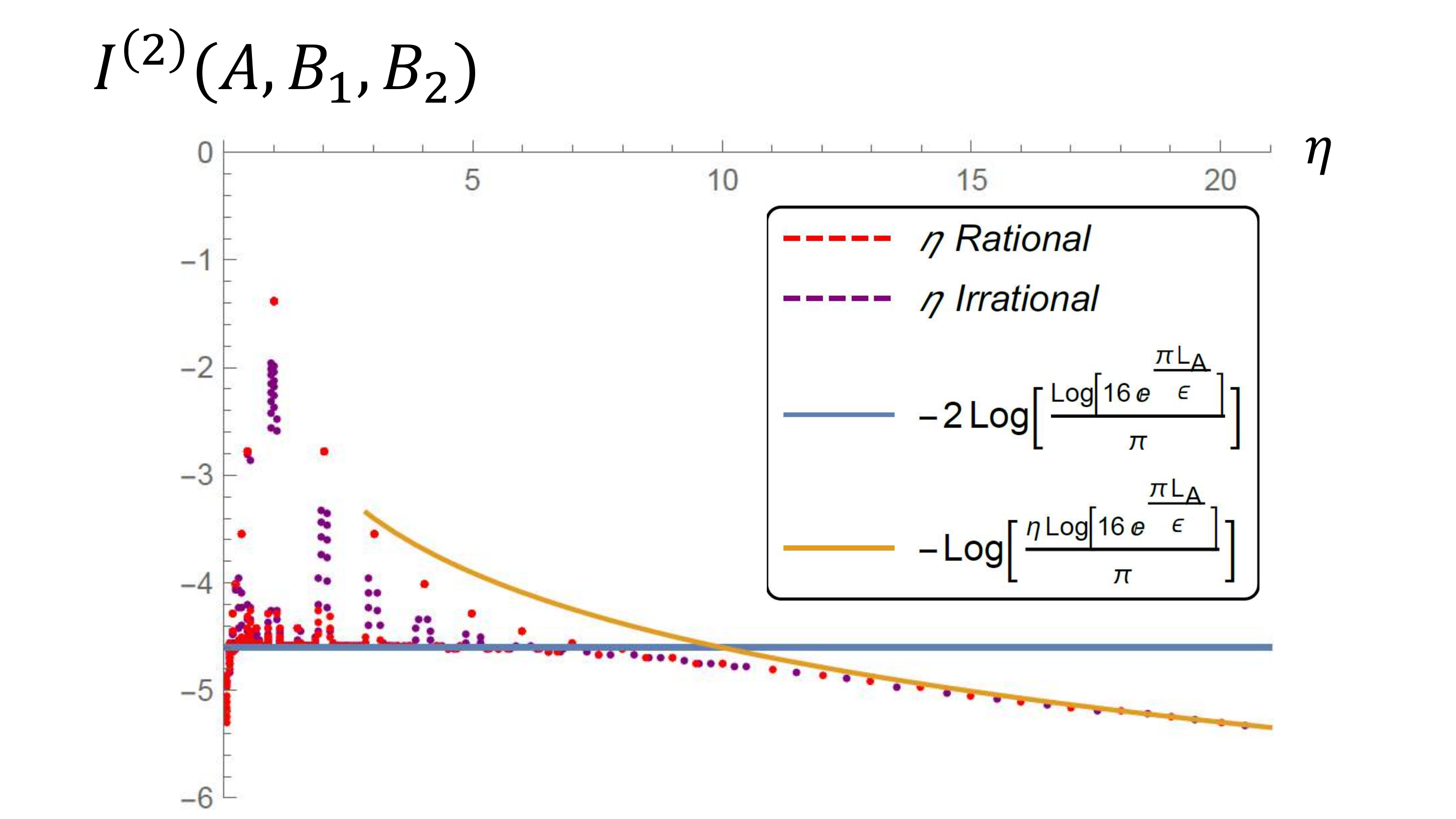}
   \includegraphics[width=100mm]{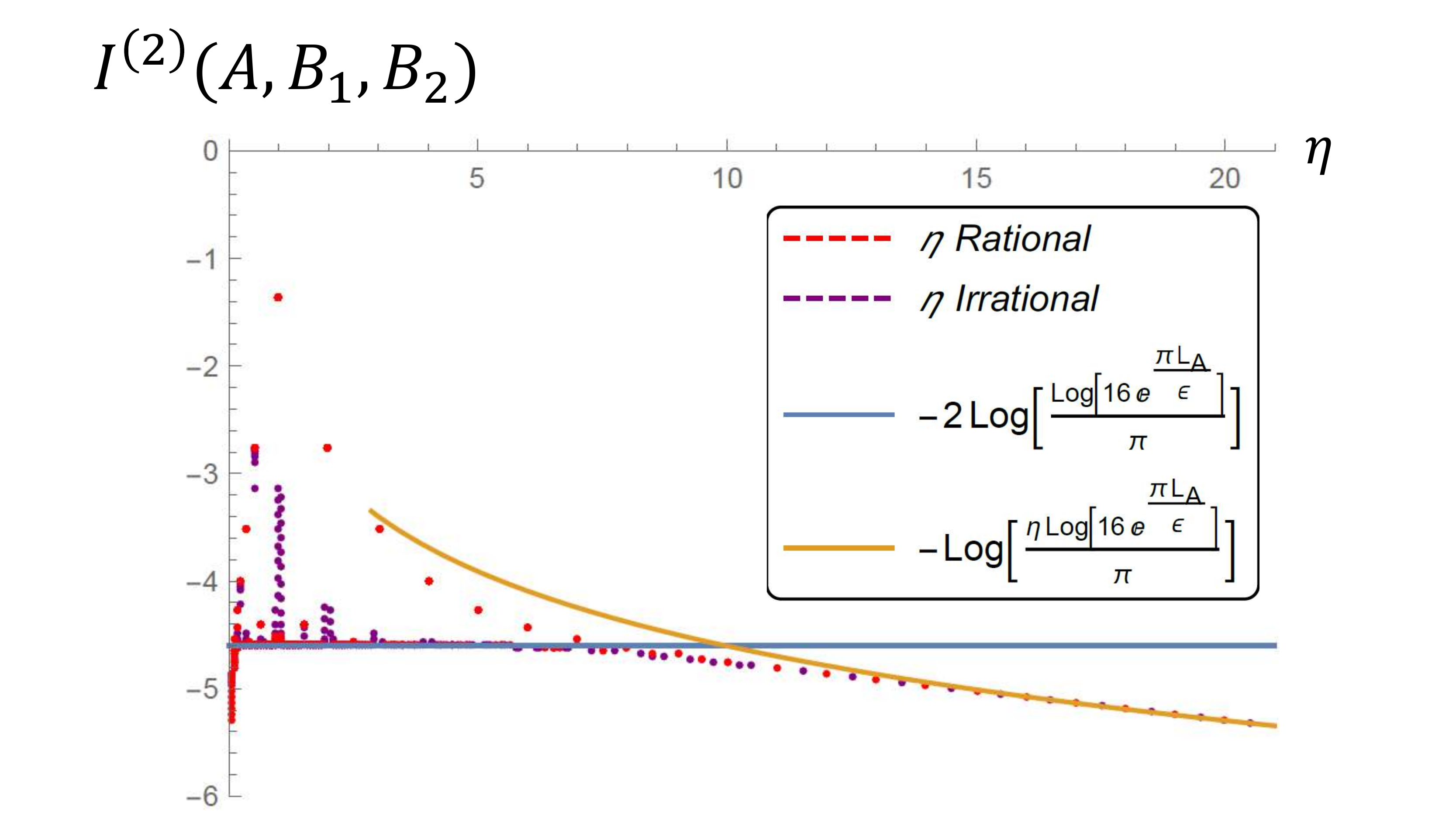}
   \caption{  
     \label{TOMI FB late time 300}
     The late-time average value obtained from averaging 
     $I^{(2)}(A,B_1,B_2)(t)$ between $t = 200$ and $300$ (Top)
     and between $t = 1000$ and $1100$ (Bottom), 
     as a function of the radius squared $\eta$.
   The red dots correspond to rational values of $\eta$ while
   the purple dots correspond to irrational values of $\eta$. The continuous
   blue curve is given by
   $-2 \log [({1}/{\pi})
     \log (16 e^{ \frac{\pi L_A}{\epsilon}}) ]$
   while the continuous orange curve is given by
   $
  - \log [({\eta}/{\pi})
        \log (16 e^{\frac{\pi L_A}{\epsilon}} ) ]$.
      }
\end{figure}


Figure \ref{TOMI FB late time 300} summarizes
the numerical extraction of the late-time behaviors of TOMI
$\lim\limits_{t \rightarrow \infty}I^{(2)}(A,B_1,B_2)(t)$ for the selected set of radii;
for rational radii we choose $p,p'  = 1, \ldots, 20$
while the irrational radii are given by $\eta = \frac{p+\sqrt{2}-9/10}{p'}$ with $p,p' = 1,\ldots, 20$.
In Fig.\ \ref{TOMI FB late time 300} (Top),
we average over $I^{(2)}(A,B_1,B_2)(t)$ at $t = 200,210,\ldots,300$,
and check if $I^{(2)}(A,B_1,B_2)(t)$ 
evaluated at these times lie within a certain range around the average
($\sim 0.2$ around the average).
This is satisfied for {\it most} radii,
i.e., for these radii
$I^{(2)}(A,B_1,B_2)(t)$ have either converged
or are close to converging at $t = 300$.
In Fig.\ \ref{TOMI FB late time 300} (Bottom),
we repeat the same procedure but with $t = 1000,1010,\ldots,1100$.
This time, $I^{(2)}(A,B_1,B_2)(t)$ evaluated at these times lie within
$0.2$ around the average
for {\it all} values of $\eta$ that we chose.
So they have either converged or are close to converging.
Those points for which the convergence is slowest
are those close to the small integer values of $\eta$ and their reciprocals,
and comparing the two graphs shows
that for most radii the data has converged by $t = 1100$.
There are also a few irrational values that are too close to the rational ones
and the numerical computation is unable to distinguish them.
For example, TOMI at
$\eta = \frac{1}{19}\left(\frac{81}{10}+\sqrt{2}\right)$ converges
to the same points as $\eta={1}/{2}$ since they are too close.
A higher precision would probably be able to distinguish these two points.

Also included in Fig.\ \ref{TOMI FB late time 300} are
the two curves,
the first of which (blue) is obtained from the (\ref{2pp'largeEtaSmall}) and sending $2 p p' \rightarrow \infty$,
$
  -2 \log \left[\frac{1}{\pi} \log
    \left(16 e^{ \frac{\pi L_A}{\epsilon}}\right) \right]
$.
The second curve (yellow) is obtained from (\ref{2pp'largeEtaLarge}) and dropping the exponenential factors with $2pp'$ and $\tilde{\eta}$ in the exponent,
$
  - \log \left[ 
        \frac{\eta}{\pi}
        \log\left(16 e^{\frac{\pi L_A}{\epsilon}} \right) \right]
$.
For small to intermediate values of $\eta$,
many points approach the blue curve as $2p p'$ increases,
which is a number independent of $\eta$.
To break out of this "barrier", $\eta$ needs to be very large or small. When $\eta$ is large, we see that they approach the orange curve.
We suspect as $L_A/\epsilon \to \infty$,
all the rational radius points ``survive'' (i.e., remain finite)
while all the irrational radius points sink to negative infinity.
(As seen from Fig.\ \ref{cross over},
in this limit, we see that $2 p p'$ and $\eta$ can always be regarded small, 
and hence when $\eta$ is rational the late-time TOMI is constant and is given by \eqref{2pp'etasmall}.)
For irrational $\eta$, on the other hand,
TOMI $\to -\infty $ as $t \to \infty$.

Before closing this section, let us make a few more comments;
First, the duality of the boson radius about $\eta = 1$ is evident in 
Fig.\ \ref{TOMI FB late time 300}. Notice that the two closest peaks to the self-dual point $\eta = 1$ occur at $2$ and ${1}/{2}$, which are reciprocals of each other. They are followed by peaks at $3$ and ${1}/{3}$ and so on and so forth. Also notice that the points close to $\eta = 0$ dip below the blue curve, just as the points that are larger than about $\eta = 6$. These points are approaching the large $\tilde{\eta} = \text{Max}\{\eta,\frac{1}{\eta}\}$ regime. We see that the large values of $\eta$ behave like the small values of $\eta$, as we expect from the duality of the boson radius about $\eta = 1$.

It is also interesting to note that the least negative late-time value of the
TOMI occurs at the self-dual radius,
where the theory acquires an emergent, larger $SU(2)$ symmetry compared with other radii 
and hence is expected to be the least chaotic compactified boson theory.

Finally,
the late-time values of $\eta = \frac{1}{6},\frac{2}{3},\frac{3}{2}$ and $6$ are
very close as seen from the graph. 
In fact, in Fig.\ \ref{TOMI FB late time 300}, 
the late-time value of $\eta = \frac{3}{2}$ and $\eta = 6$ differ by about $0.021$. This is further evidence that the late-time value of $I^{(2)}(A,B_1,B_2)(t)$ for rational radii is determined by the quantum dimension of the twist operator in the orbifolded theory. 

\subsubsection{Holographic CFTs}

\begin{figure}[t]
 \begin{minipage}{0.5\hsize}
  \begin{center}
   \includegraphics[width=70mm]{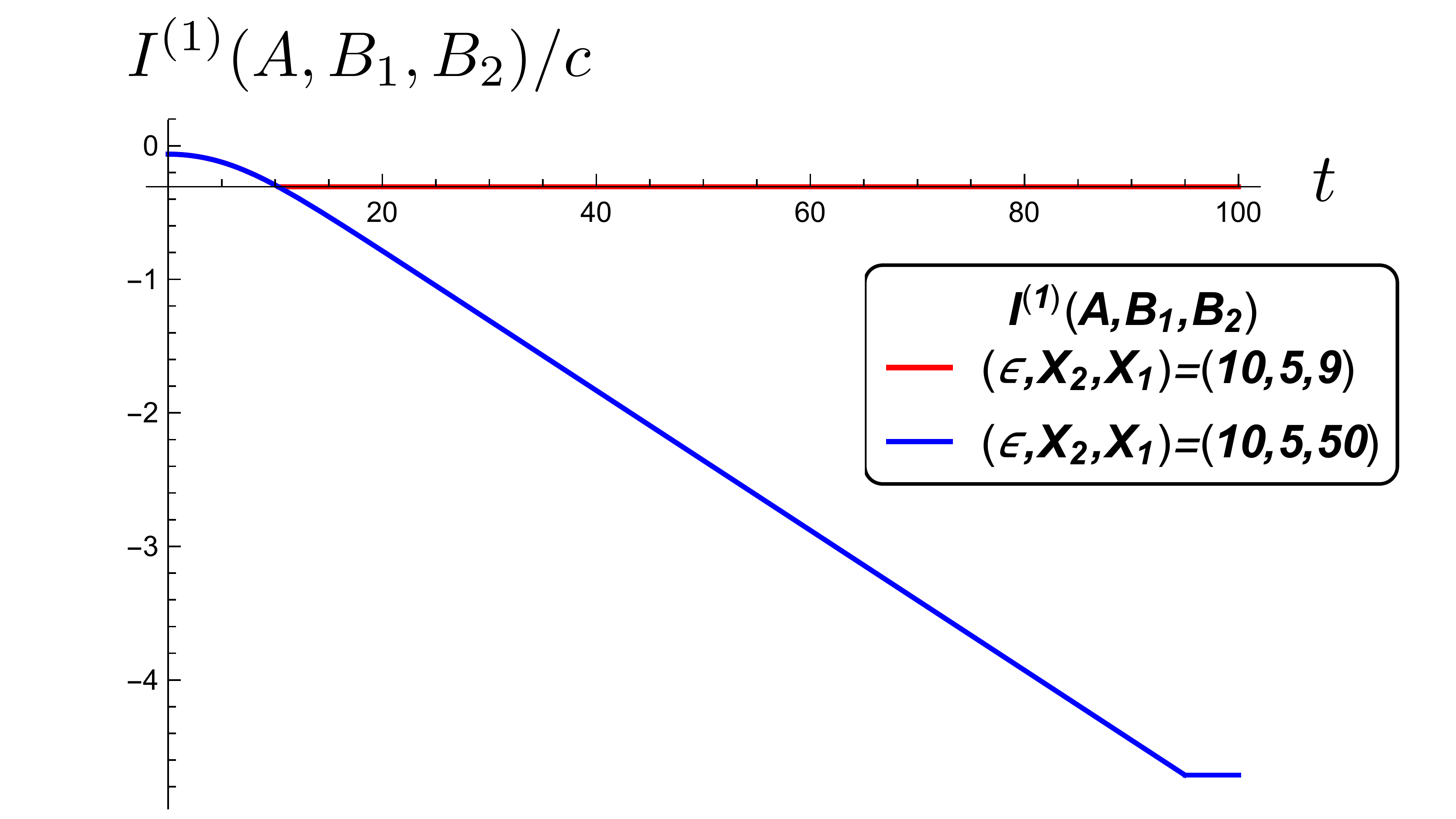}
  \end{center}
 \end{minipage}
 \begin{minipage}{0.5\hsize}
  \begin{center}
   \includegraphics[width=80mm]{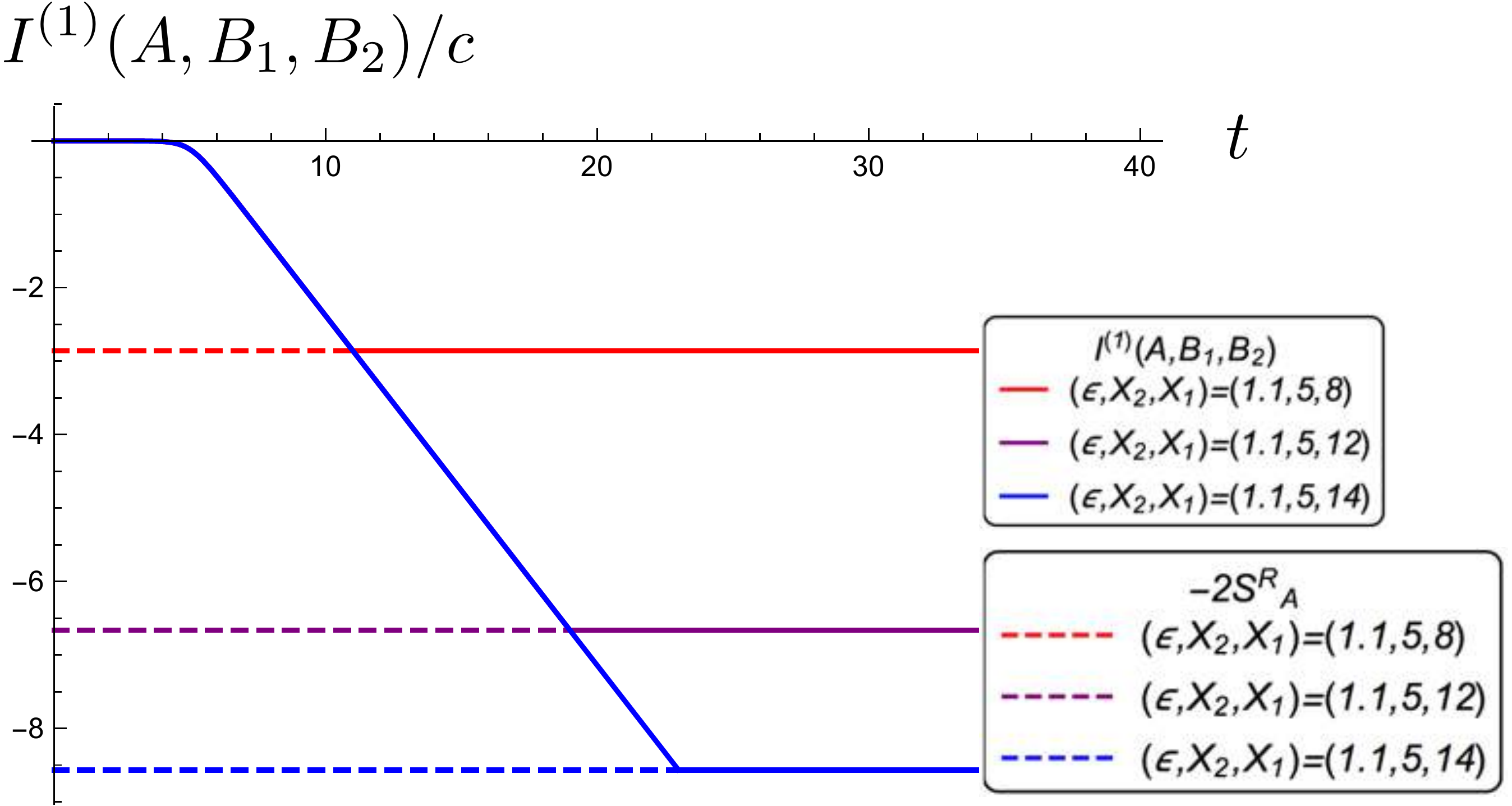}
  \end{center}
 \end{minipage}
 \caption{
   Time evolution of TOMI in holographic CFTs with infinite output subsystems. \textbf{Left:} $I^{(1)}(A,B_1,B_2)$ with $\epsilon=10$. 
   The two curves overlap with each other until $t = 10$. \textbf{Right} (same as Fig. \ref{fig:tOpMIchart} (3)): $I^{(1)}(A,B_1,B_2)$ with $\epsilon=1.1$. 
   The three curves in the left panel overlap with each other for small $t$. The dashed lines are guide to the eyes for the long time saturation values. 
   \label{comi}}
\end{figure}

TOMI for holographic CFTs is given by $ I^{(1)}(A,B_1,B_2) =$ 
$I(x_{AB_1},\bar{x}_{AB_1})+I(x_{AB_2},\bar{x}_{AB_2})-I(x_{AB},\bar{x}_{AB})$, where $x_{AB}, x_{AB_1}, x_{AB_2}$ are cross ratios defined in~\eqref{cri},
and each individual BOMI term for holographic CFTs is further given by \eqref{Holographic BOMI}.
As shown in Fig.\ \ref{comi},
TOMI $I^{(1)}(A,B_1,B_2)$ initially increases linearly in $t$,
and eventually approaches a constant which depends on the input subsystem size.
The linear behavior depends on $\epsilon$.
Here, we assume that $X_1>X_2$.
If $X_2 -Y_1' = X_2 \gg \epsilon$ ($Y_1'$ in Fig.\ \ref{fig:TOMIdetailedsetup} is taken to be zero), TOMI starts to decrease at $t= X_2$.
The late-time mutual information $I^{(1)}(A,B_1)$ and $I^{(1)}(A,B_2)$ vanish,
suggesting that we are not able to obtain information of $A$ locally from $B_1$ or $B_2$.
The late-time $I^{(1)}(A,B_1,B_2)$ approaches
to the constant,
\begin{align}\label{late time const}
  I^{(1)}(A, B_1, B_2)
  \to 
  -\f{\pi c}{3\epsilon}(X_1-X_2) =:
  -2 S_A^{R},
\end{align}
when $X_1-X_2 \gg \epsilon$.
Here, $S_A^{R}$ can be interpreted as the regulated
version of the entanglement entropy for $A$,
which is defined by
\be
\label{reg SA}
S^{R}_A = \lim_{n\rightarrow 1}
\f{1}{1-n}\log{
  \left[\f{\left \langle \sigma_n(\omega_1,
        \bar{\omega}_1)\tilde{\sigma}_{n}(\omega_2, \bar{\omega}_2)\right
      \rangle}{\left|
        {dz}/{d\omega}\right|^{4h_n}_{\omega=0}} \right]}\approx \f{\pi c}{6\epsilon} \left(X_1-X_2\right)
\ee 
in the limit $X_1 - X_2 \gg \epsilon$, where $\omega_{i}=X_i$, $z=e^{\f{\pi \omega}{\epsilon}}$. 
(The reason we have inserted $\left|{dz}/{d\omega}\right|^{4h_n}_{\omega=0}$ as the denominator is that it cancels out the factor $\big(\frac{\pi}{2\epsilon} \big)^{4h_n}$ of the numerator in~\eqref{eq:TwoPointTwist}. )
It is interesting to compare this result with the
following inequality obeyed for generic states in generic systems:
\begin{align}\label{TOMI lower bound}
I^{(1)}(A,B_1,B_2) \ge -I^{(1)}(A, B) \ge -2 Min\left[S_{A}, S_{B}\right],
\end{align}
\footnote{
  This lower boundary on TOMI can be derived as follows.
  Mutual information is written in terms of
  the relative entropy between 
$\rho_{\alpha\cup \beta}$ and $\rho_{\alpha} \otimes \rho_{\beta}$:
  \begin{align}
  I^{(1)}(\alpha,\beta)=S(\rho_{\alpha\cup \beta} ||\rho_{\alpha} \otimes \rho_{\beta})=\Tr_{\alpha\beta}\left[\rho_{\alpha\cup \beta}\left\{\log{\rho_{\alpha \cup \beta}}-\log{\left(\rho_{\alpha} \otimes \rho_{\beta}\right)}\right\}\right] \ge 0.
  \end{align}
  Due to the Araki-Lieb inequality, $S_{\alpha\cup \beta}\ge \left|S_{\alpha}-S_{\beta}\right|$, hence $I^{(1)}(\alpha,\beta)$ has a upper bound :
  \begin{align}
    I^{(1)}(\alpha,\beta) \le 2 Min\left[S_{\alpha}, S_{\beta}\right]
    \end{align}
    The inequality \eqref{TOMI lower bound} then follows.
}
where on the right hand side $S_A$ (or $S_B$) appears
instead of its regularized counterpart.
In quantum field theory context, 
the right hand side of the inequality
\eqref{TOMI lower bound}
is dependent on UV cutoff,
while the left hand side is not.
Hence, 
the bound \eqref{TOMI lower bound} itself is not 
very meaningful in quantum field theories. 
Nevertheless,
in quantum field theories
(or a suitable set of many-body quantum channels)
we conjecture that
the value $-2 S^R_A$ provides the lower bound of TOMI,
\begin{align}
I(A,B_1,B_2) \ge -2 S_A^{R},
\end{align}
and holographic channels saturate this lower bound.

\section{Discussion and future directions}
\label{Conclusion}

In this paper, 
by introducing a state dual to a unitary evolution operator $U(t)$
of $(1+1)$d CFTs (which is a time-dependent thermofield double state),
we have studied quantum correlation between input and output subsystems
of $U(t)$.
In search for specific measures of quantum correlation,
we computed the bi-partite and tri-partite operator mutual information
(BOMI and TOMI)
for various choices of input and output subsystems.
For the free fermion and compactified boson CFTs,
we found that the time-dependence of BOMI can be interpreted in terms of
the relativistic propagation of information-carrying quasi-particles.
The time-dependence of BOMI in these cases shows 
bumps or peaks that can be interpreted by the quasi-particle picture
and are indicative that we are able to obtain information of input subsystem by doing local measurements on the output subsystem.
On the other hand, for holographic CFTs, 
we have found that the time evolution of BOMI does not show any bumps and cannot be interpreted as the propagation of
local objects such as quasi-particles.
In particular, when the input and output subsystems do not overlap (disjoint case), BOMI vanishes at all times, 
indicating that one cannot mine all the information of input subsystem from merely doing local measurements on the output subsystem,   
i.e., the information scrambling effect. 

In order to measure the amount of information that is scrambled under the
unitary time evolutions, we have studied TOMI. 
It is negative when the amount of information we can get locally
is smaller than that sent from the input subsystem.
Since the free fermion channel does not scramble the information
from the initial input subsystem, TOMI vanishes.
Once again, the time-dependence of TOMI the free fermionic CFT is
interpreted in term of the relativistic propagation of local objects.

While the second R\'{e}nyi BOMI for the compactified boson theory
can be interpreted by the quasi-particle picture,
the difference between the compactified bosons and the free fermion theories appear in TOMI. 
When the output $B$ consists of the entire 1d space, 
the second R\'{e}nyi TOMI of the compactified boson theory saturates
at late times to a constant negative value
which dependes on the radius squared $\eta = R^2$, and the size of the input subsystems.
We take this to mean that the compact free boson CFT exhibits moderate scrambling.
Finally, we found that local correlation for holographic channels vanishes at late times, 
and the late-time TOMI 
approaches $-2S_A^{R}$,
which is a regulated entanglement entropy and is conjectured to be a lower bound for TOMI. We speculate that in general, the late-time value of TOMI for maximally chaotic channels saturates this bound.

\paragraph{Comparison with OTOC}

While our finding is indicative that TOMI can distinguish different
classes of systems
with different degrees of information scrambling and chaotic behaviors,
a question remains regarding to the precise connection between
BOMI and TOMI and other indicators of scrambling and chaos, such as OTOC. 
Reference \cite{2016JHEP...02..004H} shows the relation  between TOMI and 
OTOC averaged over the complete set of operators on input and output subsystems.
In CFTs, OTOC and BOMI/TOMI are both computed from four-point correlation functions.
Besides the choice of operators entering into these correlation functions,
OTOC and BOMI/TOMI differ by the behavior of the cross ratios as we send $t\to \infty$.
For OTOC, the cross ratio (during analytically continuing to the real time)
follows a closed trajectory, enclosing $x=1$.
For rational CFTs, the late-time behaviors of OTOCs are then related to the quantum dimensions
\cite{2016JHEP...08..129G,2016PTEP.2016k3B06C}.
This is also the case for four point functions entering
in the calculation of relative entropy
\cite{2018JHEP...07..002N}.
On the other hand, for BOMI/TOMI, the relevant cross ratios show rather trivial
behaviors;
they are confined on the real axis, and do not follow closed
path as $t\to \infty$.
Nevertheless, we have found that, at least for the example studied here,
i.e., the compacified free boson at rational radii, 
the late-time saturation values of TOMI are also given by the quantum dimensions. 
In addition, Ref.\ \cite{2017PhRvD..96d6020C}
studied the time-evolution of OTOCs in the orbifolded compacified
free boson theory at various radii, and identified,
for irrational radii, 
a new polynomial decay of the correlators that is a signature of an intermediate regime
between rational and chaotic models.
This is in harmony with our findings discussed in Sec.\ \ref{The compactified boson theory TOMI}.
However, again, the precise connection between OTOC and BOMI/TOMI remains elusive.

\paragraph{Toy model for holographic channel:}

While the quasi-particle picture works satisfactorily
to describe the dynamics of BOMI
for the free fermion and compacified boson CFTs,
it completely fails for holographic CFTs.
The information sent from $A$
through integrable channels is expected to
be ``localized'' (i.e., can be described by propagations
of point-like quasi-particles)
but the information for holographic channels
is expected to be delocalized due to
its complex dynamics.


\begin{figure}[htbp]
	\begin{center}
	 \begin{tabular}{l}
	 \begin{minipage}{0.50\hsize}
	  \begin{center}
	   \includegraphics[scale=0.25]{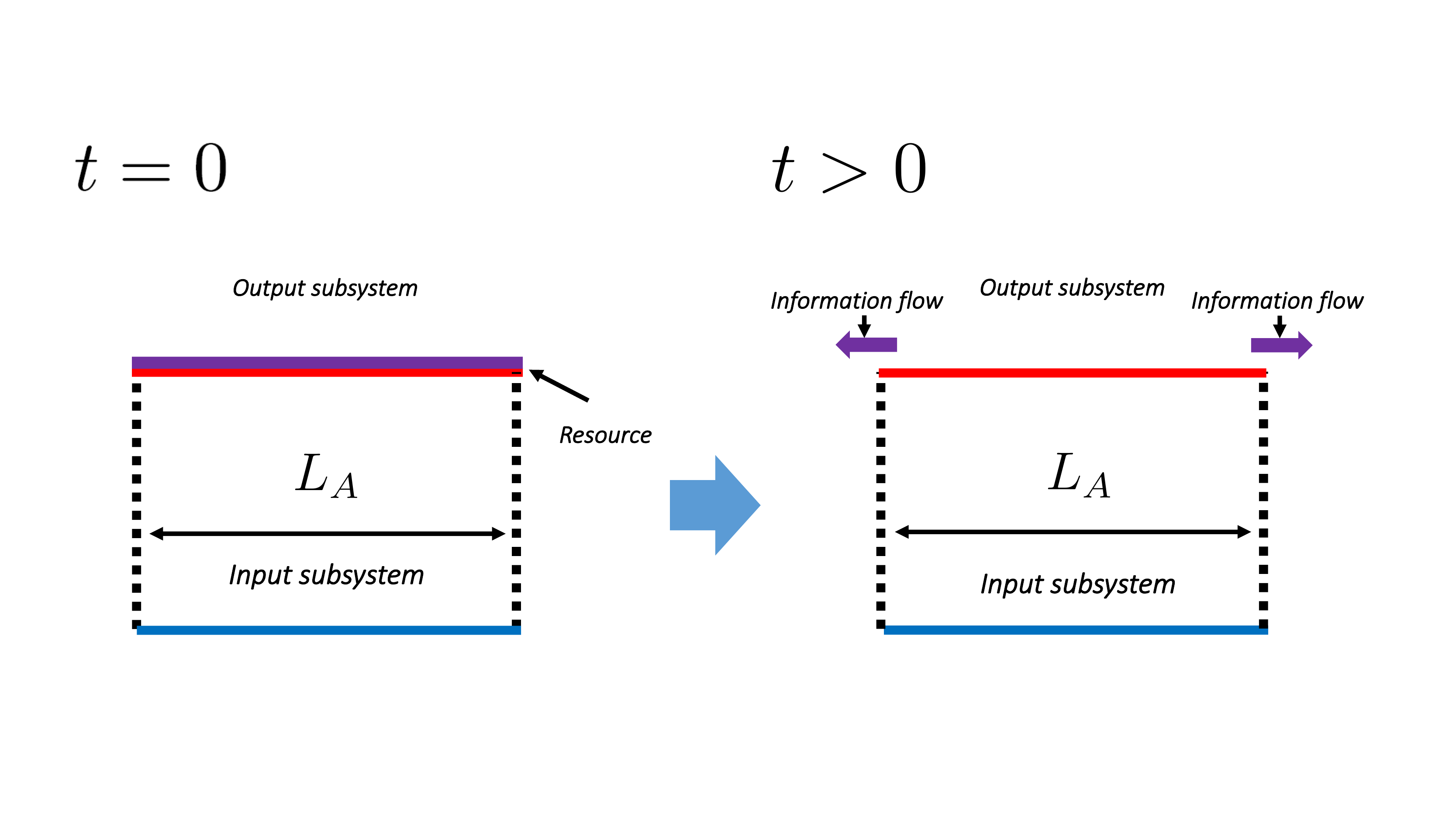}
	  \end{center}
	 \end{minipage}
   \\
   \begin{minipage}{0.5\hsize}
	  \begin{center}
	   \includegraphics[scale=0.25]{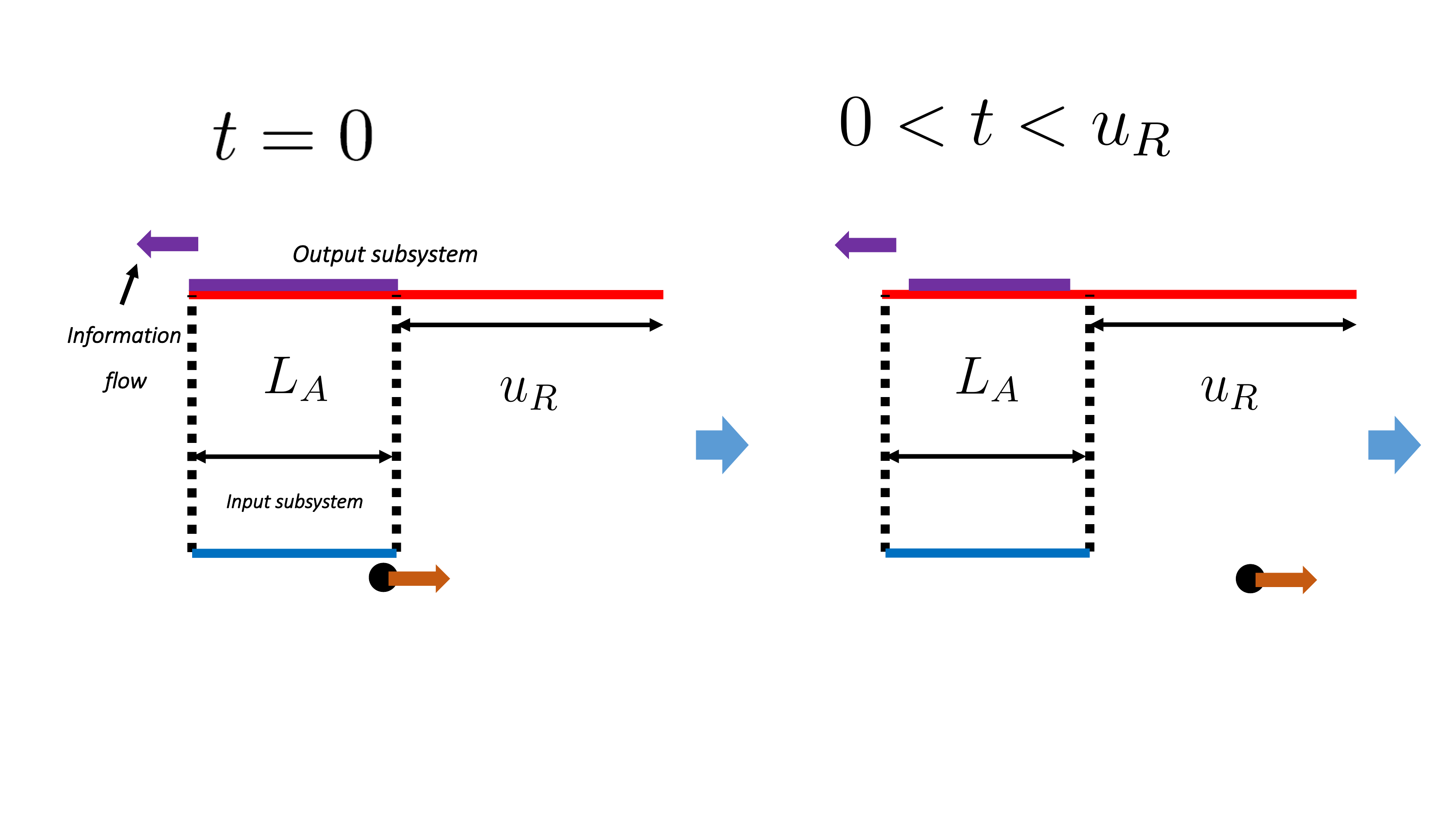}
	  \end{center}
  \end{minipage}
     \begin{minipage}{0.5\hsize}
	  \begin{center}
	   \includegraphics[scale=0.25]{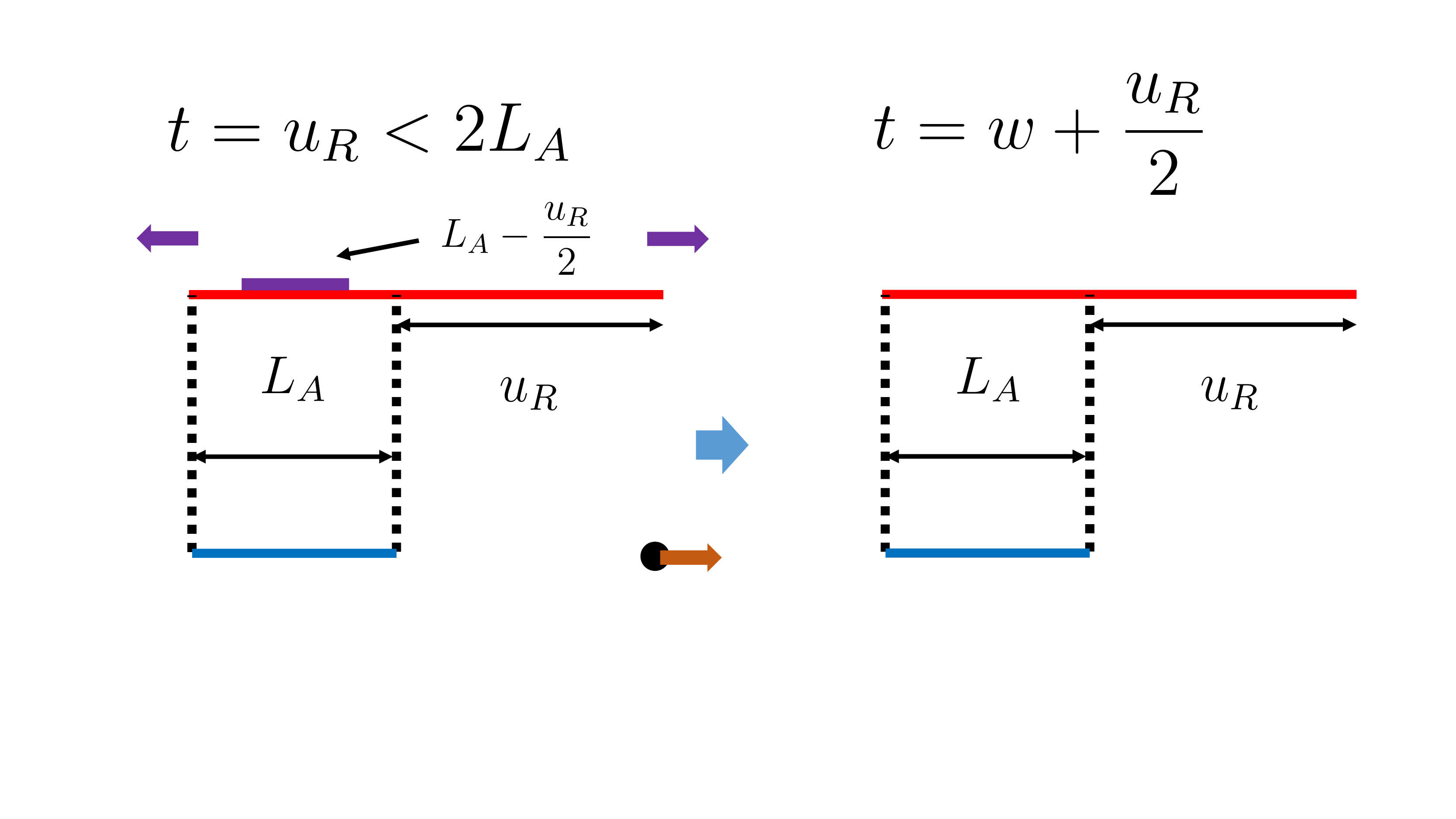}
	  \end{center}
	 \end{minipage}
	 \\
	 \begin{minipage}{0.5\hsize}
	  \begin{center}
	   \includegraphics[scale=0.25]{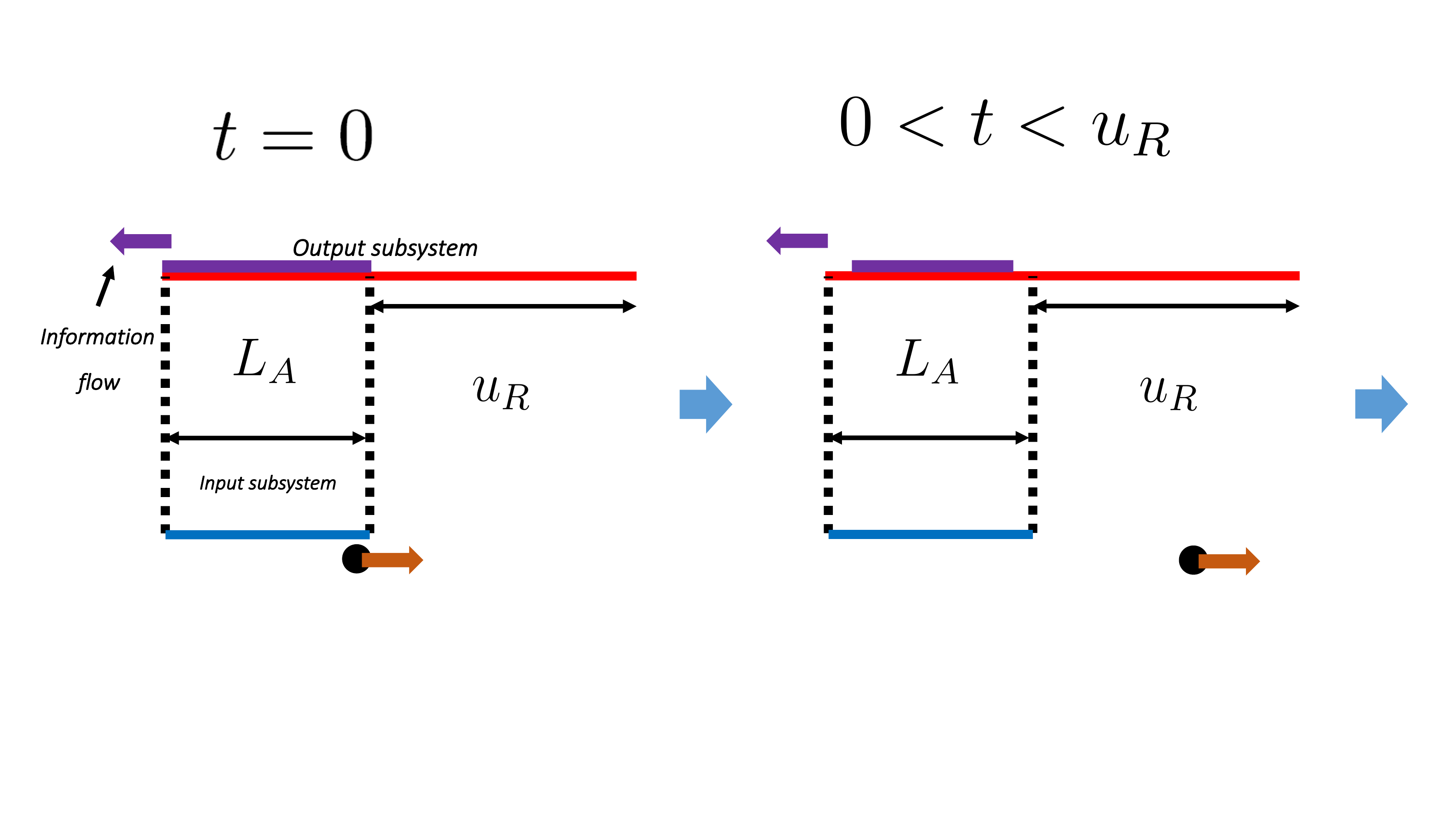}
	  \end{center}
	 \end{minipage}
	  \begin{minipage}{0.5\hsize}
	  \begin{center}
	   \includegraphics[scale=0.25]{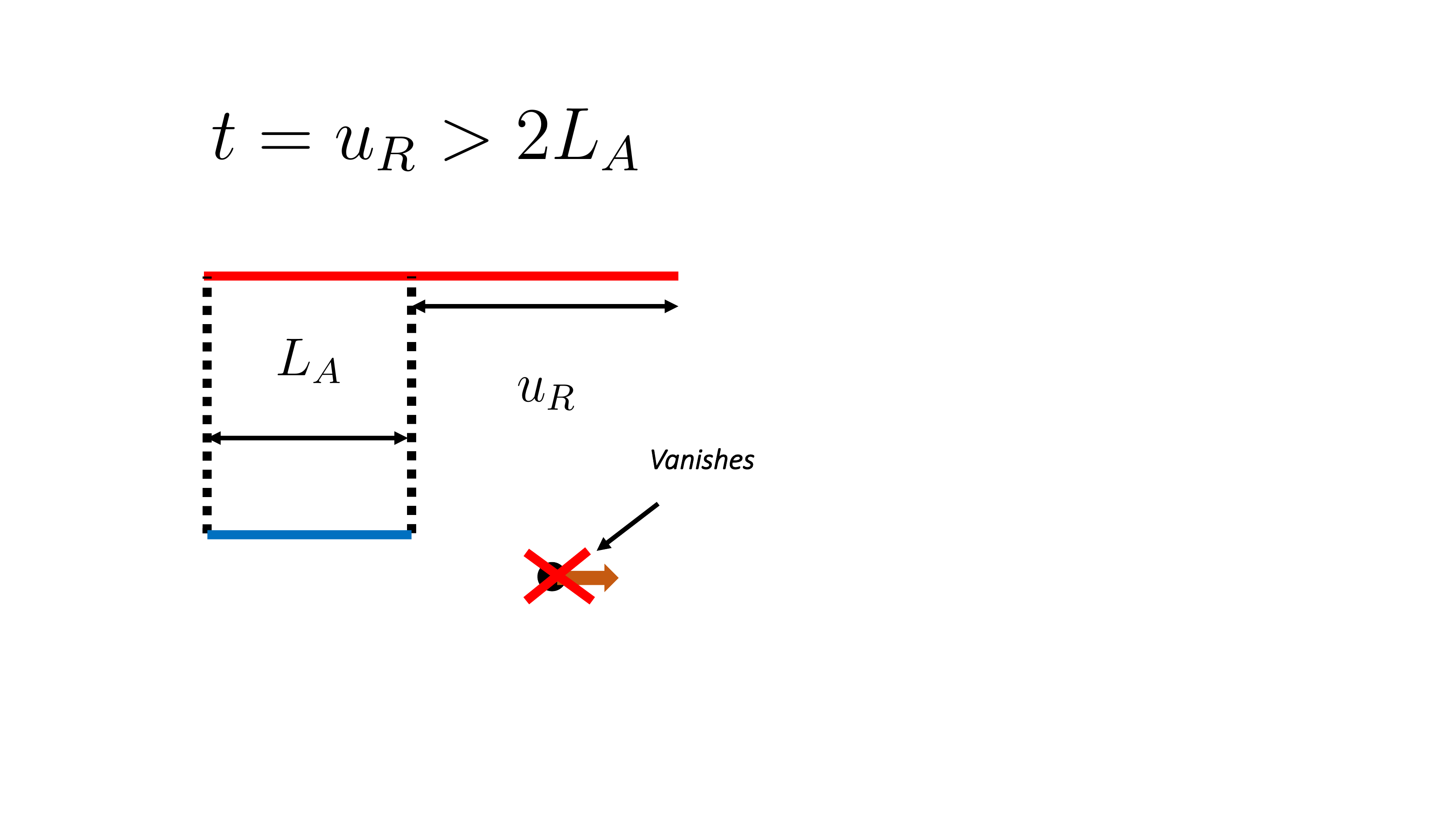}
	  \end{center}
  \end{minipage}	\end{tabular}
\caption{
  \label{pictures} 
  An empirical model describing holographic channels for the configuration corresponding to
  the purple (top row), red/blue (middle row), and green (bottom row) curves in Fig.\ \ref{BOMI chaotic vs integrable perfectly overlapping} left panel.
 As depicted in the panels of the middle row, signals are sent from the boundaries of input subsystem.
  The right and left boundaries send the right- and left-particles with the
  speed of light, respectively.
  Once the particle arrives at the boundary of output subsystem,
  information flows to out of the output subsystem until the information
  contained in the output subsystem vanishes.
 As in the right panel of the bottom row, signals disappear if the output subsystem
  empties before signals arrive at the boundary of output system.
}
\end{center}
\end{figure}

While the quasi-particle picture should not work for holographic CFTs,
one could expect that an alternative, more ``hydrodynamic" picture 
may be applicable to describe the dynamics of
quantum information in holographic channels
(See e.g., 
\cite{
  2018arXiv180300089J,
  2014PhRvL.112a1601L,
  2014PhRvD..89f6012L,
  2015PhRvD..92l6004L,
  2018arXiv180409737Z,
  2018PhRvX...8c1057K,
  2018PhRvX...8b1013V,
  2018arXiv180310244M,
  2017JHEP...05..064M,
  2017JHEP...05..065M,
  2016JHEP...07..077C,
  2018PhRvX...8b1014N, 
  2017PhRvX...7c1016N}).
Currently, we do not have a precise formalism for this.
Nevertheless, motivated by this ``delocalized'' picture
for holographic channels
we consider the following empirical/phenomenological model
where the ``delocalized'' objects carry quantum information.

Our model,
which describes how information is processed from the input
  subsystem to the output subsystem (See Fig.\ \ref{Fig:unitary}),
consists of two ingredients,
{\it resource} and {\it signal}.
We assume the input subsystem at $t=0$
has some quantum information theoretic entity.
%
%
For the lack of better name, we simply call it quantum resource.
The quantum correlation between the output and input subsystems
is given by the amount of resource stored in the output subsystem. 
In particular, if the output subsystem does not overlap with the input subsystem,
BOMI between them vanishes.
Once the resource goes out of the output subsystem, the correlation between the input and output subsystems decreases.

On the other hand,
signals are particle-like objects,
which propagate at the speed of light.
Signals by themselves
do not contribute to the quantum correlation,
but determine when the resource starts to leave (leak)
from boundaries of the output subsystem.
Once a signal hits a boundary of
the output subsystem,
the resource starts to leave from the boundary of the output subsystem.
(We will estimate the precise rate of how resource escapes from
the output subsystem below.)
If all resource in the output subsystem is lost
before the signal hits the boundary, the signal disappears.
On the other hand, if the single hits
(one of) the boundaries before all resource in
the output subsystem is lost,
resource starts to leak from the boundaries hit by the signal. 
Thus, point-like objects carrying quantum information such as quasi-particles in our toy model are not able to describe some properties of holographic channel.

Let us apply the above simple model 
for the situations depicted
in Figs.\ \ref{BOMI chaotic vs integrable perfectly overlapping},
and \ref{pictures},
where the input subsystem is included in (equal to)
the output subsystem.
\begin{itemize}


\item
  First,
  the amount of resource in the output system at $t=0$ is
  proportional to, when $L_A$ and $L_B$ are both
  sufficiently larger than
  $\epsilon$, the input subsystem size.  

\item
  The input subsystem sends signals at $t=0$.
  The left and right boundaries of input subsystem send
  left- and right-moving signals with the speed of light without costing any
  resource.
  When the signal hits the boundary of output subsystem,
  quantum information or resource flows from the boundary
  to the outside of the output subsystem with constant rate, $R_{\text{IF}}$.
  This rate can be estimated from the slope in the purple configuration
  in Fig.\ \ref{BOMI chaotic vs integrable perfectly overlapping},
  \begin{align}
    R_{\text{IF}} = \f{I^{(1)}(A, B; t=0)}{2L_A} \approx \f{\pi c}{6 \epsilon},~~L_A \gg \epsilon,
  \end{align}
  where $I^{(1)}(A,B; t=0) \approx \f{\pi c }{3 \epsilon}L_A$ in the limit, $L_A \gg \epsilon$.

\item
  For the asymmetric configuration as
  in Fig.\ \ref{pictures},
  if the other signal hits the other boundary of output subsystem before all
  quantum resource is lost, quantum information flows from the boundary to out of
  the output subsystem with $R_{\text{IF}}$.
  Therefore, once this happens, BOMI decreases twice as fast as before. 
  If all quantum resource is lost before the other signal hits the boundary of output subsystem, the signal disappears. 
  (Fig.\ \ref{pictures}.)

\item The signal exists before it hits the output boundary or all resources in the output subsystem are used.
\end{itemize}

\begin{figure}[t]
  \begin{center}
   \begin{tabular}{c}
	 \begin{minipage}{0.50\hsize}
	  \begin{center}
	   \includegraphics[scale=0.25]{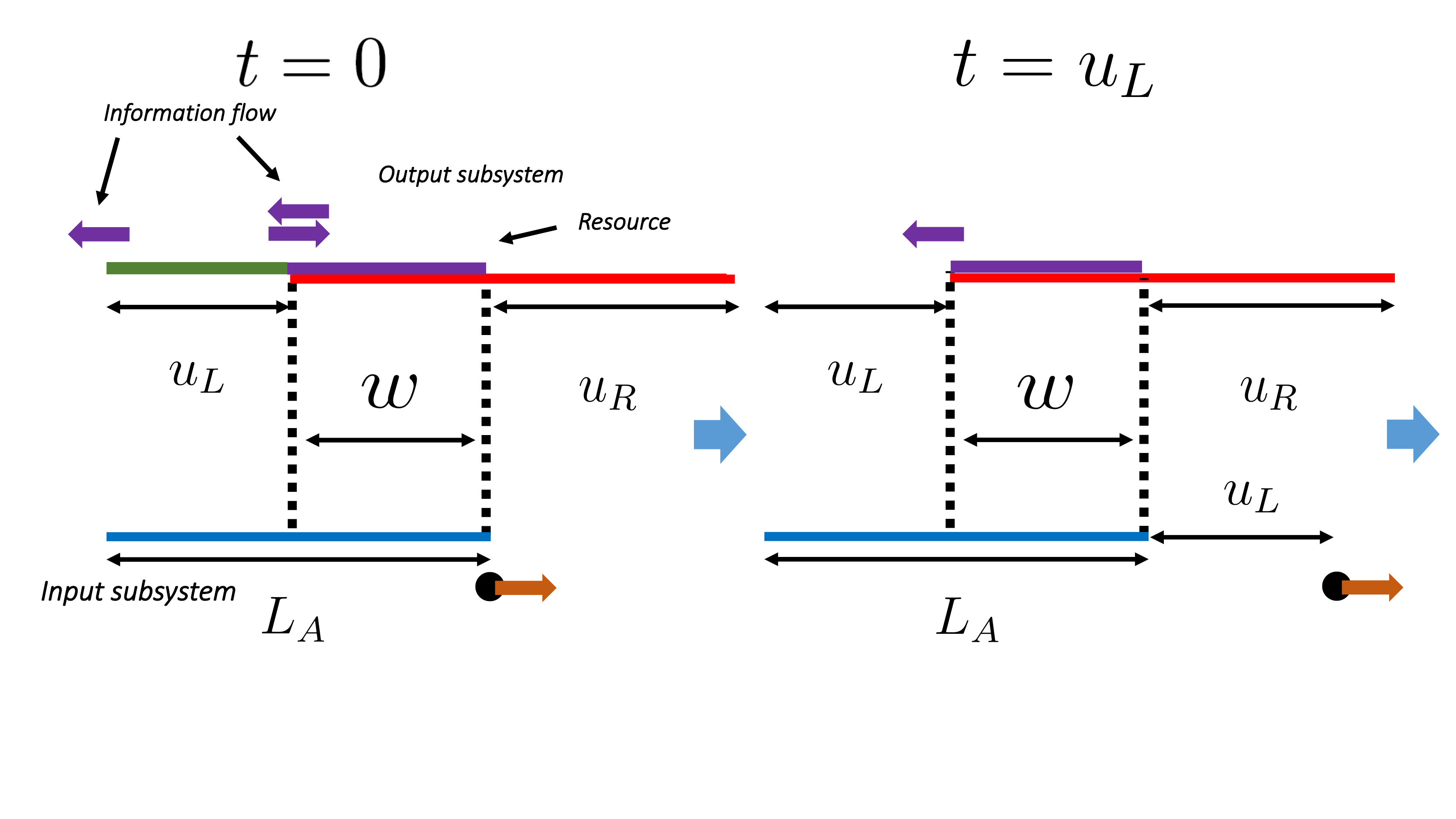}
	  \end{center}
	 \end{minipage}
	  \begin{minipage}{0.5\hsize}
	  \begin{center}
	   \includegraphics[scale=0.25]{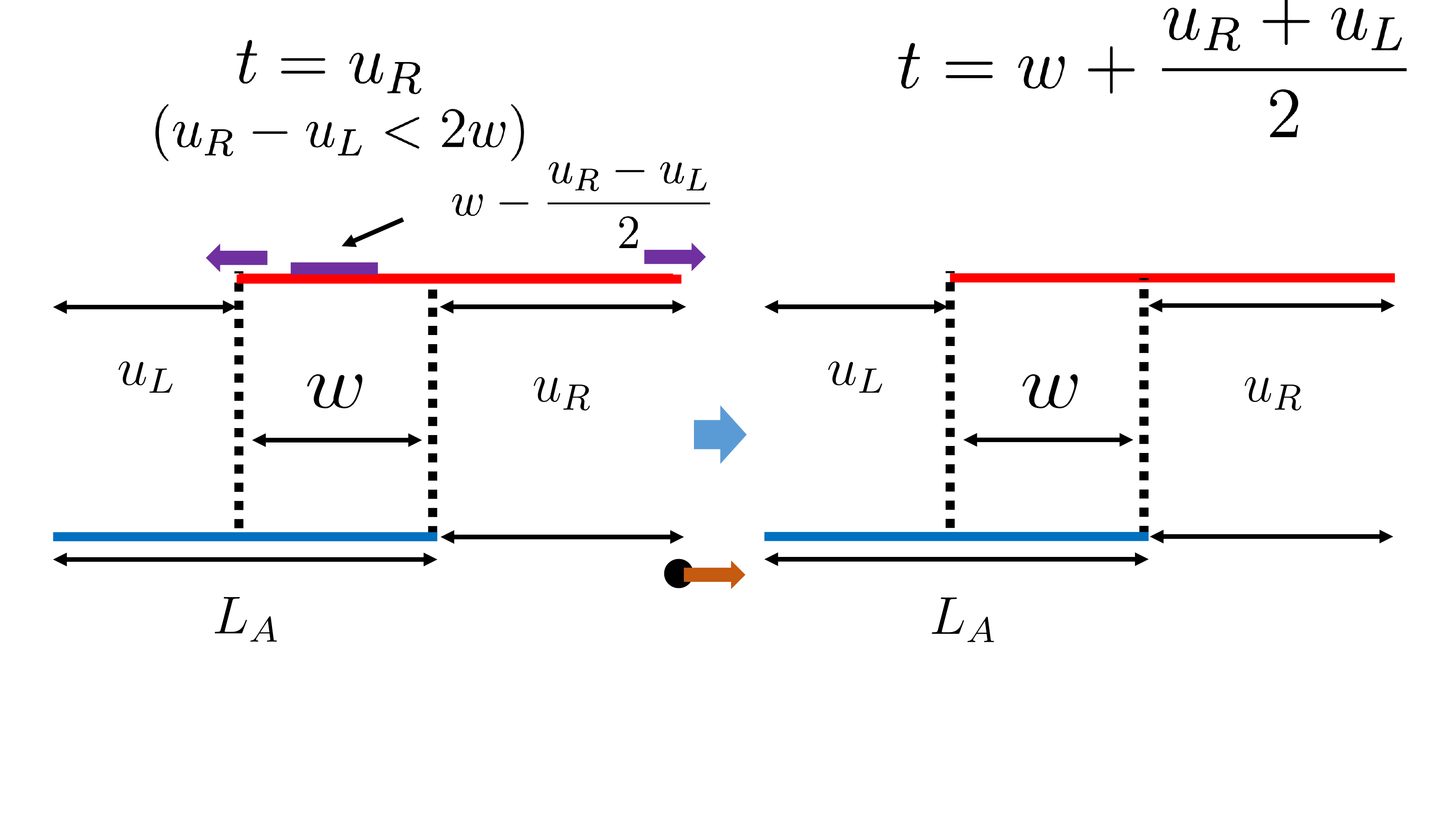}
	  \end{center}
	 \end{minipage}
	 \\
         \begin{minipage}{0.5\hsize}
	  \begin{center}
	   \includegraphics[scale=0.24]{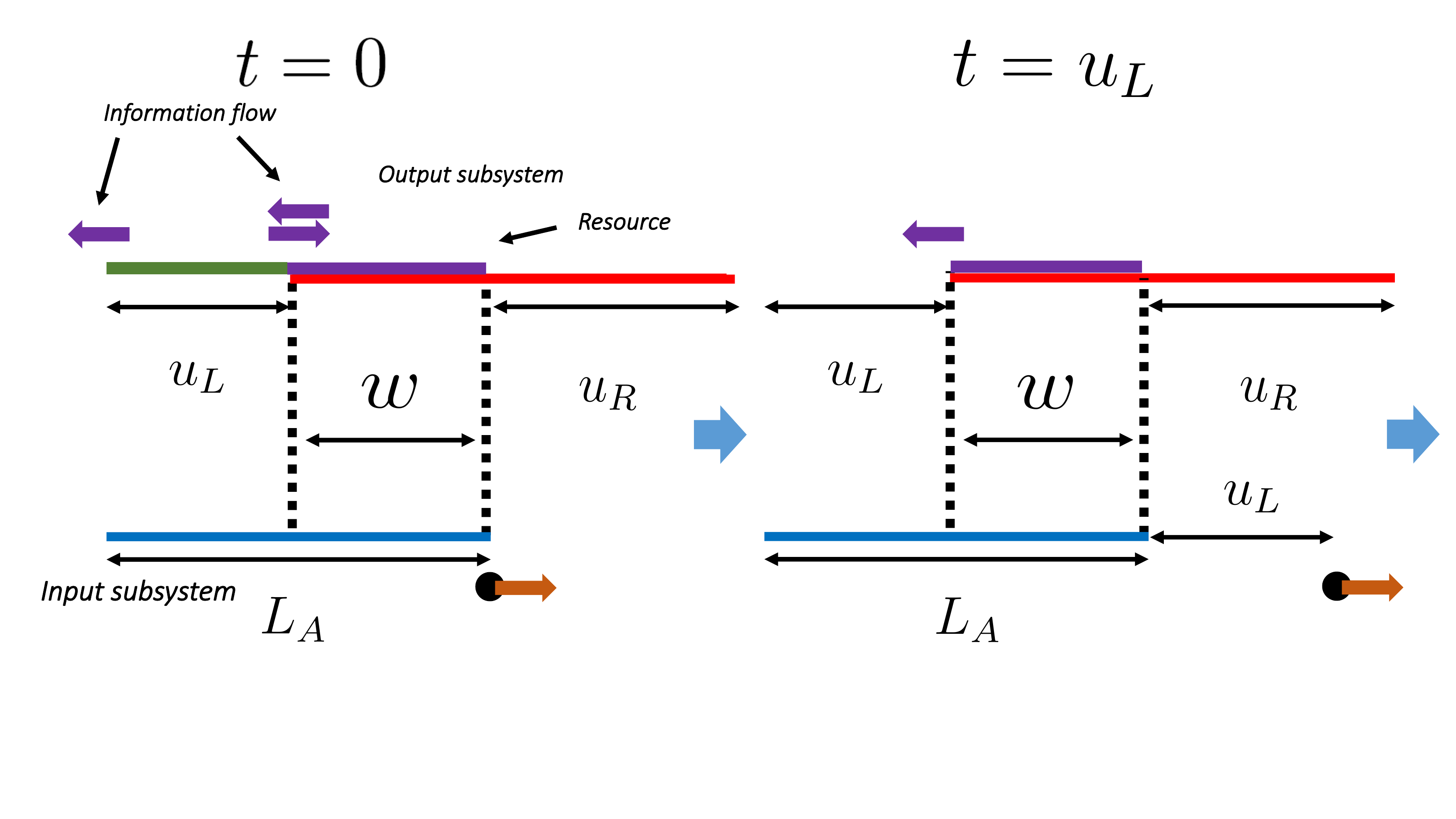}
	  \end{center}
	 \end{minipage}
	 \hspace{-3mm}
	  \begin{minipage}{0.5\hsize}
	  \begin{center}
	   \includegraphics[scale=0.24]{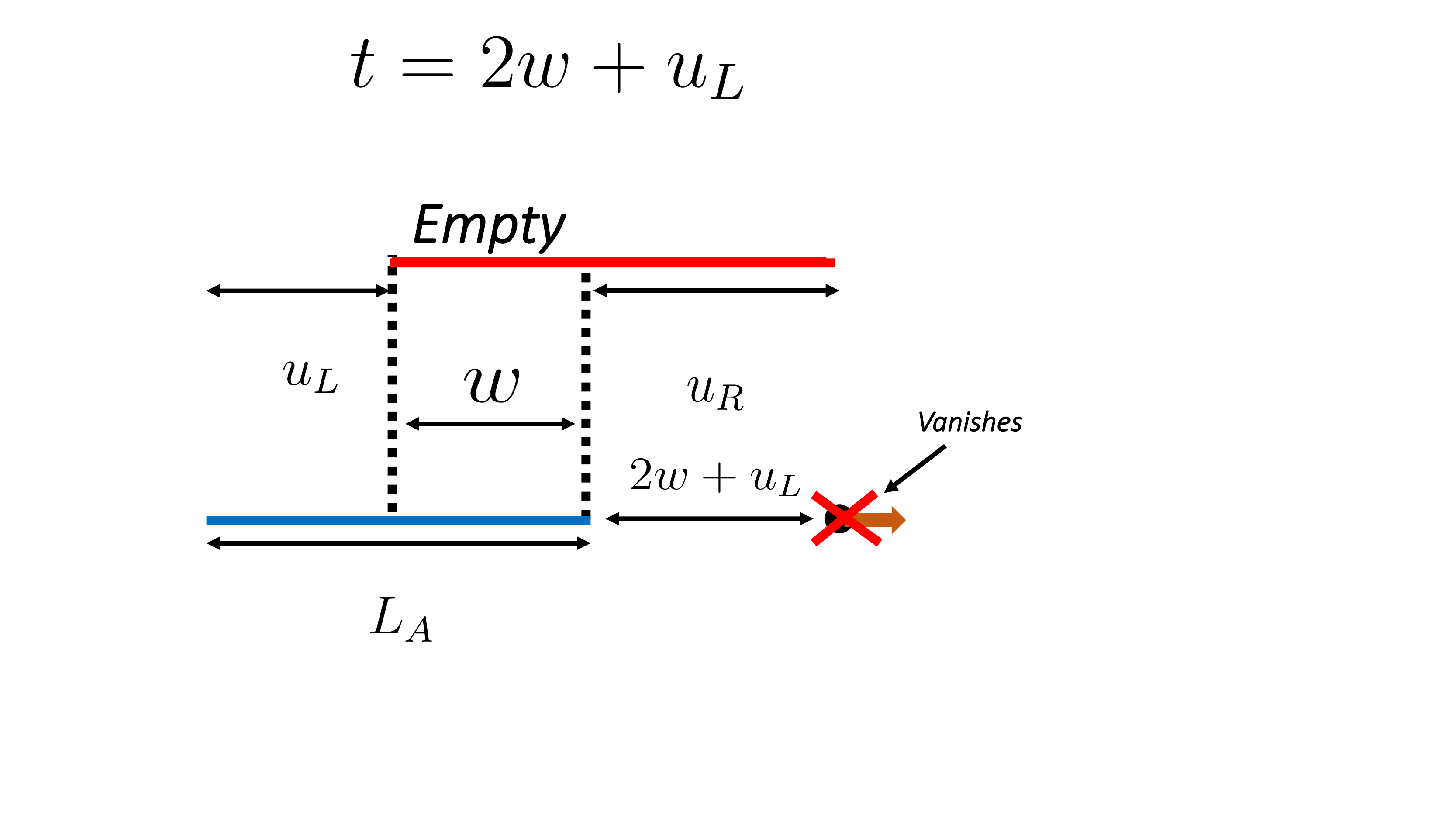}
	  \end{center}
	 \end{minipage}
\end{tabular}
   \caption{\label{Comp2}
The empirical model describing holographic channels for the configurations corresponding to the green (top row) and black (bottom row) curves in Fig.\ \ref{BOMI hol overlapping} left panel. 
The red curve ($L_A = L_B = 20, u_L = u_R = w = 10$) in Fig.\ \ref{BOMI hol overlapping} left panel is a special case of the top row due to its ``symmetric" configuration.
}
  \end{center}
\end{figure}

This model can be also applicable to 
the setups depicted in Fig.\ \ref{Comp2}.
There, we assume that the output and input subsystem are
$A =A_1 \cup A_2$ and $B=A_2 \cup B_2$, and 
$A_2$ is the region where the input subsystem partially overlaps the output subsystem.  
In this case: 
\begin{itemize}

\item
  The resources at $t=0$ are included in $A_2$ and $B_2$.
  The resources in $A_2$ contribute to BOMI at $t=0$.
  The resources in $A_2$ flow from each boundary of $A_2$. The in- and out-flows from the boundary attaching $A_1$ cancel out each others. 

\item

  Some of resources out of the output subsystem flow out of the input or output
  subsystem. Some of them flow to the output subsystem at its boundary which
  connects the subsystem to the input subsystem.
  Some of the resources in the output subsystem flow out of the subsystem.
\item
  Once the information in the output subsystem flows out of the system, it does not contribute to BOMI. A signal is created at the boundary connecting the input subsystem to the output subsystem simultaneously.  

  Since the flow to the output subsystem cancels out the flow out of the system, BOMI does not change before all resources in $A_1$  are used or the particle created at the boundary hits the other.

\item Once the resources flow out of the output subsystem,
  it cannot contribute to BOMI again.

\item
  If all resources in $A_2$ are lost before all resources in $B_2$ are lost, the residual resources cannot contribute to BOMI. Thus, BOMI between the input and output subsystem vanishes.
\end{itemize}

Finally, we observe that
BOMI even for the chaotic channel is invariant under the transformations, $t\rightarrow -t$, or $L_A \leftrightarrow L_B$, or $X_1\leftrightarrow Y_1$ combined with $X_2 \leftrightarrow Y_2$.

%
%
%

\paragraph{Future directions}

Finally, we close by listing a few future directions:
\begin{itemize}
\item
  While we have studied the operator entanglement of unitary time-evolution
  operators, it would be also interesting to study the operator entanglement of
  local operators
  \cite{2017JPhA...50w4001D,2018arXiv180804889M,2018arXiv180408655C}.
  In particular, the expectation values of local operators play
  an order-parameter-like role in eigenstate thermalizaion hypothesis (ETH).
  Since ETH is expected to be related to scrambling,
  BOMI and TOMI for local operator may shed lights on the
  connection between scrambling and ETH. 

\item
  The state dual to holographic channels is the time-dependent eternal black hole
  where holographic complexity is defined \cite{LS1,LS2,LS3,LS4}. 
The black hole is the gravity dual of the operator-dual state in (\ref{op}). Therefore, since the complexity might capture some properties of unitary time-evolution operator, we expect linear growth of complexity in $t$ to be interpreted in terms of the operator entanglement.  


\item
  The authors in \cite{tw1,tw2} have studied the relation between the
  traversability of a wormhole and the double trace deformation.
  Turning on the double trace deformation, two exteriors in the eternal black
  hole are able to communicate.
  We expect traversability of wormhole to be interpreted in terms of operator entanglement. 

\item 
  Finally,
  operator entanglement may also shed some light on
  the mechanism of holographic duality.
  For example, in Ref.\ \cite{2018arXiv180809072T},
  which proposed that arbitrary codimension one surfaces
  in AdS can be regarded as quantum circuits,
  it was argued that
  when the input and output subsystems are identical,
  the linear decrease of the operator entanglement in time 
  is (minus) the area of time-like surface in AdS space. 

  In addition,
  operator entanglement can be studied in tensor network models
  of holography. For example,  
  the tensor network structure of real-space renormalization,
  (continuous) multi-scale entanglement renormalization ansatz (cMERA),
  is expected to be related to the bulk geometry 
  \cite{CM1,CM2,CM3,CM4}.
  In cMERA, we construct the target state by acting with a series of unitary
  operators at the energy scale on the unentangled state.
  The correlation between the UV-subsystem and the subsystem at lower energy
  scale might be related to how the local information on the boundary is
  encoded in the bulk geometry.
  Therefore, it is one of the interesting future directions to study the mutual information in the state dual to the cMERA channel.

\end{itemize}

\paragraph{Acknowledgement}
We thank useful discussions with
Xiao Chen, Veronika E. Hubeny, Mukund Rangamani, Tadashi Takayanagi, Xiaoliang Qi, Pavan Hosur, Maissam Barkeshli, Meng Cheng, Anna Keselman, Victor Galitski, and Steve Shenker.
SR is supported by a Simons Investigator Grant from the Simons Foundation. LN is supported by the Kadanoff Fellowship from the University of Chicago.

\appendix
\section{Holographic computation}
\label{Holographic computation}
\begin{figure}[tbp]
 \begin{minipage}{0.5\hsize}
  \begin{center}
  \includegraphics[clip,width=6.0cm]{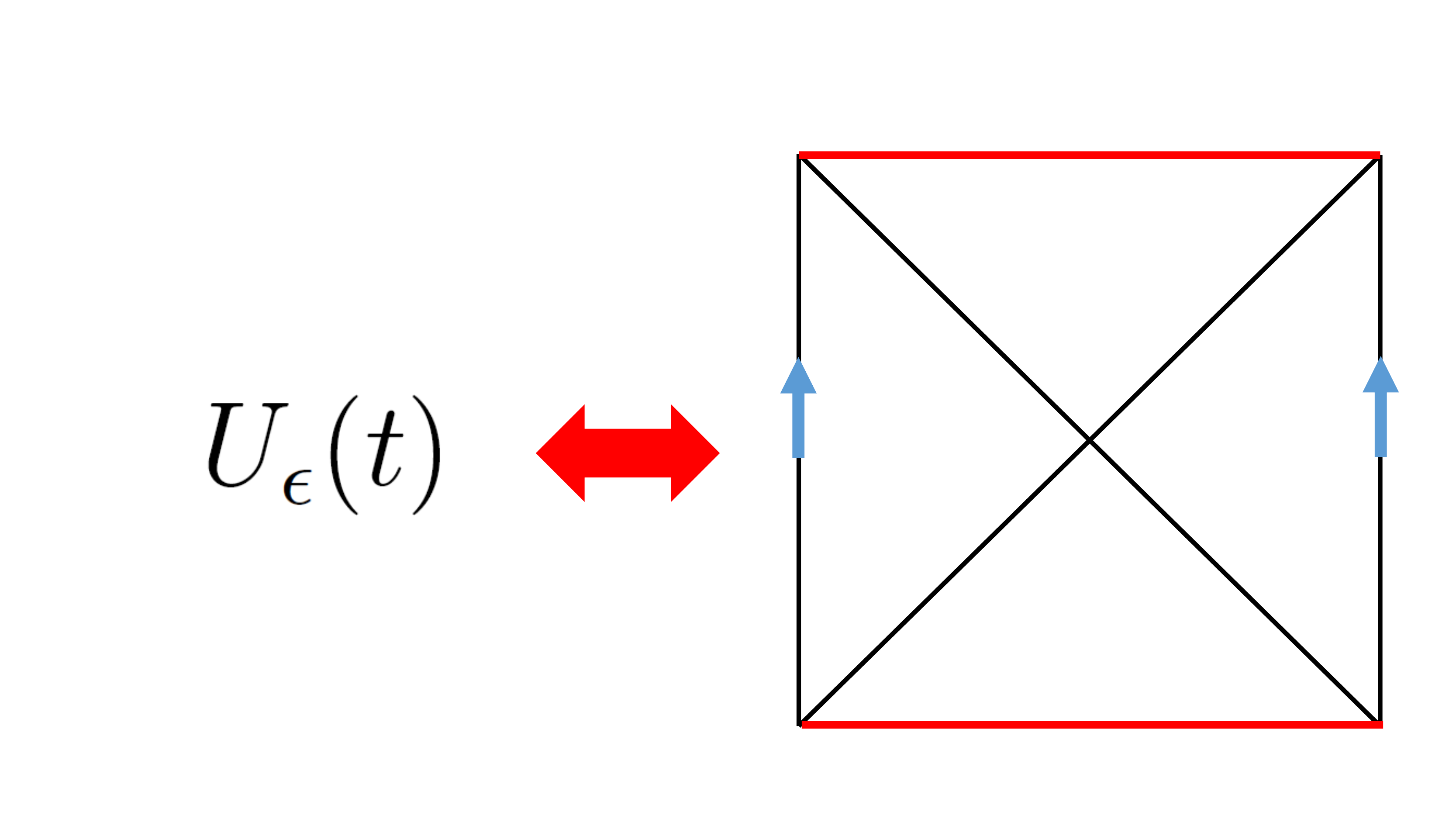}
  \end{center}
  \label{fig:one}
 \end{minipage}
 \begin{minipage}{0.5\hsize}
  \begin{center}
\includegraphics[clip,width=8.0cm]{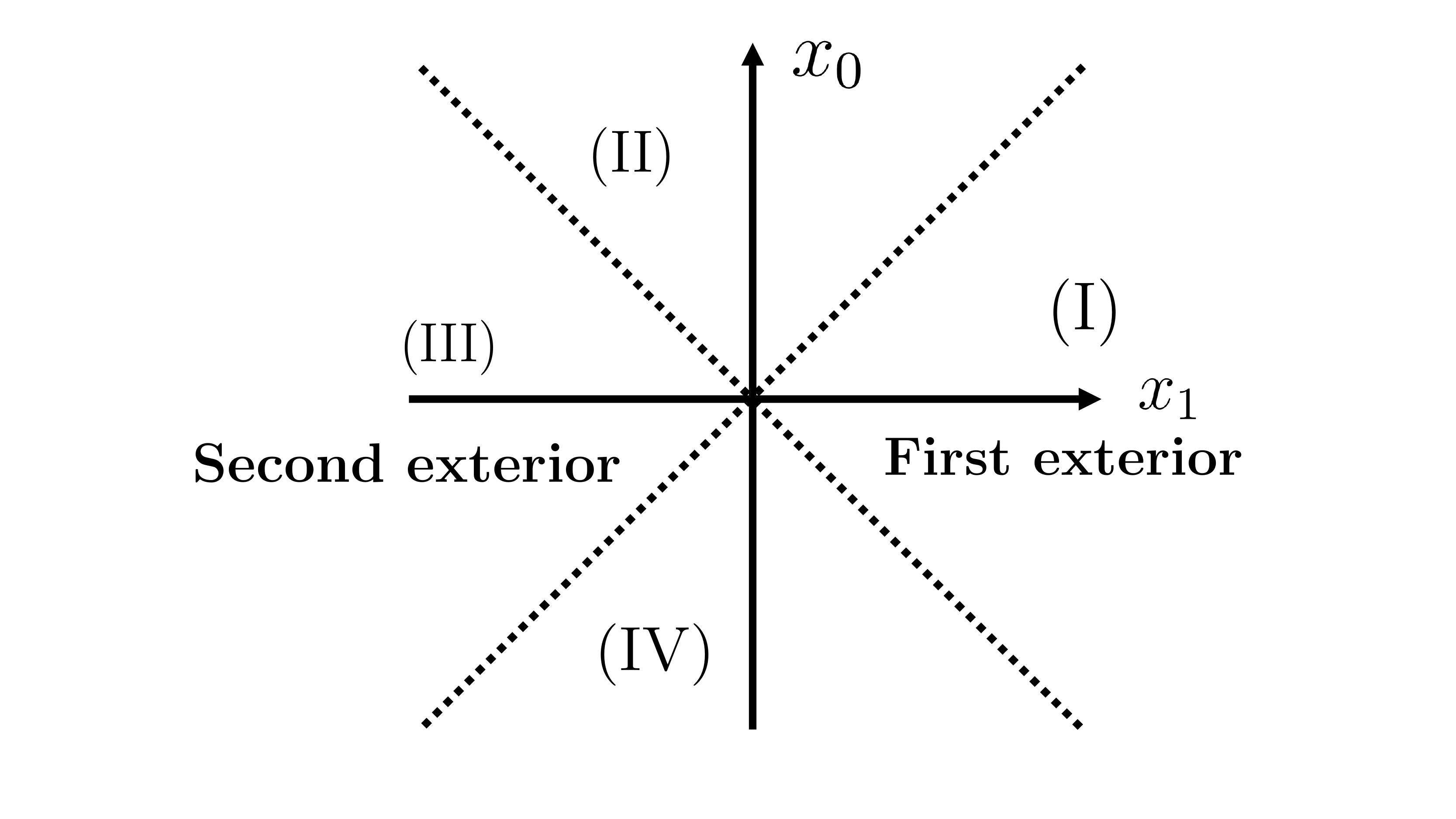}
  \end{center}
  \label{fig:two}
 \end{minipage}
  \begin{center}
        \caption{{\bf Left panel}: Gravity dual to holographic channel. {\bf Right panel:} Map from the asymptotic region of first and second exterior to the asymptotic boundary in $AdS_3$.
      }
      \label{f1}
  \end{center}
  \end{figure}

  Here, we compute BOMI and TOMI for holographic channels.
  The holographic channels has a gravity dual, which is the time-dependent eternal black hole \cite{HM}. Since the operator-dual state in (\ref{op}) is dual to the eternal black hole, BOMI and TOMI are holographically defined by the area of extremal surface in this geometry. In this paper, we study the operator entanglement of unitary time-evolution operator in $2$ dimensional CFTs. Therefore, the area of minimal surface is equal to the area of geodesic length in the $3$ dimensional time-dependent eternal black hole. 
BOMI in $(1+1)$ dimensional holographic CFTs is given by correlation function of twist and anti-twist operators. 
 BOMI for our configuration is given by the combination of two and four point functions. 
 Therefore, the channel-dependence of BOMI is given by the conformal block \cite{es}.

The gravity dual of \eqref{opeerator state U}
is given by a time-dependent eternal black
hole with two boundaries where individual CFTs live.
Let us follow the
computation in \cite{HM}.
As in Fig. \ref{f1}, 
the time direction on the left boundary is same as the direction on the right
boundary. The metric in the first exterior of time-dependent eternal black hole
 is given by
\be \label{fet}
ds^2 =d\rho^2 -\sinh^2{\rho}dt^2+\cosh^2{\rho}dx^2,
\ee 
where a horizon is at $\rho=0$, and the boundary is $\rho \rightarrow \infty$.
If we go into the horizon of black hole, we are in the interior of black hole,
whose metric is obtained by performing  an analytic continuation, $\rho=i
\alpha, t =\tilde{t}-i\f{\pi}{2}$.
The interior metric of the black hole is given by
\be 
ds^2=\sin^2{\alpha}d\tilde{t}^2+\cos^2{\alpha}dx^2-d \alpha^2
\ee
Moreover, after we take the analytic continuation, $\alpha =\f{\pi}{2}-i \tilde{\rho}, x =\tilde{x}-i\f{\pi}{2}$, its future metric  is given by
\be
ds^2=\cosh^2{\tilde{\rho}}d\tilde{t}^2-\sinh^2{\tilde{\rho}}d\tilde{x}^2+d\tilde{\rho}^2.
\ee

The time-dependent eternal black hole has two exteriors of black hole, and the metric for the other exterior is obtain by applying the analytic continuation, $t \rightarrow -t +i\pi$, to the metric in (\ref{fet}). 
This analytic continuation is slight different from
\cite{HM},
where they analytically continue
$t\to t+i\pi$.
This analytic continuation corresponds to the continuation
where the time directions in CFTs at the left and right
boundaries are opposite to each other.

Since various coordinate patches of this eternal black hole are related to Poincar\'e coordinate via the embedding space, the coordinates for the exteriors, future and interior of black hole can be mapped to Poincar\'e coordinate as follows:
\be
\begin{split}
  \text{First exterior}:\quad
  &\f{1}{2z}\left[1+(z^2-x_0^2+x_1^2)\right]
  =\cosh{x}\cosh{\rho},
  \quad
  \f{x_0}{z}=\sinh{t}\sinh{\rho},\\
  &\f{x_1}{z}=\cosh{t}\sinh{\rho},
  \quad
  \f{1}{2z}\left[1-(z^2-x_0^2+x_1^2)\right]=-\sinh{x}\cosh{\rho},\\
  \text{Interior}:\quad
  &\f{1}{2z}\left[1+(z^2-x_0^2+x_1^2)\right]=\cosh{x}\cos{\alpha},
  \quad
  \f{x_0}{z}=\cosh{\tilde{t}}\sin{\alpha},\\
  &\f{x_1}{z}=\sinh{\tilde{t}}\sin{\alpha},
  \quad
  \f{1}{2z}\left[1-(z^2-x_0^2+x_1^2)\right]=-\sinh{x}\cos{\alpha},\\
  \text{Future}:\quad
  &\f{1}{2z}\left[1+(z^2-x_0^2+x_1^2)\right]=\sinh{\tilde{x}}\sinh{\tilde{\rho}},
  \quad
  \f{x_0}{z}=\cosh{\tilde{t}}\cosh{\tilde{\rho}},\\
  &\f{x_1}{z}=\sinh{\tilde{t}}\cosh{\tilde{\rho}},
  \quad
  \f{1}{2z}\left[1-(z^2-x_0^2+x_1^2)\right]=-\cosh{\tilde{x}}\sinh{\tilde{\rho}},\\
  \text{Second exterior}:\quad
  &\f{1}{2z}\left[1+(z^2-x_0^2+x_1^2)\right]=\cosh{x}\cosh{\rho},
\quad \f{x_0}{z}=\sinh{t}\sinh{\rho},\\
&\f{x_1}{z}=-\cosh{t}\sinh{\rho},\quad
\f{1}{2z}\left[1-(z^2-x_0^2+x_1^2)\right]=-\sinh{x}\cosh{\rho}.\\
\end{split}
\ee

Near the boundary, $\rho \rightarrow \infty, z \rightarrow 0$, the asymptotic region in the exteriors, future region and interior of the black hole are mapped to the asymptotic regions of Poincar\'e coordinate as follows:
\begin{align}
  &
  \text{First exterior}: x_1^2-x_0^2 \ge 0, x_1\ge 0,
  \quad
  \text{Interior}: z^2 \ge-x_1^2+x_0^2 \ge 0,
  \nonumber \\
   & 
  \text{Future}:-x_1^2+x_0^2 \ge z^2, x_0\ge 0,
  \quad 
\text{Second exterior}: x_1^2-x_0^2 \ge 0, x_1\le 0. 
\end{align}
Fig. \ref{f1} shows that the first exterior is mapped to the region (I),
and the other exterior corresponds to (III).
Before using this map, holographic entanglement entropy is given by the area of
extremal surface for a static configuration in the time dependent background.
After the mapping, the entropy is given by the geodesic length for the time-dependent configuration, the boosted interval in the static background.

\paragraph{A holographic dual of our setup}

The normalized state dual to a unitary time evolution operator
\eqref{opeerator state U}
has the following shift symmetry:
\be \label{opstate}
\ket{U}= \mathcal{N}e^{-\left(\f{i t +\epsilon}{2}\right)\left(H_1+H_2\right)}\sum_a\ket{a}_1\ket{a}_2=\mathcal{N}e^{-\f{\left(H_1+H_2\right)\epsilon}{2}}e^{-i t_1 H_1}\sum_a\ket{a}_1e^{-i t_2 H_2}\ket{a}_2,
\ee
where $t_1+t_2=t$. 
By using this shift symmetry,
let us compute BOMI for the following, simple configuration: The subsystem, $A$, on the boundary of first exterior is an interval between $X_2 \le x \le X_1$ at $\f{t}{2}$. The subsystem, $B$, on the boundary of second exterior is a union of two intervals, $B_1$ and $B_2$ at $\f{t}{2}$. The subsystems, $B_1$ and $B_2$, are $Y_2 \le x \le Y_1$ and $Y_4 \le x \le Y_3$. 
\paragraph{Adjacent Interval}
For simplicity, we consider the case of adjacent intervals, $Y_2=Y_3$.
The map from first and second exteriors to Poincar\'e coordinate is given by
\begin{align}
 \label{mapp}
  \text{First exterior}&: x_0=e^x \sinh{t}\tanh{\rho},~~x_1=e^x \cosh{t}\tanh{\rho}, ~~z=\f{e^x}{\cosh{\rho}},
  \nonumber \\
\text{Second exterior}&: x_0=e^x \sinh{t}\tanh{\rho},~~x_1=-e^x \cosh{t}\tanh{\rho}, ~~z=\f{e^x}{\cosh{\rho}},
\end{align}
where the relation between $(\rho,t,x)$ and $(z,x_0,x_1)$ near the boundary, $\rho=\rho_{Max} \rightarrow \infty$, is given by
\be
\begin{split}
\text{First exterior}&: x^b_0 \approx e^x \sinh{t},~~x^b_1\approx e^x \cosh{t}, ~~z^b\approx2e^{x-\rho_{Max}}\\
\text{Second exterior}&: x^b_0\approx e^x \sinh{t},~~x^b_1 \approx -e^x \cosh{t}, ~~z^b\approx2e^{x-\rho_{Max}}\\
\end{split}
\ee 
The boundaries of $A$ at the boundary of the first exterior are $P_1, P_2$, and
the boundaries of $B$ at the boundary of the second exterior are $P_3, P_4$.
These points $P_i$ are mapped to $P'_i$ by (\ref{mapp}):
\be
\begin{split}
&P_1:\left(\rho_{Max}, \f{t}{2}, X_2\right) \rightarrow P'_1: \left(2e^{X_2-\rho_{Max}}, e^{X_2} \sinh{\left(\f{t}{2}\right)}, e^{X_2} \cosh{\left(\f{t}{2}\right)}\right), \\
&P_2:\left(\rho_{Max}, \f{t}{2}, X_1\right) \rightarrow P'_2:\left(2e^{X_1-\rho_{Max}}, e^{X_1} \sinh{\left(\f{t}{2}\right)}, e^{X_1} \cosh{\left(\f{t}{2}\right)}\right), \\
&P_3:\left(\rho_{Max}, \f{t}{2}, Y_4\right) \rightarrow P'_3:\left(2e^{Y_4-\rho_{Max}}, e^{Y_4} \sinh{\left(\f{t}{2}\right)}, -e^{Y_4} \cosh{\left(\f{t}{2}\right)}\right), \\
&P_4:\left(\rho_{Max}, \f{t}{2}, Y_1\right) \rightarrow P'_4:\left(2e^{Y_1-\rho_{Max}}, e^{Y_1} \sinh{\left(\f{t}{2}\right)}, -e^{Y_1} \cosh{\left(\f{t}{2}\right)}\right). \\
\end{split}
\ee
The interval between $P'_1$ and $P'_2$ corresponds to the interval in the first
exterior.
On the other hand, the interval between $P'_3$ and $P'_4$
corresponds to the interval in the second exterior. 

\subsubsection*{Geodesics between $P_1$ and $P_2$, and $P_3$ and $P_4$}
Here, let us compute the geodesic length between $P_1$ and $P_2$ and the length between $P_3$ and $P_4$. The geodesic length for $A$ is given by the length for the interval between $P'_1$ and $P'_2$, and the length for $B$ is given by the length for the interval between $P'_3$ and $P'_4$. The mapped configuration in Poincar\'e coordinate is independent of $t$. Owing to the time translational invariance in this configuration, the geodesic length is given by the one for the intervals at $t=0$.
The geodesic ending on $P_1$ and $P_2$, and the geodesic ending on $P_3$ and $P_4$ are the following semicircles:
\begin{align}
  &\text{Geodesic between $P_1$ and $P_2$}: z^2+\left(x_1-a\right)^2=C^2,
  \nonumber\\
  &\text{Geodesic between $P_3$ and $P_4$}: z^2+\left(x_1-b\right)^2=D^2, 
\end{align}
where ($X_1>X_2$ and $Y_1>Y_4$) 
\begin{align}
  &a=\left(\f{e^{X_1}+e^{X_2}}{2\tanh{\rho}}\right),
    \nonumber \\
  &b=-\left(\f{e^{Y_1}+e^{Y_4}}{2\tanh{\rho}}\right),
  \nonumber \\
  &C^2=\f{e^{2X_1}+e^{2X_2}}{2\cosh^2{\rho}}+\f{(e^{X_2}-e^{X_1})^2}{4}\tanh^2{\rho}+\f{(e^{X_2}+e^{X_1})^2}{4\sinh^2{\rho}\cosh^2{\rho}},
  \nonumber \\
  &D^2=\f{e^{2Y_1}+e^{2Y_4}}{2\cosh^2{\rho}}+\f{(e^{Y_4}-e^{Y_1})^2}{4}\tanh^2{\rho}+\f{(e^{Y_4}+e^{Y_1})^2}{4\sinh^2{\rho}\cosh^2{\rho}}.
\end{align}
The geodesic length for $A$ and $B$ are $\mathcal{L}_{12}$ and $\mathcal{L}_{34}$, which are given by
\begin{align}
  &\mathcal{L}_{12}
    =\f{1}{2}\int^{\tilde{X}_1}_{\tilde{X}_2}dx_1
    \left[\f{1}{C-x_1+a}+\f{1}{C+x_1-a}\right]
    =\f{1}{2}\log
    \left|\f{(\tilde{X}_1-a+C)(\tilde{X}_2-a-C)}{(\tilde{X}_1-a-C)(\tilde{X}_2-a+C)}\right|,
    \nonumber \\
  &\mathcal{L}_{34}=\f{1}{2}\int^{\tilde{Y}_1}_{\tilde{Y}_2} dx_1
    \left[\f{1}{D-x_1+b}+\f{1}{D+x_1-b}\right]
    =\f{1}{2}\log
    \left|\f{(\tilde{Y}_1-b+D)(\tilde{Y}_2-b-D)}{(\tilde{Y}_1-b-D)(\tilde{Y}_2-b+D)}\right|,
\end{align}
where $\tilde{X}_1=e^{X_1}\tanh{\rho_{Max}}$, $\tilde{X}_2=e^{X_2}\tanh{\rho_{Max}}$, $\tilde{Y}_1=-e^{Y_1}\tanh{\rho_{Max}}$ and $\tilde{Y}_2=-e^{Y_4}\tanh{\rho_{Max}}$. 
Near the boundaries, $e^{-\rho_{Max}}=\epsilon \ll 1$, $C, a, \tilde{X}_1$ and $\tilde{X}_2$ are given by
\begin{align}
  &C^2 \approx \left(\f{e^{X_2}-e^{X_1}}{2}\right)^2+\epsilon^2\left(e^{X_2}+e^{X_1}\right)^2,
    \quad a\approx \left(\f{e^{X_2}+e^{X_1}}{2}\right)\left(1+2\epsilon^2\right),
  \nonumber \\
  &\tilde{X}_1 \approx e^{X_1}\left(1-2\epsilon^2\right),~~\tilde{X}_2 \approx e^{X_2}\left(1-2\epsilon^2\right),
  \nonumber \\ 
  &D^2 \approx \left(\f{e^{Y_4}-e^{Y_1}}{2}\right)^2+\epsilon^2\left(e^{Y_4}+e^{Y_1}\right)^2,
    \quad b\approx -\left(\f{e^{Y_4}+e^{Y_1}}{2}\right)\left(1+2\epsilon^2\right),
  \nonumber \\
  &\tilde{Y}_1 \approx e^{Y_1}\left(-1+2\epsilon^2\right),~~\tilde{Y}_2 \approx e^{Y_4}\left(-1+2\epsilon^2\right).
\end{align}
Correspondingly, $\mathcal{L}_{12}$ and $\mathcal{L}_{34}$ are given by
\be
\begin{split}
&\mathcal{L}_{12} \approx \f{1}{2}\log{\left[\left(\f{e^{X_1}-e^{X_2}}{2\epsilon
      e^{\f{(X_1+X_2)}{2}}}\right)^4\right]}=2
\log{\left[\f{\sinh{\left[\f{X_1-X_2}{2}\right]}}{\epsilon}\right]},
\nonumber \\
&
\mathcal{L}_{34} \approx \f{1}{2}\log{\left[\left(\f{(e^{Y_1}-e^{Y_4})}{2\epsilon e^{\f{(Y_1+Y_4)}{2}}}\right)^4\right]}=2 \log{\left[\f{\sinh{\left[\f{Y_1-Y_4}{2}\right]}}{\epsilon}\right]}.\\
\end{split}
\ee
\subsubsection*{Geodesics between $P_2$ and $P_4$, and $P_1$ and $P_3$}
Here, we consider a geodesic between $P_2$ and $P_4$, and the other between $P_1$ and $P_3$. 
The length for them are given by the length for boosted intervals\footnote{(\ref{geofo}) is derived in Appendix \ref{Appc}.}:
 \be \label{geofo}
  \mathcal{L}=\log{\left[\f{-\Delta x_0^2+\Delta x_1^2}{\epsilon_{A}\epsilon_{B}}\right]},
  \ee
which is the geodesic length between $(z, x_0, x_1)=(\epsilon_A, x_0^A, x_1^A)$ and $(\epsilon_B, x_0^A+\Delta x_0, x_1^A+\Delta x_1)$.

Holographic entanglement entropy for $A \cup B$  is given by the ratio of minimal surface are to $4G_N$:
\be
\begin{split}
S_{A\cup B}=\f{\text{Min}\left[\mathcal{L}_{12}+\mathcal{L}_{34}, \mathcal{L}_{13}+\mathcal{L}_{24}\right]}{4G_N}
\end{split}
\ee
where $1/G_N$ is dimensionless Newton's constant, which is $2 c/3$ 
in the CFT language. Here, ${\mathcal{L}_{13}+\mathcal{L}_{24}}$ is given by
\be
\begin{split}
  {\mathcal{L}_{13}+\mathcal{L}_{24}}&=
  \log{\left[\f{\left(\cosh{(X_2-Y_4)}+\cosh{t}\right)\left(\cosh{(X_1-Y_1)}+\cosh{t}\right)}{2^2 \epsilon^4}\right]} \\
  &=
  {\log{\left[\f{\left(-\sinh{\left[\f{Y_1-Y_4}{2}\right]}\sinh{\left[\f{X_1-X_2}{2}\right]}+\cosh{\left[\f{(X_1-Y_4-t)}{2}\right]}\cosh{\left[\f{(X_2-Y_1-t)}{2}\right]}\right)}{
         \epsilon^2}\right]}}
   \\
&+{\log{\left[\f{\left(-\sinh{\left[\f{Y_1-Y_4}{2}\right]}\sinh{\left[\f{X_1-X_2}{2}\right]}+\cosh{\left[\f{(X_1-Y_4+t)}{2}\right]}\cosh{\left[\f{(X_2-Y_1+t)}{2}\right]}\right)}{ \epsilon^2}\right]}}. \\
\end{split}
\ee

BOMI between $A$ and $B$, $I(A, B)$ is given by $S_A+S_B-S_{A\cup B}$. Therefore, BOMI is holographically given by the linear combination of geodesic length, which is given by
\be
I(A,B)=\f{c}{6}\left(\mathcal{L}_{12}+\mathcal{L}_{34}-\text{Min}\left\{\mathcal{L}_{12}+\mathcal{L}_{34},\mathcal{L}_{13}+\mathcal{L}_{24}\right\}\right),
\label{IAB}
\ee
where if $\mathcal{L}_{12}+\mathcal{L}_{34}\le \mathcal{L}_{13}+\mathcal{L}_{24}$, then BOMI vanishes.
$\mathcal{L}_{13}+\mathcal{L}_{24}-\mathcal{L}_{12}-\mathcal{L}_{34}$ is 
\be
\begin{split}
&\mathcal{L}_{13}+\mathcal{L}_{24}-\mathcal{L}_{12}-\mathcal{L}_{34}\\
&=-\log{\left[\f{\sinh{\left[\f{Y_1-Y_4}{2}\right]}\sinh{\left[\f{X_1-X_2}{2}\right]}}{ \left(-\sinh{\left[\f{Y_1-Y_4}{2}\right]}\sinh{\left[\f{X_1-X_2}{2}\right]}+\cosh{\left[\f{(X_1-Y_4-t)}{2}\right]}\cosh{\left[\f{(X_2-Y_1-t)}{2}\right]}\right)}\right]} \\
&-\log{\left[\f{\sinh{\left[\f{Y_1-Y_4}{2}\right]}\sinh{\left[\f{X_1-X_2}{2}\right]}}{ \left(-\sinh{\left[\f{Y_1-Y_4}{2}\right]}\sinh{\left[\f{X_1-X_2}{2}\right]}+\cosh{\left[\f{(X_1-Y_4+t)}{2}\right]}\cosh{\left[\f{(X_2-Y_1+t)}{2}\right]}\right)}\right]} \\
\end{split}
\ee
Since this geometry corresponds to a time-dependent thermofield double state in (\ref{opstate}),
we need replace $t\rightarrow \f{\pi t}{\epsilon}$, $X_1\rightarrow \f{\pi X_1}{\epsilon}$, $X_2\rightarrow \f{\pi X_2}{\epsilon}$, $Y_1\rightarrow \f{\pi Y_1}{\epsilon}$ and $Y_4\rightarrow \f{\pi Y_2}{\epsilon}$ in order to compute BOMI and TOMI for (\ref{opstate}). 
 After replacing, $\mathcal{L}_{13}+\mathcal{L}_{24}-\mathcal{L}_{12}-\mathcal{L}_{34}$ is given by
\be
\begin{split}
&\mathcal{L}_{13}+\mathcal{L}_{24}-\mathcal{L}_{12}-\mathcal{L}_{34}=-\log{\left[\f{x\bar{x}}{(1-x)(1-\bar{x})}\right]},
\end{split}
\ee
where we write the linear combination of geodesic length in terms of cross ratios defined in (\ref{cora}).
Thus, $I(A,B)$ in \eqref{IAB} written in terms of cross ratios is given by
\be
\begin{split}
  I(A,B)
&=\f{c}{6}\log{\left[\f{x\bar{x}}{\text{Min}\left\{(1-x)(1-\bar{x}), x \bar{x}\right\}}\right]},
\end{split}
\ee
which agrees with \eqref{Holographic BOMI}.

\subsection{Geodesic length for a boosted interval \label{Appc}}
Here, we compute geodesic length for a boosted interval in $AdS_3$. 
The ${\it AdS}_3$ metric in Poincar\'{e} coordinate is 
\be
ds^2 =\f{dz^2-dx_0^2+dx_1^2}{z^2}.
\ee
The  geodesic ending on $(z,x_1,x_0)=(\epsilon_A, A, T)$ and $(z,x_1,x_0)=(\epsilon_B, B, T)$ is given by the semicircle,
\be
\begin{split}
&z^2+(x_1-a)^2=C^2,\\
&a=\f{A+B}{2}+\f{\epsilon_A^2-\epsilon_B^2}{2(A-B)}, \\
&C^2=\left(\f{A-B}{2}\right)^2+\f{\epsilon_A^2+\epsilon_B^2}{2}+\f{(\epsilon_A^2-\epsilon_B^2)^2}{4(A-B)^2}
\end{split}
\ee
where we assume $A>B$.
The geodesic length $\mathcal{L}$ between $(z, x_1 ,x_0)=(\epsilon_A,A,t)$ and $(\epsilon_B,B,t)$ is given as follows;
\be
\begin{split}
  \label{actionL}
\mathcal{L}
  =\int^{A}_{B}dx_1 \f{\sqrt{1+\left(\f{dz}{dx_1}\right)^2}}{z}
&=\f{1}{2}\int ^{A}_{B}dx_1 \left[\f{1}{C-a+x_1}-\f{1}{-C-a+x_1}\right] \\
&=\f{1}{2}\log{\left|\f{(A-a+C)(B-a-C)}{(A-a-C)(B-a+C)}\right|}
\end{split}
\ee
 If we take  the limit where $\epsilon_A\ll1$ and  $\epsilon_B\ll1$, $\mathcal{L}$ is given by
 \be
 \mathcal{L}\approx\log{\left[\f{(A-B)^2}{\epsilon_A\epsilon_B}\right]}.
 \ee 
 We now consider a Lorentz boost:
 \be \label{lbt}
 \begin{split}
 \begin{pmatrix}
 T \\
 X
 \end{pmatrix}=
 \begin{pmatrix}
 \cosh{\theta}&-\sinh{\theta} \\
 -\sinh{\theta}&\cosh{\theta} \\
 \end{pmatrix}
 \begin{pmatrix}
 t \\
 x\\
 \end{pmatrix}
 \end{split}
 \ee
  The map, $(t,x)^{T} \rightarrow (T,A)^{T}, ~~(t+\Delta t,x+\Delta x)^{T} \rightarrow (T,B)^{T}$, means that
  \be
  \begin{split}
  &\Delta t= \Delta x \tanh{\theta}, \\
  &A= -\sinh{\theta} t+\cosh{\theta} x\\
  &B=-\sinh{\theta} t +\cosh{\theta}x- \Delta t \sinh{\theta}+\Delta x \cosh{\theta}.
  \end{split}
  \ee
Thus, $(A-B)^2$ is given by
    $(A-B)^2
 =-\Delta t^2+\Delta x^2$.
  Thus, the geodesic length of boosted interval is 
  \be
  \mathcal{L}=\log{\left[\f{-\Delta t^2+\Delta x^2}{\epsilon_A \epsilon_B}\right]}.
  \ee

\section{Late-time behavior of TOMI for the compactified boson theory}
\label{Late time behavior of TOMI for the compactified boson theory}

In this section, we derive
analytic approximations for the late-time behavior
of TOMI for the compactified boson theory,
i.e., 
we consider the limit
$t \rightarrow \infty, X_1-X_2 \equiv L_A \gg \epsilon$.
All numerical plots in this section are made with the choice $\epsilon = 1.1$, $X_2 = 0$, $X_1 = 10$.


\subsection{Late-time behavior at self-dual radius}

As a warm-up, we consider the self-dual radius $\eta = R^2 = 1$.

Firstly, let us compute $I^{(2)}(A,B)$.
We start by noting $x_{AB}=\bar{x}_{AB}$;
For this special case of $x=\bar{x}\in [0,1]$,
the modular parameter is purely imaginary ($\bar{\tau}=-\tau$),
and the Siegel theta function factorizes into two theta functions
\begin{align}\label{ThetaFactorize}
  \Theta(0|T)&= \sum_\nu e^{i\pi \frac{\tau \eta}{2}\nu^2} \sum_\mu e^{i\pi \frac{\tau}{2\eta}\mu^2}
= \theta_3(\tau \eta)  \theta_3({\tau}/{\eta})
\end{align}
Hence,  our expression simplifies
\begin{equation}
  F_2(x_{AB},\bar{x}_{AB}) = \left[
    \theta_3(\eta \tau)\theta_3\left({\tau}/{\eta}\right)/
    \theta_3(\tau)^2 \right]^2=1
\end{equation}
when $\eta=1$. Applying (\ref{CBI2}) leads to 
\begin{equation}
  I^{(2)}(A,B)(t) = \log \frac{1}{(1-x_{AB})^{1/4}(1-\bar{x}_{AB})^{1/4}} = \log\frac{1}{\sqrt{e^{-\frac{\pi}{\epsilon}L_A}}}= \frac{\pi L_A}{2\epsilon}.
\end{equation}

As for $I^{(2)}(A,B_1)(t)$,
we note that the cross ratios (\ref{cri}) approach,
as $t\rightarrow \infty $, $X_1-X_2 \equiv L_A \gg \epsilon$
($t\gg X_1, X_2,L_A$),
\begin{align}
  x_{AB_1} &\approx e^{\frac{\pi}{\epsilon}(X_1-t)},
     \quad
  \bar{x}_{AB_1} 
  \approx 
    1 - e^{-\frac{\pi L_A}{\epsilon}}.
\end{align}
We are taking $t$ to be much larger than any spatial coordinates since we are
sending it to infinity. Since $x_{AB_1}$ goes to zero, we can use the saddle
point approximation in \cite{2017PhRvD..96d6020C} because $\tau_{AB_1} \rightarrow i
\infty$. In fact, using (3.16) in the aforementioned paper gives the following
approximations for the modular parameters:
Let $\bar{x}_{AB_1} = 1-e^{-\frac{\pi L_A}{\epsilon}}\equiv 1 - \Delta x$ where $\Delta x \ll 1$.
\begin{align}
  \tau_{AB_1} &\approx \frac{i}{\pi} \log \frac{16}{x_{AB_1}} \approx \frac{i}{\pi} \log \frac{16}{e^{\frac{\pi}{\epsilon}(X_1-t)}} = \frac{i}{\epsilon}(t-X_1+\frac{\epsilon}{\pi}\log 16),
  \\ \nonumber
\bar{\tau}_{AB_1} &= \frac{1}{i\frac{K(1-\Delta x)}{K(\Delta x)}} \approx \frac{1}{\frac{i}{\pi}\log \frac{16}{\Delta x}} = \frac{-i \pi}{\log{\frac{16}{e^{-\frac{\pi L_A}{\epsilon}}}}}.
\end{align}
We can see that $\tau_{AB_1}  \rightarrow i \infty$ as $t\rightarrow \infty$. The Siegel theta function ((3.1) in \cite{2017PhRvD..96d6020C}) can be written as
\begin{equation}
\Theta(0| T_{AB_1}) = \sum_{\mu,\nu \in \mathbb{Z}} \exp\left(-\frac{\pi |\tau_{AB_1}|}{2} (\nu+\mu)^2 \right)  \exp\left(-\frac{\pi i\bar{\tau}_{AB_1}}{2} (\nu-\mu)^2 \right)
\end{equation}
where we set $\eta = 1$. Since $|\tau_{AB_1}| \rightarrow \infty$, the leading term comes from $\nu + \mu = 0$, and
\begin{align}
  \Theta(0|T_{AB_1}) = \sum_{\nu \in \mathbb{Z}} e^{-2\pi i \bar{\tau}_{AB_1} \nu^2} = \theta_3(-2\bar{\tau}_{AB_1}),
\end{align}
where the anti-holomorphic modular parameter is given by
$\bar{\tau}_{AB_1} =- i\frac{K(1-\bar{x}_{AB_1})}{K(\bar{x}_{AB_1})}$.
To simplify this further, we will employ the doubling identity ((10.272) in \cite{YellowBook})
\begin{equation}
\Theta(0| T_{AB_1})=\sqrt{\frac{\theta_3(-\bar{\tau}_{AB_1})^2+\theta_4(-\bar{\tau}_{AB_1})^2}{2}}=\sqrt{\frac{\theta_3(-\bar{\tau}_{AB_1})^2+\sqrt{\theta_4(-\bar{\tau}_{AB_1})^4}}{2}}
\end{equation}
Next, apply the Jacobi quartic identity found on page 137 in \cite{JacobiQuartic} $\theta_4^4(\tau) = \theta_3^4(\tau)-\theta_2^4(\tau)$ and rewrite some of the Jacobi theta functions in terms of the cross ratio using (154) in \cite{CalabreseNegativity}, $x(\tau)=\left(\frac{\theta_2(\tau)}{\theta_3(\tau)} \right)^4$,
\begin{equation}
\Theta(0| T_{AB_1})= \sqrt{\frac{\theta_3(-\bar{\tau}_{AB_1})^2+\sqrt{\theta_3(-\bar{\tau}_{AB_1})^4-\theta_2(-\bar{\tau}_{AB_1})^4}}{2}}
= \sqrt{ \theta_3(-\bar{\tau}_{AB_1})^2 \frac{1+\sqrt{1-\bar{x}_{AB_1}}}{2}}
\end{equation}
The remaining Jacobi theta function can be written in terms of a hypergeometric function. In particular, apply the identity on page 13 of \cite{CalabreseTwoDisjointEE}, $ _2F_1(1/2,1/2,1,\bar{x}_{AB_1}) \equiv f_{1/2}(\bar{x}_{AB_1}) = \theta_3^2(-\bar{\tau}_{AB_1})$ so that
\begin{equation}
F_2(x_{AB_1},\bar{x}_{AB_1}) = \frac{\Theta(0| T_{AB_1})^2}{f_{1/2}(x_{AB_1}) f_{1/2}(\bar{x}_{AB_1})  } = \frac{1}{f_{1/2}(x_{AB_1})} \frac{1+\sqrt{1-\bar{x}_{AB_1}}}{2} \approx  \frac{1}{1+\frac{1}{4}x_{AB_1}} \frac{1+\sqrt{1-\bar{x}_{AB_1}}}{2}
\end{equation}
Finally, our bi-partite mutual information (\ref{CBI2}) simplfies to
\begin{align}
  I^{(2)}(A,B_1)(t)
                  &\approx \log \frac{1+e^{-\frac{\pi L_A}{2\epsilon}}}{2\left(1+\frac{1}{4}e^{\frac{\pi}{\epsilon}(X_1-t)} \right)\left(1-e^{\frac{\pi}{\epsilon}(X_1-t)} \right)^{1/4}e^{-\frac{\pi L_A}{4\epsilon}}  }
\approx \log \frac{1}{2 e^{-\frac{\pi L_A}{4\epsilon}}}.
\end{align}

Lastly, we compute $I^{(2)}(A,B_2)(t)$. As $t \rightarrow \infty$, $X_1-X_2 \equiv L_A \gg \epsilon$, the cross ratios (\ref{cri}) are approximately
\begin{align}
  x_{AB_2} &\approx 1 - e^{-\frac{\pi L_A}{\epsilon}},
             \quad
\bar{x}_{AB_2} \approx e^{-\frac{\pi}{\epsilon}(t+X_2)}
\end{align}
Since
$\bar{\tau}_{AB_2} \rightarrow -i \infty$
as $\bar{x}_{AB_2} \rightarrow 0$,
we can use the saddle point approximation as before
for $\Theta(0|T_{AB_2})$:
\begin{equation}
\Theta(0|T_{AB_2}) = \sum_{\mu,\nu} \exp \left(\frac{\pi i \tau_{AB_2}}{2}(\nu + \mu)^2 \right) \exp \left(-\frac{\pi | \bar{\tau}_{AB_2}|}{2}(\nu - \mu)^2 \right).
\end{equation}
The leading terms are those with $\nu - \mu = 0$, and hence
\begin{align}
  \Theta(0|T_{AB_2}) &= \sum_\mu e^{2 \pi i \tau_{AB_2}\mu^2}
  = \theta_3(2\tau_{AB_2})
= \sqrt{\frac{\theta_3(\tau_{AB_2})^2+\theta_4(\tau_{AB_2})^2}{2}} \\ \nonumber
                     &=  \sqrt{\frac{\theta_3(\tau_{AB_2})^2+\sqrt{\theta_3(\tau_{AB_2})^4-\theta_2(\tau_{AB_2})^4}}{2}}
= \sqrt{\theta_3(\tau_{AB_2})^2 \left(\frac{1+\sqrt{1-x_{AB_2}}}{2} \right)}.
\end{align}
Noting $\theta_3(\tau_{AB_2})^2 =  _2F_1(1/2,1/2,1,x_{AB_2})$,
and using \eqref{CBI2},
we obtain 
\begin{align}
F_2(x_{AB_2},\bar{x}_{AB_2}) &= \frac{\Theta(0|T_{AB_2})^2}{f_{1/2}(x_{AB_2})f_{1/2}(\bar{x}_{AB_2})} 
= \frac{1+\sqrt{1-x_{AB_2}}}{2f_{1/2}(\bar{x}_{AB_2})} 
\approx \frac{1+\sqrt{1-x_{AB_2}}}{2\left(1+\frac{\bar{x}_{AB_2}}{4} \right)}, 
\nonumber \\
  I^{(2)}(A,B_2)(t)
  &= \log \frac{1+\sqrt{1-x_{AB_2}} }{2\left(1+\frac{\bar{x}_{AB_2}}{4} \right) (1-x_{AB_2})^{1/4}(1-\bar{x}_{AB_2})^{1/4}}
\approx \log \frac{1}{2e^{-\frac{\pi L_A}{4\epsilon}}}.
\end{align}

Combining the three bi-partite mutual information,
we obtain the tri-partite mutual information
\begin{align}
  I^{(2)}(A,B_1,B_2)(t)
   &= \log \frac{1}{2e^{-\frac{\pi L_A}{4\epsilon}}}+\log \frac{1}{2e^{-\frac{\pi L_A}{4\epsilon}}}-\frac{\pi L_A}{2\epsilon}
= \log \frac{1}{4}.
\end{align}
This is independent of the subsystem length $L_A$ and the cylinder circumference $2\epsilon$, provided that $L_A \gg \epsilon$. The three bi-partite mutual informations and the tri-partite mutual information are plotted, along with the analytic approximations.

\begin{figure}[t]
 \begin{minipage}{0.5\hsize}
  \begin{center}
   \includegraphics[width=65mm]{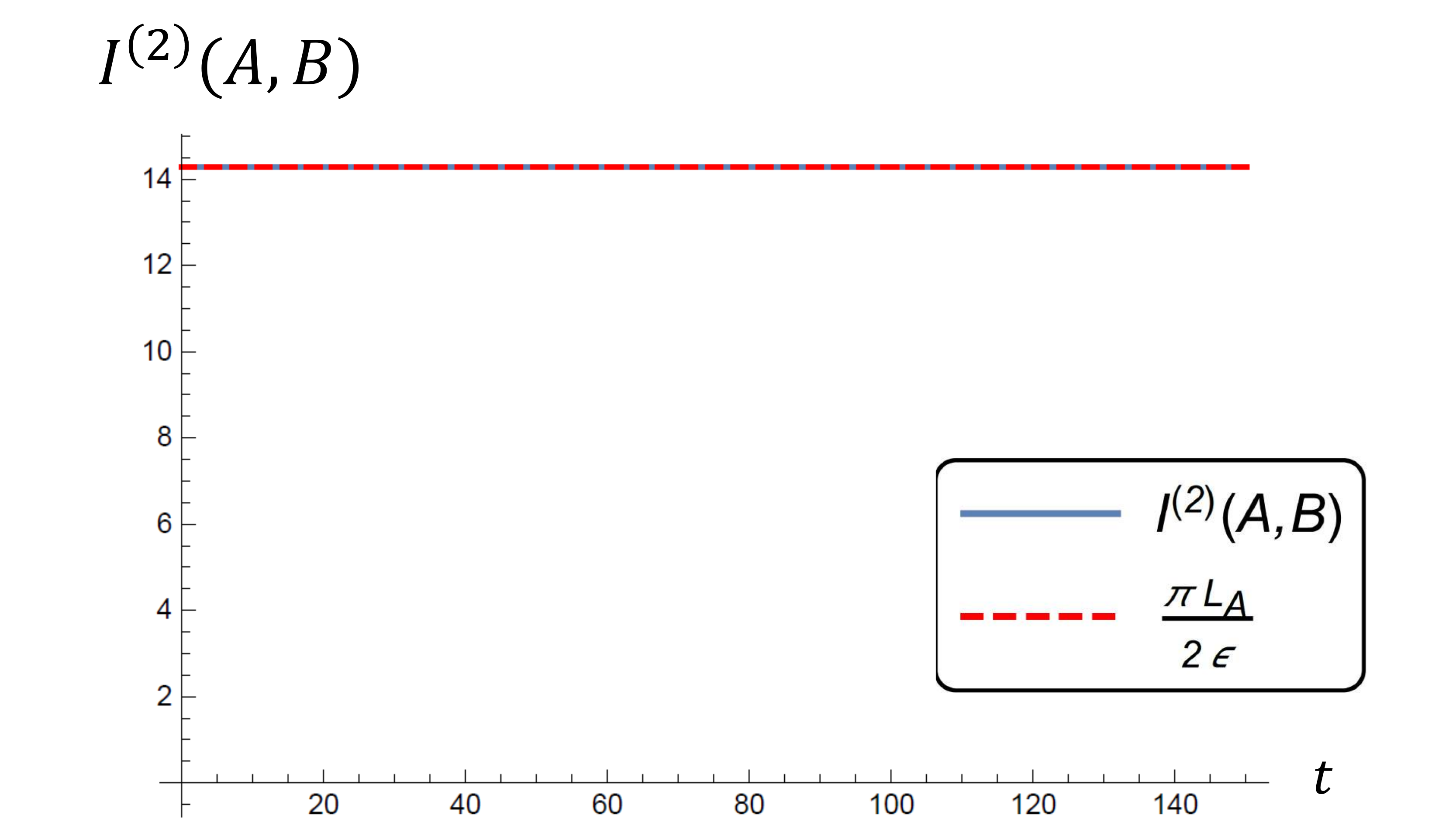}
  \end{center}
 \end{minipage}
 \begin{minipage}{0.5\hsize}
  \begin{center}
   \includegraphics[width=65mm]{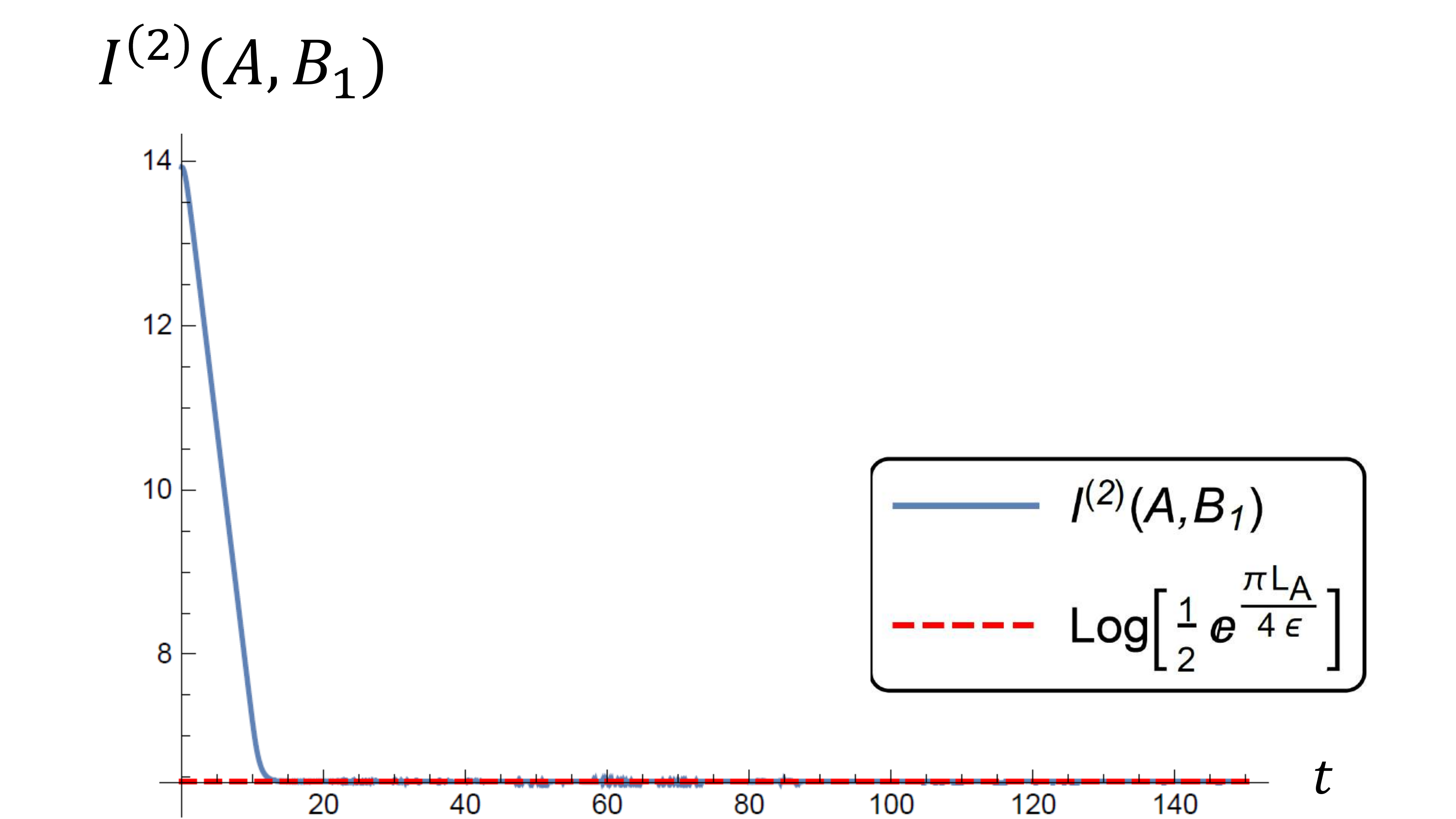}
  \end{center}
 \end{minipage}
 \begin{minipage}{0.5\hsize}
  \begin{center}
   \includegraphics[width=65mm]{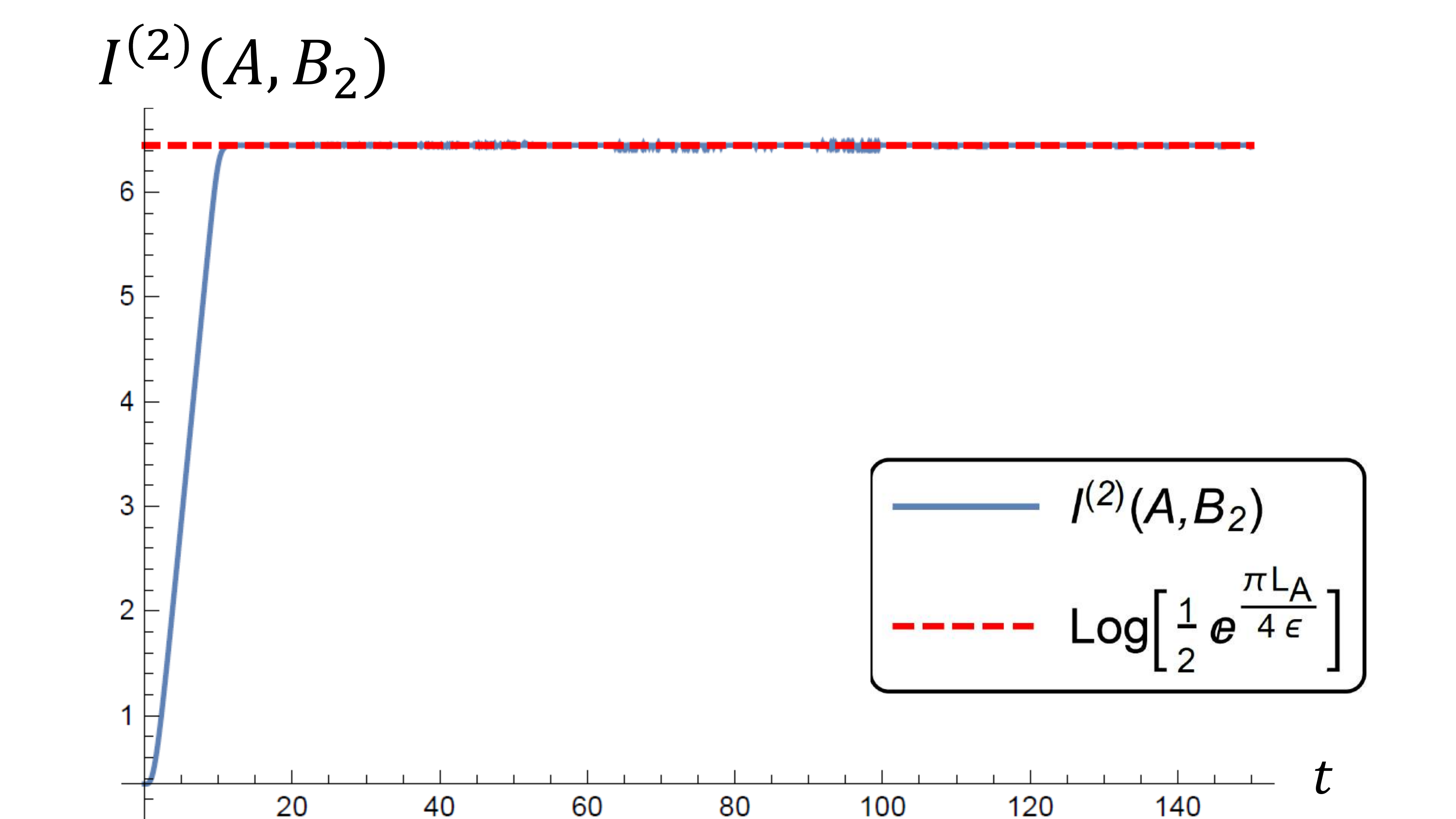}
  \end{center}
 \end{minipage}
 \begin{minipage}{0.5\hsize}
  \begin{center}
   \includegraphics[width=65mm]{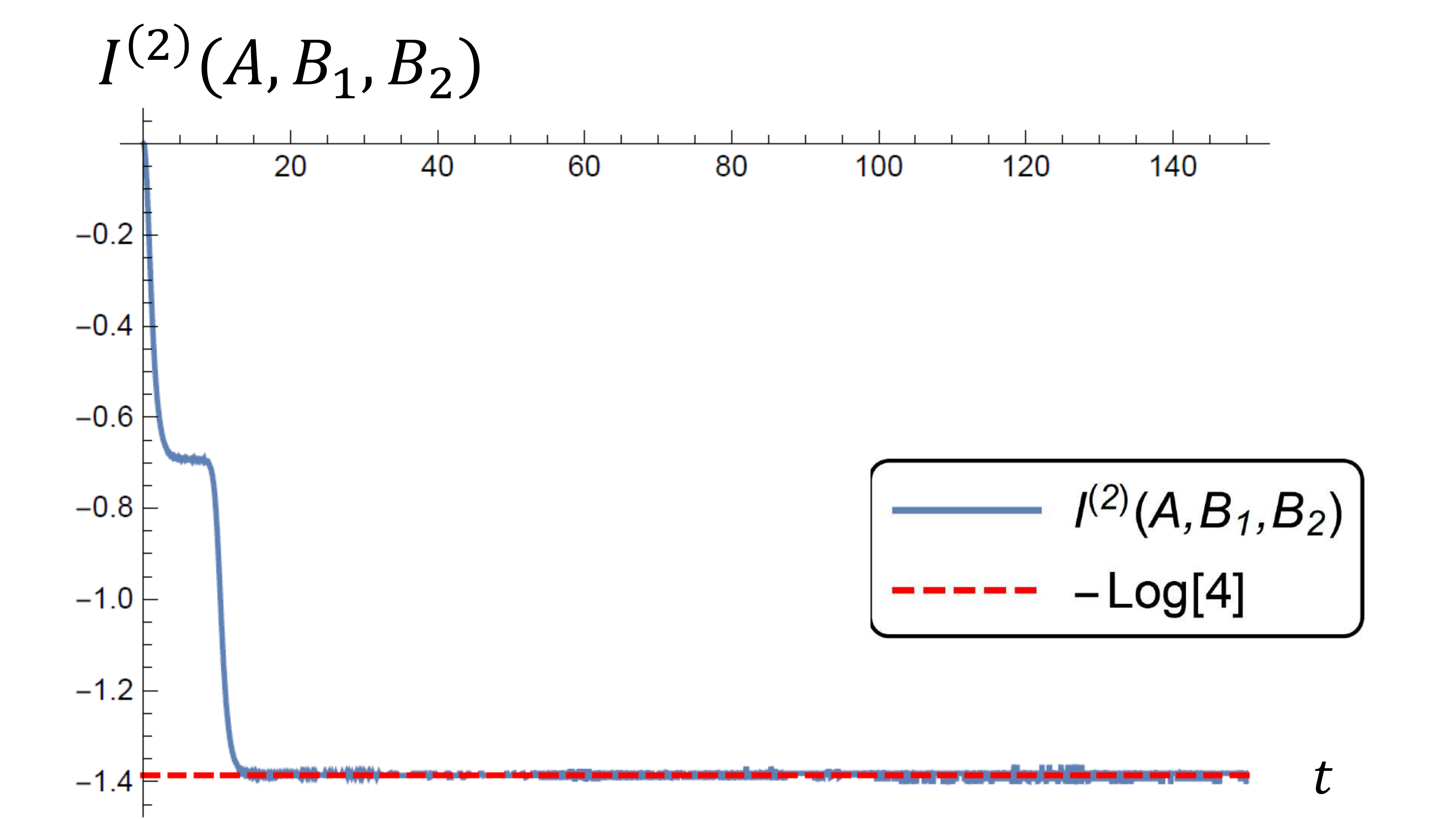}
  \end{center}
 \end{minipage}
 \caption{
 The time dependence of $I^{(2)}(A,B)$, $I^{(2)}(A,B_1)$,
   $I^{(2)}(A,B_2)$, $I^{(2)}(A,B_1,B_2)$ for the compactified boson at self-dual
   radius $\eta = R^2 = 1$.
   The subregions are chosen such that $X_1 = 10$, $X_2 = 0$, $B_1 =
   [0,\infty)$, $B_2 = (-\infty,0)$. The blue curves are obtained by using the
   full expressions derived earlier while the red dashes are the analytic
   approximations for the late-time behavior and are given by
   $\frac{\pi L_A}{2\epsilon}$,
   $\log \frac{1}{2e^{-\frac{\pi L_A}{4\epsilon}}}$,
   $\log \frac{1}{2e^{-\frac{\pi L_A}{4\epsilon}}}$
   and $\log \frac{1}{4}$, respectively.}
\end{figure}

\subsection{Arbitrary $\eta = R^2$}
\subsubsection{Approximating $\theta_3(q)$}

\begin{figure}[H]
  \begin{center}
   \includegraphics[width=80mm]{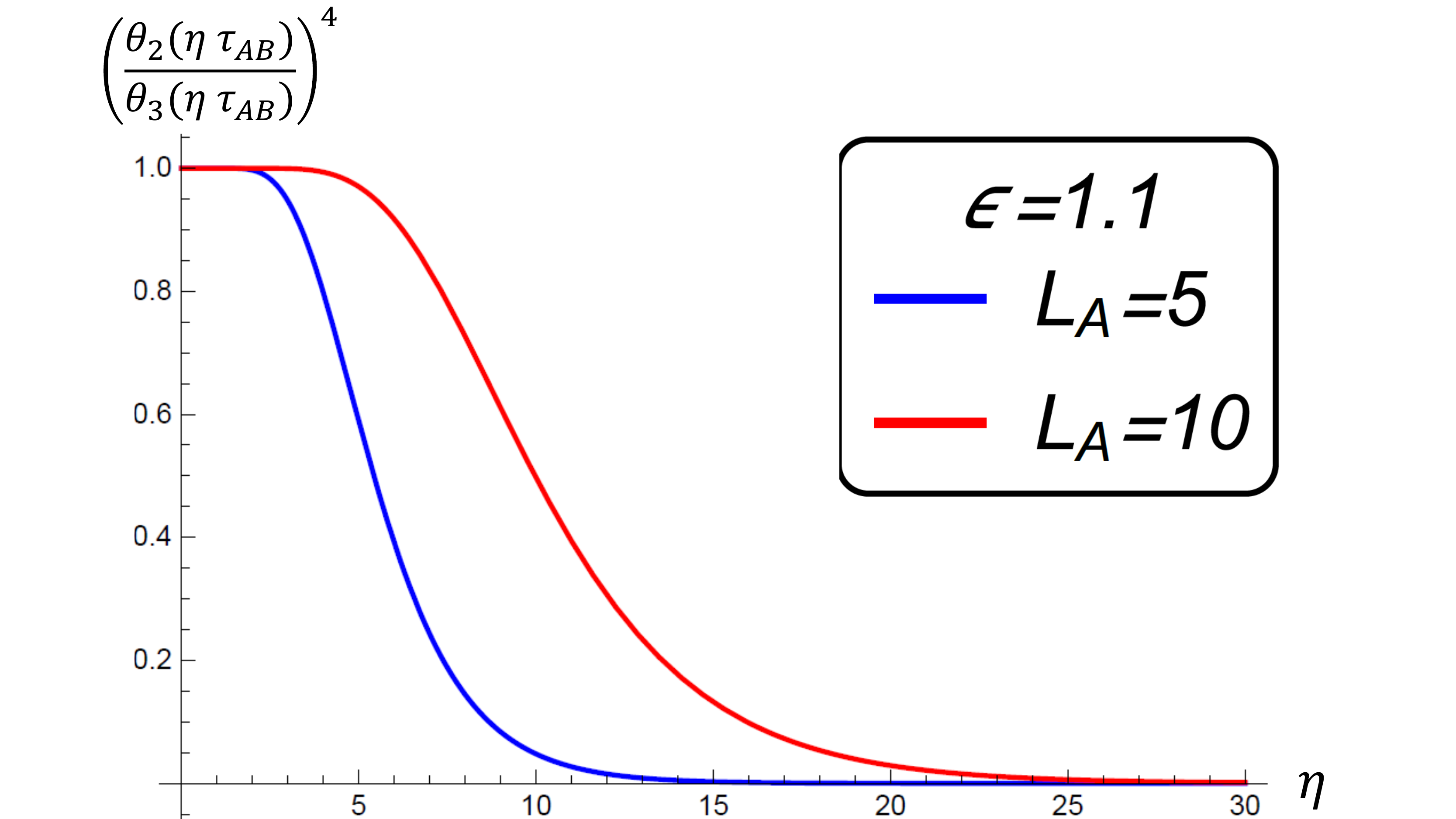}
  \end{center}
  \caption{
    The cross ratio $\left(\frac{\theta_2(\eta \tau_{AB})}{\theta_3(\eta
        \tau_{AB})} \right)^4$ as a function of $\eta$
    for $\tau_{AB} =i \frac{K(1-x_{AB})}{K(x_{AB})}$ with $L_A =5, 10,\epsilon = 1.1$. The blue curve corresponds to $L_A = 5$ while the red curve correponds to $L_A = 10$.}
      \label{cross rato v.s. radius}
\end{figure}

For arbitrary radius,
we cannot use the doubling identity for the Jacobi theta function as
we did in the previous section,
so we have to approximate $\theta_3(\eta \tau)$.
We have to consider different approximations to $\theta_3(\eta \tau)$
for two different regimes, small and large $\eta$.
This is so since for different values of $\eta$,
the corresponding cross ratio
(given by $\left(\frac{\theta_2(\eta \tau)}{\theta_3(\eta \tau)} \right)^4$)
takes different values:
For small $\eta$, the cross ratio is close to $1$,
while for large $\eta$, the cross ratio is close to $0$
(see Fig.\ \ref{cross rato v.s. radius}).
The cross ratio corresponding to $\eta \tau$ is close to $1$ for small $\eta$ and close to $0$ for large $\eta$. Below, we show the plots for $\epsilon = 1.1$ and $L_A = 5, 10$ with the time fixed at $t = 10$.
We see that as $L_A$ increases, the effective cross ratio $\left(\frac{\theta_2(\eta \tau_{AB})}{\theta_3(\eta
        \tau_{AB})} \right)^4$ stays closer to $1$ for a larger range of $\eta$.

\paragraph{Approximation for small $\eta$}
This approximation can be found in a footnote on page 28 of \cite{es}. The magnitude of $\tau$ in our set-up is small, so the cross ratio $x$ is close to 1, and we can simplify the relation between $\tau$ and $x$.
\begin{equation}
\tau = i\frac{K(1-x)}{K(x)} \approx \frac{i \pi}{\log \frac{16}{1-x}}
\end{equation}
The Jacobi Theta function can then be approximated as
\begin{align}\label{smallEta}
  \theta_3(\tau) &=\sqrt{\theta_3 (\tau)^2}
= \sqrt{_2F_1\left(\frac{1}{2},\frac{1}{2},1,x\right)} \\ \nonumber
                 &= \sqrt{\frac{1}{\pi}\log\frac{16}{1-x}+\frac{1-x}{4\pi}\left(\log\frac{16}{1-x}-2 \right)}
                   \approx \sqrt{\frac{1}{\pi}\log\frac{16}{1-x}}
= \sqrt{\frac{i}{\tau}}.
\end{align}
We can scale $\tau \rightarrow \eta \tau$ to obtain 
\begin{equation}\label{smalletathetatau}
\theta_3(\eta \tau) \approx \sqrt{\frac{i}{\eta \tau}}.
\end{equation}
As long as $\eta$ is small, the cross ratio is close to $1$
and our expansion of the Hypergeometric function remains valid.
One can show, in a similar fashion, that
\begin{equation}\label{smalletathetataubar}
\theta_3(-\eta \bar{\tau}) \approx \sqrt{\frac{1}{i \eta \bar{\tau}}}.
\end{equation}

\begin{figure}[t]
 \begin{minipage}{0.5\hsize}
  \begin{center}
   \includegraphics[width=70mm]{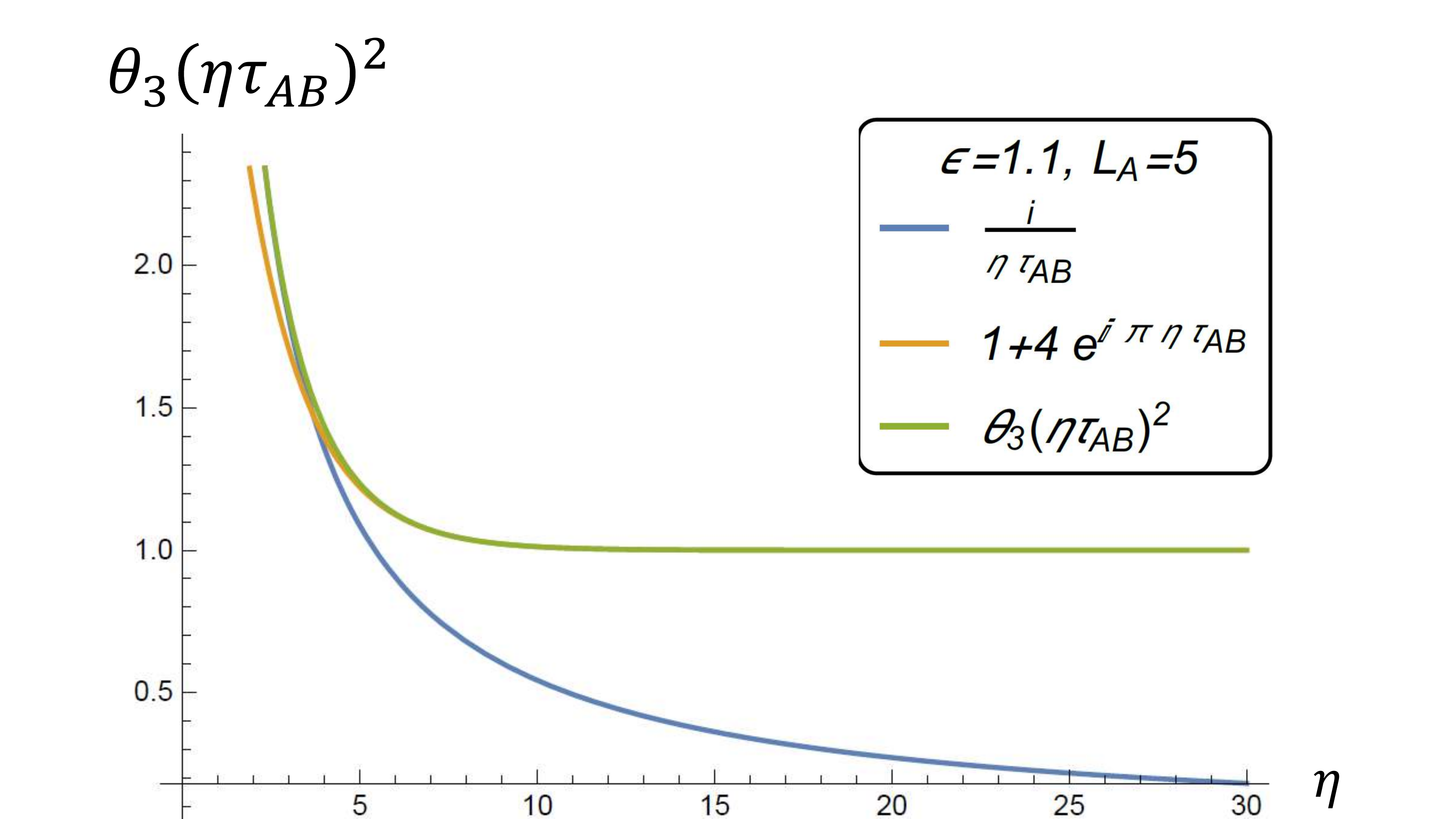}
  \end{center}
 \end{minipage}
 \begin{minipage}{0.5\hsize}
  \begin{center}
   \includegraphics[width=70mm]{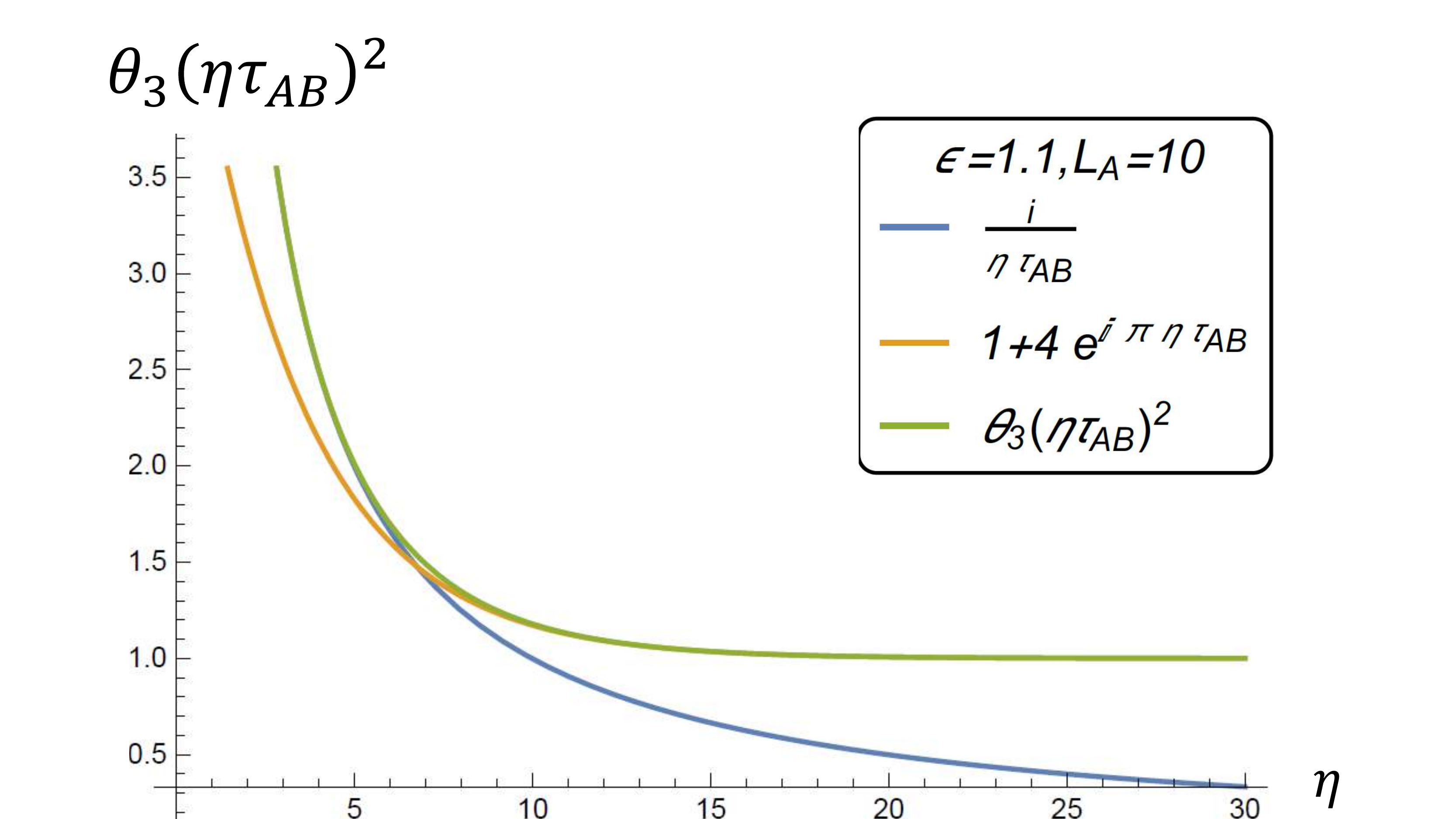}
  \end{center}
 \end{minipage}
  \caption{
    \label{cross over}
    Plot for $\theta_3(\eta \tau_{AB})^2$ (Green) against $\eta$ with
    $\tau_{AB} = i \frac{K(1-x_{AB})}{K(x_{AB})}$
    with
    $t=10$, $\epsilon=1.1$, and
    $L_A = 5$ (Left) and $L_A=10$ (Right).
    Also shown on the graph are the two approximations
    $\frac{i}{\eta \tau_{AB}}$ (Blue) and
    $1+4 e^{i \pi \eta \tau_{AB}}$ (Yellow).
    Eventually, when $\eta$ is sufficiently large, the correct approximation is given by the yellow curve.
  }
\end{figure}

\paragraph{Approximation for large $\eta$}
When the cross ratio corresponding to the modular parameter is much less than $1$, we have to expand the Hypergeometric function differently.
\begin{equation}
\theta_3(\tau) = \sqrt{\theta_3(\tau)^2} = \sqrt{_2F_1\left(\frac{1}{2},\frac{1}{2},1,x\right)} \approx \sqrt{1+\frac{x}{4}}=\sqrt{1+4e^{i \pi \tau}}
\end{equation}
For large $\eta$, $\left(\frac{\theta_2(\eta \tau)}{\theta_3(\eta \tau)} \right)^4$ is close to $0$ and we have
\begin{equation}\label{largeEta}
\theta_3(\eta \tau) = \sqrt{1+4 e^{i\pi \eta \tau}}
\end{equation}
From Fig.\ \ref{cross over}, we can see that the crossover from one approximation to the other occurs at about $\eta = 7$ for our set-up with $\tau_{AB} = i \frac{K(1-x_{AB})}{K(x_{AB})}$ with $L_A = 10,\epsilon = 1.1$.

From these two graphs, we see that as the subsystem size $L_A$ increases, the "small $\eta$" regime becomes larger as the cross over point occurs for larger values of $\eta$. For $L_A = 5$, the crossover occurs at about $\eta = 4$, while for $L_A = 10$, the crossover occurs at about $\eta = 7$.
\paragraph{When is $\eta$ large or small?}
In our approximations, we have a cross ratio of the form $x_{AB} = 1-e^{-\pi L_A/\epsilon}$. The corresponding modular parameter is approximately
\begin{equation}
\tau_{AB} = i \frac{K(1-x_{AB})}{K(x_{AB})}\approx \frac{i \pi}{\log\left(16 e^{\frac{\pi L_A}{\epsilon}}\right)}
\end{equation}
The arguments of the theta functions we want to approximate are of the form
\begin{equation}
\eta \tau_{AB} = i \frac{K(1-x')}{K(x')}
\end{equation}
where $x'=\left(\frac{\theta_2(\eta \tau_{AB})}{\theta_3(\eta \tau_{AB})} \right)^4$ is the corresponding cross ratio, a plot of which can be found in Fig. \ref{cross rato v.s. radius}. These two equations can be combined to give
\begin{equation}
\frac{\pi \eta}{\log\left(16 e^{\pi L_A/\epsilon} \right)} = \frac{K(1-x')}{K(x')}
\end{equation}
When $x'$ is close to one, this is approximately given by
\begin{equation}
\frac{\pi \eta}{\log\left(16 e^{\pi L_A/\epsilon} \right)}  \approx \frac{\pi}{\log\frac{16}{1-x'}}
\end{equation}
Since $x'$ is close to unity, we have
\begin{equation}
\eta \ll \frac{1}{\pi}\log \left(16 e^{\pi L_A/\epsilon} \right)
\end{equation}
This tells us what values of $\eta$ are considered small. When $x'$ is close to $0$, however, we have
\begin{equation}
\frac{\pi \eta}{\log\left(16 e^{\pi L_A/\epsilon} \right)} \approx \frac{1}{\pi} \log \left(\frac{16}{x'} \right)
\end{equation}
Hence, the values of $\eta$ that are considered large satisfy
\begin{equation}
\eta \gg \frac{1}{\pi}\log \left(16 e^{\pi L_A/\epsilon} \right)
\end{equation}

\subsubsection{bi-partite $I^{(2)}(A,B)$}
First, we compute $I^{(2)}(A,B)$. Because $x_{AB}=\bar{x}_{AB}$, $\bar{\tau}_{AB}=-i\frac{K(1-\bar{x}_{AB})}{K(\bar{x}_{AB})}=-i\frac{K(1-x_{AB})}{K(x_{AB})}=-\tau_{AB}$, the Siegel theta function factorizes so
\begin{equation}
  F_2(x_{AB},\bar{x}_{AB}) =
  {\theta_3(\tau_{AB}\eta)^2 \theta_3\left({\tau_{AB}}/{\eta}\right)^2}/{f_{1/2}(x_{AB}^2)^2}.
\end{equation}
We know that $\theta_3(\tau)^2 = _2F_1\left(\frac{1}{2},\frac{1}{2},1,x\right)$, where the modular parameter $\tau$ and the cross ratio are related by $\tau = i \frac{K(1-x)}{K(x)}$ and $x = \left(\frac{\theta_2(\tau)}{\theta_3(\tau)} \right)^4$. In the self-dual radius case, we used the doubling identity to work out $\theta_3(\eta\tau)$ for $\eta = 2$. However, it is harder to do this exactly for $\eta \neq 2$, so we will instead approximate it. First, write $x_{AB} = 1 - \underbrace{e^{-\frac{\pi L_A}{\epsilon}}}_{\equiv \Delta x}=1-\Delta x$. For $L_A \gg \epsilon$, $\Delta x \ll 1$ and we can make the following approximation,
\begin{equation}
\tau_{AB} = -\frac{1}{i\frac{K(1-\Delta x)}{K(\Delta x)}} \approx \frac{i\pi}{ \log \frac{16}{\Delta x}}
\end{equation}
Let us split our approximations into two cases which depend on the magnitude of $\tilde{\eta} \coloneqq \text{Max}\left\{\eta, \frac{1}{\eta}\right\}$

\begin{flushleft}
\underline{$\tilde{\eta} \ll \frac{1}{\pi}\log\left(16 e^{\pi L_A/\epsilon} \right)$}:
\end{flushleft}
In this case, we can use the small $\eta$ approximation (\ref{smalletathetatau}) for both $\theta_3$ functions
\begin{align}
  \theta_3(\tilde{\eta} \tau_{AB})^2 &= \frac{i}{\tilde{\eta}\tau_{AB}} = \frac{1}{\tilde{\eta} \pi}\log \frac{16}{\Delta x},
                                       \quad
\theta_3(\tau_{AB}/\tilde{\eta})^2 = \frac{i \tilde{\eta}}{\tau_{AB}}=\frac{\tilde{\eta}}{\pi}\log \frac{16}{\Delta x}.
\end{align}
Hence, recalling \eqref{CBI2},
\begin{align}
 \label{IABSmall}
  F_2(x_{AB},\bar{x}_{AB})
  &= \frac{\left(\frac{1}{\tilde{\eta} \pi}\log \frac{16}{1-x}\right)\left(\frac{\tilde{\eta}}{\pi}\log \frac{16}{1-x}\right)}{\left(\frac{1}{\pi}\log \frac{16}{1-x}\right)^2} = 1,
    \nonumber \\
  I^{(2)}(A,B)(t)
  &= \log \frac{1}{\sqrt{1-x_{AB}}} = \log \frac{1}{\sqrt{e^{-\frac{\pi L_A}{\epsilon}}}} = \frac{\pi L_A}{2\epsilon}.
\end{align}
Note that this is independent of $\eta$.

\begin{figure}[H]
 \begin{minipage}{0.5\hsize}
  \begin{center}
   \includegraphics[width=70mm]{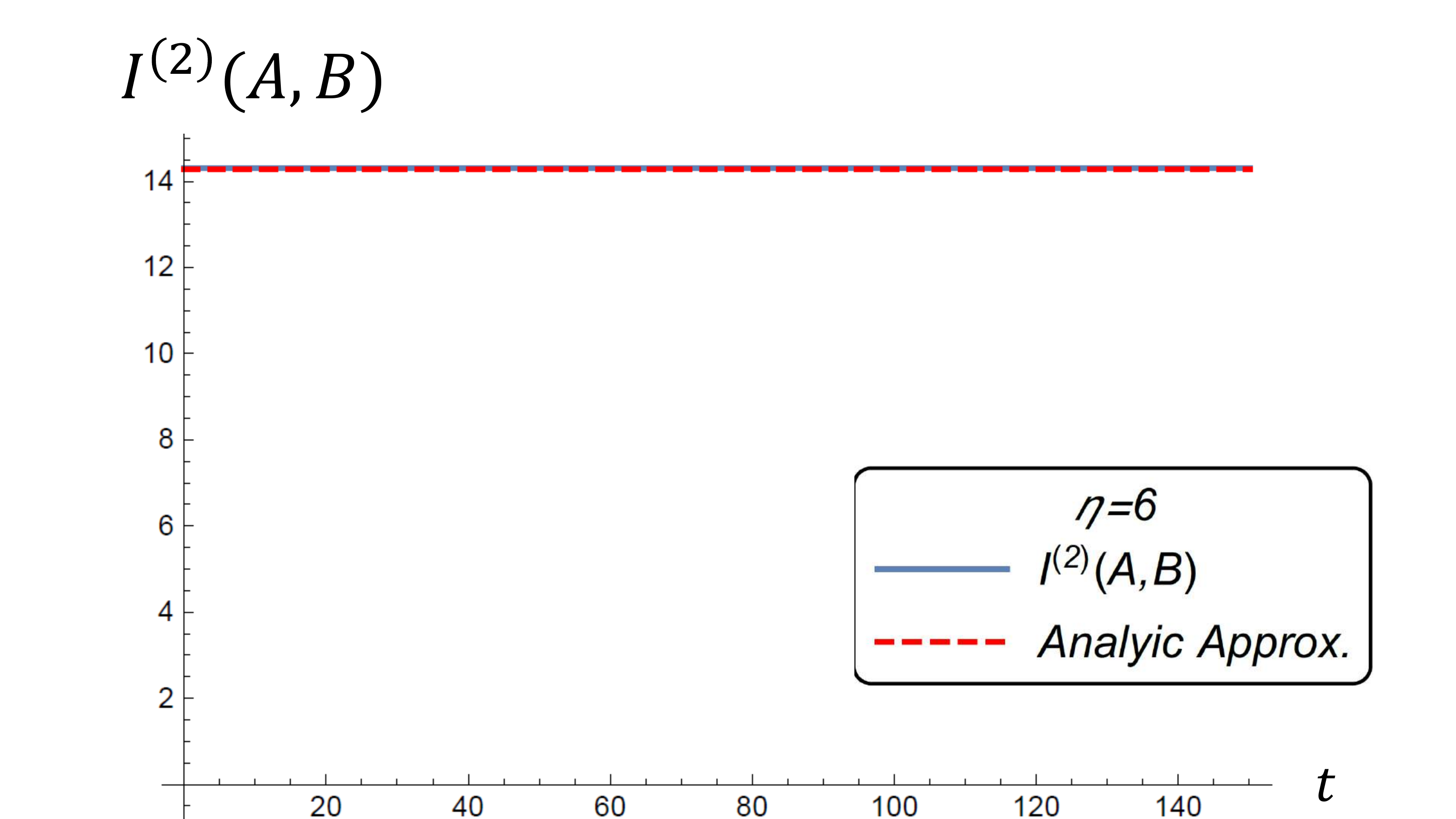}
  \end{center}
 \end{minipage}
 \begin{minipage}{0.5\hsize}
  \begin{center}
   \includegraphics[width=70mm]{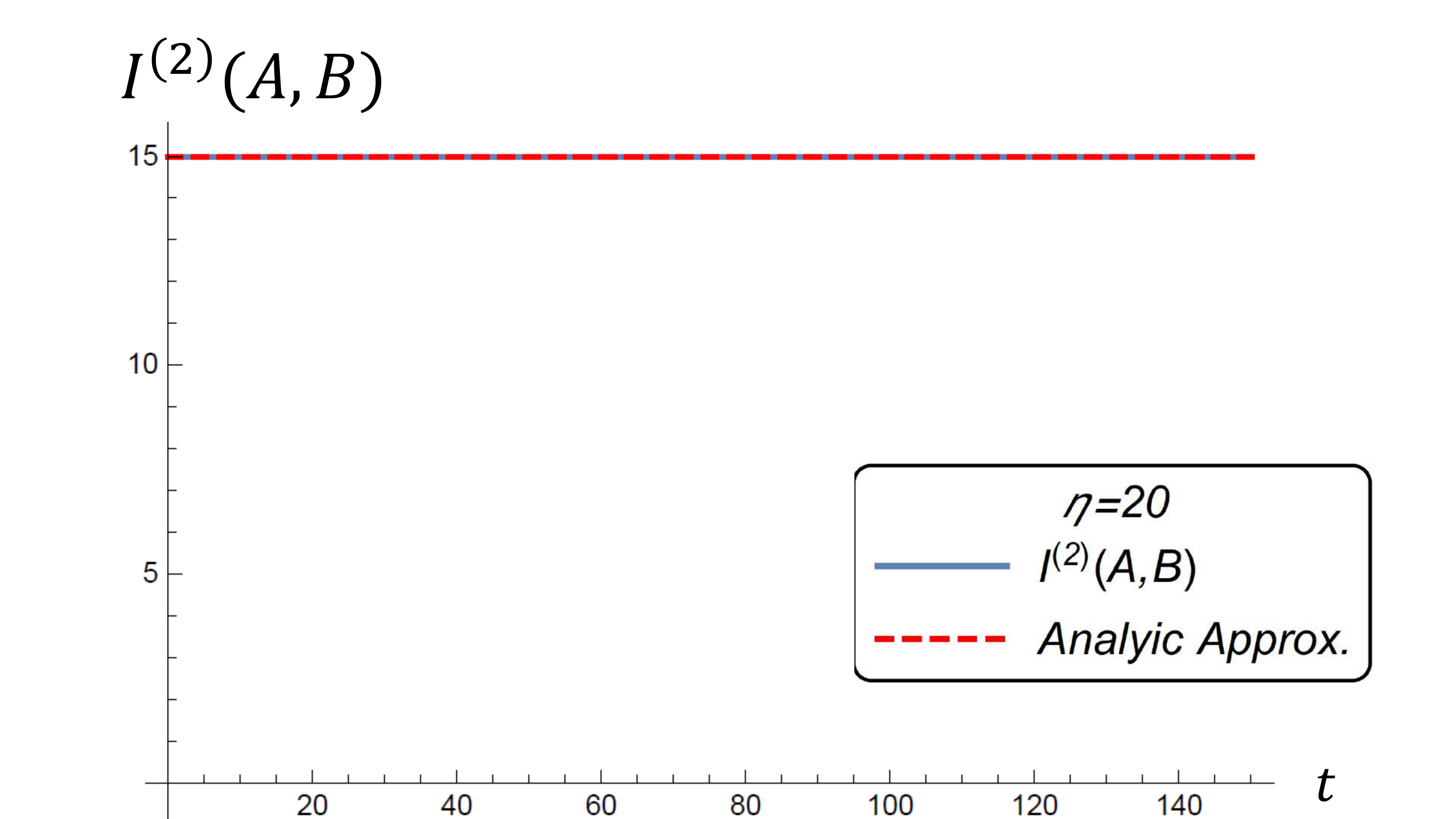}
  \end{center}
 \end{minipage}
 \caption{
   \label{plort I2AB}
   Plots of $I^{(2)}(A,B)$ for $\eta =6,20$. The blue lines are the full expression while the red dashed lines are the approximations. The $\eta = 6$ case is approximated with the $\tilde{\eta} \ll  \frac{1}{\pi}\log\left(16 e^{\pi L_A/\epsilon} \right)$ approximation, while the $\eta = 20$ case is approximated with the $\tilde{\eta} \gg  \frac{1}{\pi}\log\left(16 e^{\pi L_A/\epsilon} \right)$ approximation.}
\end{figure}

\begin{flushleft}
\underline{$\tilde{\eta} \gg  \frac{1}{\pi}\log\left(16 e^{\pi L_A/\epsilon} \right)$}:
\end{flushleft}
In this case,
we can still use the small $\eta$ approximation (\ref{smalletathetatau})
for $\theta_3(\tau_{AB}/\tilde{\eta})^2$,
while 
we have to use the other approximation (\ref{largeEta}) for $\theta_3(\tilde{\eta} \tau_{AB})^2 $:
\begin{equation}
\theta_3(\tau_{AB}/\tilde{\eta})^2 = \frac{i \tilde{\eta}}{\tau_{AB}}=\frac{\tilde{\eta}}{\pi}\log \frac{16}{\Delta x},
\quad
\theta_3(\tilde{\eta} \tau_{AB})^2 \approx 1 + 4 e^{i \pi \tilde{\eta}\tau_{AB}}.
\end{equation}
This leads to
\begin{align}
\label{IABLarge}
  F_2(x_{AB},\bar{x}_{AB})
  &\approx \frac{\left(1+4 e^{\frac{- \pi^2 \tilde{\eta}}{\log\left(16 e^{\pi L_A/\epsilon} \right)}} \right)\frac{\tilde{\eta}}{\pi}\log\left(16 e^{\frac{\pi L_A}{\epsilon}} \right)}{\left[\frac{1}{\pi}\log\left(16 e^{\frac{\pi L_A}{\epsilon}} \right)  \right]^2}
                              = \pi \tilde{\eta} \frac{1+4 e^{\frac{- \pi^2 \tilde{\eta}}{\log\left(16 e^{\pi L_A/\epsilon} \right)}}}{\log\left(16 e^{\frac{\pi L_A}{\epsilon}} \right)},
    \nonumber \\
  I^{(2)}(A,B)(t)
  &= \frac{\pi L_A}{2\epsilon}+\log \left[ \pi \tilde{\eta} \frac{1+4 e^{\frac{- \pi^2 \tilde{\eta}}{\log\left(16 e^{\pi L_A/\epsilon} \right)}}}{\log\left(16 e^{\frac{\pi L_A}{\epsilon}} \right)} \right].
\end{align}
In Fig.\ \ref{plort I2AB},
we show two plots, $\eta  = 6$ corresponding to the first case and $\eta = 20$ corresponding to the second case.

\subsubsection{bi-partite $I^{(2)}(A,B_1)$}
We now proceed to compute $I^{(2)}(A,B_1)$. As $t \rightarrow \infty$, $X_1-X_2 \equiv L_A \gg \epsilon$, the cross ratios (\ref{cri}) become 
\begin{align}
  x_{AB_1} &= e^{\frac{\pi}{\epsilon}(X_1-t)},
             \quad
\bar{x}_{AB_1} = 1 - e^{-\frac{\pi L_A}{\epsilon}}.
\end{align}
Since $x_{AB_1}\rightarrow 0$, $\tau_{AB_1}\rightarrow i \infty$ so we can use
the saddle point approximation for
\begin{equation}
\Theta(0|T_{AB_1}) = \sum_{\mu, \nu} \exp \left[-\frac{\pi |\tau_{AB_1}|}{2 p p'}(\nu p +\mu p')^2 \right]\exp \left[-\frac{\pi i \bar{\tau}_{AB_1}}{2 p p'}(\nu p -\mu p')^2 \right].
\end{equation}
The dominant terms satisfy the condition 
\begin{equation}
\nu p + \mu p' =0,
\end{equation}
I.e., $\mu = p \tilde{\mu}$,
$\nu = - p' \tilde{\mu}$ where $\tilde{\mu}\in \mathbb{Z}$.
Hence,  
\begin{equation}
\Theta(0|T_{AB_1}) = \sum_{\tilde{\mu}\in \mathbb{Z}} \exp \left[-\frac{\pi i \bar{\tau}_{AB_1}}{2 p p'}(-2p' p \tilde{\mu})^2 \right] = \theta_3(-2 p p' \bar{\tau}_{AB_1})
\end{equation}
where $-\bar{\tau}_{AB_1} = i\frac{K(1-\bar{x}_{AB_1})}{K(\bar{x}_{AB_1})}$ so the relations between $\tau$ and $x$ hold for $-\bar{\tau}_{AB_1}$ and $\bar{x}_{AB_1}$. Note that 
\begin{equation}
-\bar{\tau}_{AB_1}= - \frac{1}{i\frac{K(1-\Delta x)}{K(\Delta x)}} = \frac{i \pi}{\log \frac{16}{\Delta x}}
\end{equation}
It remains to approximate $\theta_3(-2 p p' \bar{\tau}_{AB_1})$. As before, we split this into two cases

\begin{flushleft}
\underline{$2 p p' \ll  \frac{1}{\pi}\log\left(16 e^{\pi L_A/\epsilon} \right)$}:
\end{flushleft}
We will use the small $\eta$ approximation 
\begin{align}
  \Theta (0|T_{AB_1})^2 &= \theta_3(-2\bar{\tau}_{AB_1}p p')^2
  \approx
 \frac{1}{2 p p' \pi}\log(16 e^{\frac{\pi L_A}{\epsilon}}).
\end{align}
So, $F_2$ \eqref{F2} and $I^{(2)}(A,B_1) $\eqref{CBI2} are given by
\begin{align}
\label{IAB1Small}
  F_2(x_{AB_1},\bar{x}_{AB_1})
  &= \frac{1}{2p p'\pi} \frac{\log \left(16 e^{\frac{\pi L_A}{\epsilon}} \right)}{\left[  1+\frac{1}{4}e^{\frac{\pi}{\epsilon}(X_1-t)}\right]\left[ \frac{1}{\pi}\log \left(16 e^{\frac{\pi L_A}{\epsilon}}\right)\right]}
    = \frac{1}{2 p p'\left(1+\frac{1}{4}e^{\frac{\pi}{\epsilon}(X_1-t)} \right) },
    \nonumber \\
  I^{(2)}(A,B_1)(t) &= \log \frac{1}{2p p' \left(1+\frac{1}{4}e^{\frac{\pi}{\epsilon}(X_1-t)} \right)\left(1-e^{\frac{\pi}{\epsilon}(X_1-t)} \right)^{\frac{1}{4}}e^{-\frac{\pi L_A}{4 \epsilon}} }
\xrightarrow{t \rightarrow \infty} \log \frac{e^{\frac{\pi L_A}{4\epsilon}}}{2 p p'}.
\end{align}

\begin{flushleft}
\underline{$2 p p' \gg  \frac{1}{\pi}\log\left(16 e^{\pi L_A/\epsilon} \right)$ }:
\end{flushleft}
Use the large $\eta$ approximation to obtain
\begin{align}
  \Theta (0|T_{AB_1})^2 &= \theta_3(-2\bar{\tau}_{AB_1}p p')^2
                          \approx 1+4 e^{i \pi(-2p p' \bar{\tau}_{AB_1})}
= 1+4 e^{-\frac{2 p p' \pi^2}{\log\left(16 e^{\pi L_A/\epsilon} \right)}}.
\end{align}
So, $F_2$ \eqref{F2} and $I^{(2)}(A,B_1) $\eqref{CBI2} are given by
\begin{align}
\label{IAB1Large}
  F_2(x_{AB_1},\bar{x}_{AB_1})
   &= \frac{ 1+4 e^{-\frac{2 p p' \pi^2}{\log\left(16 e^{\pi L_A/\epsilon} \right)}}}{\left(1+\frac{1}{4}e^{\frac{\pi}{\epsilon}(X_1-t)} \right)\frac{1}{\pi}\log (16 e^{\pi L_A/\epsilon})}
     \xrightarrow{t \rightarrow \infty} \frac{\pi}{\log (16 e^{\pi L_A/\epsilon})} \left( 1+4 e^{-\frac{2 p p' \pi^2}{\log\left(16 e^{\pi L_A/\epsilon} \right)}} \right),
     \nonumber \\
  I^{(2)}(A,B_1)(t)
  &= \log\left[
    \pi\left(
    1+4 e^{-\frac{2 p p' \pi^2}{\log\left(16 e^{\pi L_A/\epsilon} \right)}}
    \right)
    \right]
    - \log\left[
    \log (16 e^{\pi L_A/\epsilon})
    \left(1-e^{\frac{\pi}{\epsilon}(X_1-t)} \right)^{\frac{1}{4}}
     e^{-\frac{\pi L_A}{4\epsilon}}
     \right]
 \nonumber \\
  &\xrightarrow{t \rightarrow \infty}
    \log \left[
    \pi\left(1+4 e^{-\frac{2 p p' \pi^2}{\log\left(16 e^{\pi L_A/\epsilon} \right)}}
    \right)
    \right]
    -\log\left[
    {\log (16 e^{\pi L_A/\epsilon})e^{-\frac{\pi L_A}{4\epsilon}}}\right].
\end{align}

\begin{flushleft}
\underline{$\eta$ irrational}:
\end{flushleft}
If $\eta$ is irrational, there is no solution to $\nu \eta + \mu = 0$, so we could never get the exponent in the sum in $\Theta(0|T)$ to be zero unless $\mu=\nu=0$, so all the other terms get suppressed when $\tau_{AB_1}\rightarrow i \infty$ and the only term that survives is the $\mu = \nu = 0$ term, leading to $\Theta(0|T_{AB_1})\approx 1$ ,which has the same effect as taking $p\rightarrow \infty$ or $p' \rightarrow \infty$ in the large $p p'$ approximation, so
\begin{equation}\label{IAB1Irrational}
I^{(2)}(A,B_1) \xrightarrow{t \rightarrow \infty} \log \frac{\pi e^{\frac{\pi L_A}{4 \epsilon}}}{\log\left(16 e^{\pi L_A/\epsilon} \right)}
\end{equation}

Below we show the plots of $I^{(2)}(A,B_1)(t) $ for $\eta = 3, \frac{3}{4} , \sqrt{2}$ which are approximated with the small $2 p p'$, large $2pp'$ and irrational approximations respectively.

\begin{figure}[t]
 \begin{minipage}{0.5\hsize}
  \begin{center}
   \includegraphics[width=65mm]{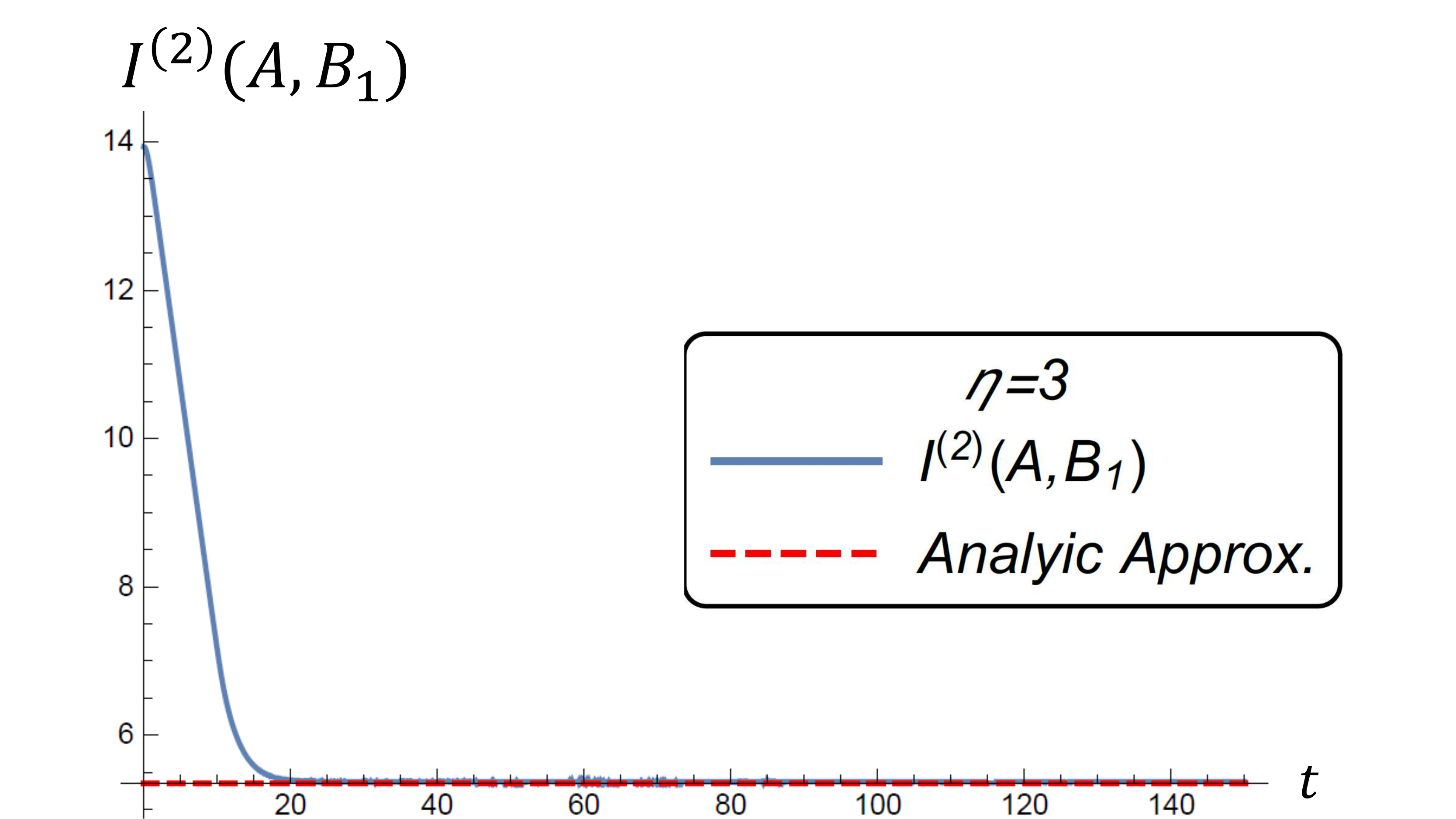}
  \end{center}
 \end{minipage}
 \begin{minipage}{0.5\hsize}
  \begin{center}
   \includegraphics[width=65mm]{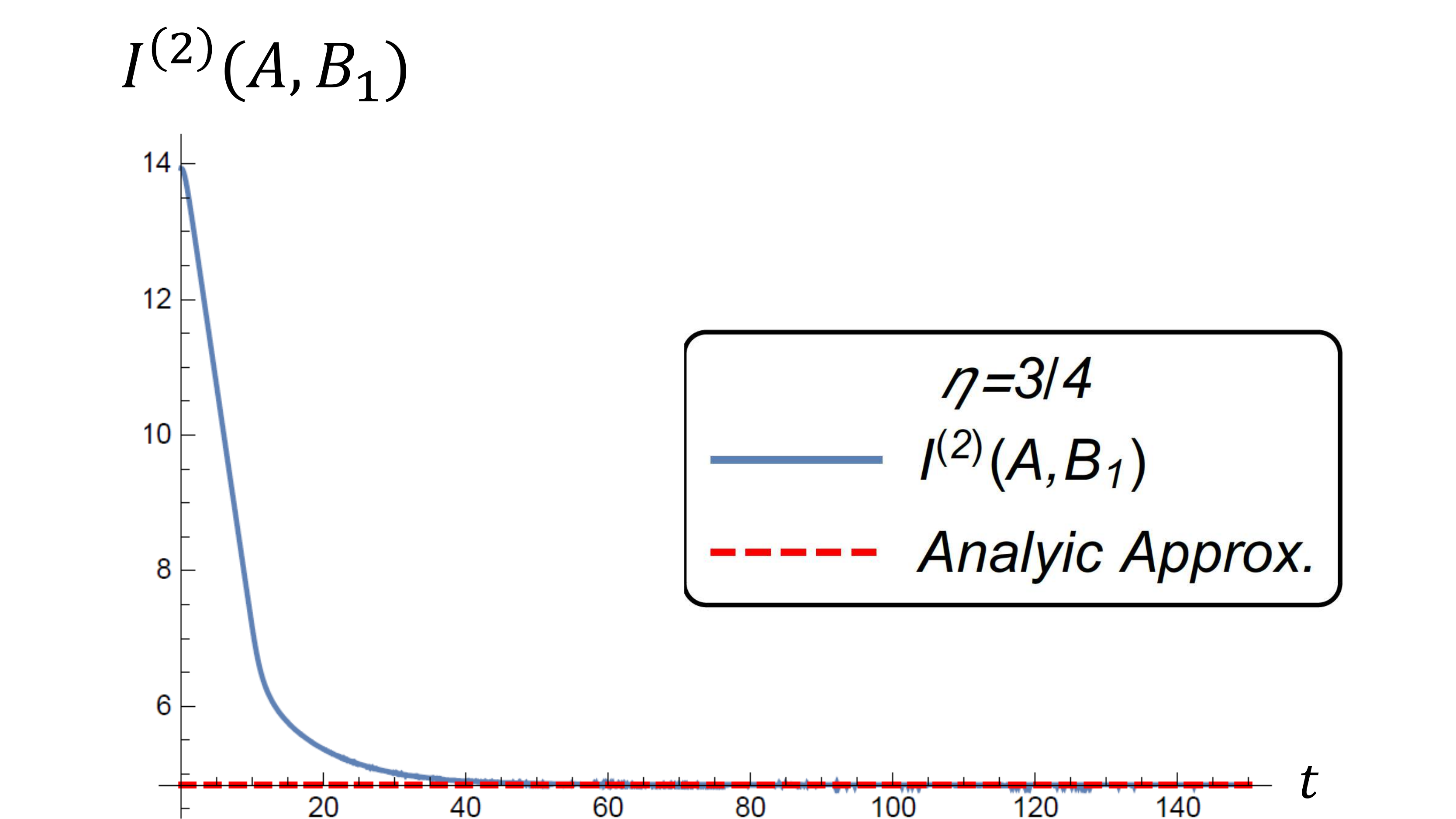}
  \end{center}
 \end{minipage}
 \begin{minipage}{0.5\hsize}
  \begin{center}
   \includegraphics[width=65mm]{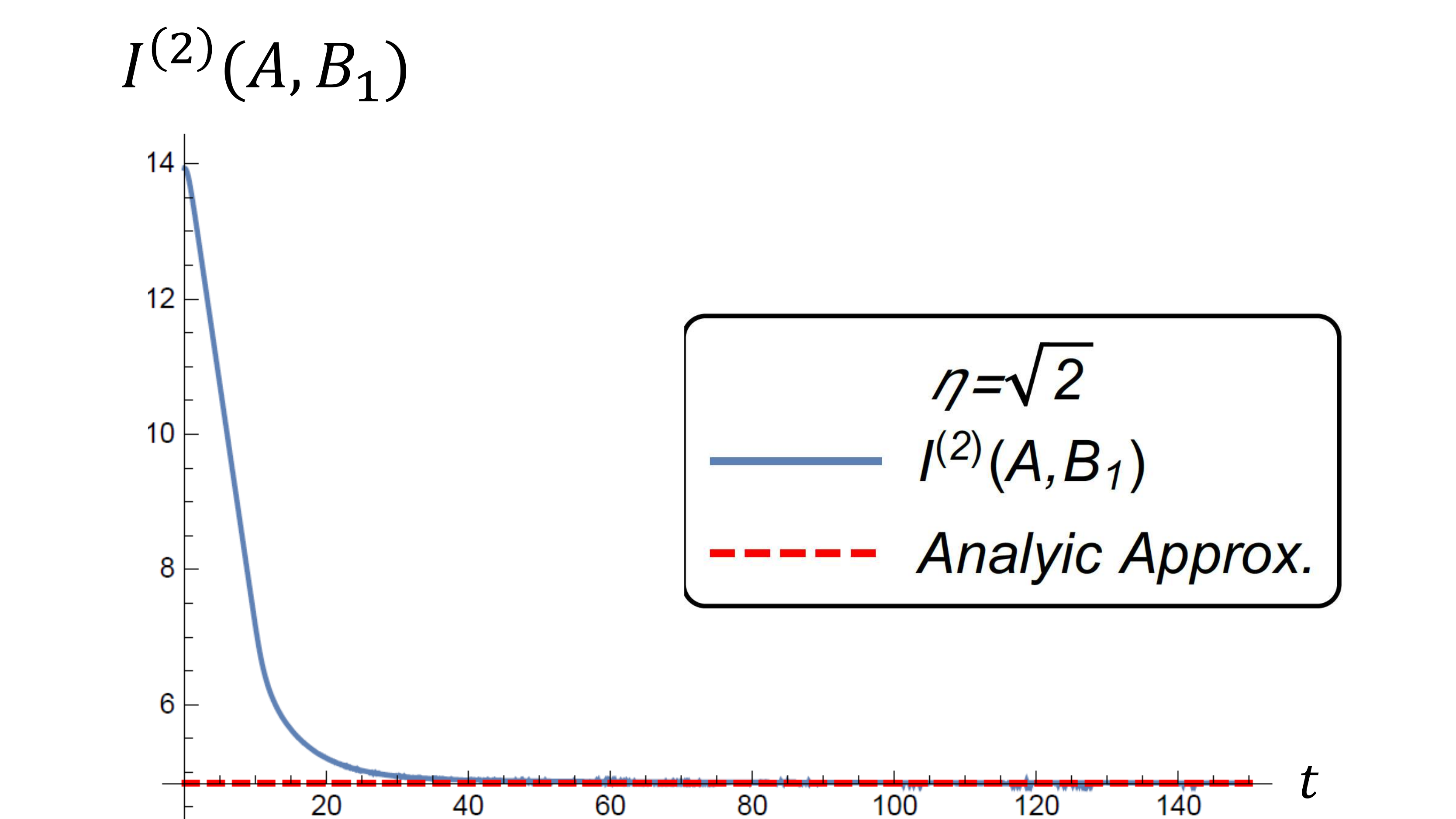}
  \end{center}
 \end{minipage}
  \caption{  Plots of $I^{(2)}(A,B_1)$ for $\eta =3,\frac{3}{4}, \sqrt{2}$. The blue lines are the full expression while the red dashed lines are the approximations. The $\eta = 3$ case is approximated with the $2pp' \ll  \frac{1}{\pi}\log\left(16 e^{\pi L_A/\epsilon} \right)$ approximation, while the $\eta = \frac{3}{4}$ case is approximated with the $2pp' \gg  \frac{1}{\pi}\log\left(16 e^{\pi L_A/\epsilon} \right)$ approximation.}
\end{figure}

The bi-partite operator mutual information decreases monotonically since quasi-particles are leaving region $B_1$.

\subsubsection{bi-partite $I^{(2)}(A,B_2)$}
The computation for $I^{(2)}(A,B_2)$ proceeds in a similar fashion. As $t \rightarrow \infty$ and for $X_1-X_2 \equiv L_A\gg \epsilon$, the cross ratios (\ref{cri}) are 
\begin{align}
  x_{AB_2} &= 1 - e^{-\frac{\pi L_A}{\epsilon}},
             \quad
\bar{x}_{AB_2} \approx e^{-\frac{\pi }{\epsilon}(t+X_2)}.
\end{align}
Because $\lim\limits_{t\rightarrow \infty}\bar{x}_{AB_2}=0$, $\lim\limits_{t\rightarrow \infty}\bar{\tau}_{AB_2}=-i \infty$ and we can apply the saddle point approximation.
\begin{equation}
\Theta(0|T_{AB_2}) = \sum_{\mu,\nu} \exp\left(\frac{\pi i \tau_{AB_2}}{2p p'}(\nu p +\mu p')^2 \right) \exp\left(-\frac{\pi |\bar{\tau}_{AB_2}|}{2p p'}(\nu p -\mu p')^2 \right)
\end{equation}
The dominant terms satisfy 
\begin{equation}
\nu p -\mu p' = 0
\quad
\Longrightarrow 
\quad 
\mu = p \tilde{\mu}\quad  \mbox{and}\quad  \nu = p' \tilde{\mu},
\end{equation}
so the Siegel theta function simplifies to
\begin{align}
  \Theta(0|T_{AB_2})  &= \sum_{\tilde{\mu}} \exp \left[\frac{\pi i \tau_{AB_2}}{2p p'}(2p' p \tilde{\mu})^2 \right]
= \theta_3(2pp'\tau_{AB_2})
\end{align}
where $\tau_{AB_2} \approx -\frac{\pi}{i \log(16 e^{\pi L_A/\epsilon})}$.
As before, we discuss the following two cases separately:
\begin{flushleft}
\underline{$2 p p' \ll  \frac{1}{\pi}\log\left(16 e^{\pi L_A/\epsilon} \right)$ }:
\end{flushleft}
Use the small $\eta$ approximation (\ref{smallEta})
\begin{equation}
\theta_3^2(2pp'\tau_{AB_2})= -\frac{1}{i \tau_{AB_2}2 p p'} =\frac{1}{2pp' \pi}\log\left(16e^{\frac{\pi L_A}{\epsilon}} \right)
\end{equation}
so the Siegel theta function becomes
\begin{equation}
\Theta(0|T_{AB_2})^2 = \frac{1}{2 pp'\pi}\log \left(16 e^{\frac{\pi L_A}{\epsilon}} \right).
\end{equation}
Correspondingly, $F_2$ \eqref{F2} and $I^{(2)}(A,B_1)$ \eqref{CBI2} are given by
\begin{align}
  \label{IAB2Small}
  F_2 (x_{AB_2},\bar{x}_{AB_2})
   &=\frac{1}{2 p p' \pi}\frac{\log\left(16 e^{\frac{\pi L_A}{\epsilon}} \right)}{\left(\frac{1}{\pi}\log \frac{16}{e^{-\frac{\pi L_A}{\epsilon}}} \right)\left(1+\frac{1}{4}e^{-\frac{\pi}{\epsilon}(t+X_2)} \right)}
= \frac{1}{2p p'\left(1+\frac{1}{4}e^{-\frac{\pi}{\epsilon}(t+X_2)} \right)},
     \nonumber \\
  I^{(2)}(A,B_2)(t) &= \log \frac{1}{2p p'\left(1+\frac{1}{4}e^{-\frac{\pi}{\epsilon}(t+X_2)} \right)e^{-\frac{\pi L_A}{4\epsilon}}\left(1-e^{-\frac{\pi}{\epsilon}(t+X_2)}\right)^\frac{1}{4}}
\xrightarrow{t \rightarrow \infty} \log \frac{e^{\frac{\pi L_A}{4\epsilon}}}{2p p'}
\end{align}

\begin{flushleft}
\underline{$2 p p' \gg  \frac{1}{\pi}\log\left(16 e^{\pi L_A/\epsilon} \right)$ }:
\end{flushleft}
Using (\ref{largeEta}), we obtain 
\begin{equation}
\Theta(0|T_{AB_2})^2 = \theta_3(2pp'\tau_{AB_2})^2 \approx 1+4 e^{i \pi (2 p p' \tau_{AB_2})} = 1+ 4 e^{-\frac{2pp'\pi^2}{\log(16e^{\pi L_A/\epsilon})}}.
\end{equation}
Correspondingly, $F_2$ \eqref{F2} and $I^{(2)}(A,B_1)$ \eqref{CBI2} are given by
\begin{align}
  \label{IAB2Large}
  F_2 (x_{AB_2},\bar{x}_{AB_2})
                              &= \frac{1+ 4 e^{-\frac{2pp'\pi^2}{\log(16e^{\pi L_A/\epsilon})}}}{\frac{1}{\pi}\log(16 e^{\pi L_A/\epsilon})(1+\frac{1}{4}e^{-\frac{\pi}{\epsilon}(t+X_2)})}
\xrightarrow{t \rightarrow \infty} \frac{\pi}{\log(16 e^{\pi L_A/\epsilon})} \left[1+ 4 e^{-\frac{2pp'\pi^2}{\log(16e^{\pi L_A/\epsilon})}} \right],
                                \nonumber \\
  I^{(2)}(A,B_2)(t)
                  &= \log \frac{\pi\left[ 1+ 4 e^{-\frac{2pp'\pi^2}{\log(16e^{\pi L_A/\epsilon})}}\right]}{\log(16 e^{\pi L_A/\epsilon})e^{-\frac{\pi L_A}{4\epsilon}}\left(1-e^{-\frac{\pi}{\epsilon}(t+x_2)} \right)^{1/4}}
\xrightarrow{t \rightarrow \infty}\log \frac{\pi\left[ 1+ 4 e^{-\frac{2pp'\pi^2}{\log(16e^{\pi L_A/\epsilon})}}\right]}{\log(16 e^{\pi L_A/\epsilon})e^{-\frac{\pi L_A}{4\epsilon}}}.
\end{align}

\begin{flushleft}
\underline{$\eta$ irrational}:
\end{flushleft}
As before, the only term that is not suppressed when $\bar{\tau}_{AB_2}\rightarrow -i \infty$ is the $\mu = \nu =0$ term, so $\Theta(0|T_{AB_2})\approx 1$
\begin{equation}\label{IAB2Irrational}
I^{(2)}(A,B_2) \xrightarrow{t \rightarrow \infty} \log \frac{\pi e^{\frac{\pi L_A}{4 \epsilon}}}{\log\left(16 e^{\pi L_A/\epsilon} \right)}
\end{equation}

Below we show the plots of $I^{(2)}(A,B_2)(t) $ for $\eta = 3,\frac{3}{4}, \sqrt{2}$, which are approximated using the small $2pp'$ approximation, large $2p p'$ approximation and irrational approximation respectively.
\begin{figure}[ht]
 \begin{minipage}{0.5\hsize}
  \begin{center}
   \includegraphics[width=65mm]{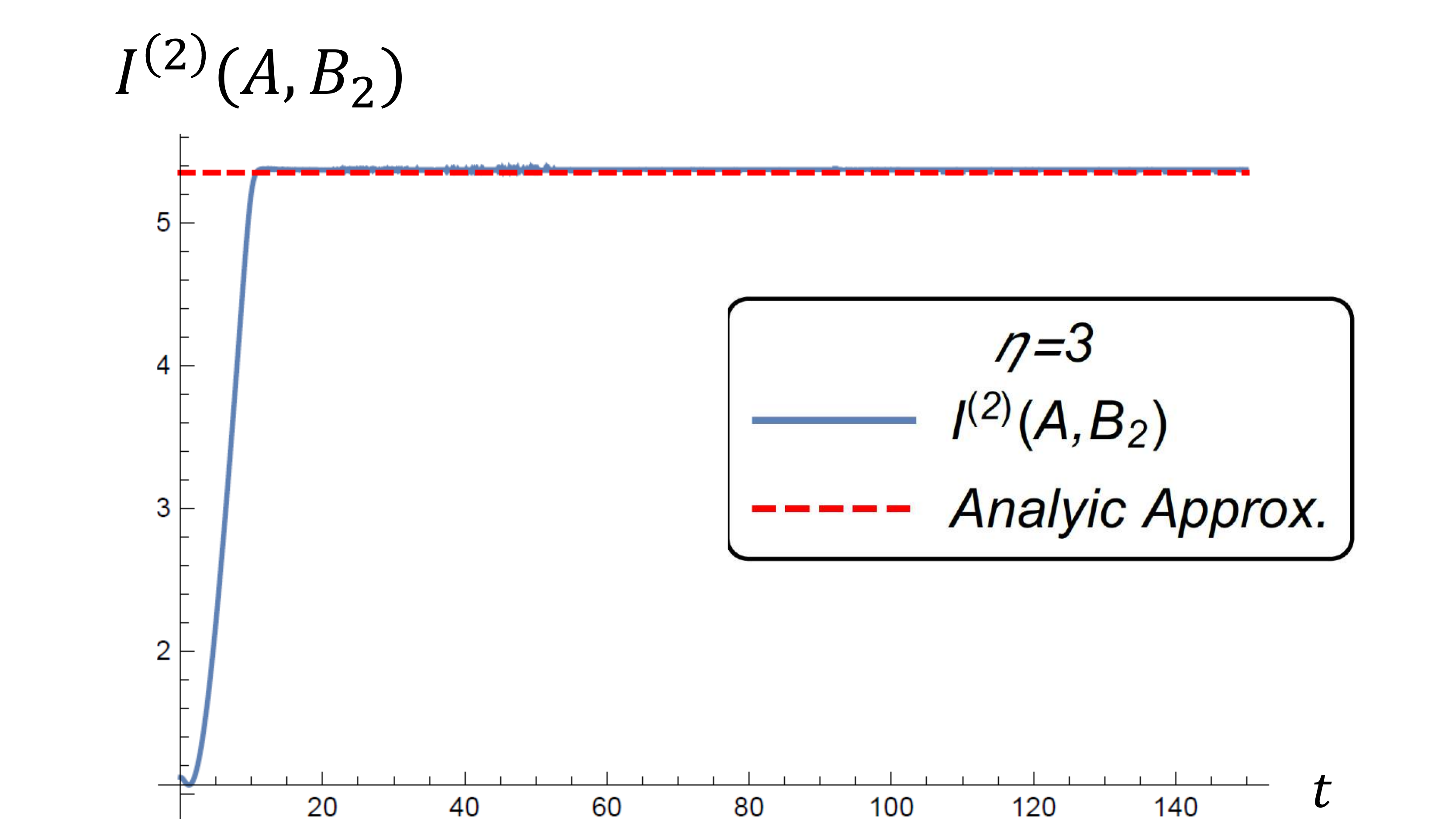}
  \end{center}
 \end{minipage}
 \begin{minipage}{0.5\hsize}
  \begin{center}
   \includegraphics[width=65mm]{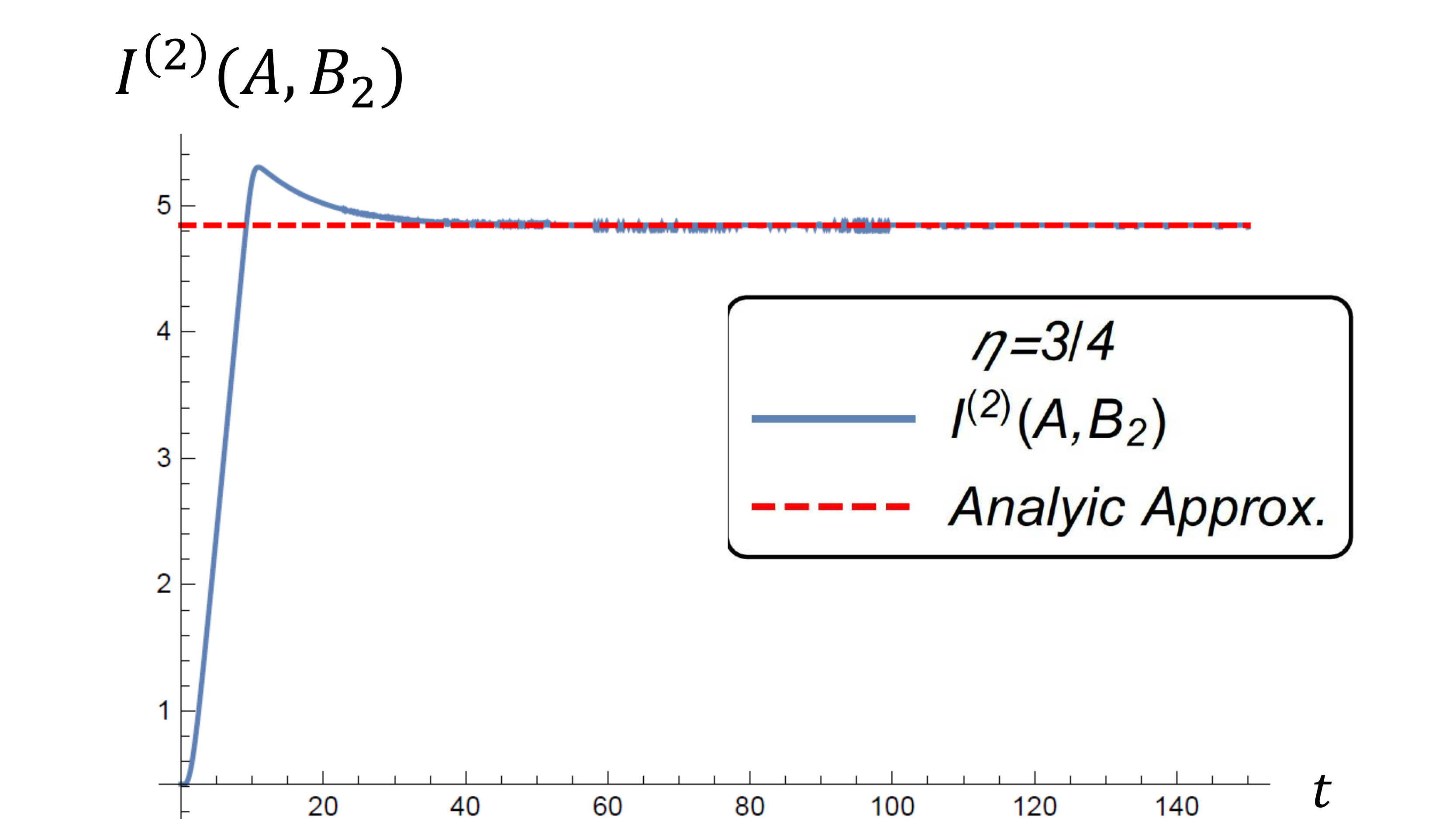}
  \end{center}
 \end{minipage}
 \begin{minipage}{0.5\hsize}
  \begin{center}
   \includegraphics[width=65mm]{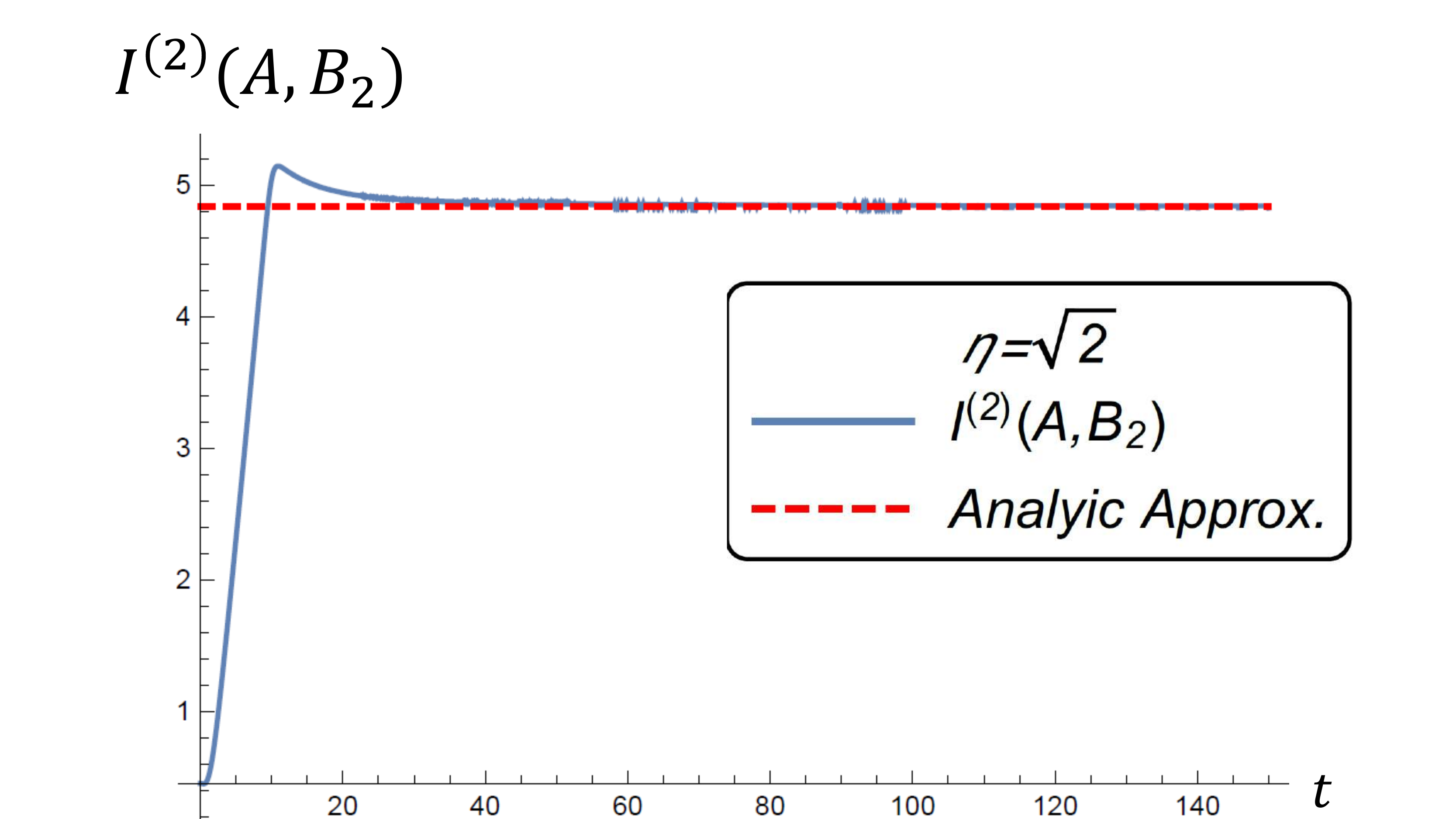}
  \end{center}
 \end{minipage}
 \caption{  The time-dependence of $I^{(2)}(A,B_2)$
   for $\eta =3,\frac{3}{4}, \sqrt{2}$.
   The blue lines are the full expression while the red dashed lines are the
   approximations.
   The $\eta = 3$ case is approximated with the $2pp' \ll  \frac{1}{\pi}\log\left(16 e^{\pi L_A/\epsilon} \right)$ approximation,
   while the $\eta = \frac{3}{4}$ case is approximated with the $2pp' \gg  \frac{1}{\pi}\log\left(16 e^{\pi L_A/\epsilon} \right)$ approximation.}
\end{figure}

The $\eta = 3$ case can be explained nicely by the quasi-particle picture. The
bi-partite mutual information saturates once the last left moving quasi-particle
from subregion $A$, which was originally at $X_1 = 10$, enters subregion $B_2$
at $t = X_1 - 0 = 10$,
$I^{(2)}(A,B_2)$ attains its final value. We see small deviations from the quasi-particle picture for $\eta =\frac{3}{4}, \sqrt{2}$. This relates to the earlier comment that the quasi-particle picture does not account for the boson radius.

\subsubsection{tri-partite $I^{(2)}(A,B_1,B_2)$}
Having compute the various bi-partite operator mutual information, we can combine them to obtain the tri-partite operator mutual information. We list the various cases below.
\begin{itemize}
\item
  For rational radius $\eta=p/p'$,
  when both $2pp',\eta \ll  \frac{1}{\pi}\log\left(16 e^{\pi L_A/\epsilon} \right)$,
  \begin{align}
    \label{2pp'etasmall}
    I^{(2)}(A,B_1,B_2) \stackrel{t\to \infty}{\longrightarrow}  -2\log d_{\sigma_2},
  \end{align}
where $d_{\sigma_2}=2p p'$ is the quantum dimension of
the twist operator for the two-sheeted Riemann surface.
Setting $p=p'=1$, we recover the self-dual case.
See Fig.\ \ref{TOMI CB different radii} for an example with $\eta = 3$.

\item
  For rational radius $\eta=p/p'$,
  when $2pp' \gg  \frac{1}{\pi}\log\left(16 e^{\pi L_A/\epsilon} \right)$ and $\eta \ll  \frac{1}{\pi}\log\left(16 e^{\pi L_A/\epsilon} \right)$,
\begin{align}\label{2pp'largeEtaSmall}
  I^{(2)}(A,B_1,B_2)
  &
\stackrel{t\to \infty}{\longrightarrow} 
      -2 \log 
    \left[ 
    \frac{1}{\pi} \log\left(16 e^{\frac{\pi L_A}{\epsilon}}\right) \right]
    +2\log
    {
    \left[ 1+ 4 e^{-\frac{2pp'\pi^2}{\log(16e^{\pi L_A/\epsilon})}}\right]
    }.
\end{align}
See Fig.\ \ref{TOMI CB different radii} for an example with $\eta = 6/5$.

\item
  For rational radius $\eta=p/p'$, when both $2pp' \gg  \frac{1}{\pi}\log\left(16 e^{\pi L_A/\epsilon} \right)$ and $\eta \gg  \frac{1}{\pi}\log\left(16 e^{\pi L_A/\epsilon} \right)$,
  \begin{align}\label{2pp'largeEtaLarge}
    I^{(2)}(A,B_1,B_2)
    &
      \stackrel{t\to \infty}{\longrightarrow} 
      -
      \log 
      \left[ 
      \frac{\tilde{\eta}}{\pi}
      \log\left(16 e^{\frac{\pi L_A}{\epsilon}} \right)
    \right]
      \nonumber \\
    &
      \qquad 
    -\log\left[
      1+4 e^{\frac{- \pi^2 \tilde{\eta}}{\log\left(16 e^{\pi L_A/\epsilon} \right)}}
      \right]
      +
      2\log\left[
      1+ 4 e^{-\frac{2pp'\pi^2}{\log(16e^{\pi L_A/\epsilon})}}
      \right].
\end{align}
See Fig.\ \ref{TOMI CB different radii} for an example with $\eta = 20$.

\item
  For $\eta$ irrational and $\eta \ll  \frac{1}{\pi}\log\left(16 e^{\pi L_A/\epsilon} \right)$:
  \footnote{
The rationality of $\eta$ only matters when applying the saddle point
approximation in $I(A,B_1)$ and $I(A,B_2)$, and the approximation is basically
the same as the large $2pp'$ approximation, but without the $4
e^{-\frac{2pp'\pi^2}{\log(16e^{\pi L_A/\epsilon})}}$ factor. }
Applying (\ref{IABSmall}), (\ref{IAB1Irrational}) and (\ref{IAB2Irrational}) gives
\begin{align}\label{EtaIrrationalSmall}
I^{(2)}(A,B_1,B_2)
  &
    \stackrel{t\to \infty}{\longrightarrow} 
     -2 \log 
    \left[
    \frac{1}{\pi}
    \log\left(16 e^{\frac{\pi L_A}{\epsilon}}\right)
    \right].
\end{align}
See Fig.\ \ref{TOMI CB different radii} for an example with $\eta = \pi$.

\item
  For $\eta$ irrational and $\eta \gg  \frac{1}{\pi}\log\left(16 e^{\pi L_A/\epsilon} \right)$:
\begin{align}\label{EtaIrrationalLarge}
  I^{(2)}(A,B_1,B_2)
  &
    \stackrel{t\to \infty}{\longrightarrow} 
    -
     \log 
    \left[ 
    \frac{\tilde{\eta}}{\pi}
    \log\left(16 e^{\frac{\pi L_A}{\epsilon}} \right)
    \right]
    -\log
    \left[
    1+4 e^{\frac{- \pi^2 \tilde{\eta}}{\log\left(16 e^{\pi L_A/\epsilon} \right)}}
 \right].
\end{align}
See Fig.\ \ref{TOMI CB different radii} for an example with $\eta = \sqrt{300}$.
\end{itemize}
Let us also make a comment on the decompactification limit. The dependence of
the late-time value on the radius squared $\eta$ only comes in through
$I^{(2)}(A,B)$. $I^{(2)}(A,B)$ is independent of $\eta$ when $\eta \ll  \frac{1}{\pi}\log\left(16 e^{\pi L_A/\epsilon} \right)$ and only depends on $\eta$ when $\eta \gg  \frac{1}{\pi}\log\left(16 e^{\pi L_A/\epsilon} \right)$.
The late-time value of $I^{(2)}(A,B)$ for large $\eta$ is given by
\begin{equation}\label{IABLarge}
  I^{(2)}(A,B)(t)
  = \frac{\pi L_A}{2\epsilon}
  +\log \left[
    \frac{1}{\pi \tilde{\epsilon}}
    \log{\left(16 e^{\frac{\pi L_A}{\epsilon}} \right)}
\right]
  +\log \left[
    1+4 e^{\frac{- \pi^2 \tilde{\eta}}{\log\left(16 e^{\pi L_A/\epsilon} \right)}}
    \right].
\end{equation}
The late-time value is obtained by expanding the expressions in terms of $\Delta x = e^{-\frac{\pi L_A}{\epsilon}} \ll 1$. The first term of (\ref{IABLarge}) is $\mathcal{O}\left(\log \frac{1}{\Delta x} \right)$ while the second term is $\mathcal{O}\left(\log\left[\frac{\tilde{\eta}}{\log\frac{1}{\Delta x}}\right] \right)$. If $\tilde{\eta}$ is sufficiently large, the second term can be comparable to the first and the expansion in $\Delta x$ might break down, so it is not entirely clear what happens in the case of the decompactified boson.

\section{Late-time behavior of TOMI for holographic CFTs}

In this section, we provide the derivation of \eqref{late time const}.
If we take the late-time limit, the cross ratios in (\ref{cri}) becomes
\eqref{cross ratio, late time}. TOMI $I(A,B_1,B_2)=I(A,B_1)+I(A,B_2)-I(A,B)$ is given by
\be \label{HTOMI}
\begin{split}
I(A,B_1,B_2)&=\f{c}{6}\log{\left[\f{x_{AB_1}\bar{x}_{AB_1}x_{AB_2}\bar{x}_{AB_2}}{x_{AB}\bar{x}_{AB}}\right]}\\
&+\f{c}{6}\log{\left[\f{\text{Min}\left\{x_{AB}\bar{x}_{AB},(1-x_{AB})(1-\bar{x}_{AB})\right\}}{\text{Min}\left\{x_{AB_1}\bar{x}_{AB_1},(1-x_{AB_1})(1-\bar{x}_{AB_1})\right\}\text{Min}\left\{x_{AB_2}\bar{x}_{AB_2},(1-x_{AB_2})(1-\bar{x}_{AB_2})\right\}}\right]}
\end{split}
\ee
The first term in (\ref{HTOMI}) is given by
\begin{align}
\f{c}{6}\log{\left[\f{x_{AB_1}\bar{x}_{AB_1}x_{AB_2}\bar{x}_{AB_2}}{x_{AB}\bar{x}_{AB}}\right]}=\f{c}{6}\log{\left[\f{4\sinh^2{\left[\f{\pi}{2\epsilon}(X_1-X_2)\right]}}{e^{2\f{\pi t}{\epsilon}}}\right]}.
\end{align}

As for the contributions
given by the second term in \eqref{Holographic BOMI},
first for $I(A,B)$,
we note that
$x_{AB}\bar{x}_{AB}
  \rightarrow \left(1-e^{-\f{\pi}{\epsilon}(X_1-X_2)}\right)^2$,
$(1-x_{AB})(1-\bar{x}_{AB} )\rightarrow e^{-\f{2 \pi}{\epsilon}(X_1-X_2)}$
at late times.
Thus
\footnote{More precisely, 
the threshold separating these two cases 
is given by $X_1-X_2=\f{\epsilon}{\pi}\log{2}$.}, 
\begin{align}
  &
\mathrm{Min}\{x_{AB}\bar{x}_{AB} ,(1-x_{AB})(1-\bar{x}_{AB}
                )\}
    \nonumber \\
  &=
  \left\{
  \begin{array}{ll}
    x_{AB}\bar{x}_{AB}=\left(1-e^{-\f{\pi}{\epsilon}(X_1-X_2)}\right)^2
    &
      \mbox{when $X_1-X_2 ~\ll~ \epsilon$}
    \\
(1-x_{AB})(1-\bar{x}_{AB})
    =
e^{-\f{2 \pi}{\epsilon}(X_1-X_2)}
    &
\mbox{when $X_1-X_2~ \gg ~\epsilon$}
    \end{array}
  \right.
\end{align}
Similarly, for $I(A,B_1)$ and $I(A,B_2)$, 
noting that
\begin{align}
&x_{AB_1}\bar{x}_{AB_1} \rightarrow \left(1-e^{-\f{\pi}{\epsilon}(X_1-X_2)}\right)\f{2\sinh{\left[\f{\pi}{2\epsilon}(X_1-X_2)\right]}}{e^{\f{\pi}{2\epsilon}(2t-X_1-X_2)}}, 
\nonumber \\
&(1-x_{AB_1})(1-\bar{x}_{AB_1} )\rightarrow e^{-\f{\pi}{\epsilon}(X_1-X_2)}\left(1-\f{2\sinh{\left[\f{\pi}{2\epsilon}(X_1-X_2)\right]}}{e^{\f{\pi}{2\epsilon}(2t-X_1-X_2)}}\right), 
\nonumber \\
&x_{AB_2}\bar{x}_{AB_2} \rightarrow \left(1-e^{-\f{\pi}{\epsilon}(X_1-X_2)}\right)\f{2\sinh{\left[\f{\pi}{2\epsilon}(X_1-X_2)\right]}}{e^{\f{\pi}{2\epsilon}(2t+X_1+X_2)}}, \nonumber \\
&(1-x_{AB_2})(1-\bar{x}_{AB_2} )\rightarrow e^{-\f{\pi}{\epsilon}(X_1-X_2)}\left(1-\f{2\sinh{\left[\f{\pi}{2\epsilon}(X_1-X_2)\right]}}{e^{\f{\pi}{2\epsilon}(2t+X_1+X_2)}}\right), 
\end{align}
at late times,
thus, 
\begin{align}
  \mathrm{Min}\{x_{AB_1}\bar{x}_{AB_1} ,(1-x_{AB_1})(1-\bar{x}_{AB_1} )\}
    &=x_{AB_1}\bar{x}_{AB_1}
      \nonumber \\
&=\left(1-e^{-\f{\pi}{\epsilon}(X_1-X_2)}\right)\f{2\sinh{\left[\f{\pi}{2\epsilon}(X_1-X_2)\right]}}{e^{\f{\pi}{2\epsilon}(2t-X_1-X_2)}},
                     \nonumber \\
\mathrm{Min}\{x_{AB_2}\bar{x}_{AB_2} ,(1-x_{AB_2})(1-\bar{x}_{AB_2}
  )\}
  &=
x_{AB_2}\bar{x}_{AB_2}
    \nonumber \\
&=\left(1-e^{-\f{\pi}{\epsilon}(X_1-X_2)}\right)\f{2\sinh{\left[\f{\pi}{2\epsilon}(X_1-X_2)\right]}}{e^{\f{\pi}{2\epsilon}(2t+X_1+X_2)}}.
\end{align}
Collecting these contributions,
$I(A,B_1,B_2)$ is given by
\be
\begin{split}
I(A,B_1,B_2)=\begin{cases}
0 & ~\text{when} ~X_1-X_2 \ll \epsilon, \\
-\f{\pi c}{3\epsilon}(X_1-X_2) &~\text{when}~ X_1-X_2 \gg \epsilon. \\
\end{cases}
\end{split}
\ee

\bibliographystyle{ieeetr}
\bibliography{reference}

\end{document}